\documentclass[a4paper,11pt]{article}
\pdfoutput=1 % if your are submitting a pdflatex (i.e. if you have
\usepackage{jheppub} % for details on the use of the package, please
\usepackage[T1]{fontenc} % if needed
\usepackage{tikz}
\usetikzlibrary{decorations.pathmorphing}
\usepackage{braket}
\usepackage{bbm}
\usepackage{mathabx}
\usepackage{mathrsfs}
\usepackage{comment}

\usepackage{enumitem}
\setlist[enumerate]{itemsep=2pt, label=(\arabic*), ref=(\arabic*)}

\newcommand{\op}[1]{\boldsymbol{#1}}

\newcommand{\Tr}{\text{Tr}}
\definecolor{indigo(dye)}{rgb}{0.0, 0.25, 0.42}

\title{Generalized Entropy is von Neumann Entropy II:
     \\The complete symmetry group and edge modes}
\newcommand{\beq}{\begin{eqnarray}}
\newcommand{\eeq}{\end{eqnarray}}
\newcommand{\beqn}{\begin{eqnarray}}
\newcommand{\eeqn}{\end{eqnarray}}

\newcommand{\defn}{\mathrel{\mathop:}=} %shrtct for definition operator

\newtheorem{theorem}{Theorem}[section]
\newtheorem{remark}{Remark}[section]

\author[a,b]{Marc S.~Klinger,}
\author[c]{Jonah Kudler-Flam,}
\author[d]{Gautam Satishchandran}

\affiliation[a]{Walter Burke Institute for Theoretic Physics, California Institute of Technology, Pasadena, CA 91125, USA}
\affiliation[b]{Illinois Center for Advanced Studies of the Universe \& Department of Physics, University of Illinois, Urbana IL 61801, USA}
\affiliation[c]{School of Natural Sciences, Institute for Advanced Study, Princeton, NJ 08540, USA}

\affiliation[d]{Princeton Gravity Initiative, Princeton University, Princeton, NJ 08544, USA}

\emailAdd{klingerm@caltech.edu}
\emailAdd{jkudlerflam@ias.edu}
\emailAdd{gautam.satish@princeton.edu}

\abstract{We consider the algebra of observables of perturbative quantum gravity in the exterior of a stationary black hole or the static patch of de Sitter spacetime. It was previously argued in \cite{Chandrasekaran:2022eqq,Chandrasekaran:2022cip,Kudler-Flam:2023qfl} that the backreaction of gravitons on the spacetime perturbs the area of the horizon at second-order which gives rise to a non-trivial constraint on the algebra of physical observables in the subregion. The corresponding "dressed" algebra including fluctuations of the total horizon area admits a well-defined trace and is Type II. In this paper we show that, at the same perturbative order at which the horizon area (and angular momentum) fluctuates, gravitational backreaction also perturbs the horizon area in an angle-dependent way. These fluctuations are encoded in horizon charges --- i.e., ``edge modes'' --- which are related to an infinite dimensional ``boost supertranslation'' symmetry of the horizon. Together, these charges impose an infinite family of nontrivial constraints on the gravitational algebra. We construct the full algebra of observables which satisfies these constraints. 
We argue that the resulting algebra is Type II and its trace is shown to take a universal form. The entropy of any ``semiclassical state'' is the generalized entropy with an additional ``edge mode'' contribution as well as a state-independent constant. 
For any black hole spacetime, the algebra has no maximum entropy state and is Type II$_{\infty}$. In de Sitter, the static patch is defined relative to the worldline of a localized ``observer''. 
We show that a consistent quantization of the static-patch algebra requires a more realistic model of the observer, in which higher multipole moments perturb the “shape” of the cosmological horizon. We argue that a proper account of the observer’s rotational kinetic energy and (non-gravitational) binding energy implies that the algebra is of Type II$_{1}$ and thereby admits a maximum entropy state.}

\begin{document} 
\maketitle
\flushbottom

\section{Introduction}

It has been long established that one can formulate a fully satisfactory, well-defined theory of local quantum physics without gravity where the quantum fields propagate on a fixed, background (globally hyperbolic\footnote{In some cases (e.g., asymptotically AdS spacetimes), if the spacetime is not globally hyperbolic on can obtain a well-posed initial value problem
by imposing appropriate boundary conditions at infinity \cite{Ishibashi:2004wx}.}) spacetime $\mathscr{M}$ \cite{Haag:1996hvx,Streater:1989vi,Brunetti:1999jn,Hollands:2001nf}. In this approach, the fundamental object to predict any local, physical phenomena in a spacetime subregion $\mathscr{R}\subset \mathscr{M}$ is the collection of all  (renormalized) quantum observables $\mathfrak{A}(\mathscr{R})$ in the region. This collection forms an {\em algebra of observables}. For an interacting theory, the expected value of any observable in $\mathfrak{A}(\mathscr{R})$ can be computed perturbatively in any physical state (see, e.g., \cite{Hollands:2014eia} for details). It is expected that this theory corresponds to the $G_{\textrm{N}}\to 0$ limit of a more fundamental quantum theory of gravity wherein the spacetime metric itself is treated in accordance with the principles of quantum theory. Quantum field theory in curved spacetime is expected to be applicable to a wide range of quantum phenomena from the early universe to the vicinity of black holes. However, this theory will fail to give sensible predictions for phenomena near singularities or observables that scale inversely in $G_{\textrm{N}}$ --- e.g., the big bang singularity or the late time behavior of black hole evaporation. On these scales, the backreaction of quantum fields on the spacetime geometry becomes important and one must necessarily include quantum gravitational effects. 

One runs into serious conceptual and practical issues in directly extending the above approach to quantum gravity. In particular, the region $\mathscr{R}$ is not a priori well-defined since it can be changed by a diffeomorphism. Even in perturbation theory this means that, at a practical level, given a region $\mathscr{R}$ in the ``background'' spacetime, one cannot identify the location of the region in the perturbed spacetime. In other words, $\mathscr{R}$ is not well-defined when one considers gravitational fluctuations. To overcome this problem, at the outset, $\mathscr{R}$ must be defined in a diffeomorphism invariant way. For some regions this can be straightforwardly accomplished by defining $\mathscr{R}$ relative  to some other invariantly defined surface such as an extremal surface or to infinity. A simple example of the latter case is the maximal ``domain of influence'' $\mathscr{R}(\Gamma)$ of an observer -- i.e., $\Gamma$ is a causal curve
on $\mathscr{M}$ --- where she can initiate and receive results from any experiments she conducts throughout her lifetime\footnote{More precisely, $\mathscr{R}(\Gamma) = I^{+}(\Gamma)\cap I^{-}(\Gamma)$.}. If the observer's experiments occur over a finite time then one must invariantly define the events when she starts and stops her experiments. This is, in general, extremely cumbersome to accomplish. However, in the limit that the experiments occur over infinite time\footnote{In perturbation theory off a fixed, background spacetime it suffices to consider observer lifetimes which scale inversely in $G_{\textrm{N}}$ which tends to infinity as $G_{\textrm{N}}\to 0$. } then $\mathscr{R}(\Gamma)$ is invariantly defined. Even in this idealistic limit, if the spacetime  contains any event horizons  --- e.g., any black holes or cosmological horizons -- then $\mathscr{R}(\Gamma)$ is a  proper subset of $\mathscr{M}$.  
Another simple way to invariantly define a spacetime region is if there exists a (partial) Cauchy surface $\Sigma$ bounded by an extremal surface. Then one can invariantly define the region $\mathscr{R}(\Sigma)$ as the domain of dependence of $\Sigma$. Outside of these examples, we know of no general procedure for defining an arbitrary subregion in an invariant way.

Given an invariantly defined subregion, $\mathscr{R}$, the quantization of the gravitational field in this subregion is now well-defined and the physical observables in $\mathscr{R}$ are those that satisfy the gravitational constraints. Since the observer's vision is fundamentally limited by an event horizon, one might expect that one can assign an ``entropy'' associated to any ``degrees of freedom'' that the observer cannot access. This expectation is perhaps best understood in the case where $\mathscr{R}$ is the outer domain of communication of a stationary black hole. If $\mathfrak{A}(\mathscr{R})$ is the algebra of gravitational fluctuations in $\mathscr{R}$ --- i.e., ``free gravitons'' with no backreaction --- it was shown by Hawking that any quantum state approaches a thermal state with a non-zero temperature  at asymptotically late times \cite{Hawking:1975vcx,Fredenhagen:1989kr} and thereby the spacetime has a finite entropy \cite{Bardeen:1973gs,Bekenstein:1973ur,Bekenstein:1974ax}. The appropriate entropy, in the sense that it obeys the second law of thermodynamics \cite{Bekenstein:1974ax,Wall:2011hj}, is the ``generalized entropy'' 
\begin{equation}
\label{eq:introSgen}
S_{\textrm{gen.}}(\mathscr{R}) = \frac{A}{4G_{\textrm{N}}} + S_{\textrm{vN.}}(\rho_{\omega})
\end{equation}
which is the sum of the area of the black hole (in Planck units) together with the von Neumann entropy of the ``density matrix'' $\rho_{\omega}$ of any quantum fields in $\mathscr{R}$ 
\begin{equation}
\label{eq:VN}
S_{\textrm{vN.}}(\rho_{\omega}) \defn -\textrm{Tr}(\rho_{\omega}\log \rho_{\omega}).
\end{equation}
We have suppressed all fundamental constants except for $G_{\textrm{N}}$. The two terms in \eqref{eq:introSgen} are individually divergent in the semi-classical limit. The second term diverges due to ultraviolet correlations across the horizon which are a universal property of any physical state in quantum field theory. These ``ultraviolet divergences'' are a characteristic feature of the algebra of observables of local quantum field theory which is a so-called Type III$_{1}$ von Neumann algebra \cite{Murray:1936gtq,ASENS_1973_4_6_2_133_0,Fredenhagen:1984dc}. These algebras have no pure states and no notion of a trace or density matrices so the expression \eqref{eq:VN} is undefined. Nevertheless, it is expected that the sum \eqref{eq:introSgen} is better defined when one includes the backreaction of quantum gravitational effects \cite{Susskind:1994sm}. 

In this paper we construct the algebra $\mathfrak{A}_{\textrm{dress.}}(\mathscr{R})$ of (perturbatively) backreacted quantum gravitational fluctuations and consider the trace and entropy of states on this algebra. As we will explain, $\mathfrak{A}_{\textrm{dress.}}(\mathscr{R})$ is constructed by imposing constraints\footnote{The present work also intersects with a growing literature concerning the role of dynamical quantum reference frames in the algebraic formulation of quantum gauge theory and gravity, see e.g. \cite{Klinger:2023auu,AliAhmad:2024wja,AliAhmad:2024vdw,Araujo-Regado:2024dpr,Carrozza:2022xut,Fewster:2024pur,Fewster:2025ijg,Janssen:2025uzf}.} on $\mathfrak{A}(\mathscr{R})$ arising from fluctuations in the ``geometry'' of $\mathscr{R}$ at second order in perturbation theory. The results of this paper build upon recent work \cite{Chandrasekaran:2022eqq,Chandrasekaran:2022cip,Kudler-Flam:2023qfl} which considered the constraint arising from the fluctuations of a {\em single} mode: the total area of the horizon. It was shown in these references that the resulting algebra is a so-called Type II algebra which implies that --- in contrast to the Type III case --- one can assign a density matrix to $\mathscr{R}$ and compute a (renormalized) von Neumann entropy. For ``semiclassical states'' this entropy was shown to be equivalent to $S_{\textrm{gen.}}$.

While the analyses of \cite{Chandrasekaran:2022eqq,Chandrasekaran:2022cip,Kudler-Flam:2023qfl} strongly suggest that the entropy of an (invariantly defined) subregion becomes better defined when one takes into account semiclassical effects, 
 there are a number of open questions that remain to be addressed. In section~\ref{sec:killinghorizon}, we show that the backreaction of gravitons will, in general, perturb the black hole horizon in an angle-dependent way. At precisely the same order in perturbation theory that the total area fluctuates, there are an {\em infinite} set of charges  which encode the fluctuations of the ``shape'' of the black hole\footnote{It was previously proposed by one of us with R. Leigh \cite{Klinger:2023tgi,Klinger:2023auu} that the complete physical algebra should include contributions from gravitational charges associated with generic region preserving diffeomorphisms. Moreover, it was conjectured that the resulting `dressed' algebra is of Type II. However, \cite{Klinger:2023tgi,Klinger:2023auu} did not fully address the problem of invariantly defining subregions, nor did it explicitly construct the associated trace on the algebra. } . 
 As we will explain below, these additional charges impose an infinite number constraints on the first-order gravitons\footnote{See \cite{CF2026Subregion} for related work that will be appearing on the arXiv simultaneously.}. Given this result, it is a priori unclear whether one can consistently quantize just a single mode of the (second-order) gravitational field. Furthermore, one cannot conclude from these analyses whether (or not!) the full algebra $\mathfrak{A}_{\textrm{dress.}}(\mathscr{R})$ is Type II. If the resulting algebra \emph{is} Type II, then in what sense is the entropy equivalent to $S_{\textrm{gen.}}$? Is there a maximum entropy state? It is clear that, at the very least, these questions must be addressed before we can definitively conclude anything regarding the entropy of gravitational subregions. 
 
 The central purpose of this paper is to investigate these questions. Since many of our arguments require a considerable amount of technical machinery, in the remainder of this section, we present a sketch of these arguments focusing primarily on the physical origin of the constraints and the resulting properties of the algebra. 

We first briefly review the arguments of \cite{Chandrasekaran:2022eqq,Chandrasekaran:2022cip,Kudler-Flam:2023qfl} which consider the algebra obtained from quantizing a single gravitational mode. The backreaction of gravitational perturbations changes the area of the horizon. The black hole area is a ``charge'' on the horizon conjugate to a symmetry which, if the affine time $U=0$ is the bifurcation surface $\mathcal{B}$ and $U=e^{\kappa u}$, acts on the horizon $\mathcal{H}$ as \cite{wald1993black}
\begin{equation}
\label{eq:translation}
u \longrightarrow u + c
\end{equation}
where $c$ is a constant. The second-order perturbed charge associated to this symmetry at affine time $U$ on $\mathcal{H}$ is \cite{Chandrasekaran:2018aop,Hollands:2024vbe} 
\begin{equation}
\delta^{2}\mathcal{Q}_{U} = \delta^{2}A_{U} - \int_{S_{U}}d\Omega~U\delta^{2}\theta 
\end{equation}
where $\delta^{2}A_{U}$ is the perturbed area of the constant $U$ cross-section and $\delta^{2}\theta$ is the perturbed expansion of the horizon which vanishes when the black hole is stationary. We note that, on $\mathcal{B}$, $\delta^{2}\mathcal{Q}_{0}=\delta^{2}A_{0}$. The constraint arises from taking the difference of the limit as $U\to \pm \infty$ which yields
\begin{equation}
\label{eq:transchargeflux}
\frac{\delta^{2}\mathcal{Q}^{\textrm{L}}}{4\beta G_{\textrm{N}}} - \frac{\delta^{2}\mathcal{Q}^{\textrm{R}}}{4\beta  G_{\textrm{N}}} = -F
\end{equation}
where $\delta^{2}\mathcal{Q}^{\textrm{L}}$ is the perturbed area in the ``left wedge'' $\mathscr{L}$ and $\delta^{2}\mathcal{Q}^{\textrm{R}}$ is the perturbed area of the ``right wedge'' $\mathscr{R}$. 
Here,  $F$ is the flux of gravitational radiation energy propagating into the black hole 
\begin{equation}
\label{eq:F}
F = -\frac{1}{2\beta G_{\textrm{N}}} \int_{-\infty}^{\infty}dU \int_{\mathbb{S}^{2}}d\Omega~U\delta \sigma_{AB}\delta \sigma^{AB} 
\end{equation}
where $\delta \sigma_{AB}(U,x^{A})$ is the perturbed shear of $\mathcal{H}$ which represents the radiation falling across the horizon. At second-order, $F$ is generically non-vanishing and so the area of the black hole fluctuates at this order. 

Eq.~\eqref{eq:transchargeflux} imposes a non-trivial constraint on $\mathfrak{A}(\mathscr{R})$ in the following way: $F$ is the ``boost energy'' of the first-order gravitons and, as one might expect, thereby generates boosts of these fields in both $\mathscr{L}$ and $\mathscr{R}$. As noted by Chandrasekharan, Penington  and Witten (CPW), the algebra of physical observables in $\mathscr{R}$ must commute with all observables in $\mathscr{L}$ including $\delta^{2}\mathcal{Q}^{\textrm{L}}$ \cite{Chandrasekaran:2022eqq}. The physical observables are those that commute with $\delta^{2}\mathcal{Q}^{\textrm{R}}-F$. However, our original algebra $\mathfrak{A}(\mathscr{R})$ includes only first-order gravitons and does not include any second-order observables such as $\delta^{2}\mathcal{Q}^{\textrm{R}}$. Considering this constraint for a Schwarzschild-AdS black hole, it was shown in \cite{Chandrasekaran:2022eqq} that one can solve this constraint by quantizing $\delta^{2}\mathcal{Q}^{\textrm{R}}$ on the Hilbert space $L^{2}(\mathbb{R})$. Due to a ``matching condition'' on perturbations, the intrinsic fluctuations of $\delta^{2}\mathcal{Q}^{\textrm{R}}$ imply fluctuations of the perturbed ADM mass\footnote{For a Kerr-Newmann black hole $\delta^{2}\mathcal{Q}^{\textrm{R}}$ can be matched to a linear combination of the perturbed ADM mass, angular moment and electromagnetic charge.}. We denote the algebra of ``dressed observables'' that commute with $\delta^{2}\mathcal{Q}^{\textrm{L}}$ as $\mathfrak{A}_{\textrm{dress.}}(\mathscr{R};\mathbb{R})$ where the label $\mathbb{R}$ denotes the fact we have only imposed the constraint arising from the one-dimensional group given by \eqref{eq:translation}. 
In contrast to the ``undressed'' algebra $\mathfrak{A}(\mathscr{R})$, it follows from a theorem of Takesaki \cite{Takesaki2003} and the thermal properties of the Hartle-Hawking vacuum that $\mathfrak{A}_{\textrm{dress.}}(\mathscr{R};\mathbb{R})$ is of the form of a ``modular crossed product'' and is thereby a Type II (von Neumann) algebra. Such algebras admit a ``renormalized,'' densely defined trace given by 
\begin{equation}
\label{eq:tracereals}
\tau_{\mathbb{R}}(\hat{a}) = \braket{0_{\mathbb{R}},\omega |e^{X/2} \hat{a} e^{X/2}|0_{\mathbb{R}},\omega} \quad \quad \quad \quad  \quad \quad \quad \quad \textrm{($G=\mathbb{R}$)}
\end{equation}
where $\ket{0_{\mathbb{R}},\omega}=\ket{0_{\mathbb{R}}}\otimes \ket{\omega}$, and $\ket{\omega}$ is the stationary vacuum of gravitons in $\mathscr{R}$ --- i.e., in Schwarzschild-AdS $\ket{\omega}$ is the Hartle-Hawking state.  We have that $X\defn \delta^{2}\mathcal{Q}^{\textrm{R}}/4G_{\textrm{N}}\beta$ is  $\ket{0_{\mathbb{R}}}$ is the (improper) vector whose wavefunction is a ``delta-function'' in the conjugate variable to $X$. Eq.~\eqref{eq:tracereals} is the key step in defining density matrices and computing entropies associated to the subregion $\mathscr{R}$. 

Using \eqref{eq:tracereals}, the entropy of ``dressed gravitons'' in the exterior of a Schwarzschild-AdS black hole can be computed (up to a state-independent constant) and was shown to be equivalent to the generalized entropy \cite{Chandrasekaran:2022eqq}. Since, in perturbation theory, the fluctuations of the area are not bounded from above, the algebra was a so-called ``Type II$_{\infty}$'' algebra.
Subsequently, Chandrasekaran, Longo, Penington, and Witten (CLPW) extended this analysis to the static patch of de Sitter spacetime \cite{Chandrasekaran:2022cip}. An important difference is that, in de Sitter space, these authors introduced a (spherically symmetric) ``observer'' whose energy is bounded from below in order to obtain a nontrivial algebra. The entropy is then bounded from above, in accordance with the lore that empty de Sitter space has maximal entropy. Such an algebra is called Type II$_{1}$ \cite{Maeda:1997fh,Bousso:2000nf,Bousso:2000md,2018JHEP...07..050D,2022arXiv220601083L}. These arguments were subsequently generalized by two of us together with S. Leutheusser (KFLS) to arbitrary spacetimes with non-degenerate Killing horizons --- including spacetimes that are ``out of equilibrium'' \cite{Kudler-Flam:2023qfl}. A similar analysis has also been extended to inflationary cosmological spacetimes \cite{Kudler-Flam:2024psh,Chen:2024rpx}.\footnote{Related work has considered an application of these ideas to general subregions \cite{Jensen:2023yxy}, although the problem of invariantly defined regions is not clearly understood in that case. General subregions and the generalized second law for gravitational operator algebras are further explored in \cite{Faulkner:2024gst,Kirklin:2024gyl,CF2026Subregion}} However, in all of these works, the dressed algebra is obtained by imposing only one constraint. To consistently quantize only one mode, one must restrict to perturbations under which the remaining modes do not fluctuate. Thus, the consistency of the above quantization strongly depends on the remaining physical degrees of freedom of the horizon at second order and the corresponding constraints that have been neglected.

What constraints have been neglected? In sec.~\ref{subsec:symmcharges}, we obtain the complete set of second-order charges on the horizon which arise from an infinite dimensional symmetry group of the horizon \cite{Chandrasekaran:2018aop}. Similar to the perturbed area, these charges encode perturbations in the geometry of the horizon at second order. However, only a subset of these charges impose constraints on the first-order gravitons in a similar manner to that of \eqref{eq:transchargeflux}. This is the case if the ``flux'' of these charges is entirely due to the backreaction of first-order gravitons -- i.e., their flux is locally constructed from only first order perturbations. In this case, commutation with the charge in $\mathscr{L}$ imposes a non-trivial constraint on the (first-order) graviton algebra. 

Any isometry of the background will generate a non-trivial constraint on the graviton algebra. Thus, in addition to the constant boosts given by \eqref{eq:transchargeflux}, we obtain non-trivial constraints due to the angular momentum flux  $F(\psi)$ through the horizon where $\psi^{A}$ is the generator of the rotational isometry group $H_{\textrm{isom.}}$ of the background horizon cross-sections. This flux satisfies a charge-flux relation 
\begin{equation}
\label{eq:chargefluxY}
\frac{\delta^{2}\mathcal{Q}^{\textrm{L}}(\psi)}{8\pi G_{\textrm{N}}} - \frac{\delta^{2}\mathcal{Q}^{\textrm{R}}(\psi)}{8\pi G_{\textrm{N}}} = - F(\psi)
\end{equation}
which imposes a constraint analogous to \eqref{eq:transchargeflux}. These charges $\delta^{2}\mathcal{Q}^{\textrm{L}/\textrm{R}}(\psi)$ directly match onto the perturbed ADM angular momenta of the left and right wedges. The work of CLPW, CPW and KFLS assumes that the relevant set of constraints at this order correspond only to the isometries of $(\mathscr{R},g)$. With this assumption, to consistently restrict to the algebra $\mathfrak{A}_{\textrm{dress.}}(\mathscr{R};\mathbb{R})$ one must restrict to spherically symmetric perturbations such that the angular momentum charges do not fluctuate. For more general perturbations, the authors speculated that the inclusion of constraints associated to any compact group does not change the type classification of the algebra. 

In section \ref{sec:isom}, we construct the algebra $\mathfrak{A}_{\textrm{dress.}}(\mathscr{R};G_{\textrm{isom.}})$ satisfying the constraints \eqref{eq:transchargeflux} and \eqref{eq:chargefluxY} associated to the isometry group where $G_{\textrm{isom.}} = \mathbb{R}\times H_{\textrm{isom.}}$ in any spacetime with a Killing horizon. Generalizing the case of $G = \mathbb{R}$, one must now allow for intrinsic fluctuations of the angular momenta $\delta^2 \mathcal{Q}^{\textrm{R}}(\psi)$, and ensure that the dressed algebra commutes with $\delta^2 \mathcal{Q}^{\textrm{L}}(\psi)$. Using a recent extension of Takesaki's theorem by one of us with S. Ali Ahmad and S. Lin (AAKL)  to include compact groups \cite{AliAhmad:2024eun}, we prove that this algebra is of Type II$_{\infty}$ with trace
\begin{equation}
\label{eq:traceGisom}
\tau_{\textrm{G}_{\textrm{isom.}}}(\hat{a}) = \braket{e_{G_{\textrm{isom.}}},\omega|e^{X/2} \hat{a} e^{X/2}|e_{G_{\textrm{isom.}}},\omega} \quad \quad  \quad \quad \quad  \textrm{($G=G_{\textrm{isom.}}$)}.
\end{equation}
For Schwarzschild, $\ket{e_{G_{\textrm{isom.}}}} = \ket{0_{\mathbb{R}}}\otimes \ket{e_{\textrm{SO}(3)}} $ where $\ket{e_{\textrm{SO}(3)}}$ is the (improper) rotationally invariant vector in the Hilbert space $L^2(SO(3))$ with $\delta$-function wavefunction with respect to the Haar measure on $H_{\textrm{isom.}}$. For any black hole spacetime, the algebra $\mathfrak{A}_{\textrm{dress.}}(\mathscr{R};G_{\textrm{isom.}})$ is  Type II$_{\infty}$ and, as an example, we explicitley construct the algebra and entropy for the exterior of a Kerr black hole in sec.~\ref{subsec:kerr}. While this result confirms the above expectations for black holes, it seems to naively give a dramatically different result from that of CLPW in the case of de Sitter. As we explain in  sec.~\ref{subsec:deSitter}, if the charges were treated as independent, then bounding the energy of the observer from below merely bounds the spectrum of $X$. The trace remains non-normalizable since the norm of the vector $\ket{0_{G_{\textrm{isom.}}}}$ is infinite. Nevertheless, in any physically realistic model of the observer, their own spin contributes to the total gravitational energy. Taking the spin-energy of the observer into account, we find that bounding the total energy of the observer from below yields a normalizable trace. For a physical observer, the isometry invariant algebra of the static patch is Type II$_{1}$. 

As we have emphasized, the isometries do not constitute the full set of second-order constraints. Of the full set of ``large gauge'' transformations of the horizon obtained in \cite{Chandrasekaran:2018aop} we find, in sec.~\ref{subsec:symmcharges}, that the complete group of symmetries that yield non-trivial constraints on the first-order gravitons is\footnote{While the full group of large gauge transformations on the horizon is $\textrm{Diff}(\mathbb{S}^{2})\ltimes \mathcal{S}$ \cite{Hopfmuller:2018fni,Chandrasekaran:2018aop} where $\textrm{Diff}(\mathbb{S}^{2})$ is the set of all smooth diffeomorphisms of the horizon cross-sections, as we explain in  sec.~\ref{subsec:symmcharges}, the only subgroup of $\textrm{Diff}(\mathbb{S}^{2})$ that imposes gravitational constraints on the first-order gravitons is the isometry group $H_{\textrm{isom.}}$. Furthermore, as we explain in sec.~\ref{subsec:symmquantalg}, $H_{\textrm{isom.}}$ is also the only subgroup of $\textrm{Diff}(\mathbb{S}^{2})$ which is unitarily implementable on the Fock space of linearized gravitons.}
\begin{equation}
G = H_{\textrm{isom.}}\ltimes \mathcal{S}
\end{equation}
where $\mathcal{S}$ is an {\em infinite} dimensional, abelian group of ``boost supertranslations'' which on the horizon act as 
\begin{equation}
\label{eq:boostsupertrans}
u\to u+f(x^{A})
\end{equation}
where $f$ is an arbitrary, smooth function on the $2$-sphere. These additional symmetries arise because, in addition to the area and angular momentum, the backreaction of gravitons will also cause fluctuations in the ``shape'' of $\mathscr{R}$ at second order. To see this we note that the corresponding flux which generates the symmetry \eqref{eq:boostsupertrans} on the graviton Hilbert space is 
\begin{equation}
F(f) \defn -\frac{1}{2\beta G_{\textrm{N}}}\int_{-\infty}^{\infty}dU\int_{\mathbb{S}^{2}}d\Omega~f(x^{A})U\delta \sigma_{AB}\delta \sigma^{AB}.
\end{equation}
We note that, as with the flux given by \eqref{eq:F}, $F(f)$ is locally constructed from the first-order gravitons and represents the angular distribution of gravitational radiation energy falling across the horizon. Indeed, the flux associated to the constant boost $F(1)=F$ is simply the total energy with no angular weighting. 
This flux causes the black hole to change its ``shape'' in a non-spherically symmetric way relative to the background, spherical cross-sections. This is encoded in non-trivial charges \cite{Hollands:2024vbe} 
\begin{equation}
\delta^{2}\mathcal{Q}_{U}(f) = \delta^{2}A_{U}(f) - \int_{S_{U}}d\Omega~f(x^{A})U\delta^{2}\theta 
\end{equation}
where $\delta^{2}A_{U}(f)$ is the integral of the area element on the cross-section weighted by $f$ and we note that $\delta^{2}\mathcal{Q}_{U}(1)=\delta^{2}\mathcal{Q}_{U}$ which we considered earlier. The limits of these charges as $U\to \pm \infty$ satisfy the global charge-flux relation 
\begin{equation}
\label{eq:QLQRfluxboostsuper}
\delta^{2}\mathcal{Q}^{\textrm{L}}(f) - \delta^{2}\mathcal{Q}^{\textrm{R}}(f) = -F(f)
\end{equation}
where, again, the subscript $\pm$ indicates the limit at $U\to \pm \infty$. Just like the constant boosts and rotations, \eqref{eq:QLQRfluxboostsuper} imposes a constraint  on the graviton algebra for each choice of $f$. Thus, at the same order in perturbation theory where the fluctuations of the area and angular momentum of the black hole are relevant, there are an infinite set of additional non-vanishing charges which impose non-trivial constraints. For asymptotically flat black holes, these charges have been conjectured to match to ``supermomentum charges'' corresponding to analogous supertranslations symmetries at infinity \cite{Hawking:2016msc}. 

We construct the full, dressed algebra $\mathfrak{A}_{\textrm{dress.}}(\mathscr{R};G)$ satisfying the complete set of constraints given by \eqref{eq:chargefluxY} and \eqref{eq:QLQRfluxboostsuper}. Furthermore, we show that the subspace of states for which the higher-harmonic boost supertranslation charges do not fluctuate consists only of the vacuum. Consequently, the algebras $\mathfrak{A}(\mathscr{R};\mathbb{R})$ or $\mathfrak{A}(\mathscr{R};G_{\textrm{isom.}})$ cannot be realized as subalgebras of $\mathfrak{A}(\mathscr{R};G)$. As such, we cannot deduce the type of the full algebra by quantizing any finite dimensional subgroup. Nevertheless, studying the dressed algebras obtained from quantizing smaller groups does allow for the identification of universal properties that can, in turn, be generalized to the infinite dimensional case. In particular, in sec.~\ref{sec:AdressH},  we obtain a ``universal'' form of the trace for the dressed algebra under $\mathbb{R}\times H$ where $H$ is an arbitrary, compact group. This universal form is given by 
\begin{equation}
\label{eq:traceRHintro}
\tau_{\mathbb{R}\times H}(\hat{a}) = \braket{e_{\mathbb{R}\times H},\omega|e^{X/2} \hat{a} e^{X/2}|e_{\mathbb{R}\times H},\omega} \quad \quad  \quad \quad \textrm{($G=\mathbb{R}\times H$ w/ $H$ compact)}.
\end{equation}
where $\ket{\omega}$ should be interpreted as a quantum field state invariant under the group $\mathbb{R}\times H$, and $X$ is the generator of translations in the group $\mathbb{R}$. The object $\ket{e_{\mathbb{R}\times H}}$ is the (improper) `neutral' state in $L^{2}(\mathbb{R}\times H)$ whose wavefunction is the delta function, $\delta_{\mu_{\mathbb{R}\times H}}(g,e_{\mathbb{R}\times H})$, compatible with the Haar measure $\mu_{\mathbb{R}\times H}$ on $\mathbb{R}\times H$. 

In sections \ref{sec:fulldressedalg} and \ref{sec:dSII1}, we consider the algebra of observables invariant under the full, infinite dimensional group. We note that all dressed algebras studied in this paper are of the form of ``crossed product'' von Neumann algebras. As opposed to crossed products with respect to (locally) compact groups, the crossed product with respect to an infinite dimensional (non locally compact) group is not well studied even in the mathematical literature. Therefore, our goal in these sections will be to  identify the necessary physical and mathematical properties so that the algebra is Type II and argue that these properties are satisfied. As we have emphasized, to obtain a non-trivial physical algebra in perturbative quantum gravity one must consider the crossed product with respect to an infinite dimensional group. We hope that the arguments of this paper provide sufficient motivation for further mathematical study of these von Neumann algebras.

As we explain in section \ref{sec:fulldressedalg},
the major difference as compared to the finite dimensional case is that there is no invariant Haar measure on an infinite dimensional group.  However, a measure $\mu_{G}$ can be chosen to be quasi-invariant under a ``sufficiently large'' subset of the group and consequently one can obtain a well-defined action of a ``sufficiently large'' set of charges $\delta^{2}\mathcal{Q}(f,\psi)$. We provide such a construction for the case where $\mu_{G}$ is an infinite dimensional Gaussian measure. Additionally, on this Hilbert space we propose a construction of an improper state $\ket{e_{\mu_{G}}}$ associated to the neutral element $e_{G}$ of the group, possessing a wavefunction $\delta_{\mu_{G}}(g,e_{G})$, which is the delta function with respect the measure $\mu_{G}$. With this vector in hand we argue that 
\begin{equation}
\label{eq:tauGintro}
\tau(\hat{a}) = \braket{e_{\mu_{G}},\omega|e^{X/2} \hat{a} e^{X/2}|e_{\mu_{G}},\omega} \quad \quad \quad \quad \quad \quad \textrm{$(G = H_{\textrm{isom.}} \ltimes \mathcal{S})$}.
\end{equation}
is a trace on the full dressed algebra $\mathfrak{A}_{\textrm{dress.}}(\mathscr{R};G)$. Consequently, the algebra of gravitational fluctuations of $\mathscr{R}$ is Type II$_{\infty}$. 

In sec.~\ref{sec: CP for NLC Groups}, we consider the inclusion of ``test'' matter fields with gauge group $G_{\textrm{LGT}}$ which give rise to additional constraints analogous to that of \eqref{eq:chargefluxY}. If the gauge group is compact, then the arguments of sec.~\ref{sec:AdressH} directly apply where now $H$ includes $G_{\textrm{LGT}}$. However, for theories with long-range fields (e.g., electromagnetic fields or unconfined Yang-Mills fields) the large gauge group is, in general, infinite dimensional and includes angle-dependent gauge transformations on the horizon \cite{Hawking:2016msc}. Since these fields are generally described by an interacting quantum field theory and the infinite dimensional group $G_{\textrm{LGT}}$ is non-abelian, the arguments that lead to the construction of \eqref{eq:tauGintro} cannot be immediately generalized. We outline a set of sufficient conditions such that the full dressed algebra is Type II. While we show that some of these properties can be straightforwardly achieved by a suitable choice of measure, others require further analysis of the large gauge invariance of interacting theories and the properties of the neutral element vector. Nevertheless, with these assumptions, the trace on the dressed algebra is of the form 
\begin{equation}
\label{eq:tauG}
\tau(\hat{a}) = \braket{e_{\mu_{G}},\omega|e^{X/2} \hat{a} e^{X/2}|e_{\mu_{G}},\omega} \quad \quad \quad \quad \quad \textrm{($G = H_{\textrm{isom.}} \ltimes (\mathcal{S}\times G_{\textrm{LGT}}))$}.
\end{equation}

In all cases, the existence of a trace allows for the computation of the entropy of any well defined quantum state on the dressed algebra. Of particular interest are the ``semiclassical states'' $\ket{\hat{\Phi}_{\alpha}} = \ket{\varphi}\otimes \ket{\alpha}$. Here, $\ket{\varphi}$ is a state over the gravitons as well as any other (if present) quantum fields in $\mathscr{R}$ and $\ket{\alpha}$ is a state in the Hilbert space of the group whose wavefunction is sharply peaked at a particular group element. For such semiclassical states, the von Neumann entropy of the corresponding density matrix $\rho_{\hat{\Phi}_{\alpha}}$ is 
\begin{equation}
\label{eq:SvNintro}
S_{\textrm{v.N.}}(\rho_{\hat{\Phi}_{\alpha}}) \simeq S_{\textrm{gen.}} + S(\rho_{\alpha}) + C 
\end{equation}
where $S(\rho_{\alpha})$ is an entropy associated to the fluctuation of the charges. For gauge fields with compact large gauge groups, we show that $S(\rho_{\alpha})$ directly matches the ``edge mode'' contributions to the entanglement entropy that have been previously computed by other methods (see, e.g., \cite{Donnelly:2014gva}). Eq.~\ref{eq:SvNintro} can be viewed as a generalization of these results to perturbative quantum gravity and quantum fields with more general gauge groups.\footnote{We note that contributions to the entanglement entropy from edge modes related to infinite dimensional large gauge groups have been explored in e.g. \cite{Ball:2024hqe,Ball:2024xhf,Ball:2024gti}. It would be interesting to draw a more direct comparison between the conclusions of these works and the results reported here.} Notably, in this construction, it is not necessary to introduce a UV cutoff to factorize the Hilbert space, as is frequently done in the discussion of edge modes. The Hilbert space of the theory remains non-factorizable and the edge mode contribution to the entanglement remains well-defined. 

In sec.~\ref{sec:dSII1}, we revisit the algebra of observables in de Sitter spacetime and incorporate the additional constraints arising from boost supertranslations. As before, if the horizon area and the higher-harmonic charges were independent, the trace would be non-normalizable and the algebra would be of Type II$_{\infty}$. To see that this is not the case, we note that the higher multipoles of the perturbed area are sourced by the mass multipole moments of the observer. Moreover, any physically realistic model of an observer defining a static patch must be bound by non-gravitational forces. These finite-size effects contribute to the observer’s total binding energy and hence to its total gravitational energy. We argue that these contributions, together with the spin energy identified in sec.~\ref{subsec:deSitter}, imply that the algebra of observables in de Sitter spacetime is of Type II$_{1}$ with entropy given by \eqref{eq:SvNintro}.

In section \ref{sec: discussion} with discussions of generalizations to generic regions bounded by extremal surfaces, and the interpretation of the boost supertranslation charges within the framework of AdS/CFT. 

While the presentation of the main text is entirely self-contained, we have included a sequence of appendices which present a parallel analysis from a more formal perspective. In Appendix \ref{App: CP and OVW}, we describe the construction of the dressed algebra using the terminology and notation of the crossed product. In Appendix \ref{App: General Trace}, we provide a general characterization of the type of this algebra and a construction of its trace when it exists by appealing to the theory of dual weights as pioneered by Digernes and Haagerup \cite{Digernes1975,HaagerupI,HaagerupII}. This analysis generalizes the results of \cite{AliAhmad:2024eun}. In Appendix \ref{App: CP Factorization}, we provide an alternative characterization of the dressed algebra for the case that the group of constraints is a Cartesian product. This provides an alternative proof of semifiniteness, and also facilitates the analysis of Type II$_1$ subalgebras which are explored in Appendix \ref{sec: Type II_1}.

 We work in natural units, $c=\hbar=1$, however we will not set $G_{\textrm{N}}$ to unity. We will use the notation and sign conventions of \cite{Wald:1984rg}. In particular, our metric signature is ``mostly positive'' and our sign convention for curvature is such that the scalar curvature of a round sphere is positive. Abstract $\ast$-algebras are denoted by $\mathscr{A}$, while von Neumann algebras are denoted by $\mathfrak{A}$. The algebra of observables supported in a subregion $\mathscr{R}$ is written as $\mathscr{A}(\mathscr{R})$ before weak closure and $\mathfrak{A}(\mathscr{R})$ after. Operators associated with classical phase space observables are denoted by boldface versions of their classical counterparts. For instance, the operator associated with the smeared graviton $\gamma(w)$ is written $\op{\gamma}(w)$. Similarly, the quantization of the gravitational fluxes, $F(f,\psi)$, and charges, $\delta^2\mathcal{Q}(f,\psi)$, are notated as $\op{F}(f,\psi)$ and $\delta^2 \op{\mathcal{Q}}(f,\psi)$, respectively. The algebra of dressed observables supported in a region $\mathscr{R}$ satisfying the constraints associated with a symmetry group $G$ is denoted by $\mathscr{A}_{\textrm{dress.}}(\mathscr{R},G)$, or $\mathfrak{A}_{\textrm{dress.}}(\mathscr{R};G)$ when closed in a suitable weak operator topology.

\section{The Algebra of Gravitational Perturbations in a Subregion}
\label{sec:killinghorizon}
In this section, we construct the ``dressed'' algebra of gravitational perturbations in $\mathscr{R}$. As outlined in the introduction, this algebra is constructed in four steps: In sec.~\ref{subsec:gravpert}, we define the subregion $\mathscr{R}$ in a diffeomorphism invariant way and locate the region in both the background and perturbed spacetime. In sec.~\ref{subsec:classphasespace}, we review the classical phase space of first-order gravitational perturbations of $\mathscr{R}$ and in sec.~\ref{subsec:quantization} we review its quantization. In sec.~\ref{subsec:symmcharges}, we show that the backreaction of these gravitational fluctuations perturbs the geometry of $\mathscr{R}$ which gives rise to an infinite set of ``second order charges'' a subset of which give rise to nontrivial constraints. Finally, in sec.~\ref{subsec:dressedalg}, we construct the complete, dressed algebra of $\mathscr{R}$. 

\subsection{The Definition of the Subregion $\mathscr{R}$}
\label{subsec:gravpert}
In this paper we consider a subregion $\mathscr{R}\subseteq \mathscr{M}$ of a one-parameter family\footnote{Formally, in gravitational perturbation theory, $\lambda=\sqrt{32\pi G_{\textrm{N}}}$. For perturbation theory on stationary black hole backgrounds we will additionally, fix the radius of the black hole as $G_{\textrm{N}}\to 0$. Therefore, for such spacetimes, we will be working in the limit of large black hole mass. } of spacetimes $(\mathscr{M},g_{ab}(\lambda))$ that solve Einstein's equation $G_{ab}(\lambda)+\Lambda g_{ab}(\lambda)=0$ for all $\lambda$. As emphasized in the introduction, in order for such a ``subregion'' to be well-defined, $\mathscr{R}$ must be defined in a diffeomorphism invariant way. In all cases of interest, the region $\mathscr{R}$ can be defined as the domain of communication of an inextendible causal curve $\Gamma$ whose affine parameter tends to infinity in the future and the past. Here, by domain of communication, we mean that $\mathscr{R} := I^{-}(\Gamma) \cap I^+(\Gamma)$, where $I^{\pm}(\Gamma)$ refer to the chronological future/past of the curve $\Gamma$. In this way, we restrict attention to the spacetime accessible to those observers who can perform any experiments or measurements of the spacetime in perpetuity and do not fall into any singularities. If $\mathscr{R}$ is a proper subset of $\mathscr{M}$ then the boundary $\partial \mathscr{R} = \mathcal{H}_{\textrm{R}}$ is called the ``event horizon'' which can occur in a cosmological spacetime as well as any spacetime where a black hole is present. If the spacetime admits a conformal boundary then $\mathscr{R}$ can be equivalently defined using the asymptotic structure. For example, for an asymptotically flat spacetime, the region $\mathscr{R}$ is the black hole ``exterior'' and the ``black hole region'' is\footnote{In maximally extended, stationary spacetimes this region will also contain the white hole. For black holes formed from collapse, $\mathscr{M}-\mathscr{R}$ is the black hole interior.} $\mathscr{M}-\mathscr{R}$.

The boundary $\mathcal{H}_{\textrm{R}}$ is a null surface \cite{Hawking:1971vc}. 
Adequately locating the region $\mathscr{R}$ relies upon characterizing the null boundary $\mathcal{H}_{\textrm{R}}$ and how it changes under perturbations in a geometric way. To characterize this region we will now review and define some basic properties of general (not necessarily stationary) null surfaces. Let $\mathcal{H}$ be a smooth, null hypersurface whose null generators are the null geodesic generators of $\xi^{a}$, which we do not assume to be affinely parameterized. The surface gravity of $\kappa$ measures the failure of $\xi^{a}$ to be affinely parameterized 
\begin{align}
\label{eq:geodkappa}
    \xi^a \nabla_a \xi^b \;\hat{=}\; \kappa \xi^b.
\end{align}
Hereafter, $\hat{=}$ denotes equality on $\mathcal{H}$. The submanifold inherits a degenerate metric $q_{ab}$ which satisfies $q_{ab}\xi^{b}=0$. If $n^{a}$ is the corresponding affinely parameterized vector field, then we define the second fundamental form of the null surface relative to $n^{a}$ as 
\begin{equation}
K_{ab} \defn \frac{1}{2}\pounds_{n}q_{ab}.
\end{equation}
Additionally, it will be convenient to choose an auxilliary null vector field $\ell^{a}$ which is transverse to $\mathcal{H}\simeq \mathbb{R}\times \mathcal{S}$ and is defined by affinely parameterized, geodesic transport off of $\mathcal{H}$ (i.e., $\ell^{a}$ satisfies $\ell^{a}\ell_{a}=0$, $\ell^{a}n_{a}=-1$ and $\ell^{a}\nabla_{a}\ell^{b}=0$). With this additional structure we may define the ``Hájiček rotation 1-form'' associated to $n^{a}$ 
\begin{equation}
w_{a} \;\hat{=}\;  - q_{a}{}^{c}\ell_{b}\nabla_{c}n^{b}
\end{equation}
on $\mathcal{H}$. 

Since the tensors $q_{ab}$, $K_{ab}$ and $w_{a}$ are orthogonal to $n^{a}$, we can view them as ``lower dimensional objects'' living on the tangent space of the horizon cross sections $S$. We will denote tensors lying in this tangent space with capital Latin indices, i.e., the tensor fields $q_{ab}$, $K_{ab}$ and $w_{a}$ will be denoted as $q_{AB}$, $K_{AB}$ and $w_{A}$. Since $\xi^{a}$ is hypersurface orthogonal, the second fundamental form satisfies $K_{[AB]}=0$ (i.e., the ``vorticity'' of the null surface vanishes) and we may decompose it as 
\begin{equation}
K_{AB} = \frac{1}{2}\theta q_{AB} + \sigma_{AB}
\end{equation}
where $\theta \defn q^{AB}K_{AB}$ is the expansion of $\mathcal{H}$ and the trace-free tensor $\sigma_{AB}$ is known the shear. In the following, it will be convenient to a choose coordinates $(v,x^{A})$ on the horizon such that $\xi^{a}=(\partial/\partial v)^{a}$ and $x^{A}$ are arbitrary coordinates on the horizon cross-sections.

\begin{figure}
    \centering
    \begin{tikzpicture}[scale=3.2]
    \tikzset{declare function={%
 penrose(\x,\c) = {\fpeval{2/pi*atan( (sqrt((1+tan(\x)^2)^2+4*\c*\c*tan(\x)^2)-1-tan(\x)^2) /(2*\c*tan(\x)^2) )}};%
 penroseu(\x,\t) = {\fpeval{atan(\x+\t)/pi+atan(\x-\t)/pi}};%
 penrosev(\x,\t) = {\fpeval{atan(\x+\t)/pi-atan(\x-\t)/pi}};%
 kruskal(\x,\c) = {\fpeval{asin( \c*sin(2*\x) )*2/pi}};% Penrose coordinates for Kruskal
}}
  
  \def\R{0.08} % size lightcone
  \def\Nlines{1} % number of world lines (at constant r/t)
  \pgfmathsetmacro\ta{1/sin(90*1/(\Nlines+1))} % constant r/t value 1
  \pgfmathsetmacro\tb{sin(90*2/(\Nlines+1))}   % constant r/t value 2
  \pgfmathsetmacro\tc{1/sin(90*2/(\Nlines+1))} % constant r/t value 3
  \pgfmathsetmacro\td{sin(90*1/(\Nlines+1))}   % constant r/t value 4
  \coordinate (-O) at (-1, 0); % center III: origin (r,t) = (0,0)
  \coordinate (-S) at (-1,-1); % south III: t=-infty, i-
  \coordinate (-N) at (-1, 1); % north III: t=+infty, i+
  \coordinate (-W) at (-2, 0); % east III:  r=-infty, i0
  \coordinate (-E) at ( 0, 0); % west III:  r=+infty, i0
  \coordinate (O)  at ( 1, 0); % center I: origin (r,t) = (0,0)
  \coordinate (S)  at ( 1,-1); % south I: t=-infty, i-
  \coordinate (N)  at ( 1, 1); % north I: t=+infty, i+
  \coordinate (E)  at ( 2, 0); % east I:  r=-infty, i0
  \coordinate (W)  at ( 0, 0); % west I:  r=+infty, i0
  \coordinate (B)  at ( 0,-1); % singularity bottom
  \coordinate (T)  at ( 0, 1); % singularity top
  \coordinate (X0) at ({asin(sqrt((\ta^2-1)/(\ta^2-\tb^2)))/90},
                       {-acos(\ta*sqrt((1-\tb^2)/(\ta^2-\tb^2)))/90}); % particle 1
  \coordinate (X1) at ({asin(sqrt((\tc^2-1)/(\tc^2-\td^2)))/90},
                       {acos(\tc*sqrt((1-\td^2)/(\tc^2-\td^2)))/90}); % particle 2
  \coordinate (X2) at (45:0.87); % particle falling in BH horizon
  \coordinate (X3) at (0.60,1.05); % particle falling in BH singularity

  \draw[thick,black] (N) -- (-S) ;
  \draw[thick,black] (-N) -- (S) ;

    \draw[thick ,  ->] (.65,-.5) to[out = 115, in = -115] (.65,.5);
    \draw[thick ,  <-] (-.65,-.5) to[out = 65, in = -65] (-.65,.5);
    \draw[thick ,  <-] (-.5,-.65) to[out = 25, in = 155] (.5,-.65);
    \draw[thick ,  ->] (-.5,.65) to[out = -25, in = -155] (.5,.65);
  % REGIONS
  \node[] at (-.75,0) {$\mathscr{L}$};
  \node[] at (.75,0) {$\mathscr{R}$};
  \node[] at (0,0.75) {$\mathscr{F}$};
  \node[] at (0,-0.75) {$\mathscr{P}$};

  \node[above=-2.5,rotate=45] at (.75,0.75) {$\mathcal{H}^+_R$};
  \node[below=-2.5,rotate=45] at (-.75,-0.75) {$\mathcal{H}^+_L$};
  \node[above=-2.5,rotate=-45] at (-.75,0.75) {$\mathcal{H}^-_L$};
  \node[below=-2.5,rotate=-45] at (.75,-0.75) {$\mathcal{H}^-_R$};
  \node[right] at (0,0) {$\mathcal{B}$};
  
\end{tikzpicture}

    \caption{A spacetime diagram depicting a bifurcate Killing horizon given by the union of null surfaces $\mathcal{H}=\mathcal{H}^{-}\cup \mathcal{H}^{+}$ where $\mathcal{H}^{-}=\mathcal{H}^{-}_{\textrm{R}}\cup \mathcal{H}^{-}_{\textrm{L}}$ and $\mathcal{H}^{+}=\mathcal{H}^{+}_{\textrm{R}}\cup \mathcal{H}^{+}_{\textrm{L}}$ with bifurcation surface  $\mathcal{B}$. The arrows correspond to orbits of the Killing field in a neighborhood of  $\mathcal{B}$. The spacetime is divided into four regions $\mathscr{L}, \mathscr{R},\mathscr{P}$ and $\mathscr{F}$ as shown.}
    \label{fig:bifurcate}
\end{figure}
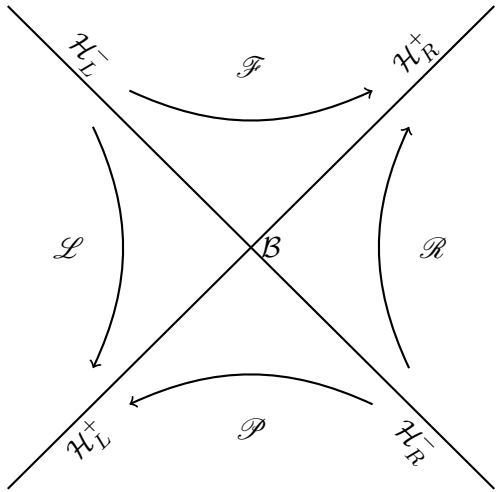

The above defines the geometry of any null surface $\mathcal{H}$. However, the boundary of $\mathscr{R}$ in $(\mathscr{M},g_{ab}(\lambda))$ is not just any null surface. It is an {\em event horizon} and, as such, is only teleologically defined. More precisely, using the fact that the spacetime satisfies Einstein's equation $G_{ab}(\lambda)+\Lambda g_{ab}(\lambda)=0$ for all $\lambda$ yields the Raychauduri equation 
\begin{equation}
\label{eq:Raychauduri}
\partial_{V}\theta(\lambda) = -\frac{1}{2}\theta(\lambda)^{2} - \sigma_{AB}(\lambda)\sigma^{AB}(\lambda)
\end{equation}
and the Damour-Navier-Stokes equation for the rotation one-form 
\begin{equation}
\label{eq:Damour}
\partial_{V}\omega_{A}(\lambda) = -\theta(\lambda)\omega_{A}(\lambda) -q^{BC}(\lambda)\mathscr{D}_{C}\sigma_{BA}(\lambda) - \frac{1}{2}\mathscr{D}_{A}\theta(\lambda),
\end{equation}
where $V$ is the affine parameter of the null generators. Here, $\mathscr{D}_{A}(\lambda)$ is the covariant derivative operator on the horizon cross-sections compatible with $q_{AB}(\lambda)$ and capital indices are raised and lowered with this metric. These equations give two relations for the three independent quantities $\theta(\lambda)$, $\omega_{A}(\lambda)$ and $\sigma_{AB}(\lambda)$ that determine the geometry of the null surface.\footnote{While there is an ``evolution'' equation for the shear (see, e.g., (9.2.33) of \cite{Wald:1984rg}), this equation is ``sourced'' by the electric Weyl tensor $C_{AcBd}n^{c}n^{d}$ on the null surface which is unconstrained by Einstein's equation. Therefore, this equation should actually be viewed as a formula for the Weyl curvature on the null surface as opposed to an equation that determines $\sigma_{AB}$.} The event horizon is uniquely defined as the outermost, outgoing null surface satisfying eqs.~\eqref{eq:Raychauduri} and~\eqref{eq:Damour} with compact cross-sections whose expansion $\theta(\lambda)$ vanishes as $V\to \infty$. The global nature of this definition means that it cannot be easily found for any $\lambda>0$. However, as we now explain, it can be straightforwardly identified at ``zeroth'' and ``first'' order in $\lambda$ \cite{2013CMaPh.321..629H}.

We first identify $\mathscr{R}$ in the 
``background'' spacetime $(\mathscr{M},g(0))$ which we assume contains a bifurcate Killing horizon $\mathcal{H}$ with bifurcation surface $\mathcal{B}$. By assumption, the spacetime admits a Killing vector $\xi^{a}$ whose integral curves are the null geodesic generators of $\mathcal{H}$ which satisfy \eqref{eq:geodkappa} with constant $\kappa$ \cite{Bardeen:1973gs}, and $\xi^{a}$ vanishes on $\mathcal{B}$. Spacetimes with Killing horizons encompass all stationary black hole spacetimes as well as the de Sitter cosmology. For any black hole formed from collapse, the exterior will rapidly settle down to an essentially stationary black hole whose event horizon is, to an excellent approximation, a Killing horizon \cite{Hawking:1971vc,Friedrich:1998wq,Alexakis:2009gi,Alexakis:2010nsf} which can always be extended to a global spacetime $(\mathscr{M},g_{ab}(0))$ with bifurcate structure \cite{Racz:1992bp,Racz:1995nh}. Any bifurcate Killing horizon globally divides the spacetime into four regions $\mathscr{F},\mathscr{P}, \mathscr{R}$ and $\mathscr{L}$ as depicted in figure \ref{fig:bifurcate} \cite{Kay:1988mu}. The subregion $(\mathscr{R},g_{ab}(0))$ corresponds to the $\lambda=0$ case of the one-parameter family of invariantly defined subregions defined above. We will denote the union $\mathscr{M}_{\textrm{R}} \defn \mathscr{F}\cup \mathscr{R}$ as the ``physically relevant region'' which, in the black hole case, corresponds to the interior and exterior of the black hole. This spacetime will, at the very least, be the smallest global extension of $\mathscr{R}$ that we will need. The physical relevance of the maximal global extension depends on the spacetime of interest. The analyses and results presented in this paper will apply equally well if the relevant global spacetime is $\mathscr{M}$ or $\mathscr{M}_{\textrm{R}}$. 

This division of $(\mathscr{M},g_{ab}(0))$ also naturally divides each horizon as depicted in fig.~\ref{fig:bifurcate}. For example, we denote the future/past horizon of $\mathscr{R}$ as $\mathcal{H}_{\textrm{R}}^{\pm}\defn \mathcal{H} \cap I^{\pm}(\mathscr{R})$ and, similarly, we define $\mathcal{H}^{\mp}_{\textrm{L}}\defn \mathcal{H} \cap I^{\pm}(\mathscr{R})$. Since $\xi^{a}$ vanishes on $\mathcal{B}$, the Gaussian null coordinates introduced above are good coordinates in a neighborhood on each component of the horizon. It will also be convenient to, for example, consider an affinely parameterized vector field $\ell^{a}=(\partial/\partial V)^{a}$ on $\mathcal{H}^{+}_{\textrm{L}}\cup \mathcal{H}^{-}_{\textrm{R}}$ with affine parameter $V$ so that $V=0$ on $\mathcal{B}$. On each component the affine parameter is exponentially related to the Killing parameter. For example, on $\mathcal{H}^{-}_{\textrm{R}}$ we have that 
\begin{align}
\label{eq:affine}
    V =        e^{\kappa v}
\end{align}
where we have chosen $V$ such that $v=0$ corresponds to $V=1$. Thus, the coordinates $(V,x^{A})$ cover $\mathcal{H}^{-}\defn \mathcal{H}^{+}_{\textrm{L}}\cup \mathcal{H}^{-}_{\textrm{R}}$. Similar statements and constructions also hold for  $\mathcal{H}^{+}=\mathcal{H}^{-}_{\textrm{L}}\cup \mathcal{H}^{+}_{\textrm{R}}$ with coordinates $(U,x^{A})$ where $U$ is the affine parameter of the null geodesic generators of $\mathcal{H}^{+}$. Finally, for a Killing horizon, we note that expansion and shear identically vanish. So, we have that on $\mathcal{H}_{\textrm{R}}$ for $\lambda=0$ 
\begin{equation}
\label{eq:Killing}
\theta(0) = 0,  \quad \sigma_{AB}(0)=0 \textrm{ and }   \quad q_{AB}(0) = q_{AB} \quad \quad \textrm{ (Killing horizon)}
\end{equation}
where $q_{AB}$ is the stationary, ``background'' metric on the horizon cross-sections. 
The rotation one-form $w_{A}(0)$, however, is generally non-vanishing --- i.e., it is non-vanishing for a stationary Kerr black hole --- and, by eqs.~\eqref{eq:Damour} and ~\eqref{eq:Killing}, is constant on the horizon. 
 
We now consider linearized perturbations of the spacetime. While we have restricted attention to the vacuum gravitational perturbations, it is entirely straightforward to extend this analysis to include matter perturbations (see sec.~\ref{sec: CP for NLC Groups}). More precisely, we consider a $1$-parameter family of classical solutions to the full vacuum Einstein equations $g_{ab}(\lambda)$ such that $g_{ab}(0)$ is a spacetime with a Killing horizon and 
\begin{equation}
\gamma_{ab} \defn\frac{dg_{ab}}{d\lambda} \bigg\vert_{\lambda =0} 
\end{equation}
is a linear perturbation which satisfies 
\begin{align}
    L[\gamma ]\defn \Box_g \gamma_{ab}+\nabla_a\nabla_b \gamma -\nabla_c \nabla_a \gamma^c{}_b-\nabla_c\nabla_b \gamma^c{}_a+2\Lambda \gamma_{ab} = 0
\end{align}
with $\gamma \defn g^{ab}\gamma_{ab}$ and $\Lambda$ a cosmological constant. In the rest of this paper we will also use the notation $\delta g_{ab} = d g_{ab}/d\lambda\vert_{\lambda=0}$ and $\delta^{2}g_{ab} = d^{2} g_{ab}/d\lambda^{2}\vert_{\lambda=0}$. We will also use this notation for any the variation of any function of the metric in the remainder of this paper.

Diffeomorphism invariance of $g_{ab}(\lambda)$ implies that any two solutions $\gamma_{ab}$ and $\gamma^{\prime}_{ab}$ are physically equivalent if they differ by a linearized gauge transformation 
\begin{equation}
\label{eq:gauge}
\gamma^{\prime}_{ab} = \gamma_{ab} + \nabla_{(a}\chi_{b)}
\end{equation}
where $\chi^{a}$ is any vector field on $\mathscr{M}$. Under such a perturbation the spacetime subregion will, in general, be perturbed as well and so we must relocate the event horizon in the perturbed spacetime. Despite the fact that the event horizon is defined only teleologically, the vanishing of the background expansion and shear implies, via \eqref{eq:Raychauduri}, that the first-order Raychaudhuri equation yields
\begin{equation}
\label{eq:deltathetaconst}
\partial_{V}\delta \theta =0\quad \quad \quad \textrm{(on $\mathcal{H}^{+}$).}
\end{equation}
 Therefore the necessary and sufficient condition for $\mathcal{H}^{+}$ to remain the event horizon  in the perturbed spacetime is that the  surface remains null with vanishing perturbed expansion. It was shown in  \cite{2013CMaPh.321..629H} that, using \eqref{eq:gauge}, the gauge condition 
\begin{equation}
\label{eq:deltatheta}
\delta \theta =0 \quad \quad \quad \textrm{(on $\mathcal{H}^{+}$)}
\end{equation}
can always be imposed. We similarly impose the identical gauge condition\footnote{Reference \cite{2013CMaPh.321..629H} imposed an additional gauge condition on the perturbed volume element of $\mathcal{B}$ and did not impose the condition that $\delta \theta$ vanish on $\mathcal{H}^{-}$. The latter condition is unnecessary (see the Remark in \cite{Sorce:2017dst}) and the former condition (i.e., that $\delta \theta=0$ on $\mathcal{H}^{-}$) together with \eqref{eq:deltatheta}  ensures that the location of $\mathcal{B}$ is not changed in the perturbed spacetime \cite{Prabhu:2015rua}. \label{foot:gauge}} $\delta \theta =0$ on $\mathcal{H}^{-}$. With these conditions we have succeeded in locating the subregion $\mathscr{R}$ in the perturbed spacetime. 
% \subsection{The Classical Phase Space of Perturbations and Its Quantization}
\subsection{The Classical Phase space of First-order perturbations of $\mathscr{R}$}
\label{subsec:classphasespace}
Having invariantly defined the subregion of interest, we can now meaningfully discuss the phase space of gravitational perturbations of this region. Observables correspond to gauge invariant functions on phase space. For example, while the perturbation $\gamma_{ab}$ is not invariant under the linearized gauge transformation given by \eqref{eq:gauge}, one can construct a gauge invariant observable by smearing 
\begin{equation}
\label{eq:gammaf}
\gamma(w) = \int_{\mathscr{M}}\sqrt{-g}d^{4}x~\gamma_{ab}(x)w^{ab}(x)
\end{equation}
where $w^{ab}$ is a smooth, symmetric and divergence-free test tensor (i.e., $w^{[ab]}=0=\nabla_{a}w^{ab}$). The phase space of linearized perturbations can be endowed with a symplectic structure with conserved, symplectic product 
\begin{equation}
\label{eq:OmegaSigma}
\mathscr{W}(\gamma_{1},\gamma_{2})= \frac{1}{16\pi }\int_{\Sigma}\sqrt{h}d^{3}x~n_{a}P^{abcdef}[\gamma_{2,bc}\nabla_{d}\gamma_{1,ef}-\gamma_{1,bc}\nabla_{d}\gamma_{2,ef}]
\end{equation}
where
\begin{equation}
P^{abcdef}\defn g^{ae}g^{fb}g^{cd}-\frac{1}{2}g^{ad}g^{be}g^{fc}-\frac{1}{2}g^{ab}g^{ae}g^{fd}+\frac{1}{2}g^{bc}g^{ad}g^{ef}.
\end{equation}
Here, $\Sigma$ is any Cauchy surface, $h_{ab}$ is the induced metric on $\Sigma$, and $n^a$ is the unit normal. We note that, for any pair of solutions $\gamma_{1},\gamma_{2}$, the symplectic product is independent of the choice of Cauchy surface. The local observable \eqref{eq:gammaf} is also an observable on phase space 
\begin{equation}
\label{eq:gammafW}
\gamma(w) = \mathscr{W}(\gamma,Ew)
\end{equation}
and generates the infinitesimal transformation $\gamma_{ab}\longrightarrow \gamma_{ab}+(Ew)_{ab}$ where $(Ew)_{ab}$ is the advanced minus retarded solution of the linearized Einstein's equation with source $w^{ab}$. 

The above construction yields the phase space of gravitational perturbations on any globally hyperbolic spacetime. We will now construct the phase space of gravitational perturbations in $\mathscr{R}$ which is the phase space of perturbations subject to the gauge conditions \eqref{eq:deltatheta}. By \eqref{eq:OmegaSigma}, the phase space of such solutions in $\mathscr{R}$ is equivalent to the space of initial data subject to our gauge conditions. Since these gauge conditions are specified on the bifurcate Killing horizon, it is extremely convenient to consider the space of initial data on $\mathcal{H}^{+}$ or $\mathcal{H}^{-}$. For definiteness we will focus on data specified on the past horizon $\mathcal{H}^{-}$. 

To identify the phase space of initial data on $\mathcal{H}^{-}$ we consider the linearized Einstein equations on the null surface. Without loss of generality, it can be shown that one can impose the additional gauge condition\footnote{This condition is compatible with working in a ``Gaussian null gauge'' which can be imposed in the neighborhood of any null surface} that $ \gamma_{ab}n^{b}\;\hat{=}\;0$ on the horizon. In this gauge, the condition \eqref{eq:deltatheta} implies that 
\begin{equation}
\label{eq:deltaexp}
\delta \theta = \frac{1}{2}q^{AB}\pounds_{n}\delta q_{AB} = 0
\end{equation}
where $\delta q_{AB}$ is the perturbed metric on the horizon cross-sections, $q_{AB}$ is the background metric on the horizon cross-sections and capital indices in \eqref{eq:deltaexp} are raised and lowered with this metric. The perturbed metric $\delta q_{AB}$ subject to \eqref{eq:deltaexp} corresponds to the ``free data'' on the horizon for the linearized Einstein equation. Indeed, by our gauge conditions, the linearized Raychaudhuri's equation is trivially satisfied and the linearized Damour-Navier-Stokes equation 
yields 
\begin{equation}
\label{eq:linearizedDNS}
\pounds_{n}\delta \omega_{A} = \mathscr{D}^{B}\delta \sigma_{BA}
\end{equation}
where, here and in the remainder of this paper, $\mathscr{D}_{A}$ is the covariant derivative compatible with  the background metric $q_{AB}$ on the horizon cross-sections. Using \eqref{eq:deltaexp}, the linearized shear is given by 
\begin{equation}
\delta \sigma_{AB} = \frac{1}{2}\pounds_{n}\delta q_{CD}.
\end{equation}
If we specify the perturbed metric $\delta q_{AB}$ on the horizon, subject to \eqref{eq:deltaexp}, $\delta \omega_{A}$ is determined by integrating \eqref{eq:linearizedDNS} with the condition that $\delta \omega_{A}$ vanishes in the far future. Without loss of generality we consider perturbations which vanish in the asymptotic past and future since we are  considering the phase space of solutions $(Ef)_{ab}$ sourced by test functions of compact support. 

The symplectic form given by \eqref{eq:OmegaSigma} with $\Sigma=\mathcal{H}^{-}$ yields a symplectic structure on the initial data \cite{2013CMaPh.321..629H} 
\begin{equation}
\mathscr{W}_{\mathcal{H}^{-}}(\gamma_{1},\gamma_{2}) = -\frac{1}{8\pi }\int_{\mathcal{H}^{-}}dUd\Omega~(\delta \sigma^{AB}_{1}\delta q_{2,AB} -  \delta q^{AB}_{1}\delta\sigma_{2,AB}). 
\end{equation}
where we take the phase space of initial data to be perturbations $\delta q_{AB}$ which decay as $U\to \pm \infty$. 
Using \eqref{eq:gammafW}, the smeared shear on $\mathcal{H}^{-}$ can be directly related to the local metric perturbation by 
\begin{equation}
\label{eq:shearsmear}
\delta \sigma(s) = \int_{\mathcal{H}^{-}_{\textrm{R}}}dVd\Omega~\delta \sigma_{AB}s^{AB} = -4\pi\mathscr{W}_{\mathcal{H}_{\textrm{R}}^{-}}(\gamma,Ew), \quad s_{ab}=(Ew)_{ab}.
\end{equation}
By \eqref{eq:gammafW}, the phase space of  perturbations in $\mathscr{R}$ is equivalent to the phase space of initial data on any Cauchy surface. In general, $\mathcal{H}^{\pm}$ will not be a Cauchy surface for $\mathscr{R}$. However, in many spacetimes (e.g., Rindler horizons in flat spacetime \cite{Unruh:1983ms}, de Sitter spacetime, black holes in AdS, etc) it is essentially a Cauchy surface for determining the evolution of solutions to the wave equation. In these cases, smoothness of the solution and the dispersive nature of waves on these spacetimes ensures that all solutions with initial data of compact support will fully ``fall through'' the horizon. Consequently, in these spacetimes, \eqref{eq:shearsmear} implies that 
\begin{equation}
\label{eq:gammasigma}
\gamma(w) =- \frac{1}{4\pi}\delta \sigma(s) 
\end{equation}
where $s_{ab} \defn (Ew)_{ab}\vert_{\mathcal{H}^{-}}$. 

In other spacetimes (e.g., black holes in de Sitter space, asymptotically flat black holes, etc.), \eqref{eq:gammasigma} is insufficient to fully reconstruct the local perturbation and one must supplement with data on another Killing horizon or at infinity. In these cases, the data can be independently specified on each asymptotic boundary and $\gamma(w)$ is simply a linear combination of data on each component of the boundary of $\mathscr{R}$ (see subsec. \ref{subsec:kerr}). 

\subsection{The Quantization of First-Order Perturbations of $\mathscr{R}$}
\label{subsec:quantization}
We will now review the quantization of first-order gravitational perturbations $\gamma$ in $\mathscr{R}$ \cite{Fewster:2012bj}. In any globally hyperbolic spacetime, the quantization algebra is a $\ast$-algebra $\mathscr{A}$ defined by starting with the free algebra generated by $\op{\gamma}(w)$, $\op{\gamma}(w)^{\ast}$ and $\op{1}$ and factoring by the relations 
\begin{enumerate}[label=(R.{\Roman*})]
\label{KGalg}
\item $ \op{\gamma}(c_{1}w_{1}+c_{2}w_{2})=c_{1}\op{\gamma}(w_{1})+c_{2}\op{\gamma}(w_{2})$ for any $w_{1},w_{2}$ and any $c_{1},c_{2}\in \mathbb{R}$ \label{A1}
\item $\op{\gamma}(L(w))=0$ for all $w$  \label{A2}
\item $\op{\gamma}(w)^{\ast}=\op{\gamma}(w)$ for all $w$ \label{A3}
\item $[\op{\gamma}(w_{1}),\op{\gamma}(w_{2})]=iE(w_{1},w_{2})\op{1}$, for any $w_{1}$ and $w_{2}$\label{A4}
\end{enumerate}
 \ref{A1} denotes the fact that $\op{\gamma}(w)$ is linear in the test tensor, \ref{A2} denotes the fact that $\op{\gamma}$ is Hermitian, \ref{A3} states that $\op{\gamma}$ satisfies the equations of motion in the distributional sense and \ref{A4} implies that  $ \op{\gamma}$ satisfies the commutation relations where $E_{abcd}(y_{1},y_{2})$ is the advanced minus retarded Green function on $\mathscr{M}$. 
 
To quantize the perturbations in $\mathscr{R}$ we must additionally impose our gauge conditions on the horizon. This can be achieved using 
\begin{equation}
\label{eq:opgammasigma}
\op{\gamma}(w) = -\frac{1}{4\pi }\delta \op{\sigma}(s) 
\end{equation}
where, again,  $s_{ab} \defn (Ew)_{ab}\vert_{\mathcal{H}^{-}}$ to define a corresponding ``initial data'' algebra on $\mathcal{H}^{-}$.  A similar algebra can be construction on the future horizon. In spacetimes where \eqref{eq:opgammasigma} does not hold, the following quantization must be supplemented by an additional quantization of degrees of freedom ``at infinity'' (see sec. \ref{subsec:kerr}). Unless otherwise stated, we will assume that \eqref{eq:opgammasigma} holds and so the quantization of the bulk field observable is equivalent to the quantization of the perturbed shear on the horizon. More precisely, $\mathscr{A}(\mathscr{R}) \cong \mathscr{A}(\mathcal{H}^{-})$ where $\mathscr{A}(\mathcal{H}^{-})$ is the $\ast$-algebra of initial data $\delta \op{\sigma}$ factored by relations equivalent to that of \ref{A1} and \ref{A3} together with the commutation relations  
\begin{equation} \label{shear commutation}
[\delta \op{\sigma}_{AB}(x_{1}),\delta \op{\sigma}_{CD}(x_{2})] = i(q_{A(C}q_{D)B}-\frac{1}{2}q_{AB}q_{CD})\delta^{\prime}(V_{1},V_{2})\delta_{\mathbb{S}^{2}}(x_{1}^{A},x_{2}^{A})\op{1}.
\end{equation}
Eqn. \eqref{shear commutation} holds in the distributional sense where the $x=(Y,x^{A})$ are points on $\mathcal{H}^{-}$ and $\delta_{\mathbb{S}^{2}}$ is the $\delta$-function on the $2$-sphere cross-sections of the horizon. We note that there is no analog of \ref{A2} since we are considering initial data. 

The algebra $\mathscr{A}(\mathcal{H}^-)$ admits a unique vacuum state\footnote{Here, we are considering a state in the algebraic sense as a positive, linear, normalized map from the algebra to the complex numbers.} $\omega$ which is invariant under the action of horizon Killing isometries. In black hole spacetimes where \eqref{eq:opgammasigma} holds, $\omega$ will be referred to as the ``Hartle-Hawking'' state\footnote{In Minkowski spacetime, $\omega$ is simply the Minkowski vacuum and in de Sitter spacetime $\omega$ is known as the Bunch-Davies state \cite{Schomblond_1976,Chernikov:1968zm,Bunch:1978yq,davies1975scalar}} \cite{Hartle:1976tp,Israel:1976ur} whereas, in cases where we must supplement \eqref{eq:opgammasigma} with data elsewhere this state will generally be referred to as the ``Unruh'' state \cite{unruh1976notes}. In either case, all such states agree on $\mathcal{H}^{-}$. Indeed, the unique state on $\mathcal{H}^{-}$ invariant under the horizon Killing isometries is a Gaussian state with vanishing 1-point function and $2$-point function given by \cite{Kay:1988mu}
\begin{equation}
\label{eq:2pt}
\omega\big(\delta \op{\sigma}_{AB}(x_{1})\delta \op{\sigma}_{CD}(x_{2})\big)= \frac{1}{\pi}\frac{(q_{A(C}q_{D)B}-\frac{1}{2}q_{AB}q_{CD})\delta_{\mathbb{S}^{2}}(x_{1}^{A},x_{2}^{A})}{(U_{1}-U_{2}-i0^{+})^{2}}.
\end{equation}
We note that the state $\omega$, is invariant under the isometries of $\mathscr{R}$. In particular, this includes the isometry group of $(q_{AB},\mathbb{S}^{2})$. If $\mathscr{R}$ is the exterior of a static black hole or the de Sitter static patch, then the cross-sections are round spheres and the isometry group $\textrm{SO}(3)$. For rotating black holes, the cross-sections are oblate spheres and the isometry group of these spheres is $U(1)$. In addition to the rotational isometries, the state $\omega$ is also invariant under the ``boost'' isometry  $U\to e^{\kappa t}U$ --- i.e., $u\to u+t$ in ``Killing time'' --- which corresponds to the action of the horizon Killing isometry of the background. On the subalgebra $\mathscr{A}(\mathcal{H}^{-}_{\textrm{R}}) \subset \mathscr{A}(\mathcal{H}^{-})$ of observables on $\mathcal{H}^{-}_{\textrm{R}}$ we can change coordinates to the ``Killing time'' $u$  which yields 
\begin{equation}
\label{eq:2ptthermal}
\omega\big(\delta \op{\sigma}_{AB}(u_{1},x_{1}^{A})\delta \op{\sigma}_{CD}(u_{2},x_{2}^{A})\big) = \frac{\kappa^{2}}{4\pi}\frac{(q_{A(C}q_{D)B}-\frac{1}{2}q_{AB}q_{CD})\delta_{\mathbb{S}^{2}}(x_{1}^{A},x_{2}^{A})}{\sinh(\frac{\kappa}{2}(u_{1}-u_{2})-i0^{+})^{2}}
\end{equation}
where $\delta_{\mathbb{S}^{2}}$ is the $\delta$-function with respect to $q_{AB}$. The state $\omega$ can then be seen to be thermal with respect to translations in $u$ in the following sense. First, we can define the translation map
\begin{equation}
\label{eq:trans}
\alpha_{t}[ \delta \op{\sigma}_{AB}(u,x^{A})] \defn \delta \op{\sigma}_{AB}(u+t,x^{A}).
\end{equation}
Then, $\omega\big(\delta \op{\sigma}_{AB}(u_{1},x_{1}^{A})\alpha_{t}[ \delta \op{\sigma}_{CD}(u_{2},x_{2}^{A})]\big)$ can be analytically continued as a distribution to complex $t$ in the strip $0<\textrm{Im} t<2\pi/\kappa$. It is bounded and continuous on this strip, and satisfies the ``KMS condition" 
\begin{equation}
\label{eq:KMS}
\omega\big(\alpha_{t+i\beta}[\delta \op{\sigma}_{AB}(u_{1},x_{1}^{A})] \delta \op{\sigma}_{CD}(u_{2},x_{2}^{A}\big)=\omega\big(\delta \op{\sigma}_{AB}(u_{2},x_{2}^{A})\alpha_{t}[ \delta \op{\sigma}_{CD}(u_{1},x_{1}^{A})]\big) 
\end{equation}
where $\beta \defn 2\pi/\kappa$. A state that obeys the above conditions is a thermal state with inverse temperature $\beta$ \cite{Kubo:1957mj,PhysRev.115.1342,Haag:1967sg}. As we will see in sec.~\ref{subsec:symmcharges}, the set of diffeomorphisms which leave $\omega$ invariant is actually considerably larger than just the isometries of $\mathscr{R}$. From the form of eq.~\ref{eq:KMS}, it is clear that $\omega$ is also invariant under a much larger group of ``boost supertranslations'' $u\to u+f(x^{A})$ as mentioned in the Introduction. This fact will play a key role in the considerations of this paper. 

Given the vacuum state \eqref{eq:2pt}, one can construct a one-particle Hilbert space $\mathscr{H}_{1}$ by defining an inner product $\braket{s_{1}|s_{2}} = \omega(\delta \op{\sigma}(s_{1})^{\ast}\delta \op{\sigma}(s_{2}))$ on the space of initial data on $\mathcal{H}^{-}$, factoring by any degenerate elements and completing this space with respect to the norm defined by the inner product. The full Fock representation $\mathscr{F} := \mathbb{C}\oplus \bigoplus_{n=1}^{\infty} \mathscr{H}_{n}$ where $\mathscr{H}_{n}$ is the n-fold symmetrized tensor product of $\mathscr{H}_{1}$. This yields the ``GNS representation'' of $\mathscr{A}(\mathcal{H}^{-})$ associated to the ``ground state'' which is represented by the vector $\ket{\omega} \in \mathscr{F}$. By construction this vector is ``cyclic,'' i.e., the action of the elements of $\mathscr{A}(\mathcal{H}^{-})$ on $\ket{\omega}$ generates a dense subspace of states.\footnote{This observation allows us to define the representation of $\mathscr{A}(\mathcal{H}^-)$ acting on $\mathscr{F}$. Given $a \in \mathscr{A}(\mathcal{H}^-)/\text{ker}(\omega)$ we define the vector $\ket{a} \equiv a\ket{\omega}$. Then, $\mathscr{A}(\mathcal{H}^-)$ simply acts by left multiplication $a \ket{b} = \ket{ab}$.} We also note that on the subalgebra $\mathscr{A}(\mathcal{H}_{\textrm{R}}^{-})$, $\ket{\omega}$ is ``separating'' in that the only element of $\mathscr{A}(\mathcal{H}_{\textrm{R}}^{-})$ that annihilates the state is the zero element. 

While the $\ast$-algebra $\mathscr{A}(\mathscr{R})$ is sufficient for most physical questions (e.g., the evolution of initial data, the expectation value of any local observable, ect.), some additional structure will be useful to discuss notions such as the ``von Neumann entropy''  of quantum states in $\mathscr{R}$.
%The initial data for states in $\mathscr{R}$ correspond (at least partially) to states on the subalgebra $\mathscr{A}_{\mathcal{H}_{\textrm{R}}^{-}}$. 
This extra structure promotes the $\ast$-algebra to a ``von Neumann algebra'' $\mathfrak{A}(\mathcal{H}_{\textrm{R}}^{-},\omega)$ which is a suitable (weak) closure of $\mathscr{A}(\mathcal{H}_{\textrm{R}}^{-})$. This closure is obtained by taking the ``double commutant'' of the original algebra. The ``commutant'' algebra $ \mathscr{A}(\mathcal{H}_R)'$  with respect to the GNS representation $\mathscr{F}$ is given by
\begin{equation}
\label{eq:commutant}
 \mathscr{A}({\mathcal{H}_{\textrm{R}}^{-}})' \defn \{b \in \mathcal{B}(\mathscr{F})~|~ ba = ab,~ \forall ~a \in \mathscr{A}(\mathcal{H}_{\textrm{R}}^{-})\} \ 
\end{equation}
where $\mathcal{B}(\mathscr{F})$ is the set of all bounded operators on $\mathscr{F}$. 
%For example, while $\delta \op{\sigma}(s)$ is not a bounded operator on $\mathscr{F}$ one can obtain a bounded operator by considering the ``Weyl operator'' 
%\begin{equation}
%\op{W}(s) \defn e^{i\delta \op{\sigma}(s)}.
%\end{equation}
The von Neumann algebra $\mathfrak{A}(\mathcal{H}^-_{\textrm{R}},\omega)$ associated the observables on $\mathcal{H}_{\textrm{R}}^{-}$ is given by  \cite{neumann1929beweis}
\begin{equation}
\mathfrak{A}(\mathcal{H}_{\textrm{R}}^{-},\omega) \defn \big[\mathscr{A}(\mathcal{H}_{\textrm{R}}^{-})\big]^{\prime \prime}.
\end{equation}
Similarly, we will denote $\mathfrak{A}(\mathcal{H}^{-},\omega)$ as the von Neumann algebra on the horizon generated by the full algebra $\mathscr{A}(\mathcal{H}^{-})$. Finally we note that, 
in spacetimes for which \eqref{eq:gammasigma} holds, we trivially have that 
\begin{equation}
\mathfrak{A}(\mathcal{H}_{\textrm{R}}^{-})  = \mathfrak{A}(\mathscr{R}) 
\end{equation}
where $\mathfrak{A}(\mathscr{R}) $ is the von Neumann algebra of gravitational perturbations in $\mathscr{R}$. Likewise, the commutant of this algebra can be identified with $\mathfrak{A}(\mathscr{R})' \simeq \mathfrak{A}(\mathscr{L})$. Hereafter, we will drop reference to the state $\omega$ in the von Neumann algebra.

The additional structure of von Neumann algebras provides useful tools for their analysis and classification. Given any state $\ket{\Psi}\in \mathscr{F}$ and a cyclic, separating state $\ket{\omega}$ one can (densely) define the relative Tomita operator $\op{S}_{\Psi|\omega}$ which satisfies 
\begin{equation}
\op{S}_{\Psi|\omega} a \ket{\omega}=a^{\ast} \ket{\Psi}.
\end{equation}
The relative Tomita operator admits a ``polar decomposition'' of the form 
\begin{equation}
\op{S}_{\Psi|\omega} = \op{J}_{\Psi|\omega}\op{\Delta}^{1/2}_{\Psi|\omega}
\end{equation}
where $\op{\Delta}_{\Psi|\omega}$ is a positive operator known as the ``relative modular operator'' and $\op{J}_{\Psi|\omega}$ is an antiunitary operator which is called the ``relative modular conjugation.'' 

The {\em modular operator} is defined as the case where both entries of the relative modular operator refer to the same state (i.e., for $\ket{\Psi}$, the modular operator is $\op{\Delta}_{\Psi} \defn \op{\Delta}_{\Psi|\Psi}$). It generates a strongly continuous, one-parameter group of automorphisms of any von Neumann algebra $\mathfrak{A}$ 
\begin{equation}
\label{eq:modflow}
 a(t)\defn \op{\Delta}_{\Psi}^{-it}a \op{\Delta}_{\Psi}^{it} \in \mathfrak{A} \quad \quad \forall a\in \mathfrak{A}, t\in \mathbb{R}
\end{equation}
where $a(t)$ is the one-parameter family of algebra elements which constitute the ``modular flow'' of $a$. Since this action is strongly continuous, there exists an unbounded (densely defined) operator $\op{K}_{\Psi}\defn -\log \op{\Delta}_{\Psi}$. The modular operator satisfies the direct analog of the ``KMS conditions'' now with respect to modular flow: 
\begin{equation}
\op{\Delta}_{\Psi}\ket{\Psi}=\ket{\Psi} \quad \quad \braket{\Psi|a(t) b|\Psi}=\braket{\Psi|a b(t+i)|\Psi} 
\end{equation}
for any $a,b\in \mathfrak{A}$. We have used the fact that  $\braket{\Psi|ab(t)|\Psi}$ satisfies appropriate analyticity conditions in $t$ (see the text above \eqref{eq:KMS}). For a general state, $\ket{\Psi}$, the modular flow \eqref{eq:modflow} will not correspond to any local, geometric flow in the spacetime. By contrast, as mentioned above, the state $\omega$ is a KMS state on $\mathcal{H}_{\textrm{R}}^{-}$ and so, by \eqref{eq:KMS}, the modular flow generated by $\op{\Delta}_{\omega}$ on that subalgebra precisely coincides with the flow generated by Killing time translations. 

The von Neumann algebras considered in this paper are ``factors'' meaning they have trivial ``centers'' (i.e., $\mathfrak{A}\cap \mathfrak{A}^{\prime} = \mathbb{C}\op{1}$). Von Neumann factors can be broadly classified into three categories: Type I, II and III. These categories are identified by the properties of their modular operators and are largely distinguished by the existence (or lack thereof) of a trace on the algebra. Such an object is a prerequisite to defining --- let alone computing --- von Neumann entropies. We now give a brief review of this classification as it pertains to the results and arguments of this paper. For further details we refer the reader to \cite{Witten:2018zxz,Sorce:2023fdx}. 

% , which, roughly speaking, characterize whether or not a trace --- and, consequently, a density matrix --- exists for the algebra. The existence of such an object is, of course, a prerequisite to defining (let alone computing!) von Neumann entropies. As we will briefly review, this classification is intimately connected to the properties of the modular operator. We now briefly review the classification of such algebras. For a more detailed review of these concepts we refer the reader to, e.g., \GSnote{[KFLS, JS,?]}. 

Type I algebras are the most familiar algebras in quantum theory. Such algebras admit irreducible Hilbert space representations and are equivalent to the algebra of all bounded operators on a given Hilbert space. These algebras admit a trace --- i.e., a densely defined, positive, linear functional $\textrm{Tr}(\cdot)$ on the algebra such that $\textrm{Tr}(ab)=\textrm{Tr}(ba)$ $\forall$ $a,b\in\mathfrak{A}$ --- which, in this case, is simply the Hilbert space trace. As such, these algebras admit pure states and density matrices. A Type I algebra on an infinite dimensional Hilbert space is denoted as Type I$_{\infty}$. A key example of such an algebra is quantum field theory on a Cauchy surface (e.g., a $t=0$ surface in Minkowski spacetime). One can assign a Hilbert space to degrees of freedom on such a surface and, unsurprisingly, the algebras on these surfaces admit pure states and density matrices. It is important to note that the Cauchy surface need not be complete for the algebra to be Type I. It need only ``sufficiently'' capture enough degrees of freedom of the quantum field so that one can assign a Hilbert space. In our case, this occurs for the algebra on the horizon $\mathcal{H}^{-}$. Despite the fact that $\mathcal{H}^{-}$ is, in general, not a complete Cauchy surface, one can construct an irreducible representation of $\mathfrak{A}(\mathcal{H}^{-})$ --- namely, $\mathscr{F}$ --- and the algebra does admit pure states --- namely, the vacuum $\omega$. 

A Type III algebra, on the other hand, has almost none of these properties. This algebra has no irreducible representations nor does it admit a trace functional. While one can define states (e.g., positive, linear maps on the algebra) these states cannot be pure and one cannot assign a density operator to them. Type III algebras arise when one considers algebras associated to subregions of the spacetime. For the purposes of this paper, an important example is the algebra $\mathfrak{A}(\mathcal{H}^{-}_{\textrm{R}})$ associated to (the boundary of) $\mathscr{R}$. The failure of assigning a density matrix to any state in $\mathscr{F}$ restricted to $\mathfrak{A}(\mathcal{H}^{-}_{\textrm{R}})$ arises from the strong entanglement across the bifurcation surface. The correlation functions of all physical states in $\mathscr{F}$ diverge at short distances, as in the vacuum state (see \eqref{eq:2pt}). This strong, universal, ``Hadamard'' divergence is the obstruction to assigning a Hilbert space to the subregion $\mathcal{H}^{-}_{\textrm{R}}$ or a density operator to any state on $\mathfrak{A}(\mathcal{H}^-_\textrm{R})$. This universal divergence also implies that the modular operator acts as an {\em outer automorphism} on $\mathfrak{A}(\mathcal{H}^{-}_{\textrm{R}})$ and cannot be split into the product of operators affiliated with $\mathfrak{A}(\mathcal{H}^-_{\textrm{R}})$ and its commutant. We will refer to the algebra $\mathfrak{A}(\mathcal{H}^{-}_{\textrm{R}})$ as a Type III$_{1}$ algebra\footnote{The specific subclassification of Type III$_{\lambda}$ algebras by Connes \cite{connes1973classification} where $0\leq \lambda \leq 1$ will not be relevant for us since the only factors that can occur in local quantum field theory are of Type III$_{1}$ \cite{Fredenhagen:1984dc}. However, the introduction of stringy physics, which is inherently non-local, has recently been argued to lead to Type III$_0$ von Neumann algebras \cite{Herderschee:2025nsb}.}. 

Finally, Type II algebras have intermediate properties between the cases of type I and III. Like Type III algebras, Type II algebras do not have irreducible representations and, as such, all states on the algebra are mixed. However, Type II algebras differ in two important ways. The first is that the modular flow is now an {\em inner automorphism}. The second is that one can, in fact, construct a trace which is unique up to a multiplicative constant.\footnote{For a discussion of this fact, see Appendix \ref{sec: Type II_1}.} As we will see, these two properties are not entirely unrelated; the von Neumann entropy is defined up to a state-independent constant and so only entropy differences are well-defined. If the algebra admits a maximum entropy state, then the algebra is called Type II$_{1}$. More technically, this means that the trace functional is normalizable. However, if the entropy has no maximum value (i.e., the trace functional is not normalizable), then the algebra is called  Type II$_{\infty}$. As mentioned in the introduction, it has recently been argued that Type II algebras are the algebras which arise in semiclassical, ``local'' quantum gravity, see, e.g., \cite{Chandrasekaran:2022cip,Chandrasekaran:2022eqq,Jensen:2023yxy,Kudler-Flam:2023qfl,Penington:2024sum,Kudler-Flam:2024psh,Chen:2024rpx,AliAhmad:2023etg}. One purpose of this paper is to analyze this claim in the case where the (invariantly defined) subregion is the region $\mathscr{R}$. 

\subsection{The Symmetries, Second-Order Charges and Constraints of $\mathscr{R}$}
\label{subsec:symmcharges}
In sec.~\ref{subsec:gravpert} we found the necessary gauge conditions to ensure that the horizons $\mathcal{H}_{\textrm{R}}^{+}$ and $\mathcal{H}^{-}_{\textrm{R}}$ remain the boundary of $\mathscr{R}$ in the perturbed spacetime. The remaining gauge freedom 
of the perturbation $h_{ab}$ is $h_{ab} \to h_{ab}+\pounds_{\chi}g_{ab}$ where $\chi^{a}$ is a smooth class of vector fields which preserve our gauge conditions on each horizon as well as any asymptotic conditions on the metric. In this subsection we will consider the group of diffeomorphisms which preserve $\mathscr{R}$. We will then find the relevant subgroup of diffeomorphisms that impose non-trivial ``constraints'' on the first-order gravitational field as described in the introduction. 
% review the group of diffeomorphisms that preserve $\mathscr{R}$, identify the subgroup of
% relevant symmetries which impose non-trivial ``constraints'' on the 
% ``symmetries'' which act non-trivially on the phase space and obtain the corresponding ``charges'' which generate the action on phase space. 

Clearly, any vector field $\chi^{a}$ which vanishes on the boundary of $\mathscr{R}$ preserves the region. The remaining gauge freedom we must thereby consider are the ``large diffeomorphisms'' --- i.e., those diffeomorphisms that act non-trivially on the boundary of $\mathscr{R}$ but preserve our gauge conditions. In this section, we will focus primarily on the symmetry group associated to diffeomorphisms that are non-trivial on the  horizon since this group is ``universal'' --- it will be common to all spacetimes considered in this paper. 
The remaining group of large diffeomorphisms is the ``asymptotic symmetry group'' which depends on the asymptotic behavior of the spacetime. We comment on the relationship of the large diffeomorphisms and charges considered here to those defined at infinity at the end of sec.~\ref{subsec:fulldressed}. 

The infinitesimal diffeomorphisms that preserve our gauge conditions are all smooth vector fields $\chi^{a}$ which generate asymptotic symmetries at infinity and, on the horizons satisfy \cite{2013CMaPh.321..629H,Chandrasekaran:2018aop,Chandrasekaran:2019ewn} 
\begin{equation}
\label{eq:symm}
\pounds_{n}\chi^{a} \;\hat{=}\; f n^{a} \quad (\textrm{on $\mathcal{H}^{+}$})\quad \quad \quad \pounds_{\ell}\chi^{a} \; \hat{=} \; f \ell^{a} \quad (\textrm{on $\mathcal{H}^{-}$})
\end{equation}
where $f$ is a smooth function on $\mathcal{H}$ that satisfies $n^{a}\nabla_{a}f=0=\ell^{a}\nabla_{a}f=0$. We recall that $\ell^{a}=(\partial/\partial U)^{a}$ and $n^{a}=(\partial/\partial V)^{a}$ are the affinely parameterized null generators of $\mathcal{H}^{+}$ and $\mathcal{H}^{-}$ respectively. To describe the symmetry generators that satisfy \eqref{eq:symm}, it is useful to extend $\ell^{a}$ and $n^{a}$ to all of $\mathcal{H}$ by requiring that the vectors remain null and orthogonal to each other (i.e. $\ell^{a}\ell_{a}=n^{a}n_{a}=0$ and $\ell^{a}n_{a}=-1$). The coordinates $(V,x^{A})$ are then extended off of $\mathcal{H}^{+}$ by holding them constant along the orbits of $\ell^{a}=(\partial/\partial U)^{a}$. Similarly, the coordinates $(U,x^{A})$ are extended off of $\mathcal{H}^{-}$ by holding them constant along the orbits of $n^{a}=(\partial/\partial V)^{a}$. This defines a ``Gaussian null coordinate system'' $(U,V,x^{A})$ in a neighborhood of $\mathcal{H}$. In these coordinates, the diffeomorphisms that satisfy \eqref{eq:symm} are given by \begin{equation}
\label{eq:chi}
\chi^{a} =  f(x^{A})\xi^{a}  + Y^{A}\bigg(\frac{\partial}{\partial x^{A}}\bigg)^{a} + UVq^{AB}\mathscr{D}_{B}f(x^{A}) \bigg(\frac{\partial}{\partial x^{A}}\bigg)^{a} + \dots 
\end{equation}
where $Y^{A}(x^{B})$ is a smooth vector field on $\mathbb{S}^{2}$, $\xi^{a}$ is the horizon Killing field 
\begin{equation}
\xi^{a}\defn \kappa\bigg[U\bigg(\frac{\partial}{\partial U}\bigg)^{a} + V\bigg(\frac{\partial}{\partial V}\bigg)^{a}\bigg],
\end{equation}
and ``$\dots$'' stand for a vector field that vanishes on $\mathcal{H}$ together with its first and second derivatives. The corresponding diffeomorphism group is an infinite dimensional group\footnote{Reference \cite{Chandrasekaran:2018aop} considered the symmetry group of an arbitrary, smooth null surface and obtained a larger group which includes, in addition to diffeomorphisms of the form \eqref{eq:chi}, an infinite dimensional, abelian group of ``affine supertranslations.'' We do not include these elements here since they would move $\mathcal{B}$ and therefore not preserve $\mathscr{R}$. The group obtained in \cite{2013CMaPh.321..629H} is slightly smaller than that generated by \eqref{eq:chi} since the authors imposed an additional (unnecessary) gauge condition --- see footnote \ref{foot:gauge}. Our symmetry group agrees with that of \cite{Chandrasekaran:2019ewn} which obtained the full symmetry group for spacetimes with a preferred ``bifurcation surface.''} whose action on $\mathcal{H}$ is $(U,V,x^{A}) \to (\bar{U},\bar{V},\bar{x}^{A})$ where
\begin{equation}
\bar{U} = e^{\kappa f(x^{A})}U, \quad \bar{V} = e^{-\kappa f(x^{A})}V, \quad \bar{x}^{A} = \bar{x}^{A}(x^{A})
\end{equation}
consisting of arbitrary diffeomorphisms of $\mathbb{S}^{2}$ together with angle-dependent rescalings of the affine parameters of the null generators. These large diffeomorphisms form the group $\textrm{Diff}(\mathbb{S}^{2})\ltimes \mathcal{S}$ where $\mathcal{S}$ is the abelian group of ``boost supertranslations'' generated by \eqref{eq:symm} with $Y^{A}=0$. The diffeomorphisms with $f=0$ and general $Y^{A}$ are referred to as ``superrotations'' which are the generators of $\textrm{Diff}(\mathbb{S}^{2})$.  

At this stage, it may be alarming that the group of ``large diffeomorphisms'' that preserve the subregion is an infinite dimensional group. We first recall that the purpose of this paper is to analyze the ``constraints'' that arise in gravitational perturbation theory. 
As we review below, for the {\em finite} dimensional group of isometries, --- i.e., $f=1$ and $Y^{A}=\psi^{A}$ is an isometry of $(q_{AB},\mathbb{S}^{2})$ --- it was shown in \cite{Chandrasekaran:2022cip,Chandrasekaran:2022eqq,Kudler-Flam:2023qfl} that the isometry group {\em does} generate non-trivial constraints at second-order in perturbation theory on the first-order gravitational perturbations. The purpose of this section is to determine (i.) the full subgroup of diffeomorphisms that yield non-trivial constraints and (ii.) the order in perturbation theory in which these constraints arise. 

The constraints arise from the ``charges'' conjugate to the large diffeomorphisms as well as their fluxes. These charges were obtained for any null boundary in \cite{Chandrasekaran:2018aop}. Therefore, we can define the charges $\mathcal{Q}(f,Y;\lambda)$ on $\mathcal{H}^{-}$ in  $(\mathscr{M},g_{ab}(\lambda))$ for any $\lambda$ and any $f,Y^{A}$. To determine (i.) and (ii.), the strategy will be to consider the ``charge-flux'' relations at every order in perturbation theory and determine which charges impose non-trivial constraints on the first-order gravitational field. 

We first consider the boost supertranslations $\mathcal{Q}_{U}(f;\lambda)\defn \mathcal{Q}(f,0;\lambda)$ which, on any constant $U$ cut $S_{U}$ of $\mathcal{H}^{-}$, are given by
\begin{equation}
\label{eq:Qboost}
\mathcal{Q}_{U}(f;\lambda) =\int_{S_{U}}d^{2}x\sqrt{q(\lambda)}f(x^{A})[1-U\theta(\lambda)].
\end{equation}
 for any $\lambda$. This is the charge ``conjugate'' to the boost supertranslations \cite{Chandrasekaran:2018aop}. By Raychauduri's equation (see \eqref{eq:Raychauduri}) it is straightforward to show that the local flux associated to this charge is 
\begin{equation}
\label{eq:boostsuperchargeflux}
\partial_{U}\mathcal{Q}_{U}(f;\lambda) = \int_{S_{U}}d^{2}x \sqrt{q(\lambda)} f(x^{A})U\bigg[\sigma_{AB}(\lambda)\sigma^{AB}(\lambda)-\frac{1}{2}\theta(\lambda)^{2}\bigg]
\end{equation}
where we recall that capital indices are raised and lowered with $q_{AB}(\lambda)$. The desired charge-flux relation is obtained by integrating the right-hand  side of \eqref{eq:boostsuperchargeflux} over the entire horizon  to obtain \cite{Wald:1999wa,Chandrasekaran:2019ewn} 
\begin{equation}
\label{eq:QRL}
\mathcal{Q}^{\textrm{R}}(f;\lambda) - \mathcal{Q}^{\textrm{L}}(f;\lambda) =-F(f;\lambda).
\end{equation}
Here, $\textrm{R}/\textrm{L}$ denotes the limit $U\to \pm$ and 
\begin{equation}
\label{eq:Flambdaf}
\quad \quad F(f;\lambda) \defn -\int_{\mathcal{H}}dUd^{2}x \sqrt{q(\lambda)} f(x^{A})U\bigg[\sigma_{AB}(\lambda)\sigma^{AB}(\lambda)-\frac{1}{2}\theta(\lambda)^{2}\bigg].
\end{equation}

We now simply evaluate the above relations at each order in perturbation theory. At zero-th order in perturbation theory, the horizon is stationary and so the only non-vanishing boost supertranslation charge is the total area which, of course, is constant
\begin{equation}
\mathcal{Q}_{U}(f;0) = \int_{S_{U}}d^{2}x\sqrt{q} = A_{\mathcal{B}} \quad \quad \partial_{U}\mathcal{Q}_{U}(f;0)  = 0 
\end{equation}
for any $f$ where we have normalized $f$ such that $\int_{\mathbb{S}^{2}}f(x^{A})=1$ and $A_{\mathcal{B}}$ is the area of the bifurcation surface. At first order in perturbation theory we find that, while now the perturbed area element need not vanish\footnote{For instance, one can consider a stationary perturbation to a ``nearby'' Schwarzschild solution. In this case the area and (by the first law) the mass would change.}, it still remains constant
\begin{equation}
\delta \mathcal{Q}_{U}(f) = \delta A_{\mathcal{B}} \quad \quad \partial_{U}\delta \mathcal{Q}_{U}(f)  = 0. 
\end{equation}
Here, $\delta A_{\mathcal{B}}$ is the perturbed area of the bifurcation surface. 

We obtain a non-trivial constraint at second order relating (second order) charges in $\mathscr{R}$ and $\mathscr{L}$ to a flux that depends only on the first-order perturbation. Taking a second-variation of \eqref{eq:Qboost} yields \cite{Hollands:2024vbe}
\begin{equation}
\delta^{2}\mathcal{Q}_{U}(f) = \delta^{2}A_{U}(f)-\int_{S_{U}}d\Omega ~U \delta^{2}\theta
\end{equation}
where $d\Omega \defn  d^{2}x\sqrt{q}$. Taking a second variation of  \eqref{eq:QRL} and \eqref{eq:Flambdaf}, we obtain 
\begin{equation}
\label{eq:QFlux1}
\frac{\delta^{2}\mathcal{Q}^{\textrm{L}}(f)}{4G_{\textrm{N}} \beta} - \frac{\delta^{2}\mathcal{Q}^{\textrm{R}}(f)}{4G_{\textrm{N}} \beta} =- \delta^{2}F(f) \quad \quad \textrm{ where } \delta^{2}F(f) \defn -\frac{1}{4G_{\textrm{N}}\beta}\int_{\mathcal{H}}dUd\Omega~f(x^{A})U\delta \sigma_{AB}\delta \sigma^{AB}.
\end{equation}
We note that $\delta^{2}\mathcal{Q}(f)\sim O(G_{\textrm{N}})$ and so we have explicitley divided by $G_{\textrm{N}}$ so that the resulting charge is finite as $G_{\textrm{N}}\to 0$. A factor of $4\beta$ was also included to simplify  later formulas of black hole entropy. Finally, we have redefined the second-order flux to absorb this factor where we recall that $\delta \sigma_{AB}\sim O(\sqrt{G_{\textrm{N}}})$ so now $\delta^{2}F(f)$ is also finite in the semiclassical limit. 

Importantly, the charges $\delta^{2}\mathcal{Q}^{\textrm{L/R}}$ depend  on the second order perturbed metric. However the flux $\delta^{2}F(f)$ depends entirely on the first-order gravitational field. Thus, while $\delta^{2}\mathcal{Q}^{\textrm{L}}(f)$ and $\delta^{2}\mathcal{Q}^{\textrm{R}}(f)$ are both independently defined, second-order observables in $\mathscr{R}$ and $\mathscr{L}$ respectively, their difference is actually an observable on the first-order gravitational phase space. Indeed, $\delta^{2}F(f)$ is the observable on phase space which generates boost supertranslations of the first-order perturbations 
\begin{equation}
\label{eq:QFlux2}
\delta^{2}F(f) = -\frac{\pi}{G_{\textrm{N}}\beta} \mathscr{W}_{\mathcal{H}}(\gamma,\pounds_{\chi}\gamma) 
%=- \int_{\mathcal{H}}dUd\Omega~f(x^{A})U\delta \sigma_{AB}\delta \sigma^{AB},
\end{equation}
where $\chi^{a} \;\hat{=}\; f \xi^{a}$ is a boost supertranslation. We note that, in the classical theory, the above ``constraint'' is fairly innocuous and is straightforward to satisfy. We will consider the analogous constraint in the quantum theory. As we shall see, satisfying this quantum constraint for the algebra of observables in $\mathscr{R}$ will not be so trivial.

We now consider the charges $\mathcal{Q}(Y)$ associated to a general superrotation 
\begin{equation}
\mathcal{Q}_{U}(Y;\lambda) = \int_{S_{U}}d^{2}x\sqrt{q(\lambda)}\omega_{B}(\lambda)Y^{B}(x^{A})
\end{equation}
and the flux of this charge is simply equal to the right-hand  side of \eqref{eq:Damour}. We will not need the full form of this flux and we will simply present the relevant fluxes at each order in perturbation theory. 
Proceeding as above, we  now identify the relevant set of superrotations that give rise to non-trivial constraints on the first-order gravitational field. 

At zero-th order in perturbation theory the only non-vanishing charges are $\mathcal{Q}_{U}(\psi;0)$ where $Y^{A}=\psi^{A}$ is an isometry of the background metric $q_{AB}$ 
\begin{equation}
\mathscr{D}_{(A}\psi_{B)}=0.
\end{equation}
These charges are simply the angular momentum of the black hole 
\begin{equation}
\frac{\mathcal{Q}_{U}(\psi;0)}{8\pi} \defn J(\psi)
\end{equation}
and are constant on the horizon. If the zero-th order charge is non-vanishing in a black hole spacetime, then the black hole is rotating in the angular direction $\psi^{A}$. We now consider the first order variation $\delta \mathcal{Q}(Y)$ which is generally non-vanishing for any superrotation $Y^{A}$. However, it directly follows from \eqref{eq:linearizedDNS} as well as the fact that $\delta q_{AB}$ vanishes at early and late affine times\footnote{The gauge invariant condition is that the integral $\int dV  \partial_{V}\delta q_{AB}=0$. This is equivalent to the initial data having no ``horizon memory'' \cite{Donnay:2018ckb,Rahman:2019bmk,Danielson:2022tdw,Danielson:2022sga}. In the quantum theory, this condition ensures that quantization of the radiative phase space corresponds to states that lie in the folium of the vacuum state $\omega$ \cite{Ashtekar:1981sf,Strominger:2017zoo,Prabhu:2022zcr,Kudler-Flam:2025pol}.} that the total flux also vanishes at first order in perturbation theory 
\begin{equation}
\delta \mathcal{Q}^{\textrm{L}}(Y) - \delta \mathcal{Q}^{\textrm{R}}(Y) = 0 
\end{equation}
for all $Y^{A}$. We now consider the second-order charges. We first consider, the second-order charges $\delta^{2}\mathcal{Q}(\psi)$ associated to the isometries $\psi^{A}$ of the background which, by varying \eqref{eq:Damour}, satisfies 
\begin{equation}
\label{eq:QFluxX}
\frac{\delta^{2}\mathcal{Q}^{\textrm{L}}(\psi)}{8\pi G_{\textrm{N}}} - \frac{\delta^{2}\mathcal{Q}^{\textrm{R}}(\psi)}{8\pi G_{\textrm{N}}} = -\delta^{2}F(\psi)
\end{equation}
where we have again introduced explicit factors of $G_{\textrm{N}}$ to ensure that the expression is well-defined in the semi-classical limit and we have redefined the flux to include these factors 
\begin{equation}
\label{eq:fluxY}
\delta^{2}F(\psi) \defn \frac{1}{4\pi G_{\textrm{N}}}\int_{\mathcal{H}^{-}} dUd\Omega~\delta \sigma^{AB}\pounds_{\psi}\delta q_{AB} = -\frac{1}{2 G_{\textrm{N}}} \mathscr{W}_{\mathcal{H}}(\gamma,\pounds_{\chi}\gamma) 
\end{equation}
and $\chi^{a}\hat{=}\psi^{A}(\partial/\partial x^{A})^{a}$ on the horizon. Therefore, for the isometries $\psi^{A}$, the second-order charges $\delta^{2}\mathcal{Q}(\psi)$ satisfy an analogous constraint whereby the difference of charges is determined entirely in terms of the radiative data. Furthermore, by the right-hand  side of \eqref{eq:fluxY}, the flux is the observable on phase space which depends locally on the perturbed metric and generates rotations along the isometry directions of the sphere. 

We note that there is no analogous relation for a general superrotation $Y^{A}$ since the corresponding flux is not locally constructed from only the first-order field. By taking a second variation of \eqref{eq:Damour} it is straightforward to show that the flux also depends on terms that are now second-order in the metric such as $\delta^{2}\sigma_{AB}\mathscr{D}^{A}Y^{B}$ as well as $\delta^{2}\theta \mathscr{D}^{A}Y_{A}$. While these terms vanish for $Y^{A}=\psi^{A}$, the flux $\delta^{2}F(Y)$, for general $Y^{A}$, depends on the second-order perturbation\footnote{We note that, for conformal Killing fields on $(q_{AB},\mathbb{S}^{2})$, the term $\delta^{2}\sigma_{AB}\mathscr{D}^{A}Y^{B}$ vanishes since $\delta^{2}\sigma_{AB}$ is trace free. While the term $\delta^{2}\theta$ can be non-locally expressed in terms of the first-order perturbation via Raychauduri's equation \eqref{eq:Raychauduri}, it appears that the corresponding flux cannot be expressed as an observable on phase space --- i.e., it cannot be expressed as $\mathscr{W}(\gamma,\pounds_{\chi}\gamma)$ for some $\chi^{a}$.}.
%\footnote{While the term proportional $\delta^{2}\theta$ can be expressed, via Raychauduri's equation, in terms of the first-order fields, the dependence on the first-order radiation is non-local and so the corresponding flux cannot be locally defined on the first-order phase space and is not a well-defined observable. This is related to the fact that the higher harmonic superrotations are not unitarily implementable in the quantum theory.  }. 
Consequently, the flux is not an observable defined entirely on the phase space of first order fields. Therefore, while the charges $\delta^{2}\mathcal{Q}(Y)$ will, in general, be non-vanishing on the horizon at second-order in perturbation theory, their charge-flux relation does not impose a constraint of the kind relevant for the considerations of this paper. These considerations do indicate, however, that the superrotation charges will will impose a non-trivial constraint on the gravitons at second-order. 

In summary, at the same order in perturbation theory in which one obtains non-trivial constraints for the isometries of the background --- e.g.,  for Schwarzschild these relations are given by eqs.~\eqref{eq:QFlux1} and \eqref{eq:QFluxX} with where $f=1$ ---  we find additional, non-trivial constraints for the $\delta^{2}\mathcal{Q}(f)$ for all $f$. The relevant symmetries that give rise to non-trivial constraints at second order in perturbation theory are 
\begin{equation}
\label{eq:groupG}
G = H_{\textrm{isom}} \ltimes \mathcal{S}
\end{equation}
where $H_{\textrm{isom}}=\textrm{SO}(3),\textrm{U}(1)$ is the rotational isometry group of $\mathscr{R}$. 

For the remainder of this paper we will restrict attention to this subgroup of the full group of large diffeomorphisms. Here, and in the remainder of this paper, we will identify any element of $G$ with their infinitesimal generators $(f,\psi)\in \mathfrak{g}$ in the Lie algebra which define a vector field through \eqref{eq:chi}. In this regard, our analysis is restricted to the identity connected component of the group $G$ with $(f,\psi)$ playing the role of exponential coordinates. 

Since the fluxes are only non-zero at second order in perturbation theory and the investigations of this paper do not go beyond second order, we will henceforth drop the $\delta$'s on the fluxes and simply refer to them as $F(f)$ and $F(\psi)$ respectively. We will not be adopting the same notation for the charges since they are generically non-vanishing at zeroth order. We conclude this subsection by noting that since the fluxes $F(f)$ and $F(\psi)$ are well-defined observables on the first-order gravitational phase space, they generate rotations and boost supertranslations on the first-order observables. Namely, they generate rotations and boost supertranslations on the gravitational phase space via the brackets 
\begin{equation}
\label{eq:Fsigms}
\{F(f),\delta \sigma(s)\} =  8\pi\delta \sigma(s^{\prime}) \quad  \quad \quad \{F(\psi),\delta \sigma(s)\} = 8\pi \delta \sigma(s^{\prime \prime}). 
\end{equation}
where 
\begin{equation}
\label{eq:sprime}
s_{AB}^{\prime} = -Uf(x^{A})\partial_{U}s_{AB}, \quad \quad s^{\prime \prime} = -\pounds_{\psi}s
\end{equation}
The total flux $F(f,\psi)=F(f)+F(\psi)$ associated to a general symmetry $(f,\psi)$ is simply the sum of the individual fluxes and their Poisson brackets are induced by the Lie brackets of $\mathfrak{g}$. If we consider two symmetries $\chi_{1}=(f_{1},\psi_{1})$ and $\chi_{2}=(f_{2},\psi_{2})$ then the Lie bracket of their associated vector fields define the following algebra:
\begin{align}
\label{eq:fX}
&[(f_{1},\psi_{1}),(f_{2},Y_{2})]= (f,\psi)  \\
&f\defn f_{1}^{A}\mathscr{D}_{A}f_{2} - \psi_{2}^{A}\mathscr{D}_{A}f_{1}  \\
&\psi^{A}\defn [\psi_{1},\psi_{2}]^{A}=\psi_{1}^{B}\mathscr{D}_{B}\psi_{2}^{A}-\psi_{2}^{B}\mathscr{D}_{B}\psi_{1}^{A}.
\end{align}

%Physically, the fact that all of these charges are non-trivial on the horizon reflects that, at second order in perturbation theory, the boundary of $\mathscr{R}$ is perturbed from a Killing horizon --- with a finite dimensional symmetry group --- to a general, smooth, bifurcate null surface whose perturbed area element is no longer spherically symmetric and whose perturbed metric $q_{AB}$ on cross-sections is an arbitrary metric on $\mathbb{S}^{2}$. The corresponding symmetry group that preserves this (much weaker) structure is $G=\textrm{Diff}(\mathbb{S}^{2})\ltimes \mathcal{S}$ . 
%\GSnote{[Include a discussion of the asymptotic symmetry group or maybe we just wait until we discuss specific cases?]}\JKFc{I think better to wait to specific cases.}
\subsection{Symmetries and Unitary Generators of the Quantization Algebra}
\label{subsec:symmquantalg}
We now comment on the inclusion of the generators of the group $G =H_{\textrm{isom}} \ltimes \mathcal{S}$ into the quantization algebra of the quantum theory. The group $G$ generates automorphisms of the algebra $\mathscr{A}(\mathscr{R})$. On the Hilbert space $\mathscr{F}$, the action of these automorphisms is represented by unitary operators whose infinitesimal generators we denote as $\op{F}(f)$ and $\op{F}(\psi)$. In the quantum theory, these operators are densely defined self-adjoint operators whose commutation relations with smeared fields and with themselves quantize the brackets \eqref{eq:fX} and \eqref{eq:Fsigms}. The vacuum state $\omega$ is invariant under $G$ and satisfies
\begin{equation}
\op{F}(f) \ket{\omega} = 0, \quad \quad \op{F}(\psi)\ket{\omega} = 0
\end{equation}
for any $f$ and $\psi$. We note that while they are well-defined operators on the Hilbert space, these operators are not elements of the algebra $\mathfrak{A}(\mathscr{R})$ since they generate a non-trivial transformation of operators in $\mathscr{L}$. Indeed, it is actually impossible to ``split'' $\op{F}(f)$ or $\op{F}(\psi)$ into an operator in $\mathfrak{A}(\mathscr{R})$ and an operator in its commutant $\mathfrak{A}(\mathscr{L})$. Such a ``half-sided'' boost supertranslation or rotation would generate singularities at the bifurcation surface and thereby yield infinite fluctuations in any physical quantum state. The unitary operators $e^{i\op{F}(f)}$ and $e^{i\op{F}(\psi)}$ generate {\em outer} automorphisms of $\mathfrak{A}(\mathscr{R})$. 

\par The failure of this splitting for $f=1$ is directly related to the fact that $\mathfrak{A}(\mathscr{R})$ is a Type III$_{1}$ von Neumann algebra. The thermality of $\omega$ with respect to Killing time translations (see eqs.~\eqref{eq:2ptthermal} - \eqref{eq:KMS}) implies that the modular Hamiltonian $\op{K}_{\omega}$ associated to $\omega$ is given by 
\begin{equation}
\label{eq:modHam}
\op{K}_{\omega} = \beta \op{F}(1) \quad \quad \quad \quad  \quad \quad \quad \quad \textrm{(on $\mathcal{H}^{-}$)}.
\end{equation}
%where we recall that the modular operator is $\op{\Delta}_{\omega}=e^{-\op{K}_{\omega}}$. 
The operator $\op{F}(1)$ generates the boost isometry on all of $\mathscr{R}$ and $\mathscr{L}$. However, we emphasize that, in general, \eqref{eq:modHam} is only guaranteed to hold on $\mathcal{H}_{\textrm{R}}^{-}$. If $\omega$ is the Hartle-Hawking state then \eqref{eq:modHam} will hold on all of $\mathscr{R}$. On the other hand, if $\omega$ is the Unruh vacuum then it is thermal only on $\mathcal{H}_{\textrm{R}}^{-}$ and the modular operator will only be geometric on the past horizon \cite{Kudler-Flam:2023qfl}. In either case, a signature property of Type III algebras is the failure of splitting the modular Hamiltonian into an operator affiliated with $\mathfrak{A}(\mathscr{R})$ and an operator affiliated with the commutant $\mathfrak{A}(\mathscr{L})$ \cite{Takesaki2003}.

As we have mentioned in the previous subsection, there exists a larger group, $\text{Diff}(\mathbb{S}^2) \ltimes \mathcal{S}$, which preserves the gravitational phase space and acts, in the quantum theory, as automorphisms of the algebra $\mathscr{A}(\mathscr{R})$. As explained there, the flux associated with a general superrotation charge cannot be written as a local integral of the first-order gravitational field. Consequently, we were not forced to include the full superrotation group among the constraints imposed on the first-order theory. Even if one attempts to incorporate superrotations as bona fide operators in the quantum theory, it is straightforward to show that the action of a generic superrotation $Y^{A}$ is not unitarily implementable on the Hilbert space\footnote{It is straightforward to show that, for a dense set of states $\psi\in \mathscr{F}$, the difference $\braket{\delta \sigma_{AB}(x_{1})\delta \sigma_{CD}(x_{1})}_{\psi} - \braket{\delta \sigma_{AB}(x_{1})\delta \sigma_{CD}(x_{1})}_{\omega} = S_{ABCD}(x_{1},x_{2})$ is a smooth function of $x_{1},x_{2}\in \mathcal{H}$. Since the action of a general superrotation on $\ket{\psi}$ does not preserve the graviton vacuum, this difference is no longer smooth and it can be shown that the resulting state does not lie in the folium of the vacuum.}. The arguments of the previous section nevertheless indicate that the superrotation charges give rise to nontrivial constraints relating first- and second-order gravitational perturbations, which must be taken into account in a complete second-order treatment. Whether superrotations continue to be “spontaneously broken’’ in the quantization of the second-order graviton field is a question that we leave for future work.\footnote{We note that it is possible to gauge symmetry actions which are not unitarily implementable provided one works from a purely algebraic perspective e.g. without prioritizing a preferred Hilbert space. We refer the reader to \cite{Klinger:2025tvg} for discussion of such a construction, including some additional comments pertaining to superrotations in perturbative quantum gravity.}

\subsection{The Complete Dressed Algebra of $\mathscr{R}$}
\label{subsec:dressedalg}
In this section we complete the gravitational algebra of observables in the region $\mathscr{R}$ by quantizing the charges $\delta^2\mathcal{Q}^L(f,\psi) \mapsto \delta^2 \op{\mathcal{Q}}^{L}(f,\psi)$ and $\delta^2 \mathcal{Q}^{R}(f,\psi) \mapsto \delta^2 \op{\mathcal{Q}}^{R}(f,\psi)$, and imposing the charge-flux constraints \eqref{eq:QFlux1} and \eqref{eq:QFluxX}. How should these charges be included into the algebra? Physically, $\delta^{2}\op{\mathcal{Q}}^{\textrm{R}}(f,\psi)$ generates boost supertranslations and superrotations of the observables in $\mathscr{R}$ and commute with all observables in $\mathscr{L}$. The charges $\delta^{2}\op{\mathcal{Q}}^{\textrm{L}}(f,\psi)$ have an analogous, non-trivial action but on observables in $\mathscr{L}$ and commute with observables in $\mathscr{R}$. 

Before providing the quantization of these charges, it is instructive to see what goes wrong if we try to implement the above considerations purely within the ``undressed'' algebras $\mathfrak{A}(\mathscr{R})$ and $\mathfrak{A}(\mathscr{L})$. Recall that the charges satisfy 
\begin{equation}
\label{eq:Gconstraints}
\frac{\delta^{2}\op{\mathcal{Q}}^{\textrm{L}}(f)}{4\beta G_{\textrm{N}}} - \frac{\delta^{2}\op{\mathcal{Q}}^{\textrm{R}}(f)}{4\beta G_{\textrm{N}}} = -\op{F}(f)\quad \quad \frac{\delta^{2}\op{\mathcal{Q}}^{\textrm{L}}(\psi)}{8\pi G_{\textrm{N}}} - \frac{\delta^{2}\op{\mathcal{Q}}^{\textrm{R}}(\psi)}{8\pi G_{\textrm{N}}} =- \op{F}(\psi). 
\end{equation}
If, for example, it were possible to ``split'' the flux into an operator affiliated with $\mathfrak{A}(\mathscr{R})$ and an operator affiliated with the commutant $\mathfrak{A}(\mathscr{L})$ then we could directly define $\delta^{2}\op{\mathcal{Q}}^{\textrm{R}}(f)$ and $\delta^{2}\op{\mathcal{Q}}^{\textrm{L}}(f)$  as generating ``half-sided'' boost supertranslations of $\mathscr{R}$ on the graviton Hilbert space $\mathscr{F}$. However, as emphasized in the previous subsection, this splitting is not possible on the algebra $\mathfrak{A}(\mathscr{R})$. Instead, to achieve the above constraints we must construct a new algebra $\mathfrak{A}_{\textrm{dress.}}(\mathscr{R};G)$ which include gravitational charges as independent fluctuating operators and in which QFT observables are ``dressed'' so as to satisfy the charge-flux constraint.

Indeed, in contrast to the fluxes, the charges are built from the second-order gravitation field and therefore are degrees of freedom which we have not yet quantized. The charges form an algebra $\mathscr{A}_{\textrm{Q}}$ which is the $\ast$-algebra generated by $\delta^{2}\op{\mathcal{Q}}^{\textrm{R}}(f,\psi)$ and $\delta^{2}\op{\mathcal{Q}}^{\textrm{L}}(f,\psi)$ for all smooth $f,Y^{A}$ with commutation relations 
\begin{flalign}
&[\delta^{2}\op{\mathcal{Q}}^{\textrm{L}}(f_1,\psi_1), \delta^{2}\op{\mathcal{Q}}^{\textrm{R}}(f_2,\psi_2)] = 0, \nonumber \\
&[\delta^2 \op{\mathcal{Q}}^{L/R}(f_1,\psi_1),\delta^2 \op{\mathcal{Q}}^{L/R}(f_2,\psi_2)] = \delta^2\op{\mathcal{Q}}^{L/R}([(f_1,\psi_1),(f_2,\psi_2)]),
\end{flalign}
where the Lie brackets are given by \eqref{eq:fX}. As we review in sec.~\ref{sec:isom}, if the group $G$ is locally compact --- as would be the case if $G$ were merely the isometry group of $(\mathscr{R},g)$ --- then there exists a canonical Hilbert space $\mathscr{H}_{\textrm{G}}=L^{2}(G,d\mu)$ on which the algebra $\mathscr{A}_Q$ naturally acts. This Hilbert space consists of square-integrable, complex functions $\Psi$ on $G$ with 
respect to the unique, invariant Haar measure $\mu_{\textrm{G}}$ with inner product 
\begin{equation}
\label{eq:innerprodG}
\braket{\Psi_{1}|\Psi_{2}} = \int_{G} d\mu_{G}(g) \overline{\Psi_{1}(g)}\Psi_{2}(g).
\end{equation}
On this Hilbert space $\delta^{2}\op{\mathcal{Q}}^{\textrm{R/L}}(g)$ generate ``left'' and ``right'' translations of $\Psi$.\footnote{To be precise, one of the two charges must also be conjugated by the unitary operator $e^{i\op{F}(\op{g})}$ described in \eqref{V unitary} to ensure that the constraint is well-posed. In particular, we take
\beq
    \big(e^{i\delta^2 \op{\mathcal{Q}}^{\textrm{L}}(g)} \hat{\Psi}\big)(h) = e^{iF(g)}\big(\hat{\Psi}(hg)\big), \qquad \big(e^{i\delta^2 \op{\mathcal{Q}}^{\textrm{R}}(g)} \hat{\Psi}\big)(h) = \hat{\Psi}(g^{-1}h).  
\eeq
}

However, the groups of interest in this paper are infinite dimensional. In this case there does not exist an invariant Haar measure. Thus, in the infinite dimensional case, one must take care to construct measures $\mu_{\textrm{G}}$ such that $\mathscr{H}_{\textrm{G}}$ with inner product (\ref{eq:innerprodG}) admits an action of all charges of physical interest.  For example, if we take the simplest case where $G=\mathcal{S}$, then we require that, at the very least, $\mathscr{H}_{\textrm{G}}$ admits a well-defined action of $\delta^{2}\op{\mathcal{Q}}^{\textrm{R}}(f)$ for all smooth $f$. The construction of a suitable measure and Hilbert space is addressed in sec.~\ref{sec:fulldressedalg}.

In the remainder of this section, we will assume
that we have a suitable Hilbert space representation $\mathscr{H}_{\textrm{G}}$ of $\mathscr{A}_{\textrm{Q}}$ and so the full Hilbert space of gravitons together with the fluctuations of the charges is 
\begin{equation}
\mathscr{H}_{\textrm{ext.}}\defn \mathscr{F}\otimes \mathscr{H}_{\textrm{G}}.
\end{equation}
We note that the Hilbert space $\mathscr{H}_{\textrm{G}}$ also admits the action of a ``multiplication operator''. If $T$ is an $L^{\infty}$ function on the group, we define the operator $T(\op{g})$ which acts as 
\beq
    \big(T(\op{g}) \Psi\big)(h) = T(h) \Psi(h).
\eeq
The commutation relation between the charges and the multiplication operator generalize the usual Heisenberg relations:
\beq
    [\delta^2\op{\mathcal{Q}}^R(h),T(\op{g})] = i \pounds_{V_h} T(\op{g}),
\eeq
where here $V_h$ is the vector field generating the group element $h$ in the group manifold.\footnote{The notation $\pounds_{V_h} T(\op{g})$ signifies the multiplication operator obtained from the function which is the derivative of $f$ in the direction of $h$.}

With these notions in place we can now construct the gravitationally dressed algebra for the region $\mathscr{R}$. This algebra, $\mathscr{A}_{\textrm{dress.}}(\mathscr{R},G)$, is the unique subalgebra of $\mathscr{A}(\mathscr{R})\otimes \mathscr{A}_{\textrm{Q}}$ which commutes with $\delta^{2}\op{\mathcal{Q}}_{\textrm{L}}(f,\psi)$ and for which $\delta^{2}\op{\mathcal{Q}}_{\textrm{R}}(f,\psi)$ generates a ``half-sided'' boost supertranslation and ``half-sided'' rotations. $\mathscr{A}_{\textrm{dress.}}(\mathscr{R},G)$ is the algebra generated by $\delta^{2}\op{\mathcal{Q}}^{\textrm{R}}(f,\psi)$ together with the ``dressed observables''
\begin{equation}
\delta\op{\sigma}(s;\op{f},\op{\psi})\defn e^{-i\op{F}(\op{f},\op{\psi})}\delta\op{\sigma}(s)e^{i\op{F}(\op{f},\op{\psi})}.
\end{equation}
Here, $\delta\op{\sigma}(s)$ is the graviton operator on $\mathscr{A}(\mathcal{H}_{\textrm{R}}^{-})$ and $e^{i\op{F}(\op{f},\op{\psi})}$ is a unitary operator on $\mathscr{H}_{\textrm{ext.}}$ in which $\op{F}$ is interpreted as a function on the group $G$ which takes values in unitary operators on the Hilbert space $\mathscr{F}$. If we treat a state in $\mathscr{H}_{\textrm{ext.}}$ as a function, $\hat{\Psi}$, from $G$ into the Hilbert space $\mathscr{F}$ we can write
\beq \label{V unitary}
    \big(e^{i\op{F}(\op{f},\op{\psi})} \hat{\Psi}\big)(f,\psi) \defn e^{i\op{F}(f,\psi)}\big(\hat{\Psi}(f,\psi)\big). 
\eeq
In the last equality, each $\hat{\Psi}(f,\psi)$ is an element of $\mathscr{F}$ for each $f,\psi$  and $e^{i\op{F}(f,\psi)}$ acts on these vectors. By construction, $\delta^{2}\op{\mathcal{Q}}^{\textrm{R}}(f,\psi)$ generates boost supertranslations and rotations on the dressed observables in $\mathscr{R}$: 
\begin{equation}
[\delta^{2}\op{\mathcal{Q}}^{\textrm{R}}(f),\delta\op{\sigma}(s;\op{f},\op{\psi})] = 8\pi i\delta \op{\sigma}(s^{\prime};\op{f},\op{\psi}) \quad \quad [\delta^{2}\op{\mathcal{Q}}^{\textrm{R}}(\psi),\delta\op{\sigma}(s;\op{f},\op{\psi})] = 8\pi i\delta \op{\sigma}(s^{\prime\prime};\op{f},\op{\psi}).
\end{equation}
where $s^{\prime}_{AB}$ and $s^{\prime \prime}_{AB}$ are the action of boost supertranslations and rotations of the test functions $s_{AB}$ respectively (see \eqref{eq:sprime}). 
Evolving this data into the bulk $\delta\op{\sigma}(s)=-4\pi \op{\gamma}(w)$ where $s_{ab}=(Ew)_{ab}$ yields the dressed bulk observable
\begin{equation}
\delta \op{\sigma}(s;\op{f},\op{\psi}) = e^{-i\op{F}(\op{f},\op{\psi})}\op{\gamma}(w)e^{i\op{F}(\op{f},\op{\psi})} \defn \op{\gamma}(w;\op{f},\op{\psi}).
\end{equation}
The von Neumann algebra of observables in $\mathscr{R}$ is 
\begin{equation} \label{Full Dressed Algebra}
\mathfrak{A}_{\textrm{dress.}}(\mathscr{R};G) \defn \{\op{\gamma}(w;\op{f},\op{\psi}), \delta^{2}\op{\mathcal{Q}}^{\textrm{R}}(f,\psi)\}''
\end{equation}
where the commutant is taken in $\mathscr{F}\otimes \mathscr{H}_{\textrm{G}}$. In the language of von Neumann algebras, $\mathfrak{A}_{\textrm{dress.}}(\mathscr{R};G)$ is known as the ``crossed product'' of $\mathfrak{A}(\mathscr{R})$ with respect to the group $G=H_{\textrm{isom.}}\ltimes \mathcal{S}$. The algebra $\mathfrak{A}_{\textrm{dress.}}(\mathscr{R};G)$ represents the complete algebra of observables in $\mathscr{R}$. The expectation value of any observable can, in principle, be computed in any state. Thus, $\mathfrak{A}_{\textrm{dress.}}(\mathscr{R};G)$ provides a fully satisfactory, well-defined theory of quantum physics \emph{with} gravity satisfying the second-order constraints. 

\subsection{The Quantization of Subgroups of $G$ and its Relation to the Full Algebra} \label{sec: subgroups}

We conclude this section by proving that it is {\em not} consistent to restrict the quantization to any finite dimensional subgroup $G_{\textrm{sub}} \subset G$. To consistently restrict to any subgroup of $G$, one must consider the subspace of perturbations where the remaining charges do not fluctuate. In order to satisfy the constraints, this implies that one must restrict to the subspace of quantum field states which are invariant under the remaining symmetries. In other words, $\mathfrak{A}_{\textrm{dress.}}(\mathscr{R};G_{\textrm{sub.}})$ is a non-trivial subalgebra of $\mathfrak{A}_{\textrm{dress.}}(\mathscr{R};G)$ only if there exists a non-trivial subspace of states in $\mathscr{F}$ invariant under $G/G_{\textrm{sub.}}$. 

For example, in \cite{Chandrasekaran:2022eqq,Chandrasekaran:2022cip,Kudler-Flam:2023qfl}, the full symmetry group was assumed to be $G_{\textrm{isom.}}=\mathbb{R}\times H_{\textrm{isom.}}$. Then, for perturbations of a Schwarzschild black hole, $\mathfrak{A}_{\textrm{dress.}}(\mathscr{R};\mathbb{R})$ is a consistent subalgebra of the isometry-invariant algebra $\mathfrak{A}_{\textrm{dress.}}(\mathscr{R};\mathbb{R}\times \textrm{SO}(3))$ since one may restrict to the subspace of spherically symmetric states\footnote{For gravitons this subspace is trivial {because it is a spin-2 field} and to obtain a non-trivial subspace one must include additional matter perturbations.} in $\mathscr{F}$. As we review in sec.~\ref{sec:isom} this restriction allowed the authors of  \cite{Chandrasekaran:2022eqq,Chandrasekaran:2022cip,Kudler-Flam:2023qfl} to directly conclude, by a theorem of Takesaki \cite{takesaki1973duality}, that the algebra $\mathfrak{A}_{\textrm{dress.}}(\mathscr{R};\mathbb{R})$ is of  Type II. 

Unfortunately, as we have emphasized, the full symmetry group is not $G_{\textrm{isom.}}$ but the substantially larger group $G=H_{\textrm{isom}}\ltimes \mathcal{S}$. Even if one ``freezes'' the rotation charges $\delta^{2}\op{\mathcal{Q}}(\psi)$  by considering only spherically symmetric or axisymmetric perturbations, there remains the full set of boost supertranslation charges $\delta^2\op{\mathcal{Q}}(f)$ for all smooth, non-constant $f$. If one wanted to restrict to the algebra $\mathfrak{A}_{\textrm{dress.}}(\mathscr{R};\mathbb{R})$, for instance, one would have to restrict to the subspace of $\mathscr{F}$ invariant under the $\mathcal{S}/\mathbb{R}$. However, the following theorem indicates that this subspace is actually trivial. 
\begin{theorem}
\label{eq:thm1}
Let $f$ be any smooth function on $\mathbb{S}^{2}$ whose support is all of $\mathbb{S}^{2}$, i.e., $f$ does not vanish identically on any open subset of $\mathbb{S}^{2}$. Suppose that the state $\Psi$ is a normalizable state in $\mathscr{F}$ invariant under the action of the boost supertranslation $u\to u+f$. Then $\Psi=\omega$, where $\omega$ is the vacuum state invariant under the full symmetry group (\ref{eq:groupG}).  
\end{theorem}
The proof of this theorem follows from a related theorem regarding the uniqueness of the (supertranslation invariant) vacuum state at null infinity \cite{Prabhu:2022zcr}. The argument of that proof uses the fact that any normalizable state $\mathscr{F}$ must decay as $u\to -\infty$. However, this decay is fundamentally incompatible with the correlation functions of the state being boost supertranslation invariant. More directly, following \cite{Prabhu:2024lmg}, it is straightforward to construct the (improper) eigenstates of $\op{F}(f)$ in $\mathscr{F}$. All such eigenstates, except the vacuum, are ``plane wave states'' which are labeled by a null momentum $p^{\mu}=\omega(1,x^{A}_{p})$ where $\omega$ is a frequency and $x^{A}_{p}$ is a point on $\mathbb{S}^{2}$. Such states on the horizon oscillate as $e^{i\omega u}\delta_{\mathbb{S}^{2}}(x^{A},x^{A}_{p})$  and are therefore non-normalizable. 
Theorem \ref{eq:thm1} tells us that we must include the full infinite-dimensional group of boost supertranslations to obtain a nontrivial algebra. We must contend with the constraints arising from the full, infinite dimensional group $G$. The remainder of this paper will be dedicated to understanding the properties of $\mathfrak{A}_{\textrm{dress.}}(\mathscr{R};G_{\textrm{sub}})$.

Of course, this does not imply that studying the algebras $\mathfrak{A}_{\textrm{dress.}}(\mathscr{R};G_{\textrm{sub}})$ is not extremely useful in gaining intuition into the structure and properties of the fundamental algebra $\mathfrak{A}_{\textrm{dress.}}(\mathscr{R};G)$. We will investigate the ``type'' of $\mathfrak{A}_{\textrm{dress.}}(\mathscr{R};G)$ by ``building up'' to the full, infinite dimensional group $G$. We will do so by picking a subgroup $G_{\textrm{sub.}}\subset G$ and simply ignoring all constraints except for those associated to $G_{\textrm{sub.}}$.
Per Theorem \ref{eq:thm1}, the algebras $\mathfrak{A}_{\textrm{dress.}}(\mathscr{R};G_{\textrm{sub.}})$ are {\em not} subalgebras of $\mathfrak{A}_{\textrm{dress.}}(\mathscr{R};G)$ so one cannot make any direct conclusions about the full algebra from the quantization of any subgroup. Nevertheless, as we will see, the quantization of these subgroups will be very useful in analyzing some universal aspects of the full algebra.

\section{The Isometry Invariant Subregion Algebra}
\label{sec:isom}
In this section we will consider the algebra of observables invariant under just the isometries of $(\mathscr{R},g)$. 
For nonextremal, stationary black holes, the rigidity theorem implies that the relevant isometry groups $G_{\textrm{isom}}$ in four dimensions\footnote{Our analysis can be straightforwardly extended to black holes in higher dimensions. The rigidity theorem in higher dimensions implies that, for rotating black holes, the isometry group is  $\mathbb{R}\times \textrm{U}(1)^{\textrm{N}}$ where $1\leq N\leq \lceil\frac{d-1}{2}\rceil$ \cite{Hollands:2006rj}.} are $\mathbb{R}\times \textrm{SO}(3)$ if the black hole is static or $\mathbb{R}\times U(1)$ for rotating black holes \cite{Hawking:1973uf,Friedrich:1998wq,Alexakis:2009gi}. In  sec.~\ref{sec:R} we first ignore the compact isometries and review the arguments of \cite{Chandrasekaran:2022cip,Chandrasekaran:2022eqq,Kudler-Flam:2023qfl} that  $\mathfrak{A}_{\textrm{dress.}}(\mathscr{R};\mathbb{R})$ is a Type II algebra. In the remainder of this section we explain how to generalize these arguments to include the full isometry group. To cover both cases we will, in section \ref{sec:AdressH}, construct an analogously defined algebra --- that we denote as  $\mathfrak{A}_{\textrm{dress.}}(\mathscr{R};\mathbb{R}\times H)$ --- invariant under the group $ \mathbb{R}\times H$ where $H$ is any compact group. In sec.~\ref{sec:trace}, we review the arguments  of \cite{AliAhmad:2024eun} that this algebra is Type II and provide a simple proof that it admits a densely defined, ``universal'' trace.  In sec.~\ref{subsec:densityops}, we obtain the corresponding density matrices and entropy of any semiclassical state on this algebra. While our arguments are valid for any region $\mathscr{R}$ bounded by Killing horizons, to illustrate our result we apply our analysis to the case where $\mathscr{R}$ is the exterior of a Kerr black hole in sec.~\ref{subsec:kerr} and the static patch of de Sitter space in sec.~\ref{subsec:deSitter}

\subsection{$\mathfrak{A}_{dress.}(\mathscr{R};\mathbb{R})$ is a Type II algebra}
\label{sec:R}
The case of $G_{\textrm{sub.}}=\mathbb{R}$ was considered in \cite{Chandrasekaran:2022eqq,Chandrasekaran:2022cip,Kudler-Flam:2023qfl} and we now briefly review the arguments and results presented in these references. At the end of this subsection we will show that ignoring all of the constraints except that of $(t,0)$ where $t$ is a constant, yields 
\begin{equation}
\mathfrak{A}_{\textrm{dress.}}(\mathscr{R};\mathbb{R}) \defn \{\op{\gamma}(w;\op{t}), \delta^{2}\op{\mathcal{Q}}^{\textrm{R}}(1)\}''
\end{equation}where $\delta^{2}\op{\mathcal{Q}}^{\textrm{R}}(1)=\delta^{2}\op{M}_{\textrm{R}}$ is equivalent to the perturbed ADM mass of the right wedge and the dressed operator is simply
\begin{equation}
\op{\gamma}(w;\op{t}) = e^{-i\op{F}(\op{t})}\op{\gamma}(w)e^{i\op{F}(\op{t})}
\end{equation}
on the Hilbert space $\mathscr{F}\otimes L^{2}(\mathbb{R})$. The structure and properties of $\mathfrak{A}_{\textrm{dress.}}(\mathscr{R};\mathbb{R})$ was analyzed in the case where $\mathscr{R}$ is the exterior of a Schwarzschild-AdS black hole \cite{Chandrasekaran:2022eqq}, the static patch of de Sitter \cite{Chandrasekaran:2022cip} and, subsequently, for any subregion $\mathscr{R}$ bounded by one or more Killing horizons (e.g., any black hole in asymptotically flat or (Anti-)de Sitter spacetimes) \cite{Kudler-Flam:2023qfl}. Remarkably, it was shown in these references that the algebra $\mathfrak{A}_{\textrm{dress.}}(\mathscr{R};\mathbb{R})$ is actually Type II and therefore admits a (densely defined) trace functional. Since it will play a distinguished role in the formulas for the trace and entropy it will be convenient, for brevity, to define 
\begin{equation}
\op{X} \defn \frac{\delta^{2}\op{\mathcal{Q}}^{\textrm{R}}(1)}{4G_{\textrm{N}}\beta}.
\end{equation}
We further note that any operator $\hat{a}\in \mathfrak{A}_{\textrm{dress.}}(\mathscr{R};\mathbb{R})$ can be expressed as a $\mathfrak{A}(\mathscr{R})$-valued function of $\op{X}$ --- i.e., $\hat{a} = a(\op{X})$ for any $\hat{a}\in \mathfrak{A}_{\textrm{dress.}}(\mathscr{R};\mathbb{R})$. If we define the (improper) eigenstates of $\op{X}$ as 
\begin{equation}
\op{X}\ket{X}=X\ket{X}
\end{equation}
then the trace on the algebra is given by 
\begin{equation}
\label{eq:TraceX}
\tau_{\mathbb{R}}(\hat{a}) =\braket{0_{\mathbb{R}}|e^{\op{X}/2}\hat{a}e^{\op{X}/2}|0_{\mathbb{R}}}= \int_{\mathbb{R}^{2}} \frac{dXdX'}{2\pi}~e^{X}\braket{\omega,X|\hat{a}|\omega,X'} \quad \quad \textrm{ for any $\hat{a}\in \mathfrak{A}_{\textrm{dress.}}(\mathscr{R};\mathbb{R})$}
\end{equation}
where $\ket{0_{\mathbb{R}}}$ is the improper eigenstate of $\op{t}$ with eigenvalue $0$ which corresponds to the ``neutral element'' of the group. Eq.~\eqref{eq:TraceX} can be shown to satisfy the properties of a trace --- i.e. $\tau_{\mathbb{R}}$ is densely defined and satisfies $\tau_{\mathbb{R}}(\hat{a}\hat{b})=\tau_{\mathbb{R}}(\hat{b}\hat{a})$ \cite{Witten:2021unn}. We note that the trace is not unique since any constant multiple of (\ref{eq:TraceX}) is also a well-defined trace on the algebra. For any state $\hat{\Phi}\in \mathscr{F}\otimes L^{2}(\mathbb{R})$ one can now define a density matrix on the algebra $\mathfrak{A}_{\textrm{dress.}}(\mathscr{R};\mathbb{R})$ defined by
\begin{equation}
\tau_{\mathbb{R}}(\rho_{\hat{\Phi}}\hat{a}) = \braket{\hat{\Phi}|\hat{a}|\hat{\Phi}}\quad \quad \textrm{ for any $\hat{a}\in \mathfrak{A}_{\textrm{dress.}}(\mathscr{R};\mathbb{R})$.}
\end{equation}
Similarly, this density matrix inherits an ambiguity of an (inverse) multiplicative constant which corresponds to an additive constant ambiguity in the entropy $S_{\textrm{vN}}(\rho_{\hat{\Phi}})$ due to the logarithm. 

In de Sitter spacetime, the spacetime region $\mathscr{R}$ is defined relative to an observer and, as we review in subsection \ref{subsec:deSitter}, the perturbed area of the horizon is bounded from above by the observer's energy. Since the observer has positive energy, this means that the relevant algebra is actually $\op{P}_{X<0}\mathfrak{A}_{\textrm{dress.}}(\mathscr{R};\mathbb{R})\op{P}_{X<0}$ where $\op{P}_{X<0}$ projects the algebra to negative values of $X$. In this case the integral in \eqref{eq:TraceX} is bounded from below and the trace can be  canonically normalized so that $\tau(\op{1})=1$. As such there exists maximum entropy state and the algebra is Type II$_{1}$ \cite{Chandrasekaran:2022cip}. However, in any spacetime with a black hole, the perturbed area of the black hole is not bounded from above (or below). In this case, $\tau(\op{1})=\infty$ and cannot be canonically normalized. The entropy is unbounded and the algebra is Type II$_{\infty}$ \cite{Chandrasekaran:2022eqq,Kudler-Flam:2023qfl}. For ``semi-classical'' states  $\ket{\hat{\Phi}_{\alpha}}=\ket{\varphi}\otimes \alpha(X)$ where $ \alpha(X)$ is slowly varying in $X$ --- so the conjugate time is sharply peaked --- it was shown that\footnote{The exact density matrix in the slowly varying limit was obtained in \cite{Jensen:2023yxy}. A rigorous derivation of $\log(\rho_{\hat{\Psi}_{\alpha}})$ in this limit as well the control of its associated errors was obtained in \cite{Kudler-Flam:2023hkl}.}
\begin{equation}
\label{eq:SvnR}
S_{\textrm{vN.}}(\rho_{\hat{\Phi}_{\alpha}}) \simeq \braket{\hat{\Phi}_{\alpha}|\beta \op{X}|\hat{\Phi}_{\alpha}} - S_{\textrm{rel.}}(\varphi|\omega) - S_{\textrm{vN.}}(\rho_{\alpha}) + \log(\beta) 
\end{equation}
where $S_{\textrm{rel.}}(\varphi|\omega)$ is the relative entropy $\rho_{\alpha}\defn |\alpha({X})|^{2}$ is the probability distribution on $L^{2}(\mathbb{R})$ and $S_{\textrm{vN.}}(\rho_{\alpha})$ is the corresponding Shannon entropy. Using an argument of Wall to relate the relative entropy to change in the generalized entropy \cite{Wall:2011hj}, one can see that the $S_{\textrm{vN.}}(\rho_{\hat{\Phi}_{\alpha}})$ is given by
\begin{equation}
\label{eq:vNSgenR}
S_{\textrm{vN.}}(\rho_{\hat{\Phi}_{\alpha}}) \simeq  S_{\textrm{gen.}}(\hat{\Phi}_{\alpha}\vert_{\mathscr{R}}) + S(\rho_{\alpha})+C
\end{equation}
where $S_{\textrm{gen.}}(\hat{\Phi}_{\alpha}\vert_{\mathscr{R}})$ is the generalized entropy to $O(1)$ in $G_{\textrm{N}}$
\begin{equation}
S_{\textrm{gen.}}(\hat{\Phi}_{\alpha}\vert_{\mathscr{R}})\defn\frac{\braket{\op{\mathcal{Q}}_{\mathcal{B}}(1)}_{\alpha}}{4G_{\textrm{N}}} + S_{\textrm{vN.}}(\varphi\vert_{\mathscr{R}})
\end{equation}
the area of the bifurcation surface at second order is $\op{\mathcal{Q}}_{\mathcal{B}}(1) = A_{\mathcal{B}}\op{1} + \delta^{2}\op{\mathcal{Q}}_{\mathcal{B}}$ and the second order perturbed area is $\delta^{2}\op{\mathcal{Q}}_{\mathcal{B}} \defn 4G_{\textrm{N}}\beta \op{X}+\int_{\mathcal{H}_{\textrm{R}}^{-}}U:\delta \op{\sigma}_{AB}\delta \op{\sigma}^{AB}:$ whose expected value is densely defined and the constant $C$ is the (divergent) state independent constant $C=-A/4G_{\textrm{N}} - S_{\textrm{vN.}}(\omega\vert_{\mathscr{R}})$. We refer the reader to, e.g., \cite{Chandrasekaran:2022eqq} or sec.~4.4 of \cite{Kudler-Flam:2023qfl} for further details. This result was originally established for Schwarzschild-AdS black holes by \cite{Chandrasekaran:2022eqq} as well as the de Sitter static patch \cite{Chandrasekaran:2022cip} and was subsequently generalized to arbitrary stationary black hole spacetimes in \cite{Kudler-Flam:2023qfl}. 

\subsection{The Algebra $\mathfrak{A}_{dress.}
(\mathscr{R};\mathbb{R}\times H)$ for $H$ Compact}
\label{sec:AdressH}
To analyze the ``type'' of the algebra $\mathfrak{A}_{\textrm{dress.}}(\mathscr{R};G_{\textrm{isom.}})$ and its trace, in this section and the next we construct an algebra $\mathfrak{A}_{\textrm{dress.}}(\mathscr{R};\mathbb{R}\times H)$ for any compact group $H$ which shares many of the mathematical properties of the isometry invariant algebra. In the case where $H=\textrm{SO}(3)$ or $\textrm{U}(1)$ then the algebras will be equivalent. In addition to covering both of the relevant isometry groups, our purpose in studying this considerably more general mathematical problem is that we will obtain, in sec.~\ref{sec:trace}, a ``universal'' form for the trace on the algebra $\mathfrak{A}_{\textrm{dress.}}(\mathscr{R};\mathbb{R}\times H)$. In addition to analyzing the properties of the isometry invariant algebra that we're actually interested in, these investigations will be very useful in section \ref{sec:fulldressedalg} when we consider infinite dimensional groups. We note that, for our more general mathematical problem, the label $\mathscr{R}$ is meaningless since $\mathfrak{A}_{\textrm{dress.}}(\mathscr{R};\mathbb{R}\times H)$ is an abstract von Neumann algebra satisfying certain properties which we delineate below. However, we will keep the label since it streamlines the notation of this paper and is suggestive of the physical applications that we have in mind. 

To construct $\mathfrak{A}_{\textrm{dress.}}(\mathscr{R};\mathbb{R}\times H)$, we start with a Type III algebra $\mathfrak{A}$ acting on a Hilbert space representation $\mathscr{F}$. We assume that the group $\mathbb{R}\times H$ admits an action as an outer automorphism on $\mathfrak{A}$. If we denote the infinitesimal generators of these automorphisms as $\op{F}(t,h)$ $t\in \mathbb{R}$ and $h\in H$ we therefore assume that the corresponding exponentiated operators are unitaries on $\mathscr{F}$ and satisfy
\begin{equation}
\label{eq:Fthautomorphism}
e^{i\op{F}(t,h)}ae^{-i\op{F}(t,h)} \in \mathfrak{A}
\end{equation}
for all $a\in \mathfrak{A}$ and $t,h\in \mathbb{R}\times H$. It follows from the structure of the group that 
\begin{equation}
\op{F}(t,h) = \op{F}(t) + \op{F}(h) \quad \textrm{ and } \quad \quad [\op{F}(t),\op{F}(h)]=0.
\end{equation}
 where $\op{F}(t)$ and $\op{F}(h)$ are the infinitesimal generators of the outer automorphisms induced by $\mathbb{R}$ and $H$ respectively. Additionally, we assume there exists a state $\ket{\omega}$ invariant under the automorphisms generated by $\mathbb{R}\times H$ meaning 
\begin{equation}
\op{F}(t)\ket{\omega} = 0, \quad \textrm{ and } \quad \quad \op{F}(h)\ket{\omega} = 0
\end{equation}
for all $t,h\in \mathbb{R}\times H$. %\MK{Should we use $t$ or $s$ instead of $c$ to match with more standard notation for the modular flow parameter? It also probably makes more sense to write $\op{F}(c,h)$, matching the notation of the previous section, and then say we can decompose this further e.g. so that $e^{i\op{F}(t,h)} = e^{i\op{K_{\omega}}t}e^{i\op{F}(h)}$ where $\op{K_{\omega}}$ is the modular Hamiltonian of the state $\omega$. This matches the notation of the next subsection. We could then do the same for the charges e.g. start with $\delta^2\op{\mathcal{Q}}(t,h)$ and then say that we decompose as $e^{i\delta^2\op{\mathcal{Q}}(t,h)} = e^{i\op{X}t} e^{i\op{Q}(h)}$ to make all our notation consistent.}
Our final condition on $\mathfrak{A}$ is that the modular flow with respect to the state $\ket{\omega}$ is generated by $e^{i\op{F}(t)}$ --- i.e. the modular operator of $\ket{\omega}$ on $\mathfrak{A}$ satisfies 
 \begin{equation}
\Delta^{it}_{\omega}=e^{i\op{F}(t)}.
 \end{equation}
We note that, the von Neumann algebra $\mathfrak{A}(\mathscr{R})$ of graviton perturbations of $\mathscr{R}$ satisfy these conditions for $H=\textrm{SO}(3)$ or $\textrm{U}(1)$.  

We now construct a von Neumann algebra of ``charges'' $\delta^{2}\op{\mathcal{Q}}^{\textrm{R}}(t,h)$ which are the infinitesimal generators of ``left'' translations on  the group. To describe this we first recall that there is a canonical Hilbert space representation which implements such an action of these charges as outlined in sec.~\ref{subsec:dressedalg}. Since $H$ is compact, there exists a unique, invariant Haar measure $\mu_{H}$ on $H$. Given such a measure one can construct a Hilbert space $\mathscr{H}_{\mathbb{R}\times H}\cong L^2(\mathbb{R}) \otimes L^{2}(H,~d\mu_H)$ where a state $\Psi$ corresponds to a specification of a complex wave function $\psi(t,h)$ on $\mathbb{R}\times H$ with finite norm 
\begin{equation}
||\Psi||^{2}=\int dtd\mu_{H}(h)~|\psi(t,h)|^{2}<\infty.
\end{equation}
Here $dt$ is the Lebesgue measure on $\mathbb{R}$. The inner product of two states $\Psi_{1},\Psi_{2}\in \mathscr{H}_{\mathbb{R}\times H}$ is then defined by 
\begin{equation}
\label{eq:Hinnerprod}
\braket{\Psi_{1}|\Psi_{2}}    = \int  dtd\mu_{H}(h)~\overline{\psi_{1}(t,h)}\psi_{2}(t,h).
\end{equation}
To describe the unitary action generated by\footnote{For brevity we drop the superscript $R$ in the charges moving forward.} $\delta^{2}\op{\mathcal{Q}}(t,h)$ it will be convenient to introduce an (improper) basis $\ket{t,h}$ with the property that for any such $\Psi$ with wavefunction $\psi(t,h)$ we have
\begin{equation}
\braket{t,h|\Psi} = \psi(t,h).
\end{equation}
The states $\ket{t,h}$ are analogous to ``position eigenstates'' of the group $\mathbb{R}\times H$ which are formally delta function normalized 
\begin{equation}
\braket{t,h|t^{\prime},h^{\prime}} = \delta(t,t^{\prime})\delta_{H}(h,h^{\prime})
\end{equation}
where $\delta_{\textrm{H}}$ is normalized so that $\int_{H} d\mu(h)\delta_{H}(h,h')=1$. It directly follows from eq.~\ref{eq:Hinnerprod} that these improper states form a basis and so any proper state $\ket{\Psi}$ can be formally expanded in this basis in the following way 
\begin{equation}
\label{eq:Psiexpand}
\ket{\Psi} = \int dtd\mu_{\textrm{H}}(h)\psi(t,h)\ket{t,h}.
\end{equation}
The unitary action of the charges $\delta^2 \op{\mathcal{Q}}(t,h)=t\op{X} +\delta^2 \op{\mathcal{Q}}(h)$ on this basis is given by 
%\footnote{Hereafter we will use the abbreviated notation $\delta^2 \op{\mathcal{Q}}(c) = \op{Q}(c)$ and $\delta^2\op{\mathcal{Q}}(h) = \op{Q}(h)$. We will also use the combined notation $\op{Q}(c,h) = \op{Q}(c) + \op{Q}(h)$. Since $c$ and $h$ commute in the group $\mathbb{R} \times H$ we will have $e^{i\op{Q}(c,h)} = e^{i\op{Q}(c)} e^{i\op{Q}(h)}$.}
\begin{equation}
e^{it^{\prime}\op{X}}\ket{t,h} = \ket{t+t^{\prime},h}\quad \textrm{and }\quad \quad e^{i\delta^{2}\op{\mathcal{Q}}(h')}\ket{t,h} = \ket{t,h^{\prime}h} 
\end{equation}
for all $t,t^{\prime}\in \mathbb{R}$ and $h,h^{\prime}\in H$. This action can be straightforwardly extended to any normalizable element of $\mathscr{H}_{\mathbb{R}\times H}$ by \eqref{eq:Psiexpand}. 
%The von Neumann algebra $\mathfrak{A}(\mathbb{R}\times H)$ generated by unitary operators $e^{i\delta^{2}\op{\mathcal{Q}}(t,h)}$ on $\mathscr{H}_{\mathbb{R}\times H}$ is called the ``group von Neumann algebra''. 
The last additional operator that will be useful to introduce on $\mathscr{H}_{\mathbb{R}\times H}$ is a ``multiplication operator''. For any  bounded function $T(t,h)$ on $\mathbb{R}\times H$ we define an operator $T(\op{t},\op{h})$ whose action on our basis elements is 
\begin{equation}
T(\op{t},\op{h}) \ket{t,h} = T(t,h)\ket{t,h}.
\end{equation}

We now have all of the ingredients in place to define the algebra $\mathfrak{A}_{\textrm{dress.}}(\mathscr{R};\mathbb{R}\times H)$ on $\mathscr{F}\otimes \mathscr{H}_{\mathbb{R}\times H}$. In direct analogy with the construction of the ``dressed'' algebras in the previous section we define the ``dressed'' von Neumann algebra with respect to $\mathbb{R}\times H$ as
\begin{equation}
\label{eq:dressRH}
\mathfrak{A}_{\textrm{dress.}}(\mathscr{R};\mathbb{R}\times H) \defn \{e^{-i\op{F}(\op{t},\op{h})}ae^{i\op{F}(\op{t},\op{h})},\delta^{2}\op{\mathcal{Q}}(t,h)\}''
\end{equation}
where, in the bracket, we are considering the set of all dressed operators for any $a\in \mathfrak{A}$ as well as the set of all charges. 

\subsection{$\mathfrak{A}_{dress.}(\mathscr{R};\mathbb{R}\times H)$ is a Type II algebra}
\label{sec:trace}

In this section we prove that the algebra $\mathfrak{A}(\mathscr{R};\mathbb{R}\times H)$ is Type II. We will accomplish this by directly constructing a densely defined trace on this algebra. The resulting trace will take a similar form to the trace constructed in sec.~\ref{sec:R} on the algebra $\mathfrak{A}(\mathscr{R};\mathbb{R})$ except now we must also take into account the dependence on group $H$.It was noted in \cite{AliAhmad:2024eun} that a suitable generalization of the trace in \eqref{eq:TraceX} is given by
% can be obtained by constructing an appropriate (improper) state $\ket{e}\in L^{2}(H;d\mu_{H})$. Given such an improper state we will show that trace on the algebra is of the form 
\begin{equation}
\label{eq:traceRH}
\tau_{\mathbb{R}\times H}(\hat{a}) = \braket{\omega,0_{\mathbb{R}},e_{H}|e^{\op{X}/2}\hat{a}e^{\op{X}/2}|\omega,0_{\mathbb{R}},e_{H}}
\end{equation}
where recall that $\op{X}\defn \delta^{2}\op{\mathcal{Q}}(1)$ and $\ket{e_{H}}$ is the improper state associated to neutral element of the group $H$. In the remainder of this section we show that \eqref{eq:traceRH} is a densely defined trace on $\mathfrak{A}_{\textrm{dress.}}(\mathscr{R};\mathbb{R}\times H)$. 

To show this, we first collect some properties of $\ket{e_{H}}$ on the von Neumann algebra $\mathfrak{A}(H)$ generated by the charges $\delta^{2}\op{\mathcal{Q}}(h)$ on $L^{2}(H,d\mu_{H})$. A general operator on $\mathfrak{A}(H)$ is given by 
\begin{equation} \label{group alg op}
\tilde{\psi}(\delta^2 \op{\mathcal{Q}}) =   \int d\mu(h) \psi(h) e^{i \delta^2 \op{\mathcal{Q}}(h)}.
\end{equation}
where $\psi(h)$ is a normalizable wavefunction on $L^{2}(H,d\mu_{H})$ and so $\tilde{\psi}(\delta^2 \op{\mathcal{Q}})$ is analogous to the ``Fourier transform'' of the wavefunction. The improper state $\ket{e_{H}}$ satisifes 
\begin{equation} 
\label{eq:neut1}
\braket{e_{H}|\tilde{\psi}(\delta^2 \op{\mathcal{Q}})|e_{H}} = \psi(e_{H}).
\end{equation}
In this sense, the ``wavefunction'' of $\ket{e_{H}}$ is equivalent to $\delta_{H}(e_{H},h)$ and, as expected, is non-normalizable. Furthermore, we note that the expected value of $\tilde{\psi}_1(\delta^2 \op{\mathcal{Q}})^{\dagger} \tilde{\psi}_2(\delta^2 \op{\mathcal{Q}})$ is simply the GNS inner product of the corresponding functions $\psi_{1}$ and $\psi_{2}$ 
\begin{equation}
\label{eq:einnerprodH}
\braket{e_{H}|\tilde{\psi}_1(\delta^2 \op{\mathcal{Q}})^{\dagger} \tilde{\psi}_2(\delta^2 \op{\mathcal{Q}})|e_{H}} = \int_{G}d\mu(h)\overline{\psi_{1}(h)}\psi(h).
\end{equation}
It directly follows that $\ket{e_{H}}$ is ``faithful'' --- i.e., $\braket{e_{H}|\tilde{\psi}(\delta^2 \op{\mathcal{Q}})^{\dagger} \tilde{\psi}(\delta^2 \op{\mathcal{Q}})|e_{H}} = 0 $ if any only if $\tilde{\psi}(\delta^{2}\op{\mathcal{Q}})=0$ --- and ``normal'' in that it is appropriately well-behaved under limits in  $\mathfrak{A}(H)$. Furthermore, from \eqref{eq:einnerprodH} it is clear that $L^{2}(H,d\mu)$ is equivalent to the GNS representation with respect to $\ket{e_{H}}$. Thus, $\ket{e_{H}}$ is a cyclic and separating vector for $\mathfrak{A}(H)$. With these ingredients, and the assumptions made on the algebra in the previous subsection, it directly follows from the ``dual weight theorem'' of Digernes \cite{Digernes1975} and Haagerup \cite{HaagerupI,HaagerupII} that $\op{X}$ generates the modular automorphism for the improper state $\ket{\omega,0_{\mathbb{R}},e_{H}}$ on $\mathfrak{A}_{\textrm{dress.}}(\mathscr{R};\mathbb{R}\times H)$. In particular, these general theorems imply that $\ket{\omega,0_{\mathbb{R}},e_{H}}$ satisfies 
\begin{equation}
\label{eq:modX}
\braket{\omega,0_{\mathbb{R}},e_{H}|\hat{a}\hat{b}|{\omega,0_{\mathbb{R}},e_{H}}} = \braket{\omega,0_{\mathbb{R}},e_{H}|\hat{b}e^{-\op{X}}\hat{a}e^{\op{X}}|{\omega,0_{\mathbb{R}},e_{H}}}
\end{equation}
Since $\op{X}\defn \delta^{2}\op{\mathcal{Q}}^{\textrm{R}}(1)/4G_{\textrm{N}}\beta$ is affiliated with the algebra, \eqref{eq:modX} implies that the modular automorphism of $\ket{\omega,0_{\mathbb{R}},e_{H}}$ is an inner automorphism. The inner implementation of the modular automorphism is a necessary and sufficient condition for the existence of a trace. Furthermore, it directly follows from \eqref{eq:modX} that \eqref{eq:traceRH} defines a trace
\begin{flalign}
\label{eq:traceabba}
    \tau_{\mathbb{R} \times H}(\hat{a} \hat{b}) 
    &=\bra{\omega,0_{\mathbb{R}},e_H} e^{\op{X}/2} \hat{a} \hat{b} e^{\op{X}/2} \ket{\omega,0_{\mathbb{R}},e_H} \nonumber
    \\
    &= \bra{\omega,0_{\mathbb{R}},e_H}( \hat{b} e^{\op{X}/2})( e^{-\op{X}} e^{\op{X}/2} \hat{a} e^{\op{X}} )\ket{\omega,0_{\mathbb{R}},e_H} \nonumber 
    \\
    &= \bra{\omega,0_{\mathbb{R}},e_H} \hat{b} \hat{a} e^{\op{X}} \ket{\omega,0_{\mathbb{R}},e_H} 
    = \tau_{\mathbb{R} \times H}(\hat{b}\hat{a}). 
\end{flalign}
where, from the first to the second line, we have invoked the KMS condition \eqref{eq:modX}. In the final line, we have used the fact that the exponential of $\op{X}$ can be commuted through in the trace \eqref{eq:traceRH}, which can be seen by expanding $\ket{0_t}$ in the $X$ basis as in \eqref{eq:TraceX}. 

Eq.~\eqref{eq:modX} is the key property that guarantees the existence of the trace $\tau_{\mathbb{R} \times H}$ on the algebra. Since this property is central to the results of our paper, we now prove \eqref{eq:modX} by a straightforward but lengthy calculation. We first note that operators of the form 
\begin{equation}
\label{eq:additiveH}
\hat{a} =  e^{-i\op{F}(\op{t},\op{h})}ae^{i\op{F}(\op{t},\op{h})}e^{i\delta^{2}\op{\mathcal{Q}}(t,{h})}
\end{equation}
form an additive basis for $\mathfrak{A}_{\textrm{dress.}}(\mathscr{R};\mathbb{R}\times H)$ where $a\in \mathfrak{A}$. Thus it suffices to prove \eqref{eq:modX} for operators of this form. We now separately compute the right-hand  side and left-hand  side of \eqref{eq:modX} for such operators and show that they are equal. For such operators, the right-hand  side of \eqref{eq:modX} is given by 
\begin{equation}
\textrm{R.H.S.}=\braket{\omega,0_{\mathbb{R}},e_{H}|e^{-i\op{F}(\op{t},\op{h})}be^{i\op{F}(\op{t},\op{h})}e^{i\delta^{2}\op{\mathcal{Q}}(t,h)}e^{-\op{X}}e^{-i\op{F}(\op{t},\op{h})}ae^{i\op{F}(\op{t},\op{h})}e^{i\delta^{2}\op{\mathcal{Q}}(t^{\prime},{h}')}e^{\op{X}}|{\omega,0_{\mathbb{R}},e_{H}}}.
\end{equation}
Using the properties of the vector $\ket{\omega,0_{\mathbb{R}},e_{H}}$ we can simplify the above expression considerably. We first recall that $\op{F}(t,h)\ket{\omega}=0$ for all $t,h$ and that  
\begin{equation}
\label{eq:modX1}
e^{i\op{F}(\op{t},\op{h})} = e^{i\op{F}(\op{t})}e^{i\op{F}(\op{h})},\quad \quad e^{i\delta^{2}\op{\mathcal{Q}}(t,h)} = e^{i\op{X}t}e^{i\delta^{2}\op{\mathcal{Q}}(h)}. 
\end{equation}
Furthermore, we have that $e^{i\delta^{2}\op{\mathcal{Q}}(h')}\ket{e_H} = \ket{h'}$ and the fluxes and charges satisfy the following commutation relations 
\begin{equation}
\label{eq:commutation}
    e^{i\op{F}(\op{t})} e^{it\op{X}}e^{-i\op{F}(\op{t})} = e^{it(\op{X}+\op{K}_{\omega})}, \quad  e^{i\op{F}(\op{h})} e^{i\delta^{2}\op{\mathcal{Q}}(h)} e^{-i\op{F}(\op{h})} = e^{i\delta^{2}\op{\mathcal{Q}}(h)} e^{i\op{F}(\op{h} h \op{h}^{-1})}
\end{equation}
where we recall that, on the horizon, $\op{K}_{\omega}\defn \beta \op{F}(1)$ is the modular Hamiltonian and we have set $\beta=1$ in the following manipulations (see the discussion around \eqref{eq:modHam} for an explanation of this fact). Using the above relations and expanding $\ket{0_{\mathbb{R}}}$ in terms an integral over the $\ket{X}$ eigenstates as in \eqref{eq:TraceX} implies that \eqref{eq:modX1} is equal to
\begin{equation}
        = \int \frac{dX dX'}{2\pi} e^{i(t+t')X^{\prime}} \bra{\omega, X, e_H}be^{it\op{K}_{\omega}}e^{i\delta^{2}\op{\mathcal{Q}}({h})}e^{i\op{F}(\op{h} {h} \op{h}^{-1})}e^{-\op{K}_{\omega}}a\ket{\omega,X', h'} 
\end{equation}
where we have used the fact that $\ket{X}$ are eigenstates of $\op{X}$ to pull out a factor of $e^{i(t+t')X^{\prime}}$. To simplify the expression further, we note that $\op{F}(\op{h}h\op{h}^{-1})$ acts diagonally on $\ket{h^{\prime}}$ and the $e^{i\delta^{2}\op{\mathcal{Q}}(h)}$ further shifts the eigenstate. So we obtain 
\begin{equation}
= \int \frac{dX dX'}{2\pi} e^{i(t+t')X^{\prime}} \bra{\omega, X, e_H}be^{it\op{K}_{\omega}}e^{-\op{K}_{\omega}}e^{i\op{F}({h}' {h} {h}'^{-1})}a\ket{\omega,X', hh'}.
\end{equation}
We now take the inner product $\braket{X|X^{\prime}}=\delta(X,X^{\prime})$ and integrate over $X^{\prime}$
\begin{equation}
= \int \frac{dX}{2\pi} e^{i(t'+t){X}} \bra{\omega,  e_H}be^{it \op{K}_{\omega}}e^{-\op{K}_{\omega}}e^{i\op{F}({h}' {h} {h}'^{-1})}a\ket{\omega, hh'}.
\end{equation}
We further note that a factor of $e^{-i\op{F}({h}' {h} {h}'^{-1})}e^{-it\op{K}_{\omega}}e^{\op{K}_{\omega}}$ can be freely added to the right-hand  side of $a$ since these operators act trivially on $\ket{\omega}$, 
\begin{equation}
 = \int \frac{dX}{2\pi} e^{i(t'+t){X}} \bra{\omega,  e_H}be^{-\op{K}_{\omega}}e^{it\op{K}_{\omega}}e^{i\op{F}({h}' {h} {h}'^{-1})}ae^{-i\op{F}({h}' {h} {h}'^{-1})}e^{-it\op{K}_{\omega}}e^{\op{K}_{\omega}}\ket{\omega, hh'}.
\end{equation}
The operator $e^{-\op{K}_{\omega}}e^{it\op{K}_{\omega}}e^{i\op{F}({h}' {h} {h}'^{-1})}ae^{-i\op{F}({h}' {h} {h}'^{-1})}e^{-it\op{K}_{\omega}}e^{\op{K}_{\omega}}$ is in the algebra $\mathfrak{A}$ since, by \eqref{eq:Fthautomorphism}, the unitaries conjugating $a$ implement algebra automorphisms. Thus we may use the KMS condition on this operator and $b$ to obtain 
\begin{align} 
        &= \int \frac{dX}{2\pi} e^{i(t'+t){X}} \bra{\omega,  e_H}e^{it\op{K}_{\omega}}e^{i\op{F}({h}' {h} {h}'^{-1})}ae^{-i\op{F}({h}' {h} {h}'^{-1})}e^{-it\op{K}_{\omega}}b\ket{\omega, hh'} 
\end{align}
which, using again the fact that $\ket{\omega}$ is annihilated by $\op{F}(t,h)$, yields
\begin{equation} \label{RHS for NLC}
     = \int \frac{dX}{2\pi} e^{i(t'+t){X}} \bra{\omega,  e_H}ae^{-i\op{F}({h}' {h} {h}'^{-1})}e^{-it\op{K}_{\omega}}b\ket{\omega, hh'}.
\end{equation}
Finally, we note that $\braket{e_{H}|hh^{\prime}}=\delta_{H}(e_{H},hh^{\prime})$ which sets $hh^{\prime}=e_{H}$ in the above equation. Since $H$ is a group it also follows that $h^{\prime}h=e_{H}$ as well. With this relation we obtain our final simplification of the right-hand  side of \eqref{eq:modX} 
\begin{equation}
\textrm{R.H.S.}= \int \frac{dX}{2\pi} e^{i(t'+t){X}} \bra{\omega,  e_H}ae^{-i\op{F}({h}'^{-1})}e^{-it\op{K}_{\omega}}b\ket{\omega, hh'}.
\end{equation}
We now consider the left-hand  side of \eqref{eq:modX} for any operator of the form of \eqref{eq:additiveH} 
\begin{equation}
\textrm{L.H.S.}=\braket{\omega,0_{\mathbb{R}},e_{H}|e^{-i\op{F}(\op{t},\op{h})}ae^{i\op{F}(\op{t},\op{h})}e^{i\delta^{2}\op{\mathcal{Q}}(t^{\prime},{h}')}e^{-i\op{F}(\op{t},\op{h})}be^{i\op{F}(\op{t},\op{h})}e^{i\delta^{2}\op{\mathcal{Q}}(t,h)}|{\omega,0_{\mathbb{R}},e_{H}}}.
\end{equation}
Again expanding the expectation value with respect to $\ket{0_{\mathbb{R}}}$ as an integral over $\ket{X}$, using the triviality of the action of unitaries on $\ket{\omega}$, that $e^{i\delta^{2}\op{\mathcal{Q}}(h)}\ket{e_{H}}=\ket{h}$, and invoking the commutation relations \eqref{eq:commutation} yields 
\begin{equation}
=\int \frac{dX dX'}{2\pi} e^{it {X}'} \bra{\omega, X, e_H} ae^{it'(\op{X}+\op{K}_{\omega})}e^{i\delta^{2}\op{\mathcal{Q}}({h}')}e^{i\op{F}(\op{h}h'\op{h}^{-1})}b\ket{\omega,X', h}. 
\end{equation}
Using the fact that $e^{it\op{X}}$ acts diagonally on $\ket{X}$, $\braket{X|X^{\prime}}=\delta(X,X^{\prime})$ and performing the $X^{\prime}$ integral we obtain 
\begin{equation} \label{LHS for NLC}
=\int \frac{dX}{2\pi} e^{i(t+t^{\prime})X}\braket{\omega,e_{H}|ae^{it^{\prime}\op{K}_{\omega}}e^{i\op{F}(hh^{\prime}h^{-1})}|\omega,h^{\prime}h}
\end{equation}
where we also used the fact that $\op{F}(\op{h}h^{\prime}\op{h}^{-1})$ acts diagonally on $\ket{h^{\prime}h}$. We note that the $X$ integral implies that we may interchange $t^{\prime}$ with $-t$ and the inner product $\braket{e_{H}|h^{\prime}h}$ implies that $\op{F}(hh^{\prime}h^{-1})$ equals $\op{F}(h^{-1})$ and $\ket{h^{\prime}h}$ can be replaced with $\ket{hh^{\prime}}$.\footnote{As we will discuss in section \ref{sec:fulldressedalg}, this point is slightly more complicated in the case that $H$ is not a compact group.} We finally obtain 
\begin{equation}
\textrm{L.H.S.}=\int \frac{dX}{2\pi} e^{i(t+t')X}\bra{\omega, e_H} ae^{-i\op{F}({h})}e^{-it\op{K}_{\omega}}b\ket{\omega, hh'}
\end{equation}
where we have freely commuted the two unitaries. This agrees with the right-hand  side computation. Thus, we have succeeded in proving \eqref{eq:modX} and so $\tau_{\mathbb{R} \times H}$ satisfies the properties of a trace.

We note that, while \eqref{eq:traceRH} is a compact expression, one can obtain a significantly more useful expression for the trace by noting that, since $H$ is compact, the  $\ket{e_{H}}$ can be  expanded in terms of a ``momentum'' basis labeled by the unitary irreducible representations of $H$ (see the end of sec. \ref{subsec:densityops}). Such a representation of the trace will be useful in determining the set of operators $\hat{a}$ with finite trace. An explicit formula for the case where $H=U(1)$ can be found in sec.~\ref{subsec:kerr}. 

\begin{remark}[Projections of Type II$_{\infty}$ algebras]
\label{rem:TypeII1}
As in the case of $\tau_{\mathbb{R}}$, the trace of the identity $\tau_{\mathbb{R}\times H}(\op{1})$ still diverges since it involves an integral which also diverges at large, positive $X$. Thus, the algebra  $\mathfrak{A}_{\textrm{dress.}}(\mathscr{R};\mathbb{R}\times H)$ is Type II$_{\infty}$. We further note that, in contrast to the case where $G=\mathbb{R}$ in sec.~\ref{sec:R}, applying the projector $\op{P}_{X<0}$ to the algebra $\mathfrak{A}_{\textrm{dress.}}(\mathscr{R};\mathbb{R}\times H)$ does not yield a normalizable trace. This is simply due to the fact that the vector $\ket{e_{H}}$ is non-normalizable and $\tau_{\mathbb{R}\times H}(\op{1})$  also includes the norm of $\ket{e_{H}}$. The projected algebra remains Type II$_{\infty}$. This fact will be relevant when we consider the algebra of observables invariant under the isometries of the de Sitter static patch in sections \ref{subsec:deSitter} and again in section \ref{sec:dSII1} when we consider the constraints arising from the larger, infinite dimensional symmetry group. In both cases, we will find that this infinite algebra can be projected to a physically relevant finite subalgebra in the spirit of \cite{Chandrasekaran:2022cip}.
\end{remark}

\subsection{Entropy and Density Matrices}
\label{subsec:densityops} 
As in sec.~\ref{sec:R}, the existence of a trace $\tau_{\mathbb{R}\times H}$ implies that any state $\hat{\Phi}\in \mathscr{F}\otimes \mathscr{H}_{\mathbb{R}\times H}$  admits an associated density matrix $\rho_{\hat{\Phi}}$ affiliated to $\mathfrak{A}_{\textrm{dress.}}(\mathscr{R};\mathbb{R}\times H)$ which satisfies 
\beq \label{Density operator condition}
	\braket{\hat{\Phi}|\hat{a}|\hat{\Phi}}  =\tau_{\mathbb{R}\times H}(\rho_{\hat{\Phi}} \hat{a}),\qquad \hat{a} \in \mathfrak{A}_{\textrm{dress.}}(\mathscr{R};\mathbb{R}\times H).
\eeq
We now explicitly compute the density matrix for the class of states
\begin{equation}
\label{eq:quantclass}
\ket{\hat{\Phi}_{\alpha}}=\ket{\varphi}\otimes \ket{\alpha}
\end{equation}
where $\ket{\varphi}\in \mathscr{F}$ and $\ket{\alpha}\in \mathscr{H}_{\mathbb{R}\times H}$. It will be useful to write the state $\ket{\alpha}$ as
\begin{equation}
\label{eq:f}
\ket{\alpha} \defn \tilde{\alpha}(\op{X},\delta^{2}\op{\mathcal{Q}})\ket{0_{\mathbb{R}},e_{H}}\quad\quad \tilde{\alpha}(\op{X},\delta^2\op{\mathcal{Q}})\defn \int dtd\mu(h) \alpha(t,h)e^{i(t\op{X} + \delta^{2}\op{\mathcal{Q}}(h))}
\end{equation}
where $\alpha(t,h)$ is square-integrable on $\mathbb{R}\times H$ and we have made the dependence of the ``Fourier transform'' on the charges $\op{X}$ and  $\delta^{2}\op{\mathcal{Q}}(h)$ explicit. In \cite{Jensen:2023yxy}, a general formula for the density operator of any state of the form of $\ket{\hat{\Phi}_{\alpha}}$ was derived for the case of $G=\mathbb{R}$ by utilizing the KMS condition and factorization of the modular operator in a semifinite algebra. Following the same procedure in our more general case, we obtain the following formula the density matrix 
\beq \label{Density operator for CQ state1}
	\rho_{\hat{\Phi}_\alpha} = \tilde{\alpha}(\op{X},\delta^{2}\op{\mathcal{Q}}) e^{-i\op{F}(\op{h})}e^{-i\op{K}_{\omega}\op{t}} e^{-\op{X}/2} \op{\Delta}_{\varphi \mid \omega} e^{-\op{X}/2} e^{i\op{K}_{\omega}\op{t}}e^{i\op{F}(\op{h})}\tilde{\alpha}(\op{X},\delta^{2}\op{\mathcal{Q}})^{\dagger}.
\eeq
where $\op{\Delta}_{\varphi \mid \omega}$ is the relative modular operator on $\mathfrak{A}$ of the states $\varphi,\omega\in\mathscr{F}$. To see that \eqref{Density operator for CQ state1} is a suitable formula for the density matrix, we now plug it back into the right-hand  side of \eqref{Density operator condition} and show that it computes the expected value $\braket{\hat{a}}_{\hat{\Phi}_{\alpha}}$. We consider the trace of the operator 
\begin{align}
	\tau(\rho_{\hat{\Phi}_\alpha} \hat{a}) &= \tau\bigg(\tilde{\alpha}(\op{X},\delta^{2}\op{\mathcal{Q}})e^{-i\op{F}(\op{h})}e^{-i\op{K}_{\omega}\op{t}} e^{-\op{X}/2} \op{\Delta}_{\varphi \mid \omega} e^{-\op{X}/2} e^{i\op{K}_{\omega}\op{t}}e^{i\op{F}(\op{h})}\tilde{\alpha}(\op{X},\delta^{2}\op{\mathcal{Q}})^{\dagger} \hat{a}\bigg). 
 \end{align}
Using cyclicity of the trace we can rewrite this expression as 
\begin{equation}
= \tau\bigg(\op{\Delta}_{\varphi \mid \omega}^{1/2} e^{-\op{X}/2} e^{i\op{K}_{\omega}\op{t}}e^{i\op{F}(\op{h})}\tilde{\alpha}(\op{X},\delta^{2}\op{\mathcal{Q}})^{\dagger} \hat{a}\tilde{\alpha}(\op{X},\delta^{2}\op{\mathcal{Q}})e^{-i\op{F}(\op{h})}e^{-i\op{K}_{\omega}\op{t}} e^{-\op{X}/2} \op{\Delta}_{\varphi \mid \omega}^{1/2}\bigg).
\end{equation}
From the definition of the trace \eqref{eq:traceRH} and expanding $\ket{0_{\mathbb{R}}}$ in terms of an integral of $\ket{X}$ we may diagonalize the factors $e^{-\op{X}/2}$ which cancel the exponential factor of $e^{\op{X}}$ coming from the trace. After completing these manipulations we obtain 
\begin{align}
	= \bra{\omega,0_t,e_H}  \op{\Delta}_{\varphi \mid \omega}^{1/2}  e^{i\op{K}_{\omega}\op{t}}e^{i\op{F}(\op{h})}\tilde{\alpha}(\op{X},\delta^{2}\op{\mathcal{Q}})^{\dagger} \hat{a}\tilde{\alpha}(\op{X},\delta^{2}\op{\mathcal{Q}})e^{-i\op{F}(\op{h})}e^{-i\op{K}_{\omega}\op{t}}  \op{\Delta}_{\varphi \mid \omega}^{1/2} \ket{\omega,0_{\mathbb{R}},e_{H}}.  \end{align}
Since $\op{\Delta}^{1/2}_{\varphi \mid \omega}\ket{\omega}=\ket{\varphi}$ we have that 
    \begin{equation}
=\bra{\varphi,0_t,e_H}   e^{i\op{K}_{\omega}\op{t}}e^{i\op{F}(\op{h})}\tilde{\alpha}(\op{X},\delta^{2}\op{\mathcal{Q}})^{\dagger} \hat{a}\tilde{\alpha}(\op{X},\delta^{2}\op{\mathcal{Q}})e^{-i\op{F}(\op{h})}e^{-i\op{K}_{\omega}\op{t}}   \ket{\varphi,0_{\mathbb{R}},e_{H}}.
    \end{equation}
The unitaries $e^{i\op{F}(\op{h})}$ and $e^{i\op{K}_{\omega}\op{t}}$ have a trivial action on $\ket{0_{\mathbb{R}},e_{H}}$ since they correspond to the neutral element of $\mathbb{R}\times H$.
    \begin{align}
=& \bra{\varphi,0_t,e_H}  \tilde{\alpha}(\op{X},\delta^{2}\op{\mathcal{Q}})^{\dagger} \hat{a}\tilde{\alpha}(\op{X},\delta^{2}\op{\mathcal{Q}})\ket{\varphi,0_t, e_H}. 
    \end{align}
Finally, using \eqref{eq:f} we obtain the expected value  $\braket{\hat{\Phi}_{\alpha}|\hat{a}|\hat{\Phi}_{\alpha}}$. 

 With the trace and density matrix defined we can now consider the von Neumann entropy which we recall, for Type II algebras, is only defined up to a multiplicative ambiguity. We first consider the case where $\ket{\varphi }=\ket{ \omega}$. In this case the density matrix of $\ket{\hat{\omega}_{\alpha}}=\ket{\omega}\otimes \ket{\alpha}$ becomes 
 \beq 
	\rho_{\hat{\omega}_{\alpha}} = |\tilde{\alpha}(\op{X},\delta^{2}\op{\mathcal{Q}})|^2e^{-\op{X}},
\eeq
since the states only differ in the wavefunction $\alpha$ on $\mathscr{H}_{\mathbb{R}\times H}$. The von Neumann entropy is given by 
\begin{equation}
S_{\textrm{vN}}(\rho_{\hat{\omega}_{\alpha}}) \defn - \tau(\rho_{\hat{\omega}_{\alpha}}\log\rho_{\hat{\omega}_{\alpha}})= \braket{\beta \op{X}}_{\hat{\omega}_{\alpha}} + S(\rho_{\alpha}) + \log(\beta) 
\end{equation}
where $\rho_{\alpha}=|\tilde{\alpha}(\op{X},\delta^{2}\op{\mathcal{Q}})|^{2}$ is the ``classical'' density matrix on $\mathscr{H}_{\mathbb{R}\times H}$ and $S(\rho_{\alpha})$ is the entropy of this density matrix 
\begin{equation}
S(\rho_{\alpha}) \defn -\tau(\rho_{\alpha}\log\rho_{\alpha}).
\end{equation}
In a moment, we will describe how this entropy can be written in a more physically transparent fashion.

In the more general case where $\varphi\neq \omega$, the logarithm of the density matrix is more complicated due to the individual terms not commuting. As we will see in the following two subsections, a particularly interesting limit is the ``semiclassical'' limit where the function $\alpha(t,h)$ is sharply peaked in its variables. In this case the logarithm of the density matrix can be approximated as\footnote{The approximation \eqref{eq:rhofH} as well as the control on of its ``errors'' can be proven using the methods of \cite{Kudler-Flam:2023hkl}.} 
\begin{equation}
\label{eq:rhofH}
\log(\rho_{\hat{\omega}_{\alpha}})\approx \log|\tilde{\alpha}(\op{X},\delta^{2}\op{\mathcal{Q}})|^{2}+\op{X}+\log\op{\Delta}_{\varphi|\omega}
\end{equation}
and so the von Neumann entropy is given by 
\begin{equation}
\label{eq:SvNRH}
S_{\textrm{vN}}(\rho_{\hat{\Phi}_{\alpha}}) \approx \braket{\beta \op{X}}_{\hat{\omega}_{\alpha}} - S_{\textrm{rel.}}(\varphi|\omega) + S(\rho_{\alpha}) + \log(\beta) .
\end{equation}
We note that the form of this expression is analogous to~\eqref{eq:SvnR} in sec.~\ref{sec:R} where now we simply get an extra contribution due to the fluctuation of the additional charges. As we will see in the following subsections, \eqref{eq:SvNRH} can be directly related to the generalized entropy mirroring the arguments of sec.~\ref{sec:R}. 

The new contribution to the entropy, $S(\rho_\alpha)$, quantifies the fluctuation of the charges in the group $G = \mathbb{R} \times H$. The Hilbert space $\mathscr{H}_{\mathbb{R} \times H}$ admits a `momentum' basis dual to its `position' basis $\ket{t,h}$. The momentum basis is spanned by $\ket{X,R,i,j}$, where here $R \in H^{\vee}$ is an irreducible unitary representation of $H$ on a Hilbert space $\mathscr{H}_R$, and $i,j = 1, ..., d_R$ are representation indices labeling an orthonormal basis for $\mathscr{H}_R$. The vector $\ket{0_t,e_H}$ admits an expansion in this basis as 
\beq
    \ket{0_t,e_H} = \int_{\mathbb{R}\times H^{\vee}} \frac{dXd\mu_H^{\vee}(R)}{\sqrt{2\pi}} \sum_{i,j = 1}^{d_R} \ket{X,R,i,j},
\eeq
where $d\mu_G^{\vee}(H)$ is the spectral measure on the space of irreducible representations of $H$. For a compact group, the set of irreps is discrete and the spectral measure is a counting measure with $\mu_H^{\vee}(R) = d_{R}$. The density operator $\rho_\alpha = |\tilde{\alpha}(\op{X},\delta^2 \op{\mathcal{Q}})|^2$ can be block diagonalized in the momentum basis as
\beq
    \rho_\alpha = \int_{\mathbb{R} \times H^{\vee}} \frac{dX d\mu_H^{\vee}(R)}{\sqrt{2\pi}} \sum_{i,j = 1}^{d_R} \rho_\alpha(X,R,i,j) \ket{X,R,i,j} \bra{X,R,i,j}. 
\eeq
Consequently, the contribution to the generalized entropy from this state is of the form
\beq \label{Entropy from group}
    S(\rho_\alpha) =  -\int_{\mathbb{R} \times H^{\vee}} \frac{dX d\mu_{H}^{\vee}(R)}{\sqrt{2\pi}} \sum_{i,j = 1}^{d_R} \rho_\alpha(X,R,i,j) \log \bigg(\rho_\alpha(X,R,i,j)\bigg).
\eeq
For reference, in the case $G = \mathbb{R}$ the density operator for a state $\ket{\alpha} \in L^2(\mathbb{R})$ can be written in the momentum basis as
\beq
    \rho_\alpha = \int_{\mathbb{R}} \frac{dX}{\sqrt{2\pi}} \; |\alpha(X)|^2 \ket{X} \bra{X}.
\eeq
The von Neumann entropy of this state with respect to the group algebra $\mathfrak{A}(\mathbb{R})$ is simply the Shannon entropy of the distribution $\rho_\alpha(X) \equiv |\alpha(X)|^2$:
\beq
    S(\rho_\alpha) = -\int_{\mathbb{R}} \frac{dX}{\sqrt{2\pi}} \; |\alpha(X)|^2 \log \bigg(|\alpha(X)|^2\bigg). 
\eeq
Eqn. \eqref{Entropy from group} has the same general structure, but with an additional contribution coming from the entropy of the density operator, $\rho_\alpha(R)$, in each block of the momentum decomposition of $\mathscr{H}_{G}$. 

Entropies of this type have appeared in the literature quantifying entanglement edge modes in general gauge theories, see e.g. \cite{Donnelly:2014gva}. To connect more directly with the analysis of \cite{Donnelly:2014gva}, one can diagonalize $\rho_\alpha(R,i,j) \mapsto p_\alpha(R) d_R^{-2} \delta_{ij}$ in each representation block, where $p(R)$ is a probability distribution over $R \in G^{\vee}$ and the factor of $d_R^{-2}$ appears to maintain normalization. Then, for a compact group $G$, the entropy of charge fluctuations can be split into the sum of a Shannon entropy for the distribution $p_\alpha$ and a term which depends only on the dimension of the various representations of $G$:
\beq \label{group ent}
    S(\rho_\alpha) = -\sum_{R \in G^{\vee}} p_\alpha(R) \log p_\alpha(R) + 2\sum_{R \in G^{\vee}} \log d_R. 
\eeq
We note that, when $G$ is a compact but infinite group, the second term in \eqref{group ent} is infinite due to the infinite number of irreducible representations appearing in the sum. This should be understood as a divergent contribution to the entropy arising from the fact that the tracial weight on the group algebra (e.g. the weight with vector representative $\ket{e}$) is not normalizable.\footnote{For compact $G$, the group algebra is of Type I$_{\infty}$ since it inherits irreducible representations from the irreps of the underlying group. Nevertheless, the fact that the algebra is of infinite type -- as can be deduced from the non-normalizability of the trace -- leads to the divergent contribution to the entropy.} This term can be absorbed into the state independent constant which appears in the generalized entropy \eqref{eq:SvNRH}. 

\subsection{Kerr Black Holes}
\label{subsec:kerr}
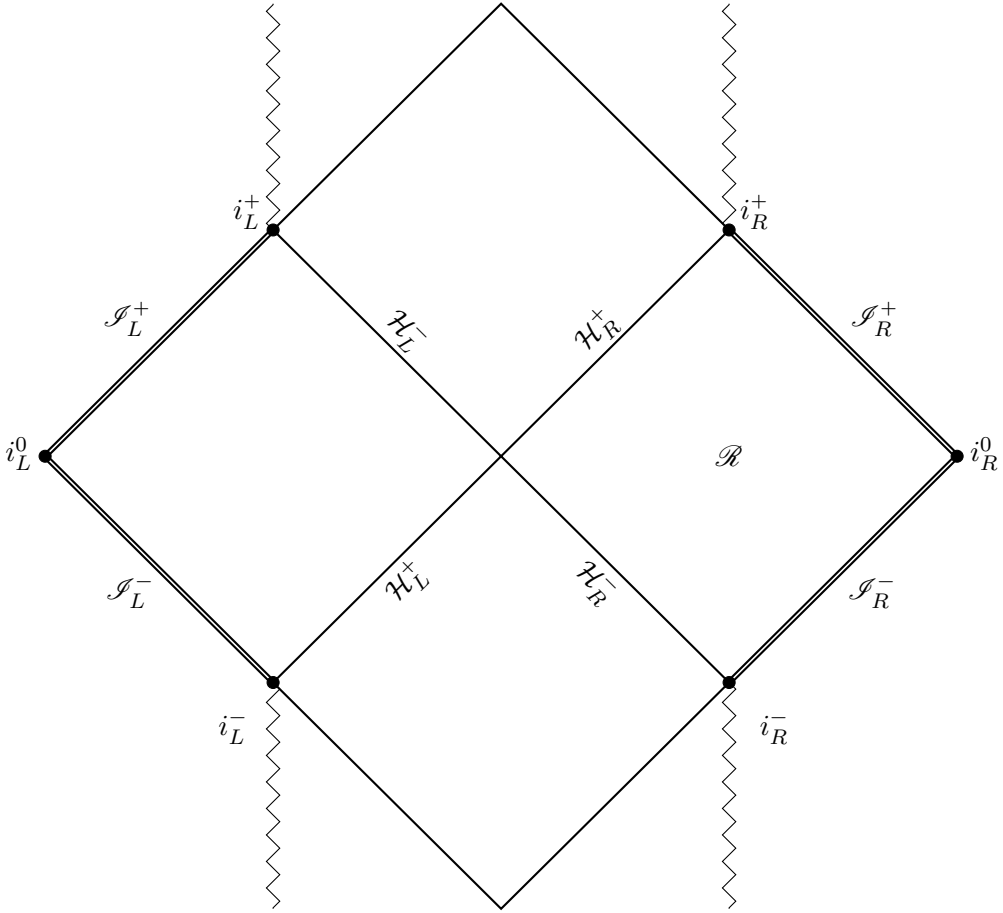
\begin{figure}
    \centering
    \begin{tikzpicture}[scale=3]
   \draw[decorate, decoration={zigzag}] (-1,1) -- node[pos=0.46,above left=-2] {\strut } (-1,2);
  \draw[decorate, decoration={zigzag}] (-1,-1) -- node[pos=0.46,above left=-2] {\strut } (-1,-2);
  \draw[decorate, decoration={zigzag}] (1,1) -- node[pos=0.46,above left=-2] {\strut } (1,2);
  \draw[decorate, decoration={zigzag}] (1,-1) -- node[pos=0.46,above left=-2] {\strut } (1,-2);
  \draw[thick,double] (1,1) -- (2,0) -- (1,-1) ;
  \draw[thick,double] (-1,1) -- (-2,0) -- (-1,-1) ;
  \draw[thick] (-1,1) -- (1,-1) -- (0,-2) -- (-1,-1) --(1,1)--(0,2)--(-1,1);
    \filldraw[] (1,1) circle (.75pt);
    \filldraw[] (-1,1) circle (.75pt);
    \filldraw[] (-1,-1) circle (.75pt);
    \filldraw[] (1,-1) circle (.75pt);
    \filldraw[] (-2,0) circle (.75pt);
    \filldraw[] (2,0) circle (.75pt);
      % INFINITY LABELS
  \node[above=1,right=1] at (2,0) {$i_R^0$};
  \node[above=1,left=1] at (-2,0) {$i_L^0$};
  \node[right=1,below right=1] at (S) {$i_R^-$};
  \node[right=1,below left=1] at (-S) {$i_L^-$};
  \node[right=1,above=1] at (1.1,.95) {$i_R^+$};
  \node[left=1,above=1] at (-1.1,.95) {$i_L^+$};
  \node[above right=-1] at (1.5,0.5) {$\mathscr{I}^+_R$};
  \node[above left=-1] at (-1.5,0.5) {$\mathscr{I}^+_L$};
  \node[below right=-2] at (1.5,-0.5) {$\mathscr{I}^-_R$};
  \node[below left=-2] at (-1.5,-0.5) {$\mathscr{I}_L^-$};
  \node[above=-2.5,rotate=45] at (.5,0.5) {$\mathcal{H}^+_R$};
  \node[below=-2.5,rotate=45] at (-.5,-0.5) {$\mathcal{H}^+_L$};
  \node[above=-2.5,rotate=-45] at (-.5,0.5) {$\mathcal{H}^-_L$};
  \node[below=-2.5,rotate=-45] at (.5,-0.5) {$\mathcal{H}^-_R$};
  \node[] at (1,0) {$\mathscr{R}$};
\end{tikzpicture}
    \caption{A portion of the maximally extended Kerr geometry. The region $\mathscr{R}$ denotes the black hole exterior which is bounded the Killing horizons $\mathcal{H}^{+}_{\textrm{R}}\cup \mathcal{H}_{\textrm{R}}^{-}$. In the conformally completed spacetime, $\mathscr{R}$ is also bounded by a boundary at infinity which consists of future/past timelike infinity $i^{+}_{\textrm{R}/\textrm{L}}$, future past null infinity $\mathscr{I}^{+}_{\textrm{R}/\textrm{L}}$ and spatial infinity $i^{0}_{\textrm{R}}$.}.
    \label{fig:kerr}
\end{figure}
While the following arguments apply to any black hole spacetime with a Killing horizon we now present, as an illustrative example, the case where $\mathscr{R}$ is the exterior of a stationary, Kerr black hole. The region $\mathscr{R}$ is an asymptotically flat, globally hyperbolic spacetime whose boundary is the portion $\mathcal{H}^{+}_{\textrm{R}}\cup \mathcal{H}^{-}_{\textrm{R}}$. The isometry group of the background spacetime $(\mathscr{R},g)$ is $\mathbb{R}\times U(1)$ corresponding to the two linearly independent Killing vectors. The cross-sections of a Kerr black hole are oblate spheres with metric $q_{AB}$ that admit only one rotational Killing vector $\psi^{A}$ which can be extended into the bulk as a spacelike Killing vector $\psi^{a}$. The $U(1)$ isometry corresponds to the orbits of the rotational Killing field $\psi^{a}$. It is conventional to identify $\mathbb{R}$ with the orbits of the asymptotically timelike Killing field $t^{a}$. However, in our analysis it will be convenient to instead identify $\mathbb{R}$ as the orbits of the horizon Killing field which is the linear combination $\xi^{a}=t^{a}+\Omega_{H}\psi^{a}$ where $\Omega_{H}$ is the angular velocity of the horizon, and $\xi^{a}$ is null on the horizon. 

The algebra $\mathscr{A}(\mathscr{R})$ of gravitatonal perturbations in $\mathscr{R}$ are quantized in manner described in sec.~\ref{subsec:gravpert}. However, in an asymptotically flat spacetime, the local graviton field is not just determined by the initial data on the horizon $\mathcal{H}^{-}$ but must be supplemented by initial data at infinity\footnote{A similar decomposition holes for any massive fields which are sufficiently non-interacting in the asymptotic past so that they behave as free fields at $i^{-}$. In that case one must supplement the data on the horizon with data at $i^{-}$ which can be straightforwardly quantized as explained in \cite{Prabhu:2022zcr} or Appendix B of \cite{Kudler-Flam:2023qfl}.}. The full algebra is given by 
\begin{equation}
\label{eq:horscriA}
\mathscr{A}(\mathscr{R}) = \mathscr{A}(\mathcal{H}^{-}_{\textrm{R}})\otimes \mathscr{A}(\mathscr{I}^{-})
\end{equation}
This decomposition is outlined 
in sec. $6.1$ and Appendix C of \cite{Kudler-Flam:2023qfl}. In the following we outline the characterization of the radiative degrees of freedom of the gravitational field and its quantization. We refer the reader to \cite{Strominger:2017zoo,Satishchandran:2019pyc} for further details on classical radiation in asymptotically flat spacetimes and \cite{Ashtekar:1981sf,Prabhu:2022zcr} for details on their quantization. 

Classically, the initial data at infinity is described by the ``radiative data'' at past infinity. In the conformally completed spacetime we can construct a coordinate system $(v,x^{A})$ at $\mathscr{I}^{-}$ with null normal $n^{a}=(\partial/\partial v)^{a}$ and $x^{A}$ are coordinates on the constant-$v$ cross-sections of $\mathscr{I}^{-}$. Choosing a null vector field $l^{a}$ transverse to $\mathscr{I}^{-}$, these coordinates are carried into the bulk of the spacetime by geodesic transport. In the physical spacetime we thereby obtain the coordinates $(r,v,x^{A})$ in a neighborhood of past null infinity where $r$ is the affine parameter of radially outgoing, past directed null geodesics. The initial data of the metric perturbation is described by the radiative field at leading order in $1/r$ in these coordinates. More precisely, the radiative data is given by the angular components 
\begin{equation}
\tilde{\gamma}_{AB}(v,x^{A}) = \lim_{r\to \infty}r\gamma_{AB}(r,v,x^{A})
\end{equation}
where we choose a ``Bondi-type'' gauge in which $q^{AB}\tilde{\gamma}_{AB}=0$ and $q^{AB}$ is the inverse of the unit sphere metric $q_{AB}$ on the cross-sections. The radiation propagating in from infinity is encoded in the Bondi News tensor which, in our coordinates, is given by 
\begin{equation}
\delta N_{AB}\defn \partial_{v}\tilde \gamma_{AB}
\end{equation}
where the ``$\delta$'' denotes that this is the radiation for the linearized metric perturbation. The quantization of the radiative data proceeds in direct parallel to the quantization of radiation on the horizon \cite{Ashtekar:1981sf,Prabhu:2022zcr}. The algebra $\mathscr{A}(\mathscr{I}^{-})$ is the $\ast$-algebra of initial data $\delta \op{N}(\tilde{s})$ factored by relations analogous to that of \ref{A1} and \ref{A3} together with commutation relations similar to that of \eqref{shear commutation} which are given by 
\begin{equation} \label{News commutation}
[\delta \op{N}_{AB}(x_{1}),\delta \op{N}_{CD}(x_{2})] = -8\pi i \bigg(q_{A(C}q_{D)B}-\frac{1}{2}q_{AB}q_{CD}\bigg)\delta^{\prime}(v_{1},v_{2})\delta_{\mathbb{S}^{2}}(x_{1}^{A},x_{2}^{A})\op{1}.
\end{equation}
The algebra admits a Gaussian vacuum state $\omega_{\mathscr{I}}$ whose two point function is also analogous to the vacuum state \eqref{eq:2pt} on the horizon 
\begin{equation}
\label{eq:2pt}
\omega_{\mathscr{I}}\big(\delta \op{N}_{AB}(x_{1})\delta \op{N}_{CD}(x_{2})\big)= -8\frac{(q_{A(C}q_{D)B}-\frac{1}{2}q_{AB}q_{CD})\delta_{\mathbb{S}^{2}}(x_{1}^{A},x_{2}^{A})}{(v_{1}-v_{2}-i0^{+})^{2}}.
\end{equation}
The GNS representation of $\mathscr{A}(\mathscr{I}^{-})$ with respect to $\omega_{\mathscr{I}}$ yields a Fock space $\mathscr{F}_{\mathscr{I}}$.  In summary, the quantization of the radiation at null infinity proceeds in direct parallel to the quantization of radiation on the horizon. Furthermore, the local graviton operator is decomposed in terms of initial data in the following way \cite{Prabhu:2022zcr}
\begin{equation}
\label{eq:gammasigmaN}
\op{\gamma}(w) = \delta\op{\sigma}(s) + \delta \op{N}(\tilde{s})
\end{equation}
where $s_{ab} \defn (Ew)_{ab}\vert_{\mathcal{H}^{-}}$, $\tilde{s}_{ab}=\lim_{r \to \infty}r(Ew)_{ab}\vert_{\mathscr{I}^{-}}$ and the restriction to $\mathscr{I}^{-}$ denotes the projection of the indices onto the tangent space of the cross-sections of $\mathscr{I}^{-}$. Furthermore, the global vacuum state is the ``Unruh state'' \cite{unruh1976notes} given by 
\begin{equation}
\ket{\omega} \cong \ket{\omega_{\mathcal{H}^{-}}}\otimes \ket{\omega_{\mathscr{I}^{-}}}.
\end{equation}
A key difference between the algebra $\mathscr{A}(\mathcal{H}^{-}_{\textrm{R}})$ and $\mathscr{A}(\mathscr{I}^{-})$ is that the vacuum state is a pure state on $\mathscr{A}(\mathscr{I}^{-})$ whereas it is mixed on $\mathscr{A}(\mathcal{H}^{-}_{\textrm{R}})$. Indeed, the corresponding von Neumann algebra $\mathfrak{A}(\mathscr{I}^{-})$ is a Type I$_{\infty}$ algebra and so admits a well-defined trace and density matrices at $\mathscr{I}^{-}$. Thus, from the point of view of the decomposition of \eqref{eq:horscriA}, the Type III nature of $\mathfrak{A}(\mathscr{R})=\mathfrak{A}(\mathcal{H}_{\textrm{R}}^{-})\otimes \mathfrak{A}(\mathscr{I}^{-})$ arises entirely from entanglement of quantum fields on the horizon. This justifies our focus in the preceding sections on the quantization of degrees of freedom at the horizon, since it is these degrees of freedom that determine the existence of a trace and of density matrices. 

Let us now consider the dressed algebra of isometry-invariant observables in $\mathscr{R}$. In this case, the relevant charges on the horizon are $\op{X}$ and $\delta^{2}\op{\mathcal{Q}}^{\textrm{R}}(\psi)$ which can be quantized on a Hilbert space $\mathscr{H}_{\mathbb{R}\times U(1)}$ as explained in sec.~\ref{sec:AdressH}.  The algebra of dressed observables on the horizon is given by the analog of \eqref{eq:dressRH} for the special case of $H=\mathbb{R}\times U(1)$ 
\begin{equation}
\mathfrak{A}_{\textrm{dress.}}(\mathcal{H}_{\textrm{R}}^{-};\mathbb{R}\times U(1)) = \{\delta \op{\sigma}(s;\op{t},\op{\psi}), \op{X},\delta^{2}\op{\mathcal{Q}}^{\textrm{R}}(\psi)\}''
\end{equation}
where we also have replaced $\mathscr{R}$ with $\mathcal{H}_{\textrm{R}}^{-}$ since the arguments of the previous section are strictly valid on the horizon and must be propagated into the bulk via \eqref{eq:gammasigmaN}. It was shown in sec. 6.1 of \cite{Kudler-Flam:2023qfl} that the algebra of isometry-invariant  observables which satisfy the constraints \eqref{eq:Gconstraints} is given by 
\begin{equation}
\mathfrak{A}_{\textrm{dress.}}(\mathscr{R};\mathbb{R}\times U(1)) \cong  \mathfrak{A}_{\textrm{dress.}}(\mathcal{H}_{\textrm{R}}^{-};\mathbb{R}\times U(1))\otimes \mathfrak{A}(\mathscr{I}^{-})
\end{equation}
Therefore, by the considerations of the previous section, the algebra is the product of a Type II$_{\infty}$ algebra with a type I algebra and so the full algebra is Type II$_{\infty}$. A general element of this algebra is a linear combination of operators of the form $\hat{a}\otimes \tilde{a}$ where $\hat{a}\in \mathfrak{A}_{\textrm{dress.}}(\mathscr{R};\mathbb{R}\times H)$ and $\tilde{a}\in \mathfrak{A}(\mathscr{I}^{-})$. The trace is then given by 
\begin{equation}
\label{eq:tracekerr}
\tau(\hat{a}\otimes \tilde{a}) = \tau_{\mathcal{H}^{-}}(\hat{a}) \cdot \tau_{\mathscr{I}^{-}}(\tilde{a})
\end{equation}
where 
\begin{equation}
\tau_{\mathcal{H}^{-}}(\hat{a}) = \braket{\omega_{\mathcal{H}^{-}},0_{\mathbb{R}},e_{U(1)}| e^{\op{X}/2} \hat{a} e^{\op{X}/2}|\omega_{\mathcal{H}^{-}},0_{\mathbb{R}},e_{U(1)}}
\end{equation}
and $\tau_{\mathscr{I}}(\tilde{a})$ is the standard Hilbert space trace over $\mathscr{F}_{\mathscr{I}}$. 

We can obtain a more useful formula for the trace by ``Fourier expanding'' the neutral element $\ket{0_{\mathbb{R}},e_{U(1)}}$. A general state in $\mathscr{H}_{U(1)}$  can be expanded in an orthonormal basis of periodic eigenfunctions $e^{im\phi}$ where $m\in \mathbb{Z}$. The Hilbert space admits the direct sum decomposition 
\begin{equation}
\mathscr{H}_{U(1)} = \bigoplus_{m}\mathscr{H}_{m}
\end{equation}
where each $\mathscr{H}_{m}$ is a one-dimensional Hilbert space of states with angular momentum $m$ --- i.e., $\delta^{2}\op{\mathcal{Q}}^{\textrm{R}}(1)\ket{m}= m\ket{m}$ with $\ket{m}\in \mathscr{H}_{m}$. In this representation, the zero element $\ket{e_{U(1)}}$ is given by 
\begin{equation}
\ket{e_{U(1)}} = \sum_{m=-\infty}^{\infty}\ket{m}.
\end{equation}
Similarly, Fourier transforming $\ket{0_{\mathbb{R}}}$ in terms of an integral over $\ket{X}$ yields the following expression for the trace 
\begin{equation}
\label{eq:TU1}
\tau_{\mathcal{H}^{-}}(\hat{a})=\int_{\mathbb{R} \times{\mathbb{R}}} \frac{dX dX'}{2\pi} e^{X} \sum_{m,m' = -\infty}^{\infty} \bra{\omega_{\mathcal{H}^{-}},X',m'} \hat{a} \ket{\omega_{\mathcal{H}^{-}},X,m}. 
\end{equation}
A general element $\hat{a}\in \mathfrak{A}_{\textrm{dress.}}(\mathcal{H}^{-}_{\textrm{R}},\mathbb{R}\times U(1))$ can be viewed as an $\mathfrak{A}$-valued function $a(\op{X},\delta^{2}\op{\mathcal{Q}}^{\textrm{R}})$ of the charges. With this representation, the dressed observables act diagonally on the basis elements --- i.e., $a(\op{X},\delta^{2}\op{\mathcal{Q}}^{\textrm{R}})\ket{X,m}=a(X,m)\ket{X,m}$. We therefore obtain 
\begin{equation}
\label{eq:TU1}
\tau_{\mathcal{H}^{-}}(\hat{a})=\int_{\mathbb{R} \times{\mathbb{R}}} \frac{dX }{2\pi} e^{X} \sum_{m = -\infty}^{\infty} \braket{\omega_{\mathcal{H}^{-}}|a(X,m)|\omega_{\mathcal{H}^{-}}}. 
\end{equation}
Consequently, in addition to sufficient decay at large negative values of $X$, the trace-class operators must also decay as $1/|m|^{1+\epsilon}$ for the sum to converge. In particular, as we have previously noted, the trace of the identity $\op{1}$ diverges and so $\mathfrak{A}_{\textrm{dress.}}(\mathcal{H}^{-}_{\textrm{R}},\mathbb{R}\times U(1))$ is a Type II$_{\infty}$ algebra. If we consider any ``semi-classical'' state $\ket{\hat{\Phi}_{\alpha}}\in \mathscr{F}\otimes \mathscr{H}_{\mathbb{R}\times U(1)}$ of the form of \eqref{eq:quantclass} where the wavefunction $\alpha(t,\psi)$ is sharply peaked in both time and angle, then, in such states, the spacetime geometry is well-described by a classical geometry even at second order. The von Neumann entropy is given by \eqref{eq:SvNRH} and, following identical manipulations that led to \eqref{eq:vNSgenR}, we obtain
\begin{equation} \label{Gen Ent Kerr}
S_{\textrm{vN.}}(\rho_{\hat{\Phi}_{\alpha}}) \simeq  S_{\textrm{gen.}}(\hat{\Phi}_{\alpha}\vert_{\mathscr{R}}) + S(\rho_{\alpha})+C.
\end{equation}
Eq.~\eqref{Gen Ent Kerr} is the generalized entropy up to a term which now involves the fluctuations of both the area $\op{X}$ and azimuthal superrotation $\delta^{2}\op{\mathcal{Q}}(\psi)$. 
 
We conclude this subsection by discussing the relationship of the quantized charges to charges defined at infinity. While this discussion is not strictly necessary for the quantization of the algebra or the construction of its trace as described above, it provides a direct connection to previous work which considered the ADM charges associated to the isometries of a black hole \cite{Chandrasekaran:2022eqq,Kudler-Flam:2023qfl}. The black hole area $\op{X}$ and the (azimuthal) superrotation charge $\delta^{2}\op{\mathcal{Q}}(\psi)$ are charges defined on the horizon. However, the corresponding horizon symmetries have a canonical extension as isometries of the background spacetime. It was shown by two of us together with S. Leutheusser  that the charges $\op{X}$ and $\delta^{2}\op{\mathcal{Q}}(\psi)$ can be ``matched'' to ADM charges at spatial infinity \cite{Kudler-Flam:2023qfl}. To see this, we first note that total perturbed ADM mass and (azimuthal) angular momentum of the black hole can be expressed in terms our initial data decomposition \eqref{eq:horscriA} as 
\begin{equation}
\label{eq:ADM}
\delta^{2}\op{M}_{i^{0}} = \op{F}^{\mathscr{I}}(t) + \delta^{2}\op{M}_{i^{-}}\quad \quad \delta^{2}\op{J}_{i^{0}}(\psi) = \op{F}^{\mathscr{I}}(\psi) + \delta^{2}\op{J}_{i^{-}}(\psi)
\end{equation}
where $\op{F}^{\mathscr{I}}(t)$ and $\op{F}^{\mathscr{I}}(\psi)$ are the generators of asymptotic time translations and azimuthal rotations of the algebra $\mathfrak{A}(\mathscr{I}^{-})$ and represents the total incoming energy and (azimuthal) angular momentum of incoming gravitons at $\mathscr{I}^{-}$. Similarly, $\delta^{2}\op{M}_{i^{-}}$ represents the perturbed incoming mass and incoming azimuthal angular momentum $\delta^{2}\op{J}_{i^{-}}(\psi)$ of the black hole. We refer the reader to sec.~6.2 of \cite{Prabhu:2022zcr} for the precise definitions of $\op{F}^{\mathscr{I}}(t)$ and $\op{F}^{\mathscr{I}}(\psi)$ in terms of radiative data at null infinity and $\delta^{2}\op{M}_{i^{-}}$ and $\delta^{2}\op{J}_{i^{-}}$ as the generators of time translations and rotations at $i^{-}$. It was argued in \cite{Kudler-Flam:2023qfl} that charges at $i^{-}$ ``match'' to the charges defined on the horizon. In particular, 
\begin{equation}
\label{eq:XQpsiiminus}
\op{X} = \delta^{2}\op{M}_{i^{-}} - \Omega_{\textrm{H}}\delta^{2}\op{J}_{i^{-}}(\psi)\quad  \textrm{ and }\quad\quad \delta^{2}\op{\mathcal{Q}}^{\textrm{R}}(\psi) = \delta^{2}\op{J}_{i^{-}}(\psi)
\end{equation}
where we recall that $\op{X} = \delta^{2}\op{\mathcal{Q}}^{\textrm{R}}(1)/4G_{\textrm{N}}\beta$. The first relation is a direct analog of the first law of black hole mechanics but for the second-order charges at $i^{-}$ and the second relation simply states that the azimuthal superrotation charge is equivalent to the black hole azimuthal angular momentum. Similar formulas hold at $i^{+}$. Since the asymptotic symmetries are inner automorphisms of $\mathfrak{A}(\mathscr{I}^{-})$, the operators $\op{F}^{\mathscr{I}}(t)$ and $\op{F}^{\mathscr{I}}(\psi)$ are in the algebra and so, by \eqref{eq:ADM} and \eqref{eq:XQpsiiminus}, including  $\op{X}$ and $\delta^{2}\op{\mathcal{Q}}^{\textrm{R}}(\psi)$ in the algebra is equivalent to including the ADM mass and and azimuthal ADM angular momentum. 

\subsection{Static Patch of de Sitter}
\label{subsec:deSitter}
We now consider the case where the background $(\mathscr{M},g)$ is de Sitter spacetime. In a closed universe, such as de Sitter, there is no asymptotic boundary by which one can invariantly define a subregion. Nevertheless, one can appropriately define such a subregion $\mathscr{R}$ relative to an ``observer'' in the spacetime. In order to sharply define a subregion we take the observer to be a localized body --- i.e., confined to a worldtube --- whose size $R$ is much smaller than the radius of curvature of the background spacetime. The appropriate classical and quantum description of the observer is an issue that we will have to contend with shortly. For now, we merely assume that the stress-energy of the observer $T_{ab}^{\textrm{obs.}}$ defines an appropriate ``center of mass'' inextendible worldline $\Gamma$ and $\mathscr{R}\defn I^{-}(\Gamma)\cap I^{+}(\Gamma)$. So that the observer does not affect the background geometry, we assume that its stress-energy is $O(1)$ in powers of $G_{\textrm{N}}$ and thus backreacts  on the spacetime only at second order in perturbation theory. In the background spacetime, we assume that $\Gamma$ is a geodesic and thereby $\mathscr{R}$ is the ``static patch'' of de Sitter spacetime which is bounded by Killing horizons $\mathcal{H}^{+}_{\textrm{R}}\cup \mathcal{H}^{-}_{\textrm{R}}$. 
The subregion background $(\mathscr{R},g)$ is a stationary spherically symmetric spacetime and so the isometry group $(\mathscr{R},g)$ is $\mathbb{R}\times SO(3)$. The subgroup $\mathbb{R}$ corresponds to time  translations along the observer's worldline and the horizon Killing field $\xi^{a}$. The subgroup $\textrm{SO}(3)$ is the orbit of the closed, spacelike Killing fields $\psi^{a}_{(i)}$ corresponding to rotations about the observer's worldline. On the horizon, the $\psi_{(i)}^{A}$ are the isometries of the round $2$-sphere cross sections of the horizon.  

Gravitational fluctuations of the subregion can be quantized to an algebra $\mathfrak{A}(\mathscr{R})\cong \mathfrak{A}(\mathscr{H}_{\textrm{R}}^{-})$ on a Fock space $\mathscr{F}$ which is equivalent to a quantization of initial data on $\mathscr{H}_{\textrm{R}}^{-}$ as described in sec.~\ref{subsec:gravpert}. The relevant charges on the horizon associated to the isometries are the perturbed horizon area $\op{X}$ and its angular momenta $\delta^{2}\op{\mathcal{Q}}(\psi_{(i)})$. As pointed out in \cite{Chandrasekaran:2022cip} and \cite{Kudler-Flam:2023qfl}, the charge $\op{X}$ ``matches'' onto the observer's energy at $i^{-}$. Similarly, as we will see, the charges $\delta^{2}\op{\mathcal{Q}}(\psi_{(i)})$ ``match'' onto the observer's angular momenta. Therefore the quantization of these charges on a Hilbert space $\mathscr{H}_{\mathbb{R}\times \textrm{SO}(3)}$ is really a quantization of the observers degrees of freedom. In order to obtain a consistent quantization of the static patch one must, at the very least, give a slightly more accurate model of an observer with these additional degrees of freedom. In this subsection we will focus on the energy and spin of the observer necessary to construct $\mathfrak{A}_{\textrm{dress.}}(\mathscr{R};\mathbb{R}\times \textrm{SO}(3))$. In section \ref{sec:dSII1} we will see that one needs a much more detailed model of an observer to construct the full dressed algebra $\mathfrak{A}_{\textrm{dress.}}(\mathscr{R};G)$. 

As previously mentioned, an observer is a ``sufficiently small body'' in the spacetime. One might imagine that a good approximation for such a body would be a ``point particle'' moving on a worldline $\Gamma$. However, this approximation runs into significant conceptual and practical issues. The first is that, due to the nonlinear nature of Einstein's equations, point particles are not well-defined as distributional sources in General Relativity \cite{Geroch:1986jjl}. In our perturbative expansion, the curvature induced by a point particle would fail to be well-defined and, in particular, produce infinite self-force effects \cite{Gralla:2009uf}. Physically, if one tries to compress a body to make it into a point particle then it should collapse into a black hole before a ``point particle'' limit can be reached. Additionally, as we consider here, if the body has spin then it also necessarily has finite size. For all of the above reasons, its far more sensible to start with a body of finite size and consider the limit where both the mass and size of the body become small.  Furthermore, if the body has finite size then it could, in principle, also have multipole moments which would affect the motion and energy of the body. Taking these additional effects into account is the main purpose of section \ref{sec:dSII1}. 

The equations of motion of a small, compact, classical body in General Relativity were obtained in a series of papers by Mathisson \cite{Mathisson:1937zz}, Pappeptrou \cite{Papapetrou:1951pa} and Dixon \cite{Dixon:1970zza,Dixon:1970zz,Dixon:1974xoz} and we now briefly summarize their main results as it pertains to this paper. In their approach, one considers a stress-energy $T_{ab}^{\textrm{obs.}}$ with support inside a worldtube of size $R$ where $R$ is much smaller than the radius of curvature. Even at leading order in $G_{\textrm{N}}$, the description and motion of the body may differ from that of a point particle simply due to the ``finite size'' effects of the body. To describe this, one defines a ``center of mass'' world line $\Gamma$, a momentum $p^{a}$ and spin tensor $S^{ab}$ which is antisymmetric, $S^{ab}=-S^{ba}$. Neglecting higher multipole moments, the equations of motion of the body are given by 
\begin{equation}
\label{eq:EOMmomentumspin}
v^{b}\nabla_{b}p^{a} = - \frac{1}{2}R_{abcd}v^{b}S^{cd}\quad \quad v^{c}\nabla_{c}S^{ab} = 2p^{[a}v^{b]}
\end{equation}
where $v^{a}$ is the four-velocity of the center of mass worldline $\Gamma$. The mass of the body is given by $m^{2}=-p^{a}p_{a}$ and we additionally note that $v^{a}$ need not be collinear with $p^{a}$. Finally, to close the equations and ensure that $\Gamma$ is indeed the center of mass worldline, one must additionally impose the supplementary condition $p^{a}S_{ab}=0$ \cite{Dixon:1970zza}. 
Eq.~\eqref{eq:EOMmomentumspin} together with this supplementary condition uniquely determines the evolution of a spinning compact body in any spacetime. These equations simplify drastically in maximally symmetric spacetimes such as de Sitter. In particular, it was shown by Dixon \cite{Dixon:1970zza} that \eqref{eq:EOMmomentumspin} simplifies to  
\begin{equation}
\label{eq:EOMdS}
v^{b}\nabla_{b}p^{a} = 0, \quad \quad v^{c}\nabla_{c}S^{ab}=0 \textrm{ and }  \quad v^{a}=p^{a}/m \quad \quad \textrm{ (de Sitter)}.
\end{equation}
Therefore $\Gamma$ is a geodesic of the background spacetime which we take to be $v^{a}=\xi^{a}$ at the ``center'' of the static patch. Given the time-like Killing vector $\xi^{a}$ and the rotational Killing vectors $\psi^{a}_{(i)}$, the conserved energy and spin of the observer are 
\begin{equation}
\label{eq:eL}
\varepsilon \defn  \int_{\Sigma}d^{3}x\sqrt{h}~n_{a}T^{ab}\xi_{b}\quad \quad L_{i} \defn  \int_{\Sigma}d^{3}x\sqrt{h}~n_{a}T^{ab}\psi_{(i)b}
\end{equation}
which, when expressed in terms of the $4$-momentum and spin of an inertial observer in de Sitter satisfying \eqref{eq:EOMdS}, are given by\footnote{For a general center of mass worldline in a spacetime with timelike and rotational Killing vectors $\xi^{a}$ and $\psi_{(i)}^{a}$, the conserved energy is $\varepsilon = -p_{a}\xi^{a} - \frac{1}{2}S^{ab}\nabla_{a}\xi_{b}$ and $L_{i}=-p_{a}\psi^{a}_{(i)} - \frac{1}{2}S^{ab}\nabla_{a}\psi_{(i)b}$} 
\begin{equation}
\varepsilon = -p_{a}\xi^{a} \quad \textrm{ and } \quad \quad L_{i} = -\frac{1}{2}S^{ab}\nabla_{a}\psi_{(i)b}.
\end{equation}

As one might expect, the angular momenta $L_{i}$ can be more directly related to the spin tensor $S^{ab}$ of our observer. To see this, we first note that the supplementary condition $p^{a}S_{ab}=0$ implies that the spin tensor is purely spatial and so we may denote it as $S^{ij}$ where $i,j$ indices refer to spatial components of the tensor. Choosing an orthonormal frame on $\Gamma$ given by $(v^{a},(e)^{a}_{i})$ we define a spin vector 
\begin{equation}
S_{i} = \epsilon_{ijk}S^{jk}
\end{equation}
where $\epsilon_{ijk} = \epsilon_{abcd}v^{a}(e)^{a}_{i},(e)^{a}_{j},(e)^{a}_{k}$. While the rotational generators $\psi^{a}_{(i)}$ vanish at $\Gamma$ since the worldline is chosen to be a fixed point of the isometry, the derivatives $\omega_{(i)ab}\defn \nabla_{a}\psi_{(i)b}\vert_{\Gamma}$ are non-vanishing, antisymmetric and also ``purely spatial'' in the sense that the only non-vanishing components are $\omega_{(i)jk}$. We define a set of three rotational directions as 
\begin{equation}
\Omega^{j}_{(i)} = \epsilon^{jkl}\omega_{(i)kl}
\end{equation}
whicher therefore defines the four-vectors $\Omega^{a}_{(i)}$ together with the relation $\Omega^{a}_{(i)}v_{a}=0$. Without loss of generality we may choose the rotational generators $\psi^{a}_{(i)}$ such that $\Omega^{a}_{(i)}=(e)_{i}^{a}$ and with this identification one can directly show that
\begin{equation}
L_{i} = S_{i}.
\end{equation}
By these relations we see that the mass and spin of the observer is precisely equivalent to the corresponding conserved quantities on the observer's wordline. By analogous arguments as presented in the previous subsection, these physical properties of the observer ``match'' onto physical charges on the horizon \cite{Kudler-Flam:2023qfl}
\begin{equation}
\label{eq:obsmatcheL}
\varepsilon = -X \quad \textrm{ and }\quad \quad L_{i} = \delta^{2}\mathcal{Q}^{\textrm{R}}(\psi_{(i)}).
\end{equation}

We now consider the dressed algebra of observables invariant under the isometry group as described in sec.~\ref{sec:AdressH}. The algebra is given by \eqref{eq:dressRH} where $H=\textrm{SO}(3)$ and the charges  correspond to the observer degrees of freedom $\op{X}\defn -\op{\varepsilon}$ and $\delta^{2}\op{\mathcal{Q}}^{\textrm{R}}(\psi_{(i)})\defn \op{L}_{i}$. In this quantization these are treated as independent degrees of freedom. This algebra and model of the observer was considered in \cite{Chandrasekaran:2022cip}. As explained in Remark \ref{rem:TypeII1}, even if we bound the spectrum of $\op{\varepsilon}$ from below the resulting algebra is {\em not} Type II$_{1}$. Therefore, naively, it would appear that the conclusions of \cite{Chandrasekaran:2022cip} do not generalize when one includes the constraints arising from the rotational isometries. 

However, this conclusion would be premature. The key point is that the kinetic energy of the rotation of the observer contributes to its total energy. As such, these degrees of freedom are not independent and we now take this into account. The relevant construction of the observer's algebra differs slightly from the construction presented in sec.~\ref{sec:AdressH} and so we will present it in a sequence of steps now.\footnote{Nevertheless, the resulting algebra is isomorphic to the dressed algebra presented in sec. \ref{sec:AdressH} -- see Appendices \ref{App: CP Factorization} and \ref{sec: Type II_1} for a discussion.}  We first impose the rotational constraints. The algebra of rotationally invariant observables are those observables which commute with the operator\footnote{As discussed in CLPW, in order to obtain a non-trivial algebra and satisfy the constraints one must also place an observer in the complementary static patch $\mathscr{L}$ \cite{Chandrasekaran:2022cip}. In our discussion, the secondary observer must also be rotating and the charges $\delta^{2}\op{\mathcal{Q}}^{\textrm{L}}$ match onto the observer's energy and angular momentum.} 
\begin{equation}
\delta^{2}\op{\mathcal{Q}}^{\textrm{L}}(\psi_{(i)}) = \op{F}(\psi_{(i)}) - \op{L}_{i}
\end{equation}
where we have used the fact that $\delta^{2}\op{\mathcal{Q}}^{\textrm{R}}(\psi_{(i)})\defn \op{L}_{i}$ and $\op{L}_{i}$ generates infinitesimal rotations in $\mathscr{H}_{\textrm{SO}(3)}$. The dressed observables that satisfy this constraint are of the form 
\begin{equation}
\op{\gamma}(w;\op{\psi}_{(i)}) \defn e^{-i\op{F}(\op{\psi}_{(i)})}\op{\gamma}(w)e^{i\op{F}(\op{\psi}_{(i)})}.
\end{equation}
 The full von Neumann algebra of rotationally invariant observables is 
\begin{equation}
\mathfrak{A}_{\textrm{dress.}}(\mathscr{R};\textrm{SO}(3)) \defn  \{\op{\gamma}(w;\op{\psi}_{(i)}),\op{L}_{i}\}''
\end{equation}
on $\mathscr{F}\otimes \mathscr{H}_{\textrm{SO}(3)}$. We note that $\mathfrak{A}_{\textrm{dress.}}(\mathscr{R};\textrm{SO}(3)) $ is a Type III algebra since we haven't yet imposed the Hamiltonian constraint.

To consider the Hamiltonian constraint we first must consider, more carefully, what we mean by a quantum mechanical observer. We note that the description of a quantum mechanical observer differs from that of a classical observer in that an initially localized quantum body will generally undergo ``wavepacket spreading'' and will thereby disperse at late times. Such an observer can, in principle, see the entire spacetime and their experience will not coincide with that of a localized, classical observer with a well-defined static patch. To obtain a similar description, one must work in the limit where the Compton wavelength of the body is much smaller than the size of the cosmological horizon --- i.e., a localized quantum mechanical observer satisfies $\lambda_{\textrm{C}}\ll \ell_{\textrm{dS}}$. In the limit as $\lambda_{C}\to 0$ we recover the description of a localized body with a well-defined static patch. To ensure this, we model the observer as having a rest mass given by  
\begin{equation}
\label{eq:restmass}
\op{m} = M\op{1} + \op{q}
\end{equation}
where $M\gg 1/\ell_{\textrm{dS}}$ and $\op{q}$ are the $O(1)$ fluctuations (in powers of $ M\ell_{\textrm{dS}}$) of the rest mass. We quantize $\op{q}$ in $L^{2}(\mathbb{R})$ with conjugate $\op{t}$, i.e. they satisfy  $[\op{q},\op{t}]=i$. Therefore, in addition to the $G_{N}\to 0$ limit we must also consider the $M\ell_{\textrm{dS}}\to \infty$ limit to obtain a localized observer. In particular, all physically properties of the observer are assumed to depend smoothly on $M$ as $M \ell_{\textrm{dS}}\to \infty$. 

With this limit in mind we now include effects due to the observers rotation. The general relationship between the angular momentum and the rotational energy depends on the precise details of the body. As an illustrative example we consider the observer to be a ``rigid body''. For a such a body one can define a symmetric ``inertia tensor'' $I_{ij}$ which is constant along $\Gamma$ \cite{Dixon:1970zza} such that the total energy of the body is given by\footnote{We assume that the size of the body scales as $O(1/\sqrt{M})$ as $M$ becomes large. Thus, $I_{ij}$ is $O(1)$ and, similarly, the rotational energy is also $O(1)$ in powers of $M$.} 
\begin{equation}
\op{\varepsilon} = M\op{1} + \op{q} + \frac{1}{2}I^{-1}_{ij}\op{S}^{i}\op{S}^{j}
\end{equation}
where $I^{-1}_{ij}$ is the inverse of $I_{ij}$. Using the fact that $\op{S}^{i}=\op{L}^{i}$, the spin of the observer is quantized in $\mathscr{H}_{\textrm{SO}(3)}$.
The total energy can be put into a positive form by considering an orthonormal basis $(\hat{e})^{i}_{\Lambda}$ ($\Lambda=1,2,3$) which diagonalizes $I_{ij}$ with real, positive\footnote{Positivity of the eigenvalues arises from the condition that $T_{ab}^{\textrm{obs.}}$ satisfies the weak energy condition.} eigenvalues $I^{\Lambda}$. This basis corresponds to the principal axes of the body which in general will not align with the angular velocities $\Omega^{a}_{(i)}$ defined above. As such the total energy of the observer is of the form 
\begin{equation}
\label{eq:totenergy}
\op{\varepsilon} = M \op{1}+ \op{q} +\frac{1}{2} \sum_{\Lambda}(I^{-1})^{\Lambda}(\op{L}\cdot e_{\Lambda} )^{2}.
\end{equation}
The Hamiltonian constraint in the quantum theory is that all physical operators in $\mathscr{R}$ commute with  
\begin{equation}
\label{eq:FqL}
\frac{\delta^{2}\op{\mathcal{Q}}^{\textrm{L}}(1)}{4G_{\textrm{N}}\beta} = -\op{F}(1) - M\op{1}-\op{q} -\frac{1}{2} \sum_{\Lambda}(I^{-1})^{\Lambda}(\op{L}\cdot e_{\Lambda})^{2}
\end{equation}
on the Hilbert space $\mathscr{F}\otimes \mathscr{H}_{\textrm{SO}(3)}\otimes \mathscr{H}_{\mathbb{R}}$ where we have have used \eqref{eq:obsmatcheL} as well as  \eqref{eq:totenergy}.
If we define the total rotational energy of the observer as 
\begin{equation} \label{eq:RotE}
\op{H}_{\textrm{rot.}}[\op{L}_{i}]\defn \sum_{\Lambda}(I^{-1})^{\Lambda}(\op{L}\cdot e_{\Lambda})^{2}
\end{equation}
then the algebra of observables which satisfy the Hamiltonian constraint is given by
\begin{flalign}
\mathfrak{A}^{\textrm{obs.}}_{\textrm{dress.}}(\mathscr{R};\mathbb{R}\times \textrm{SO}(3))=&\{e^{-i(\op{F}(\op{t}) + \op{H}_{\textrm{rot.}} \op{t})} \op{\gamma}(w,\op{\psi_{(i)}}) e^{i(\op{F}(\op{t}) + \op{H}_{\textrm{rot.}} \op{t})}, \nonumber \\
&e^{-i(\op{F}(\op{t}) + \op{H}_{\textrm{rot.}} \op{t})} \op{L}_i e^{i(\op{F}(\op{t}) + \op{H}_{\textrm{rot.}} \op{t})}, \op{q}\}''.
\end{flalign}
We have included the superscript to signify the conceptual distinction\footnote{\label{footnote:TypeII} The algebra $\mathfrak{A}^{\textrm{obs.}}_{\textrm{dress.}}(\mathscr{R},\mathbb{R} \times \textrm{SO}(3))$ is, in fact, inner unitarily equivalent to the dressed algebra constructed in sec. \ref{sec:AdressH}. This can be seen by identifying both algebras with the modular crossed product of $\mathfrak{A}_{\textrm{dress.}}(\mathscr{R};\textrm{SO}(3))$, only with respect to different reference weights. By the Connes cocycle theorem, these crossed product algebras are canonically isomorphic. The reader is referred to Appendices \ref{App: CP Factorization} and \ref{sec: Type II_1} for an in depth discussion of these facts.} of the algebra $\mathfrak{A}^{\textrm{obs.}}_{\textrm{dress.}}$ from the algebra $\mathfrak{A}_{\textrm{dress.}}$ Finally, we note that since the observer was put into the spacetime ``by hand'' we need to bound its physical properties. In particular, as in sec.~\ref{sec:R}, it is physically reasonable to bound the energy $\op{\varepsilon}$ of the observer from below. Since $\op{H}_{\textrm{rot.}}$ is positive, this amounts to bounding the spectrum of $\op{q}$ from below. The physical algebra is therefore\footnote{To get a slightly better model of the observer, it would perhaps be more sensible to bound the spectrum $\varepsilon>0$ from below. In terms of the observer's degrees of freedom this would bound the spectrum of $\op{q}$ in terms of $M$ and the total rotational kinetic energy. For example, for the isotropic body, this bound would be $q>-M - \ell(\ell+1)/2I$. While this bound is slightly more nontrivial, the algebra of observables is still Type II$_{1}$ as explained in footnote \ref{footnote:TypeII}. }
\begin{equation}
\mathfrak{A}^{+,\textrm{obs.}}_{\textrm{dress.}}(\mathscr{R};\mathbb{R}\times \textrm{SO}(3)) \defn \op{P}_{q>0}\mathfrak{A}^{\textrm{obs.}}_{\textrm{dress.}}(\mathscr{R};\mathbb{R}\times \textrm{SO}(3))\op{P}_{q>0}
\end{equation}
on $\mathscr{F}\otimes\mathscr{H}_{\textrm{SO}(3)} \otimes \op{P}_{q>0}[\mathscr{H}_{\mathbb{R}}]$ where $\op{P}_{q>0}$ is the projector onto positive values of $q$.

The algebra $\mathfrak{A}^{+,\textrm{obs.}}_{\textrm{dress.}}$ is a Type II$_{1}$ algebra. To see this we note that we first constructed the Type III algebra $\mathfrak{A}_{\textrm{dress.}}(\mathscr{R}; \textrm{SO}(3))$ and then imposed the Hamiltonian constraint. In the language of the ``crossed product'' this is equivalent to taking the crossed product of $\mathfrak{A}_{\textrm{dress.}}(\mathscr{R}; \textrm{SO}(3))$ with respect to the group $\mathbb{R}$ where the Hamiltonian on this algebra is now
\begin{equation}
\op{F}(1) + \op{H}_{\textrm{rot.}}[\op{L}_{i}].
\end{equation}
The arguments of sec.~\ref{sec:AdressH} show that  $\op{F}(1)$ on $\mathfrak{A}_{\textrm{dress.}}(\mathscr{R}; \textrm{SO}(3))$  is the modular Hamiltonian of the weight $\ket{\omega}\otimes \ket{e_{\textrm{SO}(3)}}$ on $\mathscr{F}\otimes \mathscr{H}_{\textrm{SO}(3)}$. The dressed algebra constructed in sec.~\ref{sec:AdressH} is the crossed product with respect to the modular Hamiltonian induced by this weight and the trace is given by \eqref{eq:traceRH}. In our case, the Hamiltonian is modified due to the rotation of the observer. It is straightforward to show that $\op{F}(1) + \op{H}_{\textrm{rot.}}[\op{L}_{i}]$ is the modular Hamiltonian of the weight $\ket{\omega}\otimes e^{-\op{H}_{\textrm{rot.}}[\op{L}_{i}]/2}\ket{e_{\textrm{SO}(3)}}$. It follows from identical arguments as  presented in sec.~\ref{sec:trace} that the trace on the physical algebra is given by 
\begin{equation}
\label{eq:TdS}
\tau^{\textrm{obs.}}_{\mathbb{R}\times \textrm{SO}(3)}(\hat{a}) = \braket{\omega,0_{\mathbb{R}},e_{\textrm{SO}(3)}|e^{-\frac{1}{2}(\op{q}+\op{H}_{\textrm{rot.}})}\hat{a}e^{-\frac{1}{2}(\op{q}+\op{H}_{\textrm{rot.}})}|\omega,0_{\mathbb{R}},e_{\textrm{SO}(3)}}
\end{equation}
for any $\hat{a}\in \mathfrak{A}^{+,\textrm{obs.}}_{\textrm{dress.}}(\mathscr{R};\mathbb{R}\times\textrm{SO}(3))$. We note the direct similarities between \eqref{eq:TdS} and \eqref{eq:traceRH} where $\op{X}$ is now replaced by the observer degrees of freedom $-\op{q}-\op{H}_{\textrm{rot.}}$ which acts non-trivially on the full Hilbert space $\mathscr{H}_{\textrm{SO}(3)} \otimes \op{P}_{q>0}[\mathscr{H}_{\mathbb{R}}]$. To see that the algebra is Type II$_{1}$ we can simply compute the trace of the projected identity operator 
\begin{align}
\tau^{\textrm{obs.}}_{\mathbb{R}\times \textrm{SO}(3)}(\op{1}) = &\braket{\omega,0_{\mathbb{R}},e_{\textrm{SO}(3)}|e^{-\frac{1}{2}(\op{q}+\op{H}_{\textrm{rot.}})}\op{P}_{q>0}\op{1}\op{P}_{q>0}e^{-\frac{1}{2}(\op{q}+\op{H}_{\textrm{rot.}})}|\omega,0_{\mathbb{R}},e_{\textrm{SO}(3)}}  \nonumber \\
=&\braket{0_{\mathbb{R}}|e^{-\op{q}}\op{P}_{q>0}|0_{\mathbb{R}}} \braket{e_{\textrm{SO}(3)}|e^{-\op{H}_{\textrm{rot.}}}|e_{\textrm{SO}(3)}}.
\end{align}
The first factor is finite due to the projection 
\begin{equation}
\braket{0_{\mathbb{R}}|e^{\op{-q}}\op{P}_{q>0}|0_{\mathbb{R}}} = \int_{0}^{\infty}dqe^{-q} = 1
\end{equation}
Thus, the identity is a trace-class operator if any only if $ e^{-\op{H}_{\textrm{rot.}}/2}\ket{e_{\textrm{SO(3)}}}$ is a normalizable state in $\mathscr{H}_{\textrm{SO(3)}}$. To analyze this we first consider the simple but illustrative example of an isotropic body where $I_{1}=I_{2}=I_{3}=I$. In this case, the Hamiltonian simplifies to
\begin{equation}
\op{H}_{\textrm{rot.}} = \frac{\op{L}^{2}}{2I} \quad \quad \quad  \quad \quad \quad  \textrm{(isotropic body)}
\end{equation}
To evaluate the norm of $e^{-\op{H}_{\textrm{rot.}}/2}\ket{e_{\textrm{SO(3)}}}$ we expand $\ket{e_{\textrm{SO(3)}}}$ in an orthonormal basis $\ket{\ell,n,m}$ where $\ell=0,1,\dots$ labels total angular momentum and $n,m=-\ell,\dots 0 \dots , \ell $ label the ``left'' and ``right'' azimuthal quantum numbers --- i.e., the improper basis satisfies $\braket{\ell,n,m|\ell^{\prime},n^{\prime},m^{\prime}}=\delta_{\ell,\ell^{\prime}}\delta_{n,n^{\prime}}\delta_{m,m^{\prime}}$. The non-normalizable neutral element in this basis is given by
\begin{equation}
\ket{e_{\textrm{SO}(3)}} = \sum_{\ell,n,m}\sqrt{2\ell+1}\delta_{mn}\ket{\ell,n,m}.
\end{equation}
It follows from the fact that $\op{L}^{2}\ket{\ell,n,m}=\ell (\ell+1)\ket{\ell,n,m}$ that
\begin{equation}
\braket{e_{\textrm{SO}(3)}|e^{-\op{H}_{\textrm{rot.}}}|e_{\textrm{SO}(3)}} = \sum_{\ell=0}^{\infty}(2\ell+1)e^{-\frac{\ell(\ell+1)}{2I}}\quad \quad \quad  \textrm{(isotropic body)}
\end{equation}
which is a convergent sum. By a lengthy but straightforward calculation one can indeed check that the state $e^{-\op{H}_{\textrm{rot.}}/2}\ket{e_{\textrm{SO(3)}}}$ remains normalizable if one considers a non-isotropic body simply due to the exponential suppression of the high angular momentum contributions in the sum. Up until now we have considered a rigid body, but we note that trace \eqref{eq:TdS} also applies equally well to a non-rigid body (e.g., a fluid). In this case, the total kinetic energy will receive additional contributions due to internal motions. Nevertheless, the contribution due to the rotational kinetic energy will be of the form\footnote{For a non-rigid body, the inertia tensor will, in general, be time dependent.} of $\op{H}_{\textrm{rot.}}[\op{L}_{i}]$ and the normalizability of $e^{-\op{H}_{\textrm{rot.}}/2}\ket{e_{\textrm{SO(3)}}}$  remains a necessary condition for the trace to be normalizable.  It appears likely that, for a large class of physical bodies, $\mathfrak{A}^{+,\textrm{obs.}}_{\textrm{dress.}}(\mathscr{R};\mathbb{R}\times \textrm{SO}(3))$ is a Type II$_{1}$ algebra. 

Following the arguments of sec.~\ref{subsec:densityops} we note that $\op{X}=-M\op{1}-\op{q}-\op{H}_{\textrm{rot.}}$. Thus, we find that, indeed, the entropy of any semiclassical state $\hat{\Phi}_{\alpha}$ is given by
\begin{equation}
S_{\textrm{vN.}}(\rho_{\hat{\Phi}_{\alpha}}) \simeq  S_{\textrm{gen.}}(\hat{\Phi}_{\alpha}\vert_{\mathscr{R}}) + S(\rho_{\alpha})+C
\end{equation}
which is the generalized entropy up to a term which involves the fluctuations of both the observer energy $\op{q}$ and angular momenta $\op{L}_{i}$. Finally, we note that since the algebra is Type II$_{1}$, there exists a maximum entropy state $\ket{\hat{\Phi}_{\textrm{max}}}$ whose density matrix on  $\mathfrak{A}^{+,\textrm{obs.}}_{\textrm{dress.}}(\mathscr{R};\mathbb{R}\times \textrm{SO}(3)) $ is equal to the identity operator $\op{1}$. This maximum entropy state is 
\begin{equation}
\ket{\hat{\Phi}_{\textrm{max}}} = \ket{\omega}\otimes e^{-(\op{q}+\op{H}_{\textrm{rot.}})/2}\ket{0_{\mathbb{R}},e_{\textrm{SO(3)}}}
\end{equation}
where we recall that, in de Sitter, $\ket{\omega}$ is the Bunch-Davies state.

\section{The Full Dressed Subregion Algebra} \label{sec:fulldressedalg}

We have demonstrated that the isometry invariant algebra $\mathfrak{A}_{\textrm{dress.}}(\mathscr{R};G_{\textrm{isom.}})$ is type II. However, $\mathfrak{A}_{\textrm{dress.}}(\mathscr{R};G_{\textrm{isom.}})$ is not a subalgebra of $\mathfrak{A}_{\textrm{dress.}}(\mathscr{R};G)$. In this section we consider the ``type'' of $\mathfrak{A}_{\textrm{dress.}}(\mathscr{R};G)$. 
This  quantization differs in two essential respects from the analysis presented in sec.~\ref{sec:isom}. The first is that, for an infinite dimensional group, there does not exist any translation invariant measure. The second, related issue is that, while no translation invariant measure can exist, there may exist measures, $\mu$, which are ``quasi-invariant'' under a subgroup $G_{0}\subset G$. Thus, the main issue we will have to contend with in this section is the construction of a Hilbert space $\mathscr{H}_{\textrm{G}}=L^{2}(G;\mu)$ with measure $\mu$ which is suitably quasi-invariant with respect to a ``large enough'' subgroup of $G$. To achieve this, in sec.~\ref{sec:PIforSuper}, we ignore the rotational subgroup and consider only the algebra $\mathfrak{A}_{\textrm{dress.}}(\mathscr{R};\mathcal{S})$ satisfying the boost supertranslation constraints. We argue that, for a large class of suitably quasi-invariant measures, this algebra is Type II. In sec.~\ref{subsec:fulldressed}, we combine the results of sec.~\ref{sec:PIforSuper} and sec.~\ref{sec:isom} to argue that the full dressed subregion algebra $\mathfrak{A}_{\textrm{dress.}}(\mathscr{R};G)$ is Type II and the entropy is equivalent to the generalized entropy for semiclassical states. 

Although our primary focus has been on gravitational fluctuations, in sec.~\ref{sec: CP for NLC Groups} we show that closely analogous considerations apply to matter fields. In particular, such fields may give rise to constraint equations generated by an infinite-dimensional, (possibly non-abelian) group of “large” gauge transformations on the horizon. Similar to the analysis of sec.~\ref{sec:isom}, we consider an algebra $\mathfrak{A}_{\textrm{dress.}}(\mathscr{R};\mathbb{R}\times H)$ where $H$ is a general, non-abelian, non-locally compact group and collect a set of sufficient conditions for the algebra to be Type II.

\subsection{ $\mathfrak{A}_{dress.}(\mathscr{R};\mathcal{S})$ is a Type II Algebra} \label{sec:PIforSuper}

In this section we construct the algebra $\mathfrak{A}_{\textrm{dress.}}(\mathscr{R};\mathcal{S})$ invariant under the subgroup of boost supertranslations $\mathcal{S}$. This is a simpler example than the full invariant algebra which nevertheless implicates many of the essential issues involved in quantizing the full group $G$. As explained above, the major effort in this section will be devoted to describing a suitable Hilbert space $\mathscr{H}_{\mathcal{S}}$ which admits an action of $\delta^{2}\op{\mathcal{Q}}(f)$ for a ``sufficiently large'' set of boost supertranslation functions $f$.   One can construct a Hilbert space $L^{2}(\mathcal{S},\gamma)$ given a measure $\gamma$ on the space of supertranslation functions. Since $\mathcal{S}$ is infinite dimensional, there exist many inequivalent choices of measure and none of them are a priori preferred due to the nonexistence of an invariant Haar measure. A well-studied class of measures are Gaussian measures. The choice of a centered, Gaussian measure is equivalent to that of an ``action'' on the space of boost supertranslations, as we shall now explain. 

\subsubsection{The Hilbert space $\mathscr{H}_{\mathcal{S}}$}
\label{subsubsec:HS}

Since it is more familiar, we first construct a suitable, Gaussian measure by starting with a choice of ``action'' --- i.e., a symmetric, bilinear map on $\mathcal{S}$. As a concrete example, we might consider 
\begin{equation}
\label{eq:Sf}
S[f] \defn  -\int_{\mathbb{S}^{2}}d\Omega~f\mathscr{D}^{2}f
\end{equation}
were $d\Omega$ is the measure on the $2$-sphere horizon cross-sections of the background. This particular action on the boost supertranslations was recently obtained in Euclidean quantum gravity by considering graviton fluctuations in the vicinity of a Killing horizon\footnote{Reference \cite{Law:2025ktz} was primarily interested in the case of fluctuations of the static patch in de Sitter spacetime but a similar analysis applies to any spacetime region $\mathscr{R}$ bounded by Killing horizons.} \cite{Law:2025ktz}. The arguments presented in this section will largely be agnostic to the choice of action and nearly all formulas will apply for any choice of $S[f]$. 

We note that the zero modes of the action \eqref{eq:Sf} are the space of constant boosts. In general, it will be useful to restrict attention to actions with this property. Without loss of generality, we will will assume that $S[f]$ is chosen such that its only zero modes are the constant translations.  With any action of this type, one may formally define a path integral over the group $\mathcal{S}_{0}\defn \mathcal{S}/\mathbb{R}$ given by 
\begin{equation}
\mathcal{Z}\defn \int \mathcal{D}f~e^{-S[f]}.
\end{equation}
where the ``$0$'' in $\mathcal{S}_{0}$ denotes that we have quotiented out by the constant mode.
Of course, mathematically speaking, the individual symbols in the above equation are not, a priori, well-defined. For example, the normalization $\mathcal{Z} = [\textrm{det}(-\mathscr{D}^{2})]^{-1/2}$
is formally infinite and must be regulated. In particular, the zeta-function regulation of $\mathcal{Z}$ yields $\mathcal{Z}=e^{-\frac{1}{2}\zeta^{\prime}(0)}$. Hereafter, any determinants that appear in the following manipulations are meant in the sense of such a regulation. Similarly, the objects $\mathcal{D}f$ and $e^{-S[f]}$ in the measure are not individually well-defined. However, as we will now briefly review, the measure 
\begin{equation}
\label{eq:gamma}
d\gamma(f) \defn \mathcal{Z}^{-1}\mathcal{D}fe^{-S[f]}
\end{equation}
is well-defined on $\mathcal{S}_{0}$. The object $d\gamma(f)$ on $\mathcal{S}_{0}$ can be rigorously defined as an infinite dimensional Gaussian measure on a choice of $\sigma$-algebra of measurable subsets of $\mathcal{S}$  \cite{Bogachev1998}. Starting with a topological vector space of smooth functions $f$ in $\mathcal{S}_{0}$  one can choose the $\sigma$-algebra of subsets to be generated by the ``cylindrical sets'' (see \cite{GJ,Gelfand5}). By the Bochner-Milnos theorem (see sec. A.6 of \cite{Bogachev1998}), a centered, Gaussian measure is determined by specifying any positive, symmetric billinear map $K(f_{1},f_{2})$ on the space of test functions which is to be interpreted as the covariance of the measure. For $S[f]$ given by \eqref{eq:Sf}, the covariance of $\gamma$ is the inverse of the Laplacian 
\begin{equation}
K(f_{1},f_{2}) = \int_{\mathbb{S}^{2}\times \mathbb{S}^{2}}d\Omega_{1} d\Omega_{2}~G(x_{1}^{A},x_{2}^{A})f_{1}(x^{A})f_{2}(x^{A})
\end{equation}
where 
\begin{equation}
\mathscr{D}^{2}G(x_{1}^{A},x_{2}^{A}) = \delta_{\mathbb{S}^{2}}(x_{1}^{A},x_{2}^{A})
\end{equation}
which is uniquely defined on $\mathcal{S}_{0}$. With this understanding, in the remainder of this section we freely manipulate the individual objects appearing in \eqref{eq:gamma}. As we will explicitly check, the resulting expressions will be well-defined in the above sense. Along with the Gaussian measure $\gamma$, the full measure on $\mathcal{S}$ becomes
\begin{equation}
d\mu \defn dt ~d\gamma(f) 
\end{equation}
where $f\in \mathcal{S}_{0}$ and $dt$ is the Lebesgue measure on the space $\mathbb{R}$ of constant boost supertranslations.

Given the measure $\gamma$, we now construct a corresponding Hilbert space $\mathscr{H}_{\mathcal{S}_{0}}$ defined by the space of complex wavefunctions $\Psi = \{\Psi(f)\}$ with finite norm 
\begin{equation}
||\Psi||^{2} = \mathcal{Z}^{-1}\int \mathcal{D}f \; e^{-S[f]}~\overline{\Psi(f)}\Psi(f).
\end{equation}
The inner product of two such states $\Psi_{1} = \{\Psi_{1}(f)\}$ and $\Psi_{2} = \{\Psi_{2}(f)\}$ is 
\begin{equation}
\braket{\Psi_{1}|\Psi_{2}} = \mathcal{Z}^{-1}\int \mathcal{D}f \; e^{-S[f]}
\;\overline{\Psi_{1}(f)}\Psi_{2}(f).
\end{equation}
The full Hilbert space for the boost supertranslations is given by 
\begin{equation}
\mathscr{H}_{\mathcal{S}} \defn \mathscr{H}_{\mathbb{R}}\otimes \mathscr{H}_{\mathcal{S}_{0}}
\end{equation}
As we reviewed in sec.~\ref{sec:R}, the Hilbert space $\mathscr{H}_{\mathbb{R}}$ admits a unitary action $e^{i\delta^{2}\op{\mathcal{Q}}(t)}$ by the constant translations. Similarly, $\mathscr{H}_{\mathcal{S}_{0}}$ admits a unitary action of the higher harmonic boost supertranslations. In contrast to the Lebesgue measure, the Gaussian measure is not translation invariant. Under translations $f\to f+ f^{\prime}$ the Gaussian measure changes by the Jacobian
\begin{equation}
\label{eq:Jff}
d\gamma(f) \to J(f,f^{\prime})d\gamma(f) \quad \textrm{where} \quad \quad J(f,f^{\prime}) \defn e^{-2S[f,f^{\prime}] - S[f^{\prime}]}
\end{equation}
and, for brevity, we have introduced the notation $S[f_{1},f_{2}]$ to be the billinear induced from $S[f]$. In the case where the action is given by \eqref{eq:Sf} we have that
\begin{equation}
\label{eq:Sff}
S[f_{1},f_{2}] \defn - \int_{\mathbb{S}^{2}}d\Omega~f_{1}\mathscr{D}^{2}f_{2}.
\end{equation}
Still, the translations in the space $\mathcal{S}_{0}$ may be unitarily implemented on $\mathscr{H}_{\mathcal{S}_{0}}$ as 
\begin{equation}
\label{eq:Q}
\big(e^{i\delta^{2}\op{\mathcal{Q}}(f^{\prime})}\Psi\big)[f] = J(f^{\prime},f-f^{\prime})^{-1/2}\Psi[f-f^{\prime}]
\end{equation}
where the presence of the Jacobian is due to the non-invariance of the measure. As we have indicated in the beginning of this section, not {\em all} translation operators $\delta^{2}\op{\mathcal{Q}}(f)$ are implementable on $\mathscr{H}_{\mathcal{S}_{0}}=L^{2}(\mathcal{S}_{0},\gamma)$. This is because, a translation can shift the Gaussian measure $\gamma$ to an inequivalent measure\footnote{Two measures $\gamma$ and $\gamma^{\prime}$ are equivalent if their sets of measure zero are equivalent.} $\gamma^{\prime}$. In that case, the corresponding Hilbert space $\mathscr{H}^{\prime}_{\mathcal{S}_{0}}=L^{2}(\mathcal{S}_{0},\gamma^{\prime})$ will be unitarily inequivalent as a representation of $\delta^{2}\op{\mathcal{Q}}(f)$. From the form of $J(f,f^{\prime})$ it is clear that the allowed translations must, at least, have finite action in order for the Jacobian to be well-defined. It follows from the Cameron-Martin theorem that these operators, in fact, constitute the complete set of allowed translations \cite{Bogachev1998}  --- i.e.,  the operator $e^{i\delta^{2}\op{\mathcal{Q}}(f)}$ is a unitary operator on $\mathscr{H}_{\mathcal{S}_{0}}$ if any only if $f$ has finite action. We will denote the set of boost supertranslations which are unitarily implementable on the Hilbert space $\mathscr{H}_{\mathcal{S}}$ by 
\begin{equation}
\bar{\mathcal{S}} \defn \{f\in \mathcal{S} ~|~ S[f]<\infty\}.
\end{equation}
With a judicious choice of action, the space $\bar{\mathcal{S}}$ can be so constructed as to contain all smooth boost supertranslation functions. In this sense, the Hilbert space admits an action of ``sufficiently many'' charge operators $\delta^{2}\op{\mathcal{Q}}(f)$. One can equivalently view the quantization $\mathscr{H}_{\mathcal{S}_{0}}$ as a choice of states in which the charges $\delta^{2}\op{\mathcal{Q}}(f)$ have finite fluctuations for all $f\in \bar{\mathcal{S}}$. 

Given the Hilbert space $\mathscr{H}_{\mathcal{S}}$, we can now define the group von Neumann algebra of $\mathcal{S}$. In particular, this algebra is given by
\begin{equation} \label{eq:groupalgS}
\mathfrak{A}(\mathcal{S};\gamma)
\defn \{\delta^{2}\op{\mathcal{Q}}(f)~|~ f\in \bar{\mathcal{S}}\}''
\end{equation}
which is closed in the weak operator toplogy of the Hilbert space $\mathscr{H}_{\mathcal{S}}$. This algebra admits a distinguished vector $\ket{\Omega}=\{1\}$ which corresponds to the constant wave function on $\mathcal{S}_{0}$. This state is normalized 
\begin{equation} \label{eq:gaussST}
||\Omega||^{2}= \mathcal{Z}^{-1}\int \mathcal{D}fe^{-S[f]} = 1
\end{equation}
and is a cyclic vector for the group algebra. To be precise, the action of the unitarities $e^{i\delta^{2}\op{\mathcal{Q}}(f)}$ generate the dense set of states \cite{Bogachev1998}
\begin{equation}
\label{eq:sochastic}
\ket{\Psi_{f}}\defn e^{i\delta^{2}\op{\mathcal{Q}}(f)}\ket{\Omega},\quad \quad \Psi_{f}(f^{\prime})=J(f,f^{\prime})^{1/2}
\end{equation}
for all $f\in \bar{\mathcal{S}}$.  
In the second equality in \eqref{eq:sochastic}, we have written the state as a wavefunction known as a ``stochastic exponential''. The stochastic exponentials $\ket{\Psi_f}$ are known to be dense in $L^{2}(\mathcal{S}_{0},\gamma)$ \cite{Bogachev1998}. In this sense, the unitaries $e^{i\delta^{2}\op{\mathcal{Q}}(f)}$ generate the entire Hilbert space. The Hilbert space $\mathscr{H}_{\mathcal{S}_{0}}$ also admits a ``position basis'' $\ket{f}$ which satisfies 
\begin{equation}
\braket{f|f^{\prime}} = \delta_{\gamma}(f,f^{\prime})
\end{equation}
with $\delta_{\gamma}(f,f')$ the $\delta$-function adapted to the measure $\gamma$ such that $\int_{\mathcal{S}_{0}} d\gamma(f)~ \delta_{\gamma}(f,f^{\prime}) =1$. Any state $\Psi=\{\Psi(f)\}$ in $\mathscr{H}_{\mathcal{S}_{0}}$ can be formally expanded in this basis as 
\begin{equation}
\ket{\Psi} = \int d\gamma(f)\Psi(f)\ket{f}
\end{equation}
We note that the Hilbert space also admits a multiplication operator. If $T(f)$ is any bounded function on $\mathcal{S}_{0}$, then we define the operator $T(\op{f})$ in the position basis via 
\begin{equation}
T(\op{f})\ket{f} = T(f)\ket{f}.
\end{equation}

We conclude this subsection by providing an equivalent formulation of the Hilbert space $\mathscr{H}_{\mathcal{S}_{0}}$ and the group von Neumann algebra $\mathfrak{A}(\mathscr{R};\mathcal{S}_{0})$ in terms of ``infinite tensor products''. Since the infinite tensor product of von Neumann algebras is well-studied \cite{Takesaki2003VolIII}, this re-formulation will be extremely useful for our analyses in the following sections. We first note that the space $\mathcal{S}_{0}\simeq \bigtimes_{\ell>0,m}\mathbb{R}$ since any $f\in \mathcal{S}_{0}$ can be decomposed in terms of spherical harmonics as\footnote{For rotating black holes, the analogous decomposition would be with respect to spheroidal harmonics.}
\begin{equation}
f = \sum_{\ell m}c_{\ell m} Y_{\ell m}.
\end{equation}
Under this decomposition, the Gaussian measure $\gamma$ can be decomposed in terms of an infinite tensor product of centered, one-dimensional Gaussian measures \cite{Bogachev1998}
\begin{equation}
\gamma =\bigotimes_{\ell>0,m} \gamma_{\ell m}
\end{equation}
where each $\gamma_{\ell m}$ is unit normalized on $\mathbb{R}$. Consequently, the Hilbert space $\mathscr{H}_{\mathcal{S}_{0}}= L^{2}(\mathcal{S},d\gamma)$ can similarly be decomposed as an infinite tensor product \cite{vonNeumann1939}
\begin{equation} \label{eq:STHilbITP}
\mathscr{H}_{\mathcal{S}_{0}}  \simeq \bigotimes_{\ell>0,m}L^{2}(\mathbb{R},d\gamma_{\ell m}).
\end{equation}
In this representation, any vector in this Hilbert space is defined by the specification $\Psi=\{\psi_{\ell m}\}$ of a family of wavefunctions $\psi_{\ell m}$ for each $\ell, m$ such that the product
\begin{equation}
||\Psi|| = \prod_{\ell>0, m} ||\psi||_{\ell m}^{2}
\end{equation}
converges where the norms in the right-hand  side of the above expression are computed in each $L^{2}(\mathbb{R},d\gamma_{\ell m}) $. For example, the distinguished vector $\ket{\Omega_{\gamma}}\in L^{2}(\mathcal{S},d\gamma)$ introduced above with wavefunction $\Omega_{\gamma} = 1$ now simply corresponds to the family of constant wavefunctions $\Omega_{\ell m} = 1$ for all $\ell,m$. 

The von Neumann algebra $\mathfrak{A}(\mathcal{S}_0;\gamma)$ can be similarly decomposed as an infinite tensor product of von Neumann algebras. We will now give a brief overview of this decomposition and we refer the reader to \cite{Takesaki2003VolIII} and Appendix \ref{app: ITP} for further details. Each tensor factor in \eqref{eq:STHilbITP} admits a one-dimensional translation operator $\op{\mathcal{Q}}_{\ell m}$. Following \cite{Takesaki2003VolIII}, we consider the family of von Neumann algebras $\mathfrak{A}_{\ell m}(\mathbb{R})$ corresponding to the group algebra of $\mathbb{R}$ generated by bounded functions of these $\op{\mathcal{Q}}_{\ell m}$ acting on $L^{2}(\mathbb{R},d\gamma_{\ell m})$. We then consider the infinite tensor product $\bigotimes_{\ell>0,m}\mathfrak{A}_{\ell m}(\mathbb{R})$ of such algebras which can be rigorously defined as a $C^{\ast}$-algebra \cite{Takesaki2003VolIII}. To turn it into a von Neumann algebra, we close the algebra by taking the double commutant in the Hilbert space $\mathscr{H}_{\mathcal{S}_{0}}$ to obtain
\begin{equation}
\mathfrak{A}(\mathcal{S}_0;\gamma) \simeq \big[\bigotimes_{\ell > 0,m} \mathfrak{A}_{\ell m}(\mathbb{R})\big]''.
\end{equation}
From the point of view of the infinite tensor product, the allowed set of translations $\op{\mathcal{Q}}_{\ell m}$ are those that decay sufficiently fast in $\ell$ such that the measure is mapped to an equivalent infinite tensor product measure.

\subsubsection{The Algebra $\mathfrak{A}_{dress.}(\mathscr{R};\mathcal{S})$ and its Trace}
\label{subsec:traceS}

We are now prepared to construct the dressed algebra invariant under all boost supertranslations and an associated trace on this algebra in direct analogy to the construction presented in sec.~\ref{sec:trace}. The key ingredient in that construction was the introduction of a ``neutral element'' vector for the Hilbert space. To this end, we seek an analogous vector on $\mathscr{H}_{\mathcal{S}_{0}}$. We recall that, in the Hilbert space associated to a locally compact group, such a vector was formally defined as a ``$\delta$-function'' wavefunction on the group. Our goal will be to identify an improper state on $\mathscr{H}_{\mathcal{S}_{0}}$ satisfying these properties.

The analogous vector in $\mathscr{H}_{\mathcal{S}_{0}}$ is $\ket{e_{\gamma}}$ which is the (improper) state with wavefunction $\delta_{\gamma}(0,f)$ relative to the measure $d\gamma(f)$. It is straightforward to check that $\ket{e_{\gamma}}$ has finite inner product with any state in the span of the space of the stochastic exponentials $\ket{\Psi_{f}}$ defined in the previous section. As such, the vector $\ket{e_{\gamma}}$ is a well-defined distribution on a dense set of states in $\mathscr{H}_{\mathcal{S}_0}$. To show that $\ket{e_{\gamma}}$ induces a  semifinite, normal weight on the group von Neumann algebra, it is useful to express this vector in terms of the infinite tensor product defined above as 
\begin{equation}
\label{eq:egamma}
\ket{e_{\gamma}} \defn \bigotimes_{\ell>0,m}\ket{e_{\gamma_{\ell m}}}.
\end{equation}
Here, each $\ket{e_{\gamma_{\ell m}}}$ is a vector on $L^{2}(\mathbb{R},d\gamma_{\ell m})$ whose wavefunction is a delta-function centered at $c_{\ell m}=0$ with respect to the measure $d\gamma_{\ell m}$. Each $\ket{e_{\gamma_{\ell m}}}$ moreover induces a faithful, semifinite, normal weight on the group algebra $\mathfrak{A}_{\ell m}(\mathbb{R})$ which is analogous to the neutral weight described in sec.~\ref{sec:trace}. Following the construction
of \cite{Blackadar1977Infinite}, the infinite tensor product \eqref{eq:egamma} will define a semifinite, normal weight on $\mathfrak{A}(\mathcal{S}_0;\gamma)$ provided that there exists a projector $p_{\ell m}$ on each $L^{2}(\mathbb{R},d\gamma_{\ell m})$ such that 
\begin{equation}
\label{eq:egammaplm}
\bra{e_{\gamma_{\ell m}}}p_{\ell m} \ket{e_{\gamma_{\ell m}}}=1
\end{equation}
in an almost everywhere sense relative to the indices $\ell,m$. In our case, each Hilbert space has a distinguished projector $p_{\ell m}=\ket{\Omega_{\ell m}}\bra{\Omega_{\ell m}}$ which satisfies this condition since $\braket{e_{\gamma_{\ell m}}|\Omega_{\ell m}} = 1$ for all $\ell,m$. As each $\ket{e_{\gamma_{\ell m}}}$ induces a faithful weight on $\mathfrak{A}_{\ell m}(\mathbb{R})$, and $\braket{e_{\gamma_{\ell m}}|\Omega_{\ell m}} = 1$ for each $\ell,m$ it appears likely that the weight induced by $\ket{e_{\gamma}}$ is also faithful on the infinite tensor product. However, we have not attempted prove this rigorously\footnote{If the weight induced by $\ket{e_{\gamma}}$ is not faithful, we may restrict to the corner of $\mathfrak{A}(\mathcal{S}_0;\gamma)$ induced by the support projection on which this weight is faithful. In the case of de Sitter spacetime considered in sec.~\ref{sec:dSII1}, due to the properties of the observer, the weight will be modified to a state and, as such, will manifestly be faithful, semifinite and normal. \label{foot:faithful}}. In the remainder of this section, we give a formal path integral argument that  $\ket{e_{\gamma}}$ is also faithful. 

We now explore the properties of the vector $\ket{e_{\gamma}}$ in the path integral picture.
Since the unitary operators $e^{i\delta^{2}\op{\mathcal{Q}}(f)}$ with $f\in \bar{\mathcal{S}}$, generate $\mathfrak{A}(\mathcal{S};\gamma)$ it is natural to introduce the formal expression
\begin{equation}
\label{eq:genop}
\tilde{\psi}(\delta^{2}\op{\mathcal{Q}}) = \int d\gamma(f)\psi(f)e^{i\delta^{2}\op{\mathcal{Q}}(f)}.
\end{equation}
As in the case of the path integral, the expression \eqref{eq:genop} is not to be interpreted as a literal operator-valued integral. Rather, it serves as a convenient shorthand for an operator defined indirectly, via limits of bounded operators. More precisely, $\tilde{\psi}(\delta^{2}\op{\mathcal{Q}})$ is defined as an element of the weak operator closure of the algebra generated by the unitaries $e^{i\delta^{2}\op{\mathcal{Q}}(f)}$. Concretely, one may consider a sequence of operators of the form 
\begin{equation}
\tilde{\psi}_{P}(\delta^{2}\op{\mathcal{Q}})
= \int d\gamma(f)\psi(f)e^{i\delta^{2}\op{\mathcal{Q}}(Pf)} ,
\end{equation}
where $P$ denotes a (finite-rank) projection onto $\bar{\mathcal{S}}$. For each 
$P$, the exponential defines a genuine unitary operator, and hence $\tilde{\psi}_{P}(\delta^{2}\op{\mathcal{Q}})$ is a bounded operator on the Hilbert space. The operator $\tilde{\psi}(\delta^{2}\op{\mathcal{Q}})$ is defined by taking appropriate limits of such sequences in the weak operator topology. 

We will simply ignore these more subtle issues of functional analysis and simply proceed by working formally with \eqref{eq:genop}. All manipulations will be justified by verifying that the resulting expressions are well-defined. For example, the matrix elements of \eqref{eq:genop} with respect to the dense set of states $\ket{\Psi_{f}}$ are given by
\begin{equation}
\label{eq:matrixelement}
\braket{\Psi_{f_{1}}|\tilde{\psi}(\delta^{2}\op{\mathcal{Q}})|\Psi_{f_{2}}}
    = e^{-\frac{1}{4}S[f_2-f_1]} \mathcal{Z}^{-1} \int_{\mathcal{S}} \mathscr{D} f^{\prime} \; \psi(f^{\prime}) e^{-\frac{5}{4}S[f^{\prime}]} e^{-\frac{1}{2}S[f_{2}-f_{1},f^{\prime}]},
\end{equation}
It is straightforward to show that this integral converges\footnote{As a simple example, consider the case where $\psi(f)=1$ is the constant function on all of $\mathcal{S}$. In this case, $\braket{\Psi_{f_{1}}|\tilde{\psi}(\delta^{2}\op{\mathcal{Q}})|\Psi_{f_{2}}}=(5/4)^{2/3}e^{-\frac{1}{5}S[f_{1}-f_{2}]}$ which is finite for $f_{1},f_{2}\in \bar{\mathcal{S}}$. It is also straightforward to check that the associated form is lower semibounded, with largest lower bound $\inf_{f \in \bar{\mathcal{S}}} e^{-8S[f]}$.} for any $\psi(f)$ of the form of a stochastic exponential. Such wavefunctions are dense in the full Hilbert space $\mathscr{H}_{\mathcal{S}_{0}}=L^{2}(\mathcal{S}_{0};\gamma)$. Furthermore, by considering the diagonal matrix elements it is straightforward to show that the map $\ket{\Psi}\to \braket{\Psi|\tilde{\psi}(\delta^{2}\op{\mathcal{Q}})|\Psi}$ is lower-semibounded --- i.e., the infimum of the ratio $\braket{\Psi|\tilde{\psi}(\delta^{2}\op{\mathcal{Q}})|\Psi}/||\Psi||^{2}$ is finite over the dense domain. These properties indicate that (the closure of) \eqref{eq:genop} can be defined as a good operator for $\psi(f)$ on a larger domain than $\bar{\mathcal{S}}$ \cite{Schmudgen2012}. It would be of interest to analyze the precise convergence properties of sequences of the operators $\{\tilde{\psi}_{P}(\delta^{2}\op{\mathcal{Q}})\}$ as well as identifying the full class of functions $\psi(f)$ for which the corresponding limits exist. Nevertheless, the above arguments indicate that \eqref{eq:genop} corresponds to a well-defined operator in $\mathfrak{A}(\mathcal{S};\gamma)$ for a large class of functions.

From this formal ``path integral'' picture, the vector $\ket{e_{\gamma}}$  satisfies all of the properties of the neutral weight $\ket{e_{\textrm{H}}}$ in sec.~\ref{sec:isom} 
\begin{equation} \label{eq:neutral1}
\braket{e_{\gamma}|\tilde{\psi}(\delta^{2}\op{\mathcal{Q}})|e_{\gamma}} = \psi(0)
\end{equation}
as well as 
\begin{equation}
\label{eq:innerprod}
\braket{e_{\gamma}|\tilde{\psi}(\delta^2 \op{\mathcal{Q}})^{\dagger} \tilde{\psi}(\delta^2 \op{\mathcal{Q}})|e_{\gamma}} = \int_{\mathcal{S}} d\gamma(f) \overline{\psi(f)} \psi(f).
\end{equation}
By these arguments, the Hilbert space $\mathscr{H}_{\mathcal{S}}$ is the GNS Hilbert space with respect to $\ket{e_{\gamma}}$. Consequently, in addition to being  semifinite and normal as noted above, \eqref{eq:innerprod} implies that it is also faithful. We will simply assume these properties hold without restriction (see footnote \ref{foot:faithful}) in the remainder of this paper. 

We now  construct the dressed algebra, $\mathfrak{A}_{\textrm{dress.}}(\mathscr{R};\mathcal{S})$, and its associated trace. With the Hilbert space $\mathscr{H}_{\mathcal{S}}$ defined, the dressed algebra is given by 
\begin{equation}
\mathfrak{A}_{\textrm{dress.}}(\mathscr{R};\mathcal{S}) \defn \{\delta\op{\gamma}(w;\op{f}),\delta^{2}\op{\mathcal{Q}}(f)| ~f\in\bar{\mathcal{S}}\}'',
\end{equation}
closed in the weak operator topology of $\mathscr{F}\otimes \mathscr{H}_{\mathcal{S}}$, as described in sec.~\ref{subsec:dressedalg}. Here, $\delta\op{\gamma}(w;\op{f})$ are the dressed fields in $\mathscr{R}$ invariant under any boost supertranslations $f$ such that the projection $f\vert_{\ell>0}$ to higher spherical harmonics is an element of $\bar{\mathcal{S}}$. The trace on this algebra is given by 
\begin{equation}
\label{eq:tracesupertranslations}
\tau_{\mathcal{S}}(\hat{a}) =\braket{\omega,0_{\mathbb{R}},e_{\gamma}|e^{\op{X}/2}\hat{a}e^{\op{X}/2}|\omega,0_{\mathbb{R}},e_{\gamma}}
\end{equation}
where $\hat{a}\in \mathfrak{A}_{\textrm{dress.}}$ and we recall that $\op{X}\defn \delta^{2}\op{\mathcal{Q}}(1)/4G_{\textrm{N}}\beta$. The operators 
\begin{equation}
\hat{a} = e^{-i\op{F}(\op{f})}ae^{i\op{F}(\op{f})}e^{i\delta^{2}\op{\mathcal{Q}}(f)}
\end{equation}
form an additive basis for the algebra in the sense defined above. As explained in sec.~\ref{sec:trace}, the key property that we need to check is 
\begin{equation}
\label{eq:tracecheck}
\braket{\omega,0_{\mathbb{R}},e_{\gamma}|\hat{a}\hat{b}|{\omega,0_{\mathbb{R}},e_{\gamma}}} = \braket{\omega,0_{\mathbb{R}},e_{\gamma}|\hat{b}e^{-\op{X}}\hat{a}e^{\op{X}}|{\omega,0_{\mathbb{R}},e_{\gamma}}}.
\end{equation}
By \eqref{eq:traceabba} this property directly implies that $\tau_{\mathcal{S}}$ is a trace on the dressed algebra. We first note that the following algebraic properties used in sec.~\ref{sec:trace} remain true for the algebra $\mathfrak{A}_{\textrm{dress.}}(\mathscr{R};\mathcal{S})$ 
\begin{enumerate}
\item $\ket{\omega}$ is annihilated by $\op{F}(f)$ for any $f$ and $\op{K}_{\omega}=\op{F}(1)$ is the modular Hamiltonian of this state.\label{prop1}
\item $e^{i\delta^{2}\op{\mathcal{Q}}(f)}\ket{e_{\gamma}} = \ket{f}$\label{prop2}
\item $e^{i\op{K}_{\omega}\op{t}}e^{it\op{X}}e^{-i\op{K}_{\omega}\op{t}}=e^{it(\op{X}+\op{K}_{\omega})}$ \label{prop3}
\item $e^{i\op{F}(\op{f})}e^{i\delta^{2}\op{\mathcal{Q}}(f)}e^{-i\op{F}(\op{f})}=e^{i\delta^{2}\op{\mathcal{Q}}(f)}e^{i\op{F}(\op{f})}$ \label{prop4}
\end{enumerate}
where the last relation is simplified relative to \eqref{eq:commutation} since the group $\mathcal{S}$ is abelian. Consequently, we can apply identical manipulations of the right-hand  side of \eqref{eq:tracecheck} as we did for the right-hand  side of \eqref{eq:modX} in sec.~\ref{sec:trace} up to eqn. \eqref{RHS for NLC}. These manipulations yield 
\begin{equation}
\textrm{R.H.S.} = \int \frac{dX}{2\pi} e^{i(t'+t)X} \bra{\omega, e_{\gamma}} a e^{-i\op{F}(f)}e^{-it\op{K}_{\omega}} b \ket{\omega,f+f'}.
\end{equation}
Similarly, the manipulations of the left-hand  side of \eqref{eq:tracecheck} are identical to those considered in sec.~\ref{sec:trace} of the left-hand  side of \eqref{eq:modX} up to \eqref{LHS for NLC}. This yields 
\begin{flalign} \label{LHS}
	\textrm{L.H.S.} = \int \frac{dX}{2\pi} e^{i(t+t')X} \bra{\omega,e_{\gamma}} a e^{i\op{F}(f')} e^{it'\op{K}_{\omega}} b \ket{\omega,f'+f}. 
\end{flalign}
Using the fact that the integral over $X$ yields the delta function $\delta(t+t^{\prime})$ and $\braket{e_{\gamma}|f+f^{\prime}}=\delta_{\gamma}(0,f+f^{\prime})$
we obtain
\begin{flalign}
	\textrm{R.H.S.} &= \int \frac{dX}{2\pi} e^{i(t'+t)X} \bra{\omega} a e^{i\op{F}(f^{\prime})}e^{-it\op{K}_{\omega}} b \ket{\omega}\delta_{\gamma}(0,f+f^{\prime}), \nonumber \\
	\textrm{L.H.S.} &= \int \frac{dX}{2\pi} e^{i(t+t')X} \bra{\omega} a e^{i\op{F}(f')} e^{-it\op{K}_{\omega}} b \ket{\omega}\delta_{\gamma}(0,f^{\prime}+f). 
\end{flalign}
where we permuted $f$ and $f^{\prime}$ since the group is abelian. Therefore, $\tau_{\mathcal{S}}$ is a densely defined trace on $\mathfrak{A}_{\textrm{dress.}}(\mathscr{R};\mathcal{S})$. 

Given this trace, it is straightforward to construct density matrices associated to any state of the form $\ket{\Phi_{\alpha}} = \ket{\varphi}\otimes \ket{\alpha}$ where $\ket{\varphi}\in \mathscr{F}$ and $\ket{\alpha}\in \mathscr{H}_{\mathcal{S}}$. By the same manipulations as in sec.~\ref{subsec:densityops}, the density matrix is given by  
\beq \label{Density operator for CQ state2}
	\rho_{\hat{\Psi}_\alpha} = \tilde{\alpha}(\delta^{2}\op{\mathcal{Q}}) e^{-i\op{F}(\op{f})}e^{-i\op{K}_{\omega}\op{t}} e^{-\op{X}/2} \op{\Delta}_{\varphi \mid \omega} e^{-\op{X}/2} e^{i\op{K}_{\omega}\op{t}}e^{i\op{F}(\op{f})}\tilde{\alpha}(\delta^{2}\op{\mathcal{Q}})^{\dagger},
\eeq
where 
\begin{equation}
\tilde{\alpha}(\delta^{2}\op{\mathcal{Q}}) \defn \int d\gamma(f) \alpha(f)e^{i \delta^{2}\op{\mathcal{Q}}(f)}.
\end{equation}
For any ``semiclassical'' state the von Neumann entropy is given by 
\begin{equation}
\label{eq:SvNS}
S_{\textrm{vN}}(\rho_{\hat{\Phi}_{\alpha}}) \approx \braket{\beta \op{X}}_{\hat{\omega}_{\alpha}} - S_{\textrm{rel.}}(\varphi|\omega) + S(\rho_{\alpha}) + \log(\beta).
\end{equation}

\subsection{The Full, Dressed Subregion Algebra is a Type II algebra}
\label{subsec:fulldressed}
In the previous two sections we have seperately shown that the algebra $\mathfrak{A}_{\textrm{dress}}(\mathscr{R},\mathcal{S})$ satisfying the boost supertranslation constraints as well as the algebra $\mathfrak{A}_{\textrm{dress}}(\mathscr{R},\mathbb{R}\times H_{\textrm{isom.}})$ satisfying isometry constraints are Type II algebras. In this section we combine these results to show that the full dressed subregion algebra $\mathfrak{A}_{\textrm{dress}}(\mathscr{R})$ is Type II. 

We first recall that the group $G$ is also of the form $\mathbb{R}\times H$ where now $H=H_{\textrm{isom.}}\ltimes \mathcal{S}_{0}$. Given a measure $\gamma$ on the group of boost supertranslations, we obtain a measure on the full group $G$ as 
\begin{equation}
d\mu_{G}(f,\psi)\defn dt d\mu_{H}(\psi)d\gamma(f)
\end{equation}
where $d\mu_{H}(\psi)$ is the invariant Haar measure on $H_{\textrm{isom.}}$. Following the previous section, we shall choose $\gamma$ to be a Gaussian measure on $\mathcal{S}_{0}$. The corresponding Hilbert space $\mathscr{H}_{\textrm{G}}\defn L^{2}(G;d\mu_{G})$ consists of states $\Psi(f,\psi)$ with finite norm 
\begin{equation}
||\Psi|| = \int d\mu_{G}(f,\psi)~ \overline{\Psi(f,\psi)}\Psi(f,\psi).
\end{equation}
where in the above expression, $f\in\mathcal{S}$ including the constant mode. We recall that the measure $dtd\mu_{H}(\psi)$ is invariant under the constant boosts and rotations. As we explained in the previous section, the measure $\gamma(f)$ will not be invariant under the action of higher-harmonic boost supertranslations since any choice of $S[f]$ is not invariant under the shift $f\to f+f^{\prime}$. This was the primary issue dealt with in that section in which we showed that one must more carefully consider the unitary action of boost supertranslations in the construction of the group von Neumann algebra. As we are now including rotations and, since the boost supertranslations are not rotation invariant, it appears that one might have to repeat this analysis to account for this non-trivial action. However, it is clear that $\gamma$ can be chosen such that no new complications arise. 
This follows from the fact that any choice of action $S[f]$ on $(q_{AB},\mathbb{S}^{2})$ that is local and covariant with respect to $q_{AB}$ will, in fact, be invariant under the action of $H_{\textrm{isom.}}$. An example of such an action is given by \eqref{eq:Sf}. Under this mild restriction, the measure $\gamma$ is also rotationally invariant. The full Hilbert space is simply 
\begin{equation}
\mathscr{H}_{\textrm{G}} = \mathscr{H}_{\mathbb{R}}\otimes \mathscr{H}_{\textrm{H}_{\textrm{isom.}}}\otimes \mathscr{H}_{\mathcal{S}_{0}}.
\end{equation}
For the remainder of this section, we will assume that $\gamma$ is chosen such that its associated action is local and covariant. 

We will now construct the trace, density matrices and entropy of states on the full dressed algebra. For simplicity we will first restrict to AdS black hole spacetimes where \eqref{eq:opgammasigma} holds so that $\mathfrak{A}_{\textrm{dress.}}(\mathscr{R})\cong \mathfrak{A}_{\textrm{dress.}}(\mathcal{H}^{-})$. We treat the case of de Sitter spacetime separately in sec.~\ref{sec:dSII1}. Following the arguments of sec.~\ref{sec:isom} and sec.~\ref{sec:PIforSuper} it directly follows that the algebra $\mathfrak{A}_{\textrm{dress.}}(\mathscr{R})$ on $\mathscr{F}\otimes \mathscr{H}_{\textrm{G}}$ is a Type II$_{\infty}$ algebra with trace 
\begin{equation}
\label{eq:traceG}
\tau(\hat{a}) = \braket{\omega,0_{\mathbb{R}},e_{H},e_{\gamma}|e^{\op{X}/2}\hat{a}e^{\op{X}/2}|\omega,0_{\mathbb{R}},e_{H},e_{\gamma}}
\end{equation}
for any $\hat{a}\in \mathfrak{A}_{\textrm{dress.}}(\mathscr{R})$ where we recall that $\ket{e_{H}}$ is the neutral element of the $H_{\textrm{isom.}}$ and $\ket{e_{\gamma}}$ is the neutral element of $\mathcal{S}_{0}$. For any state $\ket{\hat{\Phi}_{\alpha}}=\ket{\varphi}\otimes \ket{\alpha}$ where $\alpha$ is a wavefunction on $\mathscr{H}_{\textrm{G}}$, the density matrix of this state on $\mathfrak{A}_{\textrm{dress.}}(\mathscr{R})$ is given by \eqref{Density operator for CQ state1} where $\tilde{\alpha}$ in that expressions now corresponds to the ``Fourier transform'' of the wavefunction $\alpha$ with respect to the group $G$ and measure $\mu_{G}$ (see \eqref{eq:f} and \eqref{eq:genop}). By nearly identical manipulations as presented in sec.~\ref{sec:R} and sec.~\ref{subsec:densityops}, the entropy of any semiclassical state is given by 
\begin{equation}
\label{eq:Sgen2}
S_{\textrm{vN.}}(\rho_{\hat{\Phi}_{\alpha}}) \simeq  S_{\textrm{gen.}}(\hat{\Phi}_{\alpha}\vert_{\mathscr{R}}) + S(\rho_{\alpha})+C
\end{equation}
where $S(\rho_{\alpha})$ encodes the entropy of the fluctuations of the charges $\delta^{2}\op{\mathcal{Q}}^{\textrm{R}}(f,\psi)$. In spacetimes where \eqref{eq:opgammasigma} doesn't hold, the initial data must be further supplemented by data at past infinity. As we reviewed in sec.~\ref{subsec:kerr}, for gravitational perturbations of an asymptotically flat black hole one must further include a Type I algebra $\mathfrak{A}(\mathscr{I}^{-})$ of incoming radiation from past null infinity. 
\begin{equation}
\mathfrak{A}_{\textrm{dress.}}(\mathscr{R}) \cong \mathfrak{A}_{\textrm{dress.}}(\mathcal{H}_{\textrm{R}}^{-})\otimes \mathfrak{A}(\mathscr{I}^{-}) \quad \quad \textrm{(asymptotically flat)}.
\end{equation}
This Type I algebra simply modifies the trace \eqref{eq:traceG} by an ordinary Hilbert space trace on the degrees of freedom at $\mathscr{I}^{-}$ (see, e.g., \eqref{eq:tracekerr}). Finally, as explained in \cite{Kudler-Flam:2023qfl}, for black holes in de Sitter space, the region $\mathscr{R}$ is bounded by the black hole horizon $\mathcal{H}^{-}_{R,b}$ and the cosmological horizon $\mathcal{H}^{-}_{L,c}$. Thus, the local graviton algebra $\mathfrak{A}(\mathscr{R})$ decomposes into the product of two algebras $\mathfrak{A}(\mathcal{H}^{-}_{R,b})$ and $\mathfrak{A}(\mathcal{H}^{-}_{R,c})$ both of which are Type III. Following the arguments of sec.~$7$ of \cite{Kudler-Flam:2023qfl}, the full dressed algebra becomes 
\begin{equation}
\mathfrak{A}_{\textrm{dress.}}(\mathscr{R}) \cong \mathfrak{A}_{\textrm{dress.}}(\mathcal{H}^{-}_{R,b})\otimes \mathfrak{A}_{\textrm{dress.}}(\mathcal{H}^{-}_{L,c}) \quad \quad \textrm{(asymptotically de Sitter)}
\end{equation}
the product of two Type II$_{\infty}$ algebras and therefore is Type II$_{\infty}$. Therefore, for any stationary black hole spacetime, $\mathfrak{A}_{\textrm{dress.}}(\mathscr{R})$ is a Type II$_{\infty}$ algebra. In all cases, the generalized entropy continues to be given by \eqref{eq:Sgen2}. As indicated in sec.~\ref{subsec:deSitter} the treatment of de Sitter spacetime, being a closed universe, requires a separate analysis which we present in sec.~\ref{sec:dSII1}.

We  conclude this section with some speculative comments on the full symmetry group of $\mathscr{R}$ and the ``matching'' of the charges $\delta^{2}\mathcal{Q}^{\textrm{R}}(f,\psi)$ to other charges defined on the boundary of $\mathscr{R}$. While such a matching is not strictly necessary for the arguments presented in this paper, the conjectured relationship between symmetries on black hole horizons and those at infinity have played a significant role in discussions of “edge modes’’ in black hole spacetimes (see, e.g., \cite{Hawking:2016msc,Flanagan:2015pxa,Chandrasekaran:2018aop,Knysh:2024asf}) . We therefore take this opportunity to clarify the status of these ideas and to comment on their relevance for the considerations of the present work.

In an asymptotically flat spacetime, the complete symmetry group of $\mathscr{R}$ is comprised of the symmetries on the black hole horizon together with the symmetries at infinity. In four spacetime dimensions, the symmetries at null infinity are the BMS group which has a similar structure to the group $G=H_{\textrm{isom.}}\ltimes \mathcal{S}$ on the black hole horizon where $H_{\textrm{isom.}}$ is replaced by the Lorentz group $\textrm{SO}(1,3)$ and  $\mathcal{S}$ is now interpreted as the abelian group of supertranslations at null infinity \cite{Sachs:1962zza}. One might expect that the symmetry group defined at infinity is independent of the symmetries of the horizon --- i.e., one can perform independent actions of the symmetry group on $\mathcal{H}^{-}$ and $\mathscr{I}^{-}$. However, this is not entirely the case. As we reviewed in \ref{subsec:kerr}, the symmetry generators of the constant boosts and rotations on the horizon actually match to corresponding generators of time translations and rotations at infinity \cite{Kudler-Flam:2023qfl}. The matching of their corresponding charges at $i^{-}$ directly follows from Einstein's equation and its proof strongly uses the fact that there is a unique extension of these symmetries off of the $\mathcal{H}^{-}$ and $\mathscr{I}^{-}$ as isometries in the bulk. 

In a similar manner to how the charges $\delta^{2}\mathcal{Q}(f)$ correspond to ``angularly weighted averages'' of the area, the supertranslations at infinity generate ``supermomentum'' charges $\delta^{2}M(f)$  which correspond to ``angularly weighted averages'' of the mass at infinity (see, e.g., \cite{Prabhu:2022zcr} for a precise definition). For gravitational perturbations of a Schwarzschild black hole\footnote{For a Reissner-Nordstrom black hole the proposed analog of \eqref{eq:matchconj} presumably involves the inclusion of large gauge charges of the electromagnetic field at $i^{-}$ \cite{Prabhu:2022zcr}. The generalization of \eqref{eq:matchconj} to Kerr is far less clear since any proposed matching should generalize \eqref{eq:XQpsiiminus}. This may involve proposed extensions of the BMS group to include ``superrotations'' (or ``super-Lorentz'') charges at $\mathscr{I}$ \cite{Barnich:2011mi,Campiglia:2015yka}. However, such extensions introduce new complications in both the classical \cite{Compere:2023qoa} and the quantum theory (see sec.~\ref{subsec:symmquantalg}). }, a natural generalization of the relation   $\delta^{2}\mathcal{Q}^{\textrm{R}}(1)/4G_{\textrm{N}}\beta = \delta^{2}{M}_{i^{-}}(1)$ is
\begin{equation}
\label{eq:matchconj}
\frac{\delta^{2}\mathcal{Q}^{\textrm{R}}(f)}{4G_{\textrm{N}}\beta} = \delta^{2}M_{i^{-}}(f)
\end{equation}
for any $f$ and there is a similar matching of perturbed charges at $i^{+}$. Eq.~\eqref{eq:matchconj} is the perturbative version of a charge matching condition that was previously conjectured a decade ago by Hawking, Perry and Strominger  \cite{Hawking:2016msc} (see also \cite{Chandrasekaran:2018aop}). The matching of these symmetries is similar to an analogous problem of matching the supermomentum charges at $i^{0}$. The matching of the $f=1$ charges corresponds to conservation of the ADM mass \cite{PhysRevLett.43.181}, and the matching of the higher harmonic charges was originally proposed by Strominger \cite{Strominger:2013jfa}. Despite the fact that supertranslations are not isometries of the spacetime, the “weak-field’’ nature of spatial infinity drastically simplifies Einstein’s equations, and the matching of supermomentum charges at $i^{0}$ can be established non-perturbatively \cite{Herberthson:1992gcz,Troessaert:2017jcm,Magdy:2021rmi,Prabhu:2019fsp}. By contrast, in the presence of black holes, timelike infinity is a “strong-field’’ regime, and a comparable perturbative or non-perturbative analysis of Einstein’s equations is far more difficult to carry out\footnote{In $d>4$ dimensions, gravitational perturbations decay sufficiently fast that one can reduce the BMS group at infinity to the finite dimensional Poincaré group \cite{Hollands:2016oma}. Nevertheless, one can still define ``angular weighted averages'' of the mass which, in contrast to the situation in four dimensions are not related to any asymptotic symmetries \cite{Pate:2017fgt,Satishchandran:2019pyc}. If \eqref{eq:matchconj} holds in $d=4$, it is conceivable that these quantities may similarly match to the higher-dimensional version of $\delta^{2}\mathcal{Q}^{\textrm{R}}(f)$ at timelike infinity.}. We hope that the role played by these charges in the quantum theory of black holes, as highlighted in the present work, provides additional motivation for revisiting their possible matching in the classical theory. 

We now comment on similar matching problems for black holes in asymptotically Anti-de Sitter and de Sitter spacetimes. For black holes in de Sitter spacetime, the natural quantity that would replace the right-hand  side of \eqref{eq:matchconj} is the corresponding limit of the charge $\delta^{2}\mathcal{Q}(f)$ to $i^{-}$ along the cosmological horizon. For $f=1$, one can directly prove that the corresponding charges match \cite{Kudler-Flam:2023qfl}, and such a relation would be an obvious generalization of this result for any $f$. In sec.~\ref{sec:dSII1} we will show that, for a test body in de Sitter spacetime, a similar matching can be shown where $\delta^{2}\mathcal{Q}^{\textrm{R}}(f)$ matches onto the ``mass multipoles'' of the body. For black holes in AdS, the higher harmonic charges of the black hole do not appear to be directly related to any asymptotic symmetries of the spacetime. However, the supertranslation symmetries appear to emerge in the hydrodynamic effective field theory of the CFT (see, e.g., \cite{Knysh:2024asf,liu2025hydrocft} and sec.~\ref{disc:AdSCFT} for further details). As in the case of flat spacetime, it would be of interest to explore the matching of these charges to determine the complete symmetry group of $\mathscr{R}$. 

We emphasize that while the conjectured matching of the charges detailed above affects the nature of the global symmetry group of $\mathscr{R}$, their validity (or lack thereof) do not affect the arguments presented in this paper. As in the case of the matching of $\delta^{2}\op{\mathcal{Q}}^{\textrm{R}}(1)$, the matching of the higher harmonic charges do not affect the type of the algebra $\mathfrak{A}_{\textrm{dress.}}(\mathscr{R})$. These arguments affect only the relationship between the charges $\delta^{2}\op{\mathcal{Q}}^{\textrm{R}}(f)$ and other charges defined on the boundary of $\mathscr{R}$.

\subsection{The Dressed Algebra with Matter Fields} \label{sec: CP for NLC Groups}

Thus far we have focused on quantum gravitational perturbations of any spacetime with a bifurcate Killing horizon where we obtained constraints on the first-order gravitational field arising from the group $G=H_{\textrm{isom.}}\ltimes \mathcal{S}$.  In this section, we explore the effects of additional matter fields. The ``edge modes'' of general gauge theories (e.g., electromagnetic or Yang-Mills fields) have long been known to be related to an infinite dimensional group of ``large gauge transformations'' on the boundary. Additionally, they have played a significant role in recent computations of entanglement entropy of subregions in gauge theory \cite{Donnelly:2011hn,Donnelly:2016auv,Freidel:2020xyx,Freidel:2021cjp,Ciambelli:2021vnn,Ciambelli:2021nmv,Ciambelli:2022vot,Klinger:2023qna,Freidel:2023bnj,Ciambelli:2024swv,Klinger:2023tgi,Klinger:2023auu,AliAhmad:2024wja,AliAhmad:2024vdw,Donnelly:2022kfs,Speranza:2017gxd,Geiller:2019bti,Ball:2024hqe,Ball:2024xhf,Ball:2024gti,Araujo-Regado:2024dpr,Carrozza:2022xut}. In sec.~\ref{subsubsec:LGTmatter} we outline how these edge modes arise in our context. This motivates a further extension of our analysis to consider the crossed product with respect to infinite dimensional, non-abelian groups. In sec.~\ref{subsubsec:Hnonloc} we consider this much more general problem and obtain some sufficient conditions under which the dressed algebra remains Type II. 

\subsubsection{Large Gauge Transformations and Matter Fields}
\label{subsubsec:LGTmatter}
We assume these matter fields do not significantly backreact on the spacetime --- i.e., their stress energy is $O(1)$ in powers of $G_{\textrm{N}}$. In this case, the additional quantum fields propagate on the background spacetime and satisfy decoupled equations from that of the gravitons. The full Hilbert space is simply a tensor product Hilbert space  
\begin{equation}
\mathscr{F}^{\textrm{matt.}}\otimes \mathscr{F}^{\textrm{grav.}}
\end{equation}
where, $\mathscr{F}^{\textrm{grav.}}$ is the Fock space of gravitons and $\mathscr{F}^{\textrm{matt.}}$ is the Hilbert space of any matter fields we wish to consider. Similarly, the undressed von Neumann algebra is 
\begin{equation}
\mathfrak{A}^{\textrm{matt.}}(\mathscr{R})\otimes \mathfrak{A}^{\textrm{grav.}}(\mathscr{R})
\end{equation}
where $\mathfrak{A}^{\textrm{matt.}}(\mathscr{R})$ is the  Type III algebra of matter fields in $\mathscr{R}$. 
The gravitational constraints in the presence of matter are of the form of \eqref{eq:QFlux1} and \eqref{eq:QFluxX} where now the flux $F(f)$ receives an additional contribution due to the angular distribution of the infalling matter --- i.e., in \eqref{eq:Flambdaf} we replace $\sigma^{AB}\sigma_{AB}\to \sigma^{AB}\sigma_{AB} + 8\pi T_{UU}$ --- and, similarly, the flux $F(\psi)$ recieves an additional contribution due to the integrated flux of $T_{UA}\psi^{A}$ of matter angular momentum across the horizon. Thus, in the quantum theory, the gravitational constraints recieve contributions due to the energy and angular momentum of the matter and so the matter fields must be gravitationally dressed in an identical manner to the dressing of the gravitons considered in this paper. 

The key difference that arises in the quantization of the theory with matter is that if the region $\mathscr{R}$ contains any gauge fields then these fields will introduce additional constraints arising from ``large gauge transformations'' of the matter fields. As noted by CLPW, these constraints are of a form similar to that of \eqref{eq:QFlux1} and \eqref{eq:QFluxX}, and therefore predicate the inclusion of new charges which generate the large gauge transformations resulting in a ``crossed product'' algebra. As pointed out by several authors \cite{Donnelly:2011hn,Donnelly:2016auv,Klinger:2023qna,Klinger:2023tgi,Klinger:2023auu,Ball:2024hqe,Ball:2024xhf,Ball:2024gti}, gauge theories defined in subregions generally admit an infinite dimensional group of ``large'' gauge transformations, and so will generally give rise to an infinite set of constraints. 

As a simple but illustrative example, suppose that our matter is electromagnetically charged, in which case we must account for gauge transformations of the form 
\begin{equation}
\label{eq:GT}
A_{a}\to A_{a}+ \nabla_{a}\vartheta
\end{equation}
together with gauge transformations of any charged matter fields. 
To consider the group of large gauge transformations in $\mathscr{R}$ we first note that, in an analogous manner to conditions imposed in sec.~\ref{sec:killinghorizon}, we are free to impose the condition that $A_{a}n^{a}\vert_{\mathcal{H}}=0$ where $n^{a}$ is the null normal to the horizon. The corresponding group of large gauge transformations is then simply the group of gauge transformations where 
\begin{equation}
\vartheta\vert_{\mathcal{H}}=\vartheta(x^{A})
\end{equation}
is non-vanishing on the horizon and satisfies any additional asymptotic conditions at infinity. We note that, in asymptotically flat spacetimes, one obtains a similar group of large gauge transformations at infinity and, it was conjectured in \cite{Hawking:2016msc} that the symmetries we consider on the horizon ``match'' onto the symmetries at infinity\footnote{As in the previous section, the matching of the global gauge $\vartheta=1$ can be trivially proven by the conservation of total electric charge.}. As in the rest of this paper, our results will be independent of the precise asymptotic behavior of the spacetime and so we will simply focus on the group $G_{\textrm{LGT}}$ of large gauge transformations on the horizon. 

 The large gauge transformations on the horizon generate charges which, on any cut $S_{U}$ of the horizon, are given by \cite{Hawking:2016msc}
 \begin{equation}
 \label{eq:chargefluxalpha}
\mathcal{Q}_{U}(\vartheta) = \int_{S_{U}}d\Omega~E_{r}(U,x^{A})\vartheta(x^{A}).
 \end{equation}
Integrating Maxwell's equation along the horizon yields the ``charge-flux'' relation
\begin{equation}
\label{eq:QLQRalpha}
\mathcal{Q}^{\textrm{L}}(\vartheta) - \mathcal{Q}^{\textrm{R}}(\vartheta) = -F(\vartheta)
\end{equation}
\begin{equation}
\label{eq:fluxalpha}
 F(\vartheta) \defn- \int_{\mathcal{H}^{-}}dUd\Omega~ \vartheta(x^{A})\big[j_{U}(U,x^{A}) + \mathscr{D}^{A}E_{A}(U,x^{A})\big].
\end{equation}
The first term is the angular distribution of charge-current $j_{U}=j_{a}n^{a}\vert_{\mathcal{H}}$ crossing the horizon and is analogous to the angular distribution of gravitational wave energy-flux entering into \eqref{eq:QFlux1}. The second term is sometimes referred to as the ``soft'' contribution to the flux where $E_{A}=-\partial_{U}A_{A}$ is the pull-back of the electric field $E_{a}=F_{ab}n^{b}$ to the horizon\footnote{Since we are considering initial data on $\mathcal{H}^{-}$ we may choose initial data such that $A_{A}$ decays as $U\to \pm \infty$ so that the second term in \eqref{eq:fluxalpha} identically vanishes. We note that second term can be non-vanishing on the future horizon \cite{Hawking:2016msc,Danielson:2022sga,Danielson:2022tdw,Satishchandran:2025cfk} or at future null infinity \cite{Strominger:2017zoo,asymp-quant,Prabhu:2022zcr,Prabhu:2024lmg} and, in either case, can give rise to infrared divergences associated with the emission of an infinite number of soft quanta. The resulting quanta states lie in inequivalent sectors labeled by the value of the ``soft flux''. To accommodate these sectors one must also enlarge the von Neumann algebra in the manner described in \cite{Danielson:2025aji}.} and so the second term depends only on the ``zero frequency'' part of $A_{A}$ on the horizon. At null infinity, one obtains a very similar charge-flux relation for the infinite set of angle-dependent large gauge charges at infinity \cite{Strominger:2017zoo}. 

The key point is that the flux $F(\vartheta)$ generates large gauge transformations on the phase space and thereby yields an infinite set of non-trivial constraints. We note that, in particular, for $\vartheta=1$, the second term in $F(1)$ vanishes and \eqref{eq:QLQRalpha} is simply the constraint of global charge conservation. In the quantum theory,
to construct the full dressed algebra of $\mathscr{R}$ the physical observables must commute with $\op{\mathcal{Q}}^{\textrm{L}}(\vartheta)=\op{\mathcal{Q}}^{\textrm{R}}(\vartheta)-\op{F}(\vartheta)$. This constraint is analogous to the constraints we have considered in this paper and can be solved by $\op{\mathcal{Q}}^{\textrm{R}}(\vartheta)$ on $L^{2}(G_{\textrm{LGT}};\mu)$ where $\mu$ is a Gaussian measure. The construction of dressed operators proceeds in an identical manner. 

However, there are a couple of caveats in directly extending the arguments of our paper to more general gauge theories. The first is that, in general, the large gauge group is an infinite dimensional non-abelian group. For any Yang-Mills field, the angle-dependent large gauge transformations are now Lie algebra valued and the corresponding charges satisfy ``charge-flux'' relations similar to that of \eqref{eq:fluxalpha} where the current flux is replaced by the flux of gluons (see \cite{Strominger:2017zoo,Prabhu:2022zcr} for similar formulas at null infinity). Thus, in an unconfined phase, one will obtain constraint equations for this infinite dimensional, non-abelian group. The relevant group of symmetries that impose non-trivial constraints on the matter fields and gravitons is  
\begin{equation}
\label{eq:LGTgroup}
G=H_{\textrm{isom.}}\ltimes (\mathcal{S}\times G_{\textrm{LGT}})
\end{equation}
where, in general, $G_{\textrm{LGT}}$ is a non-abelian, infinite dimensional group. As we explain in the following section, the analysis of these groups is far more complicated than in the case of the abelian group that we have considered thus far. 

The second caveat is that, as opposed to the free gravitons considered in the majority of this paper, the algebra $\mathfrak{A}^{\textrm{matt.}}(\mathscr{R})$ is, in general, an interacting quantum field theory on a curved spacetime. In the case of a free theory of scalars, photons or gravitons one can explicitly check the invariance of the vacuum under the large gauge transformations considered in this paper as well as the action of $\op{F}(\vartheta)$ as a self-adjoint operator on the Hilbert space. These properties are far less straightforward to check in an interacting theory. We note the analogous flux operators at infinity are well-defined, self-adjoint operators \cite{Prabhu:2022zcr} so the key property one must check is that this is also the case on black hole horizons. While we think it is very likely that these symmetries are not broken in the quantum theory, we know of no proof of this statement in a general interacting quantum theory. Such a proof is outside of the scope of this paper. In the remainder of this section, we will simply assume that these properties remain when one considers an interacting theory (see \ref{ass1} of sec.~\ref{subsubsec:Hnonloc}). 

In summary, from the above discussion we have learned that the inclusion of general matter fields in $\mathscr{R}$ can introduce an additional, infinite set of constraints arising from a (possibly non-abelian) infinite dimensional large gauge group. With these applications in mind, in the following section we consider the crossed product with respect to a non-abelian, infinite dimensional group. 

\subsubsection{The Algebra $\mathfrak{A}_{dress.}(\mathscr{R},\mathbb{R}\times H)$ for $H$ non-locally compact}
\label{subsubsec:Hnonloc}

In this section, we consider the crossed product with respect to an infinite-dimensional, non-abelian group. Our strategy is to follow an approach analogous to that used in the infinite-dimensional abelian case studied in sec.~\ref{sec:PIforSuper}. We formulate a set of sufficient conditions, \ref{ass1}–\ref{ass4} below, which appropriately generalize the arguments of sec.~\ref{sec:PIforSuper}. Under these conditions, we show that the resulting dressed algebra is of Type~II. As we will see, in the general non-abelian setting, verifying some of these conditions requires a more detailed understanding of interacting quantum field theories and infinite-dimensional measures. Nevertheless, we hope that the results of this section provide a useful framework for future investigations. 

To generalize our analysis we first recall the basic properties of the dressed algebra that were essential for our construction of the trace in sec.~\ref{sec:PIforSuper}. These properties were (i.) the boost supertranslations were unitarily implementable on the Hilbert space of gravitons, (ii.) the existence of a dense subspace $\bar{\mathcal{S}}\subset \mathcal{S}$ of boost supertranslations for which the charges $\delta^{2}\op{\mathcal{Q}}(f)$ were self-adjoint operators on $L^{2}(\mathcal{S},\gamma)$, (iii.) the existence of a neutral element $\ket{e_{\gamma}}$ which induces a faithful, normal, semifinite weight on the group von Neumann algebra $\mathfrak{A}(\mathcal{S};\gamma)$. We note that even for the theory of free gravitons, properties (ii.) and (iii.) are highly non-trivial and depend on the choice of measure on the group.

In this section we consider an analogous problem where now the group is non-abelian.
To isolate the appropriate generalization of properties (i.) -- (iii.) we consider a similar mathematical problem to the one considered in sec.~\ref{sec:AdressH} where we considered the algebra of observables $\mathfrak{A}_{\textrm{dress.}}(\mathscr{R};\mathbb{R}\times H)$ invariant under the group $\mathbb{R}\times H$. The major difference is that  we will now allow $H$ to be an arbitrary non-locally compact group. We note that the large gauge group given by \eqref{eq:LGTgroup} is of this form where $H=H_{\textrm{isom.}}\ltimes (\mathcal{S}_{0}\times G_{\textrm{LGT}})$ which is an infinite dimensional group and so is non-locally compact. We now define this algebra as well as conditions \ref{ass1} --- \ref{ass3} which generalize properties (i.)---(iii.) in the abelian case.  

We denote the algebra invariant under $\mathbb{R}\times H$ as $\mathfrak{A}_{\textrm{dress.}}(\mathscr{R};\mathbb{R}\times H)$ which is constructed from a Type III algebra $\mathfrak{A}(\mathscr{R}) \subset \mathcal{B}(\mathscr{F})$ and a group von Neumann algebra which we denote as $\mathfrak{A}(\mathbb{R}\times H;\gamma_{H})$. We will assume that the algebra $\mathfrak{A}(\mathscr{R})$ has analogous properties as the one defined in section~\ref{sec:AdressH}. 
\begin{enumerate}[label=(A.{\Roman*})]
\item The group \( \mathbb{R}\times H \) acts on the algebra 
\( \mathfrak{A}(\mathscr{R})\) by outer automorphisms. 
This action is strongly continuous on the folium \( \mathscr{F} \) and is 
implemented by the unitary operators \( e^{i\op{F}(t,h)} \). 
The generator \( \op{F}(1) \) coincides with the modular 
Hamiltonian of a cyclic and separating state \( \ket{\omega}\in\mathscr{F} \) and the generators \( \op{F}(h) \) for \( h\in H \) annihilate this state.  \label{ass1}
\end{enumerate}

The major difference is in the construction of the group von Neumann algebra $\mathfrak{A}(\mathbb{R}\times H;\gamma_{H})$  which is defined relative to a choice of infinite dimensional measure $\gamma_{H}$ on $H$. The full measure on $\mathbb{R}\times H$ is given by 
\begin{equation}
d\mu_{\mathbb{R}\times H} = dt~d\gamma_{H}.
\end{equation}
While the general arguments of this section do not directly depend on whether or not $\gamma_{H}$ is a Gaussian measure, we note that if it was a Gaussian measure then $\gamma_{H}$ could be defined by specifying an ``action'' on the group $H$. Given any measure we can construct a Hilbert space $\mathscr{H}_{H}$ defined as the space of complex wave functions $\Psi=\{\Psi(h)\}$ with finite norm 
\begin{equation}
||\Psi||^{2} \defn \int d\gamma_{H}(h) \overline{\Psi(h)}\Psi(h)
\end{equation}
and the inner product of two such states $\Psi=\{\Psi_{1}(h)\}$ and $\Psi = \{\Psi_{2}(h)\}$ is given by 
\begin{equation}
\braket{\Psi_{1}|\Psi_{2}} \defn \int d\gamma_{H}(h) \overline{\Psi_{1}(h)}\Psi_{2}(h).
\end{equation}
The full Hilbert space is 
\begin{equation}
\mathscr{H}_{\mathbb{R}\times H} \defn \mathscr{H}_{\mathbb{R}}\otimes \mathscr{H}_{H}.
\end{equation}
The Hilbert space $\mathscr{H}_{H}$ contains a basis of improper states $\ket{h}$ which satisfy 
\begin{equation}
\braket{h|h^{\prime}} = \delta_{\gamma_{H}}(h,h^{\prime}),
\end{equation}
where $\delta_{\gamma_{H}}$ is the $\delta$-function adapted to the measure $\gamma_{H}$. A general vector on $\mathscr{H}_{H}$ can be formally decomposed as 
\begin{equation}
\ket{\Psi}=\int d\gamma_{H}(h)\Psi(h)\ket{h}
\end{equation}
and, for any bounded function, $T(h)$ on $H$ we can define a multiplication operator 
\begin{equation}
T(\op{h})\ket{h} = T(h)\ket{h}.
\end{equation}

With these ingredients we can now embark upon the task of defining the group von Neumann algebra $\mathfrak{A}(H;\gamma_{H})$. The full group algebra will be given by 
\begin{equation}
\mathfrak{A}(\mathbb{R}\times H;\gamma_{H})=\mathfrak{A}(\mathbb{R},dt)\otimes \mathfrak{A}(H;\gamma_{H}) 
\end{equation}
where $\mathfrak{A}(\mathbb{R},dt)$ is simply the algebra generated by $e^{i\op{X}t}$ on $\mathscr{H}_{\mathbb{R}}$ where $\op{X}=\delta^{2}\mathcal{Q}(1)/4G_{\textrm{N}}\beta$. The relevant properties of the algebra $\mathfrak{A}(H;\gamma_H)$ depend crucially on the choice of measure $\gamma_{H}$. Following the discussion in the previous section, we will now consider the appropriate generalization of properties (ii.) and (iii.). To generalize property (ii.), the key issue we have to deal with is the unitary action of left translations on the Hilbert space $\mathscr{H}_{H}$.  Since $H$ is, in general, infinite dimensional, the measure $\gamma_{H}$ cannot be invariant under all translations in $H$. The best one can hope for is that $\gamma_{H}$ is ``quasi-invariant'' under a ``large enough'' subgroup $\bar{H}\subset H$. In other words, for such translations $\bar{h}\in H$,  the measure transforms as 
\begin{equation}
d\gamma_{H}(\bar{h}h) = J(\bar{h},h)d\gamma_{H}(h)
\end{equation}
where $J(\bar{h},h)$ is the Jacobian of the measure under a left translation and $h\in H$. The space $\bar{H}$ is the set of elements $\bar{h}$ such that the Jacobian $J(\bar{h},h)$ is well-defined.
We note that even if $\gamma_{H}$ was a Gaussian measure defined with respect to an ``action'' $S[h]$, the space $\bar{H}$ will generally not be the space of ``finite action'' elements. This property was special to the previous section where the group action was simply a shift of the Gaussian measure. For general $H$, the transformation will be non-linear and so $\bar{H}$ must be characterized on a case-by-case basis. Therefore, we assume that 
\begin{enumerate}[label=(A.{\Roman*})]
\setcounter{enumi}{1}
\item There exists a measure $\gamma_{H}$ on $H$ which is quasi-invariant with respect to left translations by a ``large enough'' subgroup $\bar{H}\subset H$. \label{ass2}
\end{enumerate}
By ``large enough'' we mean that the set of translations is big enough to include all physical charges one is interested in. For example, in the case where $G_{\textrm{LGT}}\subset H$, a reasonable condition is that $\bar{H}$ includes, at least, all smooth large gauge transformations. With this assumption, the left translations are unitarily implementable with the following action 
\begin{equation}
(e^{i\delta^{2}\op{\mathcal{Q}}(\bar{h})}\Psi)[h]= J(\bar{h},\bar{h}^{-1}h)^{-1/2}\Psi[\bar{h}h]
\end{equation}
for any $\Psi\in \mathscr{H}_{H}$. The resulting group von Neumann algebra is then defined as 
\begin{equation}
\mathfrak{A}( H;\gamma_{H})
\defn \{\delta^{2}\op{\mathcal{Q}}(\bar{h})~|~ \bar{h}\in \bar{H}\}''.
\end{equation}

Moving toward our next assumption, we may again define an (improper) ``neutral element'' vector $\ket{e_{\gamma_{H}}}$ with 
``wavefunction'' $\delta_{\gamma_{H}}(e_{H},h)$ with respect to $\gamma_{H}$ where $e_{H}$ is the neutral element of the group $H$. Following the previous section, the key assumption we will make of this vector is

\begin{enumerate}[label=(A.{\Roman*})]
\setcounter{enumi}{2}
\item The weight induced by $\ket{e_{\gamma_{H}}}$ is faithful, semifinite and normal on $\mathfrak{A}( H;\gamma_{H})$. \label{ass3}
\end{enumerate}
For an arbitrary infinite dimensional measure, this property is highly non-trivial and we know of no general argument for the existence of a vector of this kind. In the following section, we outline an argument for the particular case of $H=G_{\textrm{LGT}}$. 

With these ingredients we now consider the dressed algebra $\mathfrak{A}_{\textrm{dress.}}(\mathscr{R};\mathbb{R}\times H)$. The dressed algebra is given by the analog of \eqref{eq:dressRH} where $H$ is non-locally compact and the algebra is defined on the Hilbert space $\mathscr{F}\otimes \mathscr{H}_{\mathbb{R}\times H}$ with respect to the infinite dimensional measure $\gamma_{H}$. We refer the reader to Appendix \ref{App: CP and OVW} for a general construction of the dressed algebra for arbitrary non-locally compact groups admitting quasi-invariant measures.

\subsubsection{The Trace, Entropy and Density Matrices}

Assuming  \ref{ass1} -- \ref{ass3} we now consider the construction of the trace on the dressed algebra. With these assumptions, an obvious generalization of the trace given by \eqref{eq:traceRH} and \eqref{eq:tracesupertranslations} to the case where the group is $\mathbb{R}\times H$ is 
\begin{equation}
\label{eq:trace}
\tau_{\mathbb{R}\times H}(\hat{a}) = \braket{\omega,0_{\mathbb{R}},e_{\gamma_{H}}|e^{\op{X}/2}\hat{a}e^{\op{X}/2}|\omega,0_{\mathbb{R}},e_{\gamma_{H}}}.
\end{equation}
We now investigate the conditions under which \eqref{eq:trace} is a trace on the algebra. As in sections~\ref{sec:trace} and \ref{subsec:traceS} the key identity that we need  to prove is 
\begin{equation}
\label{eq:traceident}
\braket{\omega,0_{\mathbb{R}},e_{\gamma_{H}}|\hat{a}\hat{b}|{\omega,0_{\mathbb{R}},e_{\gamma_{H}}}} = \braket{\omega,0_{\mathbb{R}},e_{\gamma_{H}}|\hat{b}e^{-\op{X}}\hat{a}e^{\op{X}}|{\omega,0_{\mathbb{R}},e_{\gamma_{H}}}}.
\end{equation}
Since the operators  
\begin{equation}
\hat{a} = e^{-i\op{F}(\op{t},\op{h})}ae^{i\op{F}(\op{t},\op{h})}e^{i\delta^{2}\op{\mathcal{Q}}(t,h)},
\end{equation}
with $a\in \mathfrak{A}$, form an additive basis for the dressed algebra in the sense defined above, we may check this property directly on such elements. We first note that \ref{ass1} -- \ref{ass3} imply that the algebra satisfies the algebraic properties \ref{prop1} -- \ref{prop3} of the sec.~\ref{subsec:traceS}. Property \ref{prop4} is modified due to the fact that the group is non-abelian. The algebra satisfies  
\begin{equation}
e^{i\op{F}(\op{t},\op{h})} e^{i\delta^2 \op{\mathcal{Q}(h)} }e^{-i\op{F}(\op{t},\op{h})} = e^{i \delta^2 \op{\mathcal{Q}}(h)} e^{i \op{F}(\op{h}h\op{h^{-1}})},
\end{equation}
which is the same property that we used in the locally compact case (see sec.~\ref{sec:trace}). With these identities the manipulations of the right-hand  side precisely mirror the manipulations performed in sec.~\ref{sec:trace}. The right-hand  side yields 
\begin{equation}
\textrm{R.H.S.} = \int \frac{dX}{2\pi} e^{i(t'+t)X} \bra{\omega, e_{\gamma_{H}}} a e^{-i\op{F}(h'hh'{}^{-1})} e^{-it\op{K}_{\omega}} b \ket{\omega,hh'},
\end{equation}
where we recall that $\op{K}_{\omega}\defn \op{F}(1)$ is the modular Hamiltonian of the state $\omega$ on $\mathfrak{A}$. Similarly, identical manipulations of the left-hand  side yield 
\begin{equation}
\textrm{L.H.S.} = \int \frac{dX}{2\pi} e^{i(t+t')X} \bra{\omega,e_{\gamma_{H}}} a e^{i\op{F}(hh'h^{-1})} e^{it'\op{K}_{\omega}} b \ket{\omega,h'h}.
\end{equation}
As in the previous section we use the fact that the integral over $X$ yields a $\delta(t+t^{\prime})$ the fact that 
\begin{equation}
\braket{e_{H}|hh^{\prime}} = \delta_{\gamma_{H}}(h^{-1},h^{\prime}) \quad \quad \braket{e_{H}|h^{\prime}h} = \delta_{\gamma_{H}}(h^{\prime -1 },h)
\end{equation}
to obtain 
\begin{flalign} 
	\textrm{R.H.S.} &= \int \frac{dX}{2\pi} e^{i(t'+t)X} \bra{\omega} a e^{-i\op{F}(h'{}^{-1})}) e^{-it\op{K}_{\omega}} b \ket{\omega} \delta_{\gamma_{H}}(h^{-1},h^{\prime}), \label{eq:LHS and RHS 21} \\
	\textrm{L.H.S.} &= \int \frac{dX}{2\pi} e^{i(t'+t)X} \bra{\omega} a e^{-i\op{F}(h'{}^{-1})} e^{-it\op{K}_{\omega}} b \ket{\omega} \delta_{\gamma_{H}}(h^{\prime -1 },h).\label{eq:LHS and RHS 22}
\end{flalign}
However, since the measure is not the Haar measure and the group is neither Abelian nor compact, the  $\delta$-functions in the above two expressions are not necessarily equal. These distributions can be related by an ``inversion''. Let $i:H\to H$ be the inverse map which, for any function $T(h)$ on $H$ acts as 
\begin{equation}
T \circ i(h) = T(h^{-1}).
\end{equation}
We note that the measure $\gamma_{H}(h)$ is not necessarily inversion invariant. If the measure is quasi-invariant under inversion then, by definition, the measure will transform as 
\begin{equation}
\label{eq:inverse}
d\gamma_{H}(h^{-1}) = I(h) d\gamma_{H}(h)
\end{equation}
where $I(h)$ is the Jacobian of the measure under inversion.  Integrating the $\delta$-function in  \ref{eq:LHS and RHS 21} against any test functions $\psi_{1}$ and $\psi_{2}$ yields 
\begin{equation}
\int d\gamma_{H}(h)\int d\gamma_{H}(h^{\prime})\delta_{\gamma_{H}}(h^{-1},h^{\prime})\psi_{1}(h)\psi_{2}(h^{\prime}) =\int d\gamma_{H}(h) \psi_{1}(h)\psi_{2}(h^{-1}).
\end{equation}
Similarly, integrating the $\delta$-function  in \eqref{eq:LHS and RHS 22} against any  pair of test functions yields 
\begin{equation}
\int d\gamma_{H}(h)\int d\gamma_{H}(h^{\prime})\delta_{\gamma_{H}}(h^{\prime-1},h)\psi_{1}(h)\psi_{2}(h^{\prime}) = \int d\gamma_{H}(h^{\prime})\psi_{1}(h^{\prime -1})\psi_{2}(h^{\prime}).
\end{equation}
Relabeling $h^{\prime}\to h^{-1}$, we obtain 
\begin{equation}
=\int d\gamma_{H}(h^{-1})\psi_{1}(h)\psi_{2}(h^{-1})  = \int d\gamma_{H}(h)I(h)\psi_{1}(h)\psi_{2}(h^{-1}).
\end{equation}
Thus, we find that  
\begin{equation}
\delta_{\gamma_{H}}(h^{\prime -1 },h)=I(h)\delta_{\gamma_{H}}(h^{-1},h^{\prime}).
\end{equation}
For the right-hand  side and left-hand  sides of \eqref{eq:traceident} to be equal in the non-abelian case we require that $I(h)=1$. In other words, we additionally assume that 
\begin{enumerate}[label=(A.{\Roman*})]
\setcounter{enumi}{3}
\item If $H$ is non-abelian, the measure $\gamma_{H}$ is invariant under inversion, i.e.,  $\gamma_{H} \circ i(h) = \gamma_{H}(h)$.  \label{ass4}
\end{enumerate}
We note that \ref{ass3} and \ref{ass4} imply that $\ket{e_{\gamma_{H}}}$ is a tracial weight on the group algebra $\mathfrak{A}(\mathbb{R}\times H;\gamma_{H})$. 

Given assumptions \ref{ass1} -- \ref{ass4} we therefore conclude that \eqref{eq:trace} indeed defines a trace on the algebra $\mathfrak{A}_{\textrm{dress.}}(\mathscr{R}, \mathbb{R} \times H)$. Given this trace we may now construct the corresponding density matrices and entropy for any a semi-classical state. By precisely the same arguments as in sec.~\ref{subsec:densityops} and sec.~\ref{subsec:traceS}, the density matrix for any state of the form $\ket{\hat{\Phi}_{\alpha}} = \ket{\varphi}\otimes \ket{\alpha}$ is given by 
\begin{equation}
\rho_{\hat{\Phi}_{\alpha}} = \tilde{\alpha}(\op{X},\delta^{2}\op{\mathcal{Q}}) e^{-i\op{F}(\op{h})}e^{-i\op{K}_{\omega}\op{t}} e^{-\op{X}/2} \op{\Delta}_{\varphi \mid \omega} e^{-\op{X}/2} e^{i\op{K}_{\omega}\op{t}}e^{i\op{F}(\op{h})}\tilde{\alpha}(\op{X},\delta^{2}\op{\mathcal{Q}})^{\dagger}.
\end{equation}
where $\ket{\alpha}\in \mathscr{H}_{\mathbb{R}\times H}$ and $\tilde{\alpha}(\op{X},\delta^{2}\op{\mathcal{Q}})$ is given by \eqref{eq:f} with the measure $d\mu(h)$ now replaced by $d\gamma_{H}(h)$ in that definition. For any semiclassical state, $\tilde{\alpha}$ is a slowly varying function of its arguments and the von Neumann entropy is given by 
\begin{equation}
\label{eq:SvNRH2}
S_{\textrm{vN}}(\rho_{\hat{\Phi}_{\alpha}}) \approx \braket{\beta \op{X}}_{\hat{\omega}_{\alpha}} - S_{\textrm{rel.}}(\varphi|\omega) + S(\rho_{\alpha}) + \log(\beta) 
\end{equation}
The entropy of charge variations $S(\rho_{\alpha})$ can be seen to include the well documented contribution of entanglement edge modes which appear in the study of general gauge theories defined on local subregions. 

We conclude this section by considering whether the above assumptions plausibly hold for the case where $H=H_{\textrm{isom.}}\ltimes (\mathcal{S}_{0}\times G_{\textrm{LGT}})$ and $G_{\textrm{LGT}}$. As discussed at the end of sec.~\ref{subsubsec:LGTmatter}, \ref{ass1} depends on properties of the vacuum in the interacting theory and we will simply assume it is satisfied for the algebra $\mathfrak{A}(\mathscr{R})$ in $\mathscr{R}$. To consider the remaining assumptions \ref{ass2} -- \ref{ass4}, we note that if $G_{\textrm{LGT}}$ is compact\footnote{If $H$ was compact then it admits left invariant Haar measure which is also always right invariant. A bi-invariant measure is clearly invariant under inversion since an inversion can be obtained by a composition of left and right translations, $g \mapsto g^{-1} g g^{-1}$.} or an infinite dimensional abelian group then the arguments of sections \ref{sec:trace} and \ref{sec:PIforSuper} apply. Therefore, we will restrict attention to the case where $G_{\textrm{LGT}}$ is an infinite dimensional, non-abelian group. 
In this case, \ref{ass4} can straightforwardly be achieved in the following way. We first choose a Haar measure $\mu_{H}(h)$ on the group $H_{\textrm{isom.}}$. On the group $\mathcal{S}_{0}\times G_{\textrm{LGT}}$ we may choose an action
\begin{equation}
S[f,\vartheta] = S[f] + S[\vartheta]
\end{equation}
where $S[f]$ is a symmetric, billinear on $\mathcal{S}_{0}$ and $S[\vartheta]$ and is a symmetric, billinear  on $G_{\textrm{LGT}}$. Furthermore, we choose both actions to be rotationally invariant. As explained in \ref{subsubsec:HS}, this defines a centered, rotationally invariant, Gaussian measure $\gamma(f,\vartheta)$. Since $\mathcal{S}_{0}\times G_{\textrm{LGT}}$ is the identity connected sector of the group, we identify its elements with its Lie algebra elements via the exponential map. As such, the action of inversion is simply 
\begin{equation}
\gamma \circ i(f,\vartheta) =  \gamma(-f,-\vartheta) 
\end{equation}
and since $\gamma$ is a centered, Gaussian measure it is inversion invariant. Since $\mu_{H}$ is the Haar measure on $H_{\textrm{isom.}}$, the full measure
\begin{equation}
d\gamma_{H}  = d\mu_{H}(\psi)d\gamma(f,\vartheta)
\end{equation}
is also inversion invariant. To consider \ref{ass2} we note that the measure is not invariant under left translations of $G_{\textrm{LGT}}$ since the action $S[\vartheta]$ will not be invariant under the group action. The measure transforms by a Jacobian 
\begin{equation}
\label{eq:Jalpha}
d\gamma(f,\vartheta) \to J(\vartheta,\vartheta^{\prime})d\gamma(f,\vartheta).
\end{equation}
and the subgroup $\bar{G}_{\textrm{LGT}} \subset G_{\textrm{LGT}}$ for which the Jacobian $J(\vartheta,\vartheta^{\prime})$ is well-defined is the subgroup which implements a unitary action on the Hilbert space. This space is difficult to characterize in general even if $\gamma_{H}$ is a Gaussian measure. In the abelian case, the group action on the measure was to simply ``shift'' the Gaussian. These ``shifts'' are well-studied transformations of infinite dimensional Gaussian measures \cite{Bogachev1998} and the relevant subgroup that generate unitary actions on the Hilbert space are simply the space of elements with ``finite action''. The group action by a general element of $\bar{G}_{\textrm{LGT}}$ will not be a simple shift of the Gaussian if the group is non-abelian and so the set of group elements where the Jacobian $J(\vartheta,\vartheta^{\prime})$ is well-defined will not be as straightforward to characterize. In the theory of Gaussian measures, the transformation  \eqref{eq:Jalpha} is known as a ``non-linear transformation'' and a precise characterization of $\bar{G}_{\textrm{LGT}}$ requires a more complete analysis of the unitary maps associated to such transformations \cite{Ramer1974}. Nevertheless, while the precise specification of $\bar{G}_{\textrm{LGT}}$ is difficult in general, it is clear that the action $S[\vartheta]$ can be chosen such that $\bar{G}_{\textrm{LGT}}$ contains, at the very least, all smooth large gauge transformations. Thus, it would appear that we can satisfy \ref{ass2} as well. 

The last condition is \ref{ass3} which concerns the properties of the neutral weight. If $\gamma_{\textrm{H}}$ is a Gaussian measure then, following sec.~\ref{subsubsec:HS},  we can expand the measure in terms of spherical harmonics. The Hilbert space $\mathscr{H}_{H}$ can written as an infinite tensor product of finite-dimensional Hilbert spaces in a similar manner to \eqref{eq:STHilbITP}. We can thereby equivalently construct construct the weight $\ket{e_{\gamma_{H}}}$ as an infinite tensor product of faithful, semifinite normal weights. By identical arguments as below \eqref{eq:egammaplm}, $\ket{e_{\gamma_{H}}}$ can be shown to be semifinite and normal. As in the infinite dimensional abelian case, the faithfulness of $\ket{e_{\gamma_{H}}}$ is more difficult to check. However by analogous ``path integral''-style manipulations as presented in sec.~\ref{subsubsec:HS}, it can be shown that it satisfies the analog of \eqref{eq:innerprod}. By these arguments, $\ket{e_{\gamma_{H}}}$ is also faithful. In total, these arguments suggest that the algebra including the large gauge constraints is also Type II$_{\infty}$.

\section{A Type II$_1$ algebra for de Sitter} \label{sec:dSII1}

A notable result of CLPW is that the de Sitter algebra they obtain admits a normalized state of maximal entropy: it is a Type II$_{1}$ algebra \cite{Chandrasekaran:2022cip}, rather than a Type II$_{\infty}$ algebra, which has no maximum entropy state. This lends support to the idea the there are no states with greater entropy than empty de Sitter space and that the entropy is given by the area of the cosmological horizon in Planck units \cite{Maeda:1997fh,Bousso:2000nf,Bousso:2000md,2018JHEP...07..050D,2022arXiv220601083L}. In the notation of this paper, the algebra they considered was $\mathfrak{A}(\mathscr{R};\mathbb{R})$ which includes just the spherically symmetric, second-order perturbed area $\delta^{2}\op{\mathcal{Q}}(1)$. This charge ``matched'' onto (minus) the energy of the observer. Therefore, imposing the physical condition that the energy of the observer be bounded from below bounds the area from above yielding a Type II$_{1}$ algebra. However, graviton fluctuations of the static patch perturb the horizon in a non-spherically symmetric way. As we will explain, these perturbations ``turn on'' an infinite set of charges which match onto other physical properties of the observer. A central question is whether the Type II$_{1}$ property survives upon incorporating the full set of charges. In this section we present an argument that a more complete, physical model of the observer yields a Type II$_{1}$ algebra. 

\subsection{An Observer in de Sitter Space}
\label{sec:observer}

We first recall, in section \ref{subsec:deSitter}, we considered a model of an observer as a sufficiently small body in de Sitter space moving on a worldline $\Gamma$. As explained in that section, the observer cannot be  point-like and a consistent description of the observer as well as its corresponding algebra must account for ``finite size'' effects of the body. At the level of the algebra, graviton fluctuations perturb the total area of the horizon as well as the angular momenta $\delta^{2}\op{\mathcal{Q}}(\psi_{(i)})$ where $\psi^{a}_{(i)}$ are the rotational isometries of the static patch. These charges ``match'' onto the energy and spin of the body and therefore a consistent description of the static patch must, at the very least, incorporate the finite size effects of a spinning observer. If the observer has finite size then it will, in principle, have multipole moments. In this section we will take these degrees of freedom of the observer into account and explain how they are related to the higher harmonic boost supertranslations. 

We first focus on the boost supertranslation charges which, as we will now explain, are related to the mass multipole moments of the observer. We will, for now, neglect the contributions due to the ``spin multipoles'' of the body. The classical description of the motion of a small body in a general curved spacetime was obtained by Dixon \cite{Dixon:1970zza,Dixon:1970zz,Dixon:1974xoz}. A point-like body follows a geodesic of the background spacetime. 
The ``finite size'' corrections to the body will in general cause a deviation from geodesic motion. We saw this explicitly in \eqref{eq:EOMmomentumspin} when we took into account the effects of spin. Similarly, the $\ell$-th multipole moment of the body couples to the gradients $\nabla_{a_{1}}\dots \nabla_{a_{\ell-1}}R_{bcde}$ of the Riemann tensor (see, e.g.,  (1.33) and (1.34) of \cite{Dixon:1974xoz}). However, in any maximally symmetric spacetime the curvature is covariantly constant and, as such, these ``force terms'' are identically zero. The center of mass motion of the body is again described by 
\begin{equation}
\label{eq:EOMdS2}
v^{b}\nabla_{b}p^{a} = 0, \quad \quad \textrm{ and }  \quad v^{a}=p^{a}/m \quad \quad \textrm{ (de Sitter)}
\end{equation}
where we recall that we are neglecting contributions from the spin of the body.
Thus, the center of mass worldline $\Gamma$ of our classical observer is a geodesic of the background spacetime. For any small body in a general curved spacetime, Dixon obtained covariant definitions for the multipole moments of such a body as spatial integrals of the stress-energy tensor \cite{Dixon:1974xoz}. These definitions simplify significantly for an observer at the ``center'' of the de Sitter static patch $\mathscr{R}$.

To compare to Dixon's definitions, it will be convenient to endow the observer with an orthonormal frame $(v^{a}, e_{(i)}^{a})$ along $\Gamma$, where $v^{a}$ is the unit tangent to $\Gamma$ and the spatial frame vectors satisfy
$v^{b}\nabla_{b} e_{(i)}^{a} = 0 $. This frame defines a system of ``Fermi normal coordinates'' as follows \cite{Manasse:1963zz}. Along $\Gamma$, let $\tau$ denote proper time, so that $v^{a} = (\partial/\partial \tau)^{a}$. At each point on $\Gamma$, consider the spacelike geodesics that emanate orthogonally from $\Gamma$ with initial tangent $X^{a} = x^{i} e_{(i)}^{a}(\tau)$. The spatial coordinates $x^{i}$ label these geodesics and coincide with the components of the initial tangent vector in the transported spatial frame. Finally, we extend the proper-time coordinate $\tau$ off of $\Gamma$ by declaring it to be constant along these orthogonal spacelike geodesics,  $X^{a}\nabla_{a}\tau = 0$. Together, $(\tau, x^{i})$ define a Fermi normal coordinate system in a neighborhood of $\Gamma$.

For our inertial observer in the ``center'' of the static patch, it is straightforward to show that the coordinates $(\tau,x^{i})$ can be extended to all of $\mathscr{R}$. Indeed, the coordinate $\tau$ coincides with the static Killing time and the Killing field is $\xi^{\mu}=(\partial/\partial \tau)^{\mu}$. The surfaces $\Sigma_{\tau}$ are the static slices which foliate $\mathscr{R}$. Additionally, since the spacetime is spherically symmetric about the observer's worldline it is convenient to define coordinates $R=\sqrt{\delta_{ij}X^{i}X^{j}}$ as the proper length along spacelike geodesics. By spherical symmetry, surfaces of constant $R$ are $2$-spheres on which we define angular coordinates $x^{A}$. In these coordinates $(\tau,R,x^{A})$, we define the multipoles\footnote{The quantities $M_{\ell m}^{\mu \nu}$ are related to the multipole moments $I^{i_{1}\dots i_{\ell}\mu \nu}$ defined by Dixon by $I^{i_{1}\dots i_{\ell}\mu \nu} = \sum_{m=-\ell}^{\ell}\mathcal{Y}^{(\ell m)i_{1}\dots i_{\ell}}M_{\ell m}^{\mu \nu}$ where  $\mathcal{Y}^{(\ell m)i_{1}\dots i_{\ell}}$ correspond to spin-$\ell$ tensor spherical harmonics.} of the observer as 
\begin{equation}
M^{\mu \nu}_{\ell m} = \int_{\Sigma_{\tau}}\sqrt{h}d^{3}x~ R^{\ell}T^{\mu \nu}Y_{\ell m}(x^{A})
\end{equation}
where $T^{\mu \nu}$ are the components of the stress tensor in Fermi normal coordinates, $h$ is the determinant of the induced metric $h_{ij}$ on $\Sigma_{\tau}$ and $\sqrt{h}d^{3}x$ is the volume element on $\Sigma_{\tau}$. To relate these quantities to the ``mass multipoles'' we define
\begin{equation}
q_{\ell m}\defn M_{\ell m}^{\mu \nu}\xi_{\mu}n_{\nu}
\end{equation}
where $\xi^{\mu}$ is the timelike Killing field in $\mathscr{R}$ and $n_{\mu}$ is the unit normal to $\Sigma_{\tau}$. 

The quantity $q_{00}=\varepsilon$ is the energy of the observer which, if the stress-energy of the observer enters Einstein's equation at second-order in perturbation theory, perturbs the area of the bifurcation surface $\delta^{2}A_{\mathcal{B}}(1)$. Since the body is a stationary perturbation of the spacetime there are no incoming gravitons and by Raychaudhuri's equation (see, e.g., \eqref{eq:QFlux1} and \eqref{eq:QFlux2}) the observer's energy ``matches'' onto the perturbed charge $\delta^{2}\mathcal{Q}^{\textrm{R}}(1)$. The quantity $q_{1 m}$ encodes the spatial momentum of the body relative to a static observer and therefore vanishes for a classical, stationary body with definite momentum. The quantities $q_{\ell m}$ for $\ell>1$ encode the mass multipoles of the observer. It was recently shown that a stationary multipole moment in de Sitter spacetime will change the shape of the horizon in an angle-dependent way \cite{Fischler:2024cgm,Fischler:2024idi}. More precisely, if the energy of the observer enters at second-order in Einstein's equation then the static mass multipole $q_{\ell m}$ perturb the $\ell,m$ spherical harmonic of the bifurcation surface area $\delta^{2}A_{\mathcal{B}}(Y_{\ell m})$ which by Raychaudhuri's equation, implies that  
\begin{equation}
\frac{\delta^{2}\mathcal{Q}^{\textrm{R}}(Y_{\ell m})}{4G_{\textrm{N}}\beta} = -q_{\ell m}.
\end{equation}

In the quantum theory, we wish to obtain an algebra of observables of the static patch. The charges $\delta^{2}\op{\mathcal{Q}}^{\textrm{R}}(Y_{\ell m})$ are physical operators which match onto the multipole moments of the observer. If we attempt to neglect these charges by, for example, setting higher multipole moments to vanish then the constraint implies that the physical states are the graviton states in $\mathscr{F}$ invariant under the action of $\op{F}(Y_{\ell m})$. However, by theorem \ref{eq:thm1}  there are no states, besides the vacuum, that are invariant under the action of all of the horizon boost supertranslations. Similarly, there are no operators that are invariant under all supertranslations except the identity. Thus we must, at the very least, include the multipole moment fluctuations of the observer in the quantum theory to obtain a non-trivial algebra.

To obtain such an algebra we now consider the description of a quantum mechanical observer. As discussed in sec.~\ref{subsec:deSitter}, a quantum mechanical system will generally disperse and will not have a well-defined static patch. To obtain a quantum mechanical observer with a similar experience as that of a classical observer we consider the limit where the observers rest mass is given by $M\op{1}+\op{q}$ where $M\ell_{\textrm{dS}}\gg 1$ and $\op{q}$ are $O(1)$ fluctuations of the observer's energy in powers of $M$. All physical quantities of the observer similarly must scale in powers of $M$. For example, we take the 3-velocity $\op{v}^{i}\sim O(1/M)$ relative to the frame of $\Gamma$ to scale inversely in $M$ so that the trajectory of the quantum body is equivalent to $\Gamma$ in the limit. We note that, by \eqref{eq:EOMdS2}, this implies that the relative momentum of the body has $O(1)$ fluctuations and, consequently, the $\ell=1$ charge fluctuates as  $\op{q}_{1m}\sim O(1)$.  Similarly, we assume that all multipole moments fluctuate as $\op{q}_{\ell m}\sim O(1)$ so they also contribute non-trivially in the limit as $M\ell_{\textrm{dS}}\gg 1$. These charge fluctuations can straightforwardly quantized on $\mathscr{H}_{\mathcal{S}}$ as described in \ref{subsubsec:HS}. Likewise, one can construct the corresponding dressed algebra $\mathfrak{A}_{\textrm{dress.}}(\mathscr{R};\mathcal{S})$. However, if the perturbations $\op{q}_{\ell m}$ are independent, the multipole moments $\op{q}_{\ell m}$ do not directly perturb the total area of the black hole. In this sense, the $\op{q}_{\ell m}$'s appear to be ``zero modes''. They can fluctuate freely without any suppression from decreasing the horizon area. In this case,the algebra is Type II$_{\infty}$ and there would be no maximum entropy state. 

The situation is reminiscent of the considerations of sec.~\ref{subsec:deSitter} where, if we assumed that the energy and angular momentum of the body were independent, then the algebra was Type II$_{\infty}$. As we explained in that section, for any physical model of the observer the energy and angular momentum are {\em not} independent. The angular momentum of the observer contributes to its energy which is thereby suppressed at large values of the angular momentum. With this in mind, we must re-ask the question: what is a good model of an observer? 

An observer is a compact object in the spacetime. In order for the observer to be compact it must be held together by some binding energy. This binding energy cannot be due to gravity because as, $G_{\textrm{N}}\to 0$, the object would become unbound. Thus any object, e.g., a star or a galaxy, whose binding energy is dominated by gravitational effects is not a good semiclassical observer. Therefore, a reasonable observer, e.g., an atom or nanoparticle, must involve some binding energy due to other forces\footnote{We thank Edward Witten for clarifying this point.} ---, e.g., electromagnetic, nuclear forces. This is important because the gravitational binding is $O(G_{\textrm{N}})$ and so will not be relevant at leading order in perturbation theory. A generic observer will have a mass distribution supported on all multipole moments. Due to the non-gravitational forces that support the observer, it takes a finite amount energy to change their multipole structure --- e.g., it costs energy for the observer to, say, stretch out her arms. The precise dependence of the energy on the multipole moments depends on the details of the observer. We can obtain a simple model as follows. Consider an observer whose state is at a local minimum of its total energy. When expanding about this minimum, the energy is quadratic in the displacement. For small displacement changes, the multipole moments are linear in the displacement and so the energy of the body is given by 
\begin{equation}
\label{eq:totalenergyobs}
\op{\varepsilon} = M\op{1} + \op{H}_{\textrm{rot.}} + \op{q} + \frac{1}{2}\sum_{\ell >0, m}c_{\ell m}\op{q}^{2}_{\ell m}
\end{equation}
where the $c_{\ell m}$ are model-dependent coefficients that depend on the response of the body and $\op{H}_{\textrm{rot.}}$ is the rotational energy of the body (see \ref{subsec:deSitter}). Crucially, all terms on the right-hand  side are of $O(G_{\textrm{N}}^{0})$. 

\subsection{The Algebra of Observables in de Sitter}
\label{subsec:algdS}
The construction of the physical algebra proceeds as in sec.~\ref{subsec:deSitter}. We first consider the constraints for all boost supertranslations $f\in \mathcal{S}_{0}$. For $f\in \mathcal{S}_{0}$ we denote the corresponding multipole charge as 
\begin{equation} \label{eq:sphHarm}
\op{q}(f) \defn \sum_{\ell>0, m}c_{\ell m}\op{q}_{\ell m} \quad \textrm{where } \quad f = \sum_{\ell>0, m}c_{\ell m}Y_{\ell m}.
\end{equation}
All physical observables must commute with 
\begin{align}
    \delta^{2}\op{\mathcal{Q}}^{\textrm{L}}(f) = \op{F}(f) + \op{q}(f) 
\end{align}
for $f\in \mathcal{S}_{0}$.  Along with $\op{q}(f)$, the observables commuting with this constraint are of the form 
\begin{equation}
\op{\gamma}(w;\op{f}) = e^{-i\op{F}(\op{f})}\op{\gamma}(w)e^{i\op{F}(\op{f})}.
\end{equation}
We denote the algebra invariant under the higher harmonic boost supertranslations as $\mathfrak{A}_{\textrm{dress.}}(\mathscr{R};\mathcal{S}_{0})$. As explained in sec.~ \ref{subsubsec:HS}, this algebra is defined with respect to choice of measure $\gamma$, and may equivalently be regarded as an infinite tensor product of one-dimensional group von Neumann algebra $\mathfrak{A}_{\ell m}(\mathbb{R})$. For the purposes of the present discussion, the latter point of view will be most useful.

We may additionally impose the rotational constraints arising from the group $\textrm{SO}(3)$ as explained in sec.~\ref{subsec:deSitter} to obtain the dressed algebra $\mathfrak{A}_{\textrm{dress.}}(\mathscr{R};\textrm{SO}(3)\ltimes \mathcal{S}_{0})$. All that remains is to impose the Hamiltonian constraint that all physical operators must commute with 
\begin{equation}
\op{F}(1) + \op{q} + \op{H}_{\textrm{rot.}} + \op{H}_{\textrm{BE}}, 
\end{equation}
where 
\beq \label{eq:BE}
    \op{H}_{\textrm{BE}} \defn \frac{1}{2}\sum_{\ell m} \op{H}_{\ell m}, \qquad \op{H}_{\ell m} \defn c_{\ell m} \op{q}_{\ell m}^2
\eeq
is the binding energy of the observer. This yields the algebra  $\mathfrak{A}^{\textrm{obs.}}_{\textrm{dress.}}(\mathscr{R};G)$ which, as in sec.~\ref{subsec:deSitter}, is the ``crossed product'' of $\mathfrak{A}_{\textrm{dress.}}(\mathscr{R},\textrm{SO}(3)\ltimes \mathcal{S}_0)$ by the automorphism generated on this algebra by the operator $\op{F}(1) + \op{H}_{\textrm{rot.}}+ \op{H}_{\textrm{BE}}$. This operator is the modular Hamiltonian of the vector
\beq \label{eq:ObsState}
    \ket{\omega} \otimes e^{-\frac{1}{2}\op{H}_{\textrm{rot.}}}\ket{e_{\textrm{SO(3)}}} \otimes e^{-\frac{1}{2}\sum_{\ell > 0, m}(\op{H}_{\ell m} + \log Z_{\ell m})} \ket{e_{\gamma}}
\eeq
As defined in sec.~\ref{subsec:traceS}, the vector $\ket{e_{\gamma}}$ is the (improper) ``neutral-element" state on the group algebra $\mathfrak{A}(\mathcal{S}_{0},\gamma)$. 
The constants $Z_{\ell m}$ are defined by the relation 
\begin{equation}
Z_{\ell m} \defn \braket{e_{\gamma_{\ell m}}|e^{-\op{H}_{\ell m}}|e_{\gamma_{\ell m}}}.
\end{equation}
Provided $Z_{\ell m} < \infty$ for each $\ell, m$ the vector 
\begin{equation}
e^{-\frac{1}{2} \sum_{\ell > 0,m}\tilde{\op{H}}_{\ell m} } \ket{e_{\gamma}}
\end{equation}
has finite norm and defines a proper state on the group algebra.\footnote{See Appendix~\ref{app: ITP} for a more detailed construction of this state.} For brevity, we have defined $\tilde{\op{H}}_{\ell m} \defn \op{H}_{\ell m}+\log Z_{\ell m}\op{1}$. Consequently, the algebra 
\begin{equation}
\mathfrak{A}^{\textrm{obs.},+}_{\textrm{dress.}}(\mathscr{R};G)=\op{P}_{q>0}\mathfrak{A}^{\textrm{obs.}}_{\textrm{dress.}}(\mathscr{R};G)\op{P}_{q>0}
\end{equation}
admits the trace
\begin{equation}
\tau^{\textrm{obs.}}(\hat{a})= \bra{\omega,0_t,e_{\textrm{SO}(3)},e_{\gamma}} e^{-\frac{1}{2}(\op{q} + \op{H}_{\textrm{rot.}}+ \sum_{\ell > 0,m}\tilde{\op{H}}_{\ell m} )} \hat{a} e^{-\frac{1}{2}(\op{q} +\op{H}_{\textrm{rot.}}+ \sum_{\ell > 0,m}\tilde{\op{H}}_{\ell m} )} \ket{\omega, 0_t, e_{\textrm{SO}(3)},e_{\gamma}},
\end{equation}
for any $\hat{a}\in \mathfrak{A}^{\textrm{obs.},+}_{\textrm{dress.}}(\mathscr{R};G)$. If
the observer is well-approximated as a rigidly rotating body with multipole moments satisfying \eqref{eq:totalenergyobs}, then it follows from the above discussion as well as the dicussion in sec.~\ref{subsec:deSitter} that the trace  $\tau^{\textrm{obs.}}(\op{1})$ is finite and the algebra is Type II$_{1}$. For a more general body, the dependence on the multipole fluctuations may be more complicated. Nevertheless, it appears likely that the norm of $e^{-\frac{1}{2}(\op{H}_{\textrm{rot.}}+\sum_{\ell m}\tilde{\op{H}}_{\ell m})}\ket{e_{\textrm{SO}(3)},e_{\gamma}}$ is finite so that the algebra $\mathfrak{A}^{\textrm{obs.},+}_{\textrm{dress.}}(\mathscr{R};G)$ is Type II$_{1}$ for a large class of physical observers in de Sitter. In all such cases, the observer algebra admits a maximum entropy state  given by 
\begin{equation}
\ket{\hat{\Phi}_{\textrm{max}}} = \ket{\omega} \otimes e^{-\frac{1}{2}(\op{q}+\op{H}_{\textrm{rot.}}+\sum_{\ell m}\tilde{\op{H}}_{\ell m})} \ket{0_{\mathbb{R}},e_{\textrm{SO(3)}},e_{\gamma}}
\end{equation}

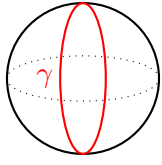
\begin{figure}
    \centering
    \begin{tikzpicture}
  \draw[thick] (0,0) circle (1);
  \draw[dotted] (0,0) ellipse (1 and 0.3);
  \draw[red,thick] (0,0) ellipse (0.3 and 1);
  \node[red] at (-.5,0) {$\gamma$};
    \end{tikzpicture}
    \caption{The sphere path integral with an world-line observer following geodesic $\gamma$. Cutting the path integral along the dotted line prepares the Hartle-Hawking state with two anti-podal, entangled observers.}
    \label{fig:sphere}
\end{figure}

    We conclude this section by briefly commenting on the relation to the Euclidean approach to the entropy of de Sitter space, which was where it was first discovered \cite{Gibbons:1978ac}. The Euclidean sphere is the dominant solution to Einstein's equations with positive cosmological constant in Euclidean signature. Following \cite{Chandrasekaran:2022cip}, we can augment the path integral to include an action for the observer, now including the energy contributions from higher charges.
    Taking the (unstable) saddle where the worldline is great circle geodesic, the path integral gives a factor of $\Tr e^{-\beta (H_{\textrm{BE}}+H_{\textrm{rot.}} +q) }$, where this is the type I trace over the Hilbert space of the charges. This is just a number. The Hartle-Hawking state can be contructed by cutting the sphere path integral in half (see figure \ref{fig:sphere}). This gives, up to the previously mentioned divergent factors, the same tracial state discussed above.

\section{Discussion} \label{sec: discussion}

In this paper, we have constructed the complete algebra of observables subject to all constraints on the first-order graviton field in the region $\mathscr{R}$ bounded by any Killing horizon. From the full, infinite dimensional group of horizon symmetries, we found that the subgroup $G = H_{\textrm{isom}}\ltimes \mathcal{S}$
imposes a non-trivial set of constraints entirely on the graviton field. In addition to the isometries $H_{\textrm{isom.}}\times \mathbb{R}$, this group includes an infinite dimensional, abelian group of ``boost supertranslations''. We have shown that, for any black hole spacetime, the full algebra of observables satisfying these constraints is Type II$_{\infty}$. In de Sitter spacetime, we argued that the spin-energy and binding energy of the observer renders the algebra Type II$_{1}$. In the remainder of this section we will comment on some extensions of the present work.

\paragraph{Generic Extremal Surfaces in Open Spacetimes}

The exterior regions of Killing horizons are of course just one of many examples of spacetime subregions that we would like to understand in quantum gravity. What then is the algebra of observables and entropy in perturbative quantum gravity associated to a more generic subregion, $\mathscr{R}$, and what are the relevant symmetries? As emphasized in the introduction, it is difficult to even define what one means by a subregion once we allow spacetime to fluctuate, even perturbatively. The two cases that we know of that are universally well-defined are (i) the region causally accessible to a worldline observer and (ii) a region bounded by an extremal surface. Case (i) is well-defined because it manifestly makes no reference on the metric, only the existence of an observer that has its endpoints chosen at infinity. Case (ii) is well-defined in perturbative quantum gravity because the ``location'' of extremal surfaces is robust to perturbations. Note that the exterior of Killing horizons are special cases that are simultaneously (i) and (ii). 

Focusing for the moment on case (ii), we review the proposal of \cite{Chen:2024rpx} for the construction of gravitational algebra that incorporates the homogeneous boost mode about generic extremal surfaces. See also \cite{Jensen:2023yxy} for related discussion. We first focus on regions bounded by extremal surfaces that extend all the way out to spatial infinity, and hence are in an open spacetime. We then consider the causal complements of these regions which may have bounded size. At present, we do not understand how to consider generic subregions bounded by extremal surfaces in closed spacetimes using the technology of this paper, though these can be analyzed in the algebraic context using no-boundary state technology developed in \cite{Blommaert:2025bgd}.

In \cite{Chen:2024rpx}, it was noted that because the surface is extremal, one can consider a one-parameter family of spacetimes, labeled by a relative boost parameter $T_0$ between the interior and exterior of the extremal surface, and that all such spacetimes satisfy the constraint equations. Denoting the Hilbert space of fluctuations about each background by $\mathscr{H}_T$, one may construct the direct integral Hilbert space
\begin{align}
    \mathscr{H} = \int^\oplus dT_0 \mathscr{H}_{T_0} .
\end{align}
The physical algebra acting on this Hilbert space was then argued to take the form of a crossed product, namely generated by the QFT algebra acting on each $\mathscr{H}_{T_0}$ in the obvious way along with an operator $i\partial_{T_0}$, given by the perturbed area operator on the horizon over $4G$ plus the one-sided modular Hamiltonian for some state. The divergences at the boundary of the region for these two operators cancel one another. Rather than taking the modular Hamiltonian for some state, we simply consider the one-sided flux
\begin{equation}
F_{\mathscr{R}}=- \int_0^\infty dU\int_{S} d^2x\sqrt{q_{U}}~U\bigg[\delta \sigma_{AB}\delta \sigma^{AB}-\frac{1}{2}\delta \theta^{2}\bigg]
\end{equation}
 where $U = 0$ is the codimension-two boundary of $\mathscr{R}$ and $U$ is an affine parameter on the null boundary. This one-sided flux has the same effect of removing the divergence as the one-sided modular Hamiltonian and is more straightforward to generalize. It is now clear, following the results of this paper, that we should also consider angle dependent boosts of the extremal surface as well as the diffeomorphism of the surface. We do not include diffeomorphisms that move the position of the extremal surface. This comprises the physical (in our context) subgroup of the so-called corner symmetry group \cite{Donnelly:2016auv,Freidel:2020xyx,Freidel:2021cjp,Ciambelli:2021vnn,Ciambelli:2021nmv,Ciambelli:2022vot,Klinger:2023qna,Freidel:2023bnj,Ciambelli:2024swv}. Taking a boost that smoothly depends on the angles also leads to new solution to Einstein's equations \cite{Chandrasekaran:2023vzb}. To each of these solutions, we can associate a Hilbert space of quantum fields on the background $\mathscr{H}_f$ and a Hilbert space can be written as a direct integral 
\begin{align}
    \mathscr{H} = \int^\oplus d\gamma(f) \mathscr{H}_f.
\end{align}
We then consider the operator on this Hilbert space which is the sum of the perturbed angle dependent area operators and angle-dependent one-sided flux operators
\begin{equation}
F_{\mathscr{R}}(f)=- \int_{0}^{\infty}dU\int_{S}d^2x\sqrt{q_U}~f(x^{A})U\bigg[\delta \sigma_{AB}\delta \sigma^{AB}-\frac{1}{2}\delta \theta^{2}\bigg].
\end{equation}
These combinations are well-defined operators on the Hilbert space and are the analogs to the charges at infinity defined in section \ref{subsec:symmcharges}. 
Together with the quantum field theory operators, this is the full gravitational algebra associated to the subregion bounded by an extremal surface. The von Neumann entropy is given by the generalized entropy of the subregion that includes a fluctuation term associated to the additional charges, which may be interpreted as the entanglement edge modes. 

It was important in the construction so far to have $\mathscr{R}$ be semi-infinite because if the $U$ integral in the one-sided flux was cut off at some finite value, there could be additional divergences introduced at these points. In order to consider the algebra associated to genuinely finite regions in open universes, we can first construct the algebra as above for the causal complement, and subsequently take the commutant. The construction and properties of these local subregion algebras will be explored in forthcoming work 

In case (i), when the boundary of the causal diamond is not an extremal surface, the symmetries of the subregion are obscure. While similar symmetry groups have been considered for general causal diamonds \cite{Chandrasekaran:2019ewn}, these will not generically describe the same region as the one defined by an infinitely extended observer. Indeed, the only universal large diffeomorphism to gauge appears to be the time translation symmetry along the observer's worldline, as was done in the background-independent proposal of \cite{Witten:2023xze}.
 
\paragraph{AdS/CFT}
\label{disc:AdSCFT}
The focus of this work has been on a large class of gravitational charges that generate symmetries of horizons. These charges led to an algebra with an entropy that had a contribution from fluctuations of these charges, which have the same form as entanglement from edge modes that appear in gauge theory. In the case of black holes in AdS, the horizon charges should have an interpretation in the dual CFT at the asymptotic boundary. For definiteness, consider an asymptotically AdS$_4$ Schwarzschild black hole, with asymptotic boundary $S^2 \times \mathbb{R}$ where the CFT lives. The homogeneous charge that generates an angle-independent boost on the horizon corresponds to the Hamiltonian of the CFT
\begin{align}
    \delta^2 H_{CFT} = \int_{S^2} d\Omega~ \delta^2 T^{CFT}_{ab}(x)n^{a}t^{b},
\end{align}
where $n^{a}$ is the timelike normal to $\mathbb{S}^{2}$, $t^{a}$ is the asymptotic timelike Killing field and the perturbation of the CFT stress tensor, $T^{CFT}_{ab}$, is about the thermal state at the Hawking temperature.
The natural extension of this correspondence to angle-dependent boosts is angle-dependent time translations of the boundary, generated by
\begin{align}
    \delta^2H_{CFT}(f) = \int_{S^2} d\Omega~ f(x^A)\delta^2 T^{CFT}_{ab}(x^A)n^{a}t^{b}.
\end{align}
From the perspective of the CFT, these may appear to be strange charges. Interestingly, a larger class\footnote{These included diffeomorphisms that move the location of the horizon, which were not relevant for our discussion.} of symmetries of black hole horizons were considered in \cite{Knysh:2024asf} and given a CFT interpretation in terms of gauge symmetries of the hydrodynamic effective field theory. The matching of symmetry transformations of the horizon and the effective field theory arose from mapping coordinates on the horizon to the boundary via null geodesics. Indeed, during the completion of this work, we learned of an upcoming paper \cite{liu2025hydrocft} that clarifies this by analyzing the connection of these hydrodynamic modes of the boundary effective field theory to entanglement edge modes that have a similar form to the type II entropy. 
\acknowledgments

We thank Shadi Ali Ahmad, Chang-Han Chen, Luca Ciambelli, Eanna Flanagan, Temple He, David Kastor, Robert Leigh, Hong Liu, Juan Maldacena, Manu Srivastava, Jennie Traschen, Edward Witten and Kathryn Zurek for discussions. JKF is supported by the Marvin L. Goldberger
Member Fund at the Institute for Advanced Study and the National Science Foundation under Grant PHY-2514611. The work of MSK was supported by the Heising-Simons foundation ``Observable Signatures of Quantum Gravity" collaboration, the Walter Burke Institute for Theoretical Physics, and the U.S. Department of Energy, Office of Science, Office of High Energy Physics, under Award Number DE-SC0011632. G.S. is supported by the Princeton Gravity Initiative at Princeton University.

\appendix

\section{Crossed Products and their Operator Valued Weights} \label{App: CP and OVW}

In this appendix, we provide an alternative construction of the crossed product of a von Neumann algebra by a non-locally compact group admitting a quasi-invariant measure. This analysis runs parallel to the discussion of subsections \ref{sec:AdressH} and \ref{sec: CP for NLC Groups}. 

Let $\mathfrak{A} \subset B(\mathscr{H})$ be a von Neumann algebra acted upon by a group automorphism $\alpha: G \rightarrow \text{Aut}(\mathfrak{A})$, with $G$ a (possibly non locally compact) group admitting a quasi-invariant measure $\nu$. A measure on $G$ is (left) quasi-invariant if there exists a cocycle $J: G_0 \times G \rightarrow \mathbb{R}$ such that\footnote{One can interpret $J(g,h)$ as the Radon-Nikodym derivative of the pushforward of $\nu$ under left translation by $g$ and $\nu$ itself. The second equality follows immediately from the chain rule for Radon-Nikodym derivatives.}
\beq
	d\nu(gh) = J(g,h) d\nu(h), \qquad J(gh,k) = J(g,hk)J(h,k), \qquad g,h \in G_0, k \in G. 
\eeq 
We may assume for simplicity that the automorphism $\alpha$ is unitarily implemented in $\mathscr{H}$ by the unitary representation $U: G \rightarrow U(\mathscr{H})$, however we emphasize that this is not a prerequisite for the definition of the crossed product. 

Using the measure $\nu$ we define a Hilbert space $L^2(G,d\nu)$ consisting of square integrable functions on $G$ with respect to the measure $\nu$. On the Hilbert space $L^2(G,d\nu)$ we can define a unitary representation $\ell_{\nu}: G_0 \rightarrow U(L^2(G,d\nu))$ where here $G_0 \subset G$ is the dense subset of group elements for which the measure $\nu$ is quasi-invariant. Explicitly,
\beq \label{ell nu}
	\bigg(\ell_{\nu}(g) f\bigg)(h) = J(g,g^{-1}h)^{-1/2} f(g^{-1} h). 
\eeq
The group algebra of $G$ with respect to the measure $\nu$ is defined to be the weak closure of this representation: $\mathfrak{A}(G;\nu) \equiv \ell_{\nu}(G_0)''$. 

The crossed product of $\mathfrak{A}$ by $G$ with respect to the action $\alpha$ is a von Neumann subalgebra of bounded operators on the extended Hilbert space $\mathscr{H} \otimes L^2(G,d\nu)$. It is generated by $\ell_{\nu}(g)$ for $g \in G_0$ and $\pi_{\alpha}(x)$ for $x \in \mathfrak{A}$, where
\beq
    \big(\pi_{\alpha}(x) \psi\big)(g) \equiv \alpha_{g^{-1}}(x) \big(\psi(g)\big), \qquad \psi \in \mathscr{H} \otimes L^2(G,d\nu).
\eeq
That is,
\beq
    \mathfrak{A} \times_{\alpha,\nu} G \equiv \{\pi_{\alpha}(x),\ell_{\nu}(g) \; | \; x \in \mathfrak{A}, g \in G_0\}''. 
\eeq
If $\alpha$ is unitarily implemented on the Hilbert space $\mathscr{H}$ we can write $\pi_{\alpha}(x) = V^{\dagger} x V$, where the unitary operator $V$ acts on $\psi \in \mathscr{H} \otimes L^2(G,d\nu)$ as $\big(V \psi)(g) = U(g)\big(\psi(g))$.  

To match with the approach of the main text, we can write $\ell(g) = e^{i \op{\mathcal{Q}}(g)}$, $U(g) = e^{i\op{F}(g)}$, and $V = e^{i \op{F}(\op{g})}$. Here, $\op{\mathcal{Q}}(g)$ can be thought of as charges, $\op{F}(g)$ as fluxes, and $f(\op{g})$ is a multiplication operator. The representation $\ell_{\nu}$ is obtained from $\ell$ by premultiplying with the appropriate Jacobian of the measure $\nu$ as described in eqn. \eqref{ell nu}, and we can use the notation $\ell_{\nu}(g) = e^{i\op{\mathcal{Q}_{\nu}}(g)}$. A dense set of (possibly unnormalizable) vectors in the Hilbert space $\mathscr{H} \otimes L^2(G,d\nu)$ is given by $\ket{x,g}_{\nu}$ where $x \in \mathfrak{A}$, and $g \in G_0$. Here, $\ket{x,g}_{\nu} = \pi(x) \ket{\omega,g}_{\nu}$ with $\ket{\omega}$ the cyclic-separating vector representative of some faithful state $\omega \in P(\mathfrak{A})$. The action of the various operators appearing in the crossed product are given by
\begin{flalign}
    & e^{i \op{\mathcal{Q}_{\nu}}(g)} \ket{x,h}_{\nu} = \ket{x,gh}_{\nu}, \nonumber \\
    & e^{i \op{F}(g)} \ket{x,h}_{\nu} = \ket{\alpha_g(x),h}_{\nu}, \nonumber \\
    & e^{i \op{F}(\op{g})} \ket{x,h}_{\nu} = \ket{\alpha_h(x),h}_{\nu}, \nonumber \\
    & \pi(y) \ket{x,h}_{\nu} = \ket{yx, h}_{\nu}, \nonumber \\
    & \pi_{\alpha}(y) \ket{x,h}_{\nu} = \ket{\alpha_{h^{-1}}(y) x,h}_{\nu}
\end{flalign}

As we have addressed in the main text, for analyzing the type of the crossed product algebra it is useful to construct a densely defined `neutral' weight on the group algebra. We will now propose a construction of such a weight, though we emphasize that the existence of such a weight is not necessary for the algebra $\mathfrak{A} \times_{\alpha,\nu} G$ to be well defined. In the next appendix we will state a theorem which describes the sufficient conditions which must be satisfied by this weight to ensure that the algebra $\mathfrak{A} \times_{\alpha,\nu} G$ is semifinite. 

A general element in $\mathfrak{A}(G;\nu)$ is given by the weak limit of a sequences of `sums'
\beq \label{General op in group alg}
    \tilde{\psi}(\ell_{\nu}) = \sum_{g \in G_0} \psi(g) \ell_{\nu}(g).
\eeq
By the density of $G_0$, the map\footnote{This map should be Cauchy continuous for the extension to be unique.} $\psi: G_0 \rightarrow \mathbb{C}$ can be extended to a map on all of $G$ which we also denote by $\psi$. We can then define a positive, linear functional
\beq
    e_{\nu}(\tilde{\psi}(\ell)^{\dagger} \tilde{\psi}(\ell)) \equiv \int_{G} d\nu(g) \overline{\psi(g)} \psi(g). 
\eeq
Depending upon various properties of the measure $\nu$, $e_{\nu}$ may define a weight on the algebra $\mathfrak{A}(G,\nu)$. In the event that it is moreover faithful, semifinite, and normal, one may regard $L^2(G,d\nu)$ as a GNS representation for $\mathfrak{A}(G;\nu)$ with respect to this weight. For example, this will always be the case whenever $G$ is a locally compact group. Consequently, we can take this Hilbert space to be densely generated by the improper states $\ket{g}_{\nu} \equiv \ell_{\nu}(g) \ket{e_\nu}$, where $\ket{e_\nu}$ is the vector representative of $e_{\nu}$. We can then formally write a general element of the group algebra as
\beq
    \tilde{\psi}(\ell_{\nu}) = \int_{G} d\nu(g) \psi(g) \ell_{\nu}(g),
\eeq
with the understanding that $\tilde{\psi}(\ell_{\nu}) \ket{e_\nu} \equiv \ket{\psi}_{\nu}$ by the process of extending the function $\psi$ appearing in \eqref{General op in group alg}. 

Given a von Neumann algebra $M \subset B(H)$, we say that a possibly unbounded operator $\mathcal{O}$ is affiliated with $M$ if it commutes with all unitary operators in $M'$. We denote the set of operators affiliated with $M$ by $\hat{M}$. Given an inclusion of von Neumann algebras $N \subset M$ we can define an extension of the notion of a weight called an operator valued weight. An operator valued weight is a map $T: M^+ \rightarrow \hat{N}^+$ which is additive and positively homogeneous with respect to $N^+$:
\beq
	T(m_1 + m_2) = T(m_1) + T(m_2), \;\; T(n^* m n) = n^* T(m) n, \;\; \forall m_1,m_2 \in M^+, n \in N^+. 
\eeq
Clearly, an ordinary weight can be regarded as an operator valued weight from $M$ to the subalgebra of complex numbers. We denote by $M_T^+$ the set of all elements $m \in M^+$ which map to elements of $\hat{N}^+$ with finite operator norm. The domain of $T$ is then defined to be the linear span of these elements, which we denote by $M_T$. An operator valued weight is called semifinite if its domain is ultraweakly dense, faithful if $T(m) = 0 \iff m = 0$ and normal if $T$ is lower semi-continuous in the ultraweak operator topology. We will denote the set of all faithful, semifinite, normal operator valued weights from $M$ to $N$ by $P(M,N)$. An operator valued weight which is unital, $T(\op{1}) = \op{1}$, is called a conditional expectation. 

Given a faithful, semifinite, normal weight $\omega$ on $N$ and a faithful, semifinite, normal operator valued weight $T: M_T \rightarrow \hat{N}$, we can define a faithful, semifinite, normal weight $\omega_T \equiv \omega \circ T$ on $M$. Denoting the inclusion of $N$ into $M$ by $i: N \hookrightarrow M$, the modular automorphisms of $\omega$ and $\omega_T$ obey the relation
\beq
	\sigma^{\omega_T}_t \circ i(n) = i \circ \sigma^{\omega}_t(n). 
\eeq
In fact, Falcone and Takesaki showed that the existence of a weight $\omega_T$ satisfying the above relation is a necessary and sufficient condition for the existence of an operator valued weight. Thus, if there exist weights $\tilde{\omega} \in P(M)$ and $\omega \in P(N)$ such that
\beq
    \sigma^{\tilde{\omega}}_t \circ i = i \circ \sigma^{\omega}_t, \qquad \forall t \in \mathbb{R}
\eeq
then there will exist an operator valued weight $T \in P(M,N)$ such that $\tilde{\omega} = \omega \circ T$. 

Generalizing eqn. \eqref{General op in group alg}, we can write a general operator in the crossed product as the weak limit of a sequence of `sums' of the form
\beq
    \tilde{a}(\ell) = \sum_{g \in G_0} \ell_{\nu}(g) \pi_{\alpha}(a(g)).
\eeq 
Again, by the density of $G_0$ we can extend the map $a: G_0 \rightarrow \mathfrak{A}$ to a map on all of $G$ which we continue to denote by $a$. We can then define a positive, linear map from $\mathfrak{A} \times_{\alpha,\nu} G$ to $\mathfrak{A}$:
\beq
    T_{\nu}(\hat{a}^{\dagger} \hat{a}) \equiv \int_{G} d\nu(g) a(g)^* a(g). 
\eeq
It can be seen to satisfy the homogeneity property of operator valued weights by direct computation. First,
\beq
    \hat{a} \pi_{\alpha}(x) = \sum_{g \in G_0} \ell_{\nu}(g) \pi_{\alpha}(a(g) x). 
\eeq
Then, we can easily write
\begin{flalign}
    T_{\nu}\bigg(\pi_{\alpha}(x)^{\dagger} \hat{a}^{\dagger} \hat{a} \pi_{\alpha}(x)\bigg) &= T_{\nu}\bigg( (\hat{a} \pi_{\alpha}(x))^{\dagger} (\hat{a} \pi_{\alpha}(x))\bigg) \nonumber \\
    &= \int_{G} d\nu(g) x^* a(g)^* a(g) x = x^* T_{\nu}(\hat{a}^{\dagger} \hat{a}) x. 
\end{flalign}
In the event that $e_{\nu}$ is a weight, $T_{\nu}$ defines an operator valued weight. If $e_{\nu}$ is moreover faithful, semifinite, and/or normal, $T_{\nu}$ will likewise be faithful, semifinite, and/or normal.

In this case, we can use the operator valued weight $T_{\nu}$ to extend any weight $\varphi$ on $\mathfrak{A}$ to a weight $\varphi \circ T_{\nu}$ on $\mathfrak{A} \times_{\alpha,\nu} G$. If $\varphi$ has vector representative $\ket{\varphi}$ in the Hilbert space $\mathscr{H}$, the dual weight $\tilde{\varphi}$ will have vector representative $\ket{\tilde{\varphi}} = \ket{\varphi,e_\nu}$ in $\mathscr{H} \otimes L^2(G,d\nu)$. In the case that $G$ is locally compact we can move through the above discussion with $\nu = \mu$ the unique left invariant Haar measure. In this case the vector $\ket{e_\mu} \equiv \ket{e}$ is the neutral element of the group and $T_{\nu} \equiv T$ is Haagerup's standard operator valued weight \cite{HaagerupI,HaagerupII}.

\section{Trace for Crossed Products by General Groups} \label{App: General Trace}

In this appendix, we state a set of sufficient conditions for the crossed product algebra $\mathfrak{A} \times_{\alpha,\nu} G$ to be semifinite. This analysis runs parallel to sections \ref{sec:trace} and \ref{subsec:traceS} in the main text. 

\begin{theorem}[Semifiniteness] \label{thm. semifinite}
    Let $\mathfrak{A}$ be a von Neumann algebra acted upon by a group $G$ via the automorphism $\alpha: G \rightarrow \text{Aut}(\mathfrak{A})$. We do not require the group $G$ to be locally compact, only that it admits a left quasi-invariant measure $\nu$ which is also quasi-invariant under inversion with module function (e.g. Jacobian under inversion) $\delta_{\nu}: G \rightarrow \mathbb{C}$. The following is a sufficient condition for the crossed product $\mathfrak{A} \times_{\alpha,\nu} G$ to be semifinite:
    \begin{enumerate}
        \item The positive, linear functional $e_{\nu}$ defines a faithful, semifinite, normal weight on the group algebra $\mathfrak{A}(G;\nu)$,
        \item There exists a faithful, semifinite, normal weight $\omega$ on $\mathfrak{A}$ and an embedding $\eta: \mathbb{R} \hookrightarrow G$ such that
        \begin{enumerate}[label=(\alph*)]
            \item The automorphism $\alpha \circ \eta: \mathbb{R} \rightarrow \text{Aut}(\mathfrak{A})$ is KMS with respect to $\omega$,
            \item The weight $\omega$ is quasi-invariant under the action of $\alpha$ e.g. $\omega \circ \alpha_g = \delta_{\nu}(g)^{-1}\omega$ for all $g \in G_0$.
        \end{enumerate}
    \end{enumerate}
\end{theorem}

This theorem can be proven by direct construction of a trace on $\mathfrak{A} \times_{\alpha,\nu} G$:
\beq \label{Super general trace}
    \tau(\hat{a}) = \omega \circ T_{\nu}(\ell_{\nu} \circ \eta(i/2) \; \hat{a} \; \ell_{\nu} \circ \eta(i/2)). 
\eeq
In the case that $G = \mathbb{R} \times H$ with $\mathbb{R}$ the modular automorphism group of $\omega$ we can write
\beq \label{Super general trace Cartesian product}
    \tau(\hat{a}) = \bra{\omega,0_t,e_{\nu_H}} e^{\op{X}/2} \; \hat{a} \; e^{\op{X}/2} \ket{\omega,0_t, e_{\nu_H}},
\eeq
where $\nu_H$ is a quasi-invariant measure on $H$ and $\ell_{\nu} \circ \eta(t) = e^{-it\op{X}}$. When $H$ is locally compact we can take $\nu = \mu$, the unique left invariant Haar measure. 

By hypothesis (1), the positive, linear map $T_{\nu}$ defines a faithful, semifinite, normal operator valued weight. The weight defined in \eqref{Super general trace} is therefore faithful, semifinite, and normal because $\omega$ and $T_{\nu}$ are. To determine its modular automorphism, it suffices to consider the modular flow of the generating elements $\pi_{\alpha}(x)$ and $\ell_{\nu}(g)$. By the Falcone-Takesaki theorem, the modular automorphism of the dual weight on operators from $\mathfrak{A}$ is the same as the modular automorphism of the weight $\omega$:
\beq
    \sigma^{\omega \circ T_{\nu}}_t(\pi_{\alpha}(x)) = \pi_{\alpha} \circ \sigma^{\omega}_t(x).
\eeq
By hypothesis $(2a)$, $\sigma^{\omega}_t = \alpha_{\eta(t)}$. The automorphism $\alpha$ is unitarily implemented by $\ell_{\nu}(g)$ in the crossed product, $\pi_{\alpha} \circ \alpha_g(x) = \ell_{\nu}(g) \pi_{\alpha}(x) \ell_{\nu}(g^{-1})$. Thus, we can write
\beq
    \sigma^{\omega \circ T_{\nu}}_t(\pi_{\alpha}(x)) = \ell_{\nu} \circ \eta(t) \pi_{\alpha}(x) \ell_{\nu} \circ \eta(-t). 
\eeq

In general, the modular automorphism of the dual weight applied to $\ell_{\nu}(g)$ is given by
\beq
    \sigma^{\omega \circ T_{\nu}}_t(\ell_{\nu}(g)) = \sigma^{\omega_{\nu}}_t(\ell_{\nu}(g)) \pi_{\alpha}(u^{\omega \circ \alpha_g | \omega}_t). 
\eeq
The first term can be derived by considering the KMS condition for products $\ell_{\nu}(g) \ell_{\nu}(h)$ and the second for products $\ell_{\nu}(g) \pi_{\alpha}(x) = \pi_{\alpha} \circ \alpha_{g}(x) \ell_{\nu}(g)$. Applying hypothesis $(2b)$ we find
\beq
    \sigma^{\omega \circ T_{\nu}}_t(\ell_{\nu}(g)) = \ell_{\nu}(g),
\eeq
where the modular flow of $e_{\nu}$ has canceled with the Connes' cocycle comparing the weight $\omega \circ \alpha_g$ with $\omega$. Hypotheses $(2a)$ and $(2b)$ together imply that $\eta(\mathbb{R})$ is a central subgroup of $G$, e.g. $\eta(t) g \eta(-t) = g$ for all $g \in G$ \cite{AliAhmad:2024eun}. Thus, we can write
\beq
    \sigma^{\omega \circ T_{\nu}}_t(\ell_{\nu}(g)) = \ell_{\nu} \circ \eta(t) \ell_{\nu}(g) \ell_{\nu} \circ \eta(-t). 
\eeq

We have therefore demonstrated that the hypotheses of our theorem imply that the modular flow of the weight $\omega \circ T_{\nu}$ is given by
\beq \label{Inner modular flow in general}
    \sigma^{\omega \circ T_{\nu}}_t(\hat{a}) = \ell_{\nu} \circ \eta(t) \; \hat{a} \; \ell_{\nu} \circ \eta(-t),  
\eeq
for all operators $\hat{a}$ in the algebra $\mathfrak{A} \times_{\alpha,\nu} G$. For the Cartesian product case, in the notation of the main text, we have
\beq
    \sigma^{\omega \circ T_{\nu}}_{t}(\hat{a}) = e^{-i t \op{X}} \hat{a} e^{i t \op{X}}. 
\eeq

In fact, eqn. \eqref{Inner modular flow in general} already implies that $\mathfrak{A} \times_{\alpha,\nu} G$ is semifinite since we have constructed a faithful, semifinite, normal weight with inner unitarily implemented modular flow. But we can also complete our proof that \eqref{Super general trace} defines a trace by a simple application of the KMS condition:
\begin{flalign}
    \tau(\hat{a} \hat{b}) &= \omega \circ T_{\nu}(\ell_{\nu} \circ \eta(i/2) \hat{a} \hat{b} \ell_{\nu} \circ \eta(i/2)) \nonumber \\
    &= \omega \circ T_{\nu}( \hat{b} \ell_{\nu} \circ \eta(i/2) \sigma^{\omega \circ T_{\nu}}_{-i}( \ell_{\nu} \circ \eta(i/2) \hat{a})) \nonumber \\
    &= \omega \circ T_{\nu}(\hat{b} \ell_{\nu} \circ \eta(i/2) \ell_{\nu} \circ \eta(-i) \ell_{\nu} \circ \eta(i/2) \hat{a} \ell_{\nu} \circ \eta(i)) \nonumber \\
    &= \omega \circ T_{\nu}(\hat{b} \hat{a} \ell_{\nu} \circ \eta(i)) = \omega \circ T_{\nu}(\ell_{\nu} \circ \eta(i/2) \hat{b} \hat{a} \ell_{\nu} \circ \eta(i/2)) = \tau(\hat{b} \hat{a}). 
\end{flalign}

Let $\hat{\varphi}_f$ be the state on $\mathfrak{A} \times_{\alpha,\nu} G$ which is induced from the vector $\ket{\varphi} \otimes \ket{f} \in \mathscr{H} \otimes L^2(G,d\nu)$. The density operator of this state with respect to the above trace is given by
\beq
    \rho_{\hat{\varphi}_f} = V^{\dagger} \tilde{f}(U \otimes \ell_{\nu}) \ell_{\nu} \circ \eta(-i/2) \Delta_{\varphi \mid \omega} \ell_{\nu} \circ \eta(-i/2) \tilde{f}(U \otimes \ell_{\nu})^{\dagger} V,
\eeq
where
\beq
    \tilde{f}(U \otimes \ell_{\nu}) = \int_{G} d\nu(g) f(g) \; U(g) \otimes \ell_{\nu}(g). 
\eeq
Using charge notation and in the Cartesian product case:
\beq
    \rho_{\tilde{\varphi}_f} = \tilde{f}(\op{\mathcal{Q}}_{\nu}) e^{-i \op{F}(\op{g})} e^{-\op{X}/2} \Delta_{\varphi \mid \omega} e^{-\op{X}/2} e^{i \op{F}(\op{g})} \tilde{f}(\op{\mathcal{Q}}_{\nu})^{\dagger},
\eeq
where
\beq
    \tilde{f}(\op{\mathcal{Q}}) \equiv \int_{G} d\nu(g) f(g) e^{i \op{\mathcal{Q}}(g)}, 
\eeq
with $\ell_{\nu}(g) = e^{i \op{\mathcal{Q}}_{\nu}(g)}$. Whenever $\ell_{\nu}(g)$ or $\op{\mathcal{Q}}_{\nu}(g)$ appear in an integral we must apply the extension procedure described in Appendix \ref{App: CP and OVW}.

\section{Semifiniteness for Cartesian Product Groups} \label{App: CP Factorization}

In this appendix, we demonstrate that, under the assumptions stated in Appendix \ref{App: General Trace} in the case where $G = \mathbb{R} \times H$, we have the isomorphism $\mathfrak{A} \times_{\alpha} (H \times \mathbb{R}) \simeq (\mathfrak{A} \times_{\alpha^H} H) \times_{\sigma^{\omega_H}} \mathbb{R}$. For simplicity, we will concentrate on the case that $H$ is locally compact, but a generalization to the non-locally compact case is straightforward (provided the existence of a suitable quasi-invariant measure). This analysis can be used to establish the isomorphism between the algebra $\mathfrak{A}_{\textrm{dress.}}(\mathscr{R},\mathbb{R} \times SO(3))$ and $\mathfrak{A}_{\textrm{dress.}}^{\textrm{obs.}}(\mathscr{R},\mathbb{R} \times SO(3))$ described in the main text.

Establishing our claim is a simple exercise in matching the generators of the algebra $\mathfrak{A} \times_{\alpha} (H \times \mathbb{R})$ and $(\mathfrak{A} \times_{\alpha^H} H) \times_{\sigma^{\omega_H}} \mathbb{R}$. Let $\mathfrak{A} \subset B(\mathscr{H})$ be a type III von Neumann algebra with faithful, semifinite, normal weight $\omega$ acted upon by an automorphism $\alpha: \mathbb{R} \times H \rightarrow \text{Aut}(\mathfrak{A})$. Moreover, let us assume that $\alpha$ satisfies the assumptions of our theorem, e.g.
\beq \label{Assumptions}
	\alpha_{(t,e)}(x) = \sigma^{\omega}_t(x), \qquad \omega \circ \alpha_{(t,h)} = \omega, \qquad \forall t \in \mathbb{R} , h \in H. 
\eeq
Note that $\alpha_{(t,h)} = \alpha_{(t,e)} \circ \alpha_{(0,h)} = \alpha_{(0,h)} \circ \alpha_{(t,e)}$. We shall denote the restriction of $\alpha$ to $H$ by $\alpha^H_{h} \equiv \alpha_{(0,h)}$.  

From our discussion above, the algebra $\mathfrak{A} \times_{\alpha} G$ is generated by the representations $\pi_G: \mathfrak{A} \rightarrow B(\mathscr{H} \otimes L^2(G))$ and $\ell_G: G \rightarrow B(\mathscr{H} \otimes L^2(G))$:
\begin{flalign}
	&\bigg( \pi_G(x) \xi_G\bigg)(t,h) \equiv \alpha_{(-t,h^{-1})}(x)\bigg(\xi_G(t,h)\bigg), \nonumber \\
    &\bigg(\ell_G(s,k) \xi_G\bigg)(t,h) \equiv \xi_G(-s+t,k^{-1}h). 
\end{flalign}
Here, we have used the notation $\xi_G \in L^2(G)$ to signify that the vector $\xi_G$ is a $\mathscr{H}$-valued function on $G$.

Likewise, the algebra $\mathfrak{A} \times_{\alpha^H} H$ is generated by the representations $\pi_H: \mathfrak{A} \rightarrow B(\mathscr{H} \otimes L^2(H))$ and $\ell_H: H \rightarrow B(\mathscr{H} \otimes L^2(H))$:
\beq
	\bigg(\pi_H(x) \xi_H\bigg)(h) \equiv \alpha_{(0,h^{-1})}(x)\bigg(\xi_H(h)\bigg), \qquad \bigg(\ell_H(k) \xi_H(h)\bigg) \equiv \xi_H(k^{-1}h). 
\eeq
We have used the notation $\xi_H$ to signify that $\xi_H$ is a $\mathscr{H}$-valued function on $H$. As we have discussed, the dual weight theorem constructs a faithful, semifinite, normal weight $\omega_H$ on $\mathfrak{A} \times_{\alpha^H} H$ with vector representative $\ket{\omega,e_H}$. The assumptions \eqref{Assumptions} imply that the modular automorphism of this weight acts on the generators of the algebra as
\beq \label{Mod aut of omega H}
	\sigma^{\omega_H}_t(\pi_H(x)) = \pi_H \circ \sigma^{\omega}_t(x), \qquad \sigma^{\omega_H}_t(\ell_H(h)) = \ell_H(h). 
\eeq	

Using \eqref{Mod aut of omega H}, let us form the algebra $(\mathfrak{A} \times_{\alpha^H} H) \times_{\sigma^{\omega_H}} \mathbb{R}$. It is the algebra generated by the representations $\pi_{\mathbb{R}}: \mathfrak{A} \times_{\alpha^H} H \rightarrow B(\mathscr{H} \otimes L^2(H) \otimes L^2(\mathbb{R}))$ and $\ell_{\mathbb{R}}: \mathbb{R} \rightarrow B(\mathscr{H} \otimes L^2(H) \otimes L^2(\mathbb{R}))$. Note that we have an isomorphism of Hilbert spaces $\mathscr{H} \otimes L^2(H) \otimes L^2(\mathbb{R}) \simeq \mathscr{H} \otimes L^2(H \times \mathbb{R})$. Of course, the operators in $\mathfrak{A} \times_{\alpha^H} H$ are themselves generated by $\pi_H(x)$ and $\ell_H(h)$. Thus, using the usual formulas, we find that the algebra $(\mathfrak{A} \times_{\alpha^H} H) \times_{\sigma^{\omega_H}} \mathbb{R}$ has three kinds of generators: 
\begin{flalign}
	\bigg(\pi_{\mathbb{R}}(\pi_H(x)) \xi_{G}\bigg)(t,h) &= \sigma^{\omega_H}_{-t}\bigg(\pi_H(x)\bigg)\bigg(\xi_G(t,\cdot)\bigg)(h) \nonumber \\ 
	&= \alpha_{(-t,h^{-1})}(x)\bigg(\xi_G(t,h)\bigg) = \bigg(\pi_G(x) \xi_G\bigg)(t,h),
\end{flalign}
\begin{flalign}
	\bigg(\pi_{\mathbb{R}}(\ell_H(k)) \xi_G\bigg)(t,h) &= \sigma^{\omega_H}_{-t}\bigg(\ell_H(k)\bigg) \bigg(\xi_G(t,\cdot)\bigg)(h) \nonumber \\
	&= \xi_G(t,k^{-1}h) = \bigg(\lambda_G(0,k) \xi_G\bigg)(t,h),
\end{flalign}	
\beq
	\bigg(\ell_{\mathbb{R}}(s) \xi_G\bigg)(t,h) = \xi_G(-s+t,h) = \bigg(\ell_G(s,e) \xi_G\bigg)(t,h). 
\eeq
In summary, we see that
\beq
	\pi_{\mathbb{R}} \circ \pi_H(x) = \pi_G(x), \qquad \pi_{\mathbb{R}} \circ \ell_H(h) \ell_{\mathbb{R}}(t) = \ell_{\mathbb{R}}(t) \pi_{\mathbb{R}} \circ \ell_H(h) = \ell_G(t,h),
\eeq
and thus we conclude
\begin{flalign} \label{alt semifinite cp}
	(\mathfrak{A} \times_{\alpha^H} H) \times_{\sigma^{\omega_H}} \mathbb{R} &\equiv \{\pi_{\mathbb{R}} \circ \pi_H(x), \pi_{\mathbb{R}} \circ \ell_H(h) \ell_{\mathbb{R}}(t) \; | \; x \in \mathfrak{A}, (t,h) \in G\}'' \nonumber \\
	&= \{\pi_G(x), \ell_G(t,h) \; | \; x \in \mathfrak{A}, (t,h) \in G\}'' = \mathfrak{A} \times_{\alpha} G, 
\end{flalign}
as desired.

\section{Finite subfactors for general type II$_{\infty}$ algebras} \label{sec: Type II_1} 

In this appendix, we consider the existence of type II$_1$ subalgebras of $\mathfrak{A} \times_{\alpha,\nu} G$ in the event that this algebra is semifinite. This analysis runs parallel to sections \ref{subsec:deSitter} and \ref{sec:dSII1} in the main text. 

Let $A$ be a type II$_{\infty}$ factor. Takesaki's structure theorem for type III von Neumann algebras establishes that there will exist a type III von Neumann algebra $B$ such that $A = B \times_{\sigma} \mathbb{R}$, e.g. $A$ is the modular crossed product of $B$ \cite{takesaki1973duality}. This can also be understood by appealing to Landstad's top down construction of the crossed product \cite{landstadDuality}. In particular, $B$ is the invariant subalgebra of $A$ under an automorphism $\theta: \mathbb{R} \rightarrow \text{Aut}(A)$ which is Pontryagin dual to $\sigma: \mathbb{R} \rightarrow \text{Aut}(B)$.\footnote{See also \cite{AliAhmad:2024wja,AliAhmad:2024vdw,AliAhmad:2025oli} for a discussion in the physical context of gravitational algebras.} In the case we are primarily interested in, $A = \mathfrak{A} \times_{\alpha,\nu} (H \times \mathbb{R})$, and $B = \mathfrak{A} \times_{\alpha^H,\nu^H} H$, see Appendix \ref{App: CP Factorization}.

On a related note, Pedersen and Takesaki have shown that a von Neumann algebra $C$ is semifinite (e.g. admits a faithful, semifinite, normal, tracial weight) if and only if the modular automorphism of any faithful, semifinite, normal weight on $C$ is innerly implemented \cite{PedersenTakesaki1973}. That is, for each $w \in P(C)$ there exists a one parameter family of unitary operators $U^{w}_t \in C$ such that
\beq \label{Pedersen Takesaki}
	\sigma^{w}_t(x) = U^{w}_t x U^{w}_{-t}, \qquad \forall x \in C. 
\eeq
By Stone's theorem, there exists a positive self adjoint operator $h_{w}$, affiliated with $C$, which generates this unitary group i.e. $U^w_t = h_{w}^{it}$. Using this fact, and appealing to Connes' cocycle theorem \cite{CONNES1980153}, Pedersen and Takesaki further demonstrated that, starting from \emph{any} $w \in P(C)$, we can construct a tracial weight $\tau_{w} \in P(C)$ given by
\beq \label{Pedersen Takesaki Trace}
	\tau_{w}(x) \equiv w(h_{w}^{-1/2} \; x \; h_{w}^{-1/2}). 
\eeq
The fact that this weight is faithful, semifinite, and normal follows from the faithful, semifinite, normal qualities of the original weight $w$. The fact that it is tracial follows from \eqref{Pedersen Takesaki} and the KMS condition. 

Returning to our analysis of the algebra $A$, the dual weight theorem tells us that -- given any $\psi \in P(B)$ -- there exists a canonically defined weight $\tilde{\psi} \in P(A)$, as discussed in Appendix \ref{App: CP and OVW}. If $B \subset B(\mathscr{K})$, and $\psi$ has vector representative $\ket{\psi}$, then $A \subset B(\mathscr{K} \otimes L^2(\mathbb{R}))$ and the dual weight is given by
\beq
	\tilde{\psi}(a) = \bra{\psi,0} a \ket{\psi,0}. 
\eeq
If we take $A = B \times_{\sigma^{\psi}} \mathbb{R}$, generated by dressed operators from $B$ and the translations $\ell_{\psi}(t) = e^{-it \op{X}_{\psi}}$, then we can also show that
\beq
	\sigma^{\tilde{\psi}}_t(a) = e^{-it \op{X}_{\psi}} a e^{it\op{X}_{\psi}}. 
\eeq
Here, we have carefully labeled the representation $\ell_{\psi}$ and the generator $\op{X}_{\psi}$ to emphasize their dependence on the choice of reference weight $\psi$. Invoking the Pedersen-Takesaki construction, we conclude that the algebra $A$ admits a faithful, semifinite, normal, tracial weight:
\beq
	\tau_{\tilde{\psi}}(a) = \bra{\psi,0} e^{\op{X}_{\psi}/2} a e^{\op{X}_{\psi}/2} \ket{\psi,0} = \int_{\mathbb{R} \times \mathbb{R}} \frac{dx_{\psi} dx_{\psi}'}{2\pi} e^{x_{\psi}} \bra{\omega,x_{\psi}} a \ket{\omega,x_{\psi}'}. 
\eeq

The choice of reference weight has no bearing on the overall structure of the modular crossed product. However, as we shall now demonstrate, it plays an important role in the interpretation of finite subalgebras therein. Consider the algebras $B \times_{\sigma^{\psi}} \mathbb{R}$ and $B \times_{\sigma^{\varphi}} \mathbb{R}$ where $\psi,\varphi \in P(B)$ are two distinct faithful, semifinite, normal weights on $B$. Both algebras act on $\mathscr{K} \otimes L^2(\mathbb{R})$. The algebra $B \times_{\sigma^{\psi}} \mathbb{R}$ is generated by the dressed representation
\beq
    \bigg(\pi_{\psi}(b) \xi\bigg)(t) \equiv \sigma^{\psi}_{-t}(b)(\xi(t))
\eeq
and $\ell_{\psi}: \mathbb{R} \rightarrow U(\mathscr{K} \otimes L^2(\mathbb{R}))$ such that 
\beq
    \pi_{\psi} \circ \sigma^{\psi}_t(b) = \ell_{\psi}(t) \pi_{\psi}(b) \ell_{\psi}(-t). 
\eeq
We can always write $\ell_{\psi}(t) \equiv e^{-it \op{X}_{\psi}}$ for some operator $\op{X}_{\psi}$. At the same time, the algebra $B \times_{\sigma^{\varphi}} \mathbb{R}$ is generated by the dressed representation
\beq
    \bigg(\pi_{\varphi}(b) \xi \bigg)(t) \equiv \sigma^{\varphi}_{-t}(b) (\xi(t))
\eeq
and $\ell_{\varphi}: \mathbb{R} \rightarrow U(\mathscr{K} \otimes L^2(\mathbb{R}))$ such that
\beq
    \pi_{\varphi} \circ \sigma^{\varphi}_t(b) = \ell_{\varphi}(t) \pi_{\varphi}(b) \ell_{\varphi}(-t).
\eeq
Likewise, we can write $\ell_{\varphi}(t) = e^{-it\op{X}_{\varphi}}$. 

The modular automorphisms of $\psi$ and $\varphi$ are related by the cocycle derivative as
\beq
    \sigma^{\varphi}_t(b) = u^{\varphi \mid \psi}_t \sigma^{\psi}_t(b) u^{\varphi \mid \psi}_{-t},
\eeq
Using this fact, we can relate the dressed representations:
\begin{flalign}
    \bigg(\pi_{\varphi}(b) \xi\bigg)(t) &= \sigma^{\varphi}_{-t}(b) (\xi(t)) \nonumber \\
    &= u^{\varphi \mid \psi}_{-t} \sigma^{\psi}_{-t}(b) u^{\varphi \mid \psi}_{t} (\xi(t)) \nonumber \\
    &= \bigg(U_{\varphi \mid \psi}^{-1} \pi_{\psi}(b) U_{\varphi \mid \psi} \xi\bigg)(t),
\end{flalign}
where here
\beq
    \bigg(U_{\varphi \mid \psi} \xi\bigg)(t) \equiv u^{\varphi \mid \psi}_t(\xi(t))
\eeq
is a unitary operator on $\mathscr{K} \otimes L^2(\mathbb{R})$. At the same time, if we take $\ell_{\varphi}(t) = U_{\varphi \mid \psi}^{-1} \ell_{\psi}(t) U_{\varphi \mid \psi}$, then we find that $\ell_{\varphi}(t)$ and $\pi_{\varphi}(b)$ satisfy the desired algebraic relation. We therefore conclude that $B \times_{\sigma^{\varphi}} \mathbb{R} = U_{\varphi \mid \psi}^{-1} (B \times_{\sigma^{\psi}} \mathbb{R}) U_{\varphi \mid \psi}$, and thus these algebras are isomorphic. Crucially, this also tells us that the modular `time translation' generators are unitarily equivalent
\beq \label{Relation between generators}
    \op{X}_{\varphi} = U_{\varphi \mid \omega}^{-1} \op{X}_{\omega} U_{\varphi \mid \omega},
\eeq
but distinct. 

The (equivalent) algebras $B \times_{\sigma^{\psi}} \mathbb{R}$ and $B \times_{\sigma^{\varphi}} \mathbb{R}$ both have natural traces which can be obtained by the Pedersen-Takesaki construction \eqref{Pedersen Takesaki Trace}. In particular
\beq
    \tau_{\tilde{\psi}}(a) = \bra{\psi,0} e^{\op{X}_{\psi}/2} a e^{\op{X}_{\psi}/2} \ket{\psi,0}, \qquad \tau_{\tilde{\varphi}}(a) = \bra{\varphi,0} e^{\op{X}_{\varphi}/2} a e^{\op{X}_{\varphi}/2} \ket{\varphi,0}. 
\eeq
To make a proper comparison between these traces, let's write them exclusively in terms of the generator $\op{X}_{\varphi}$ by using \eqref{Relation between generators}. Then we find
\begin{flalign} \label{omega trace in terms of varphi energy}
    \tau_{\tilde{\psi}}(a) &= \bra{\psi,0} e^{\op{X}_{\varphi}/2} U_{\varphi \mid \psi} a U_{\varphi \mid \psi}^{-1} e^{\op{X}_{\varphi}/2} \ket{\psi,0}
\end{flalign}
where we have concluded $U_{\varphi \mid \psi} \ket{\psi,0} = \ket{\psi,0}$ by direct computation. In fact, it's not hard to show that \eqref{omega trace in terms of varphi energy} exactly reproduces $\tau_{\tilde{\varphi}}$:\footnote{It's worth noting that \eqref{RN for Dual Weights} allows us to deduce the Connes' cocycle derivative between dual weights:
\beq
    u^{\tilde{\varphi} \mid \tilde{\psi}}_{t} = e^{it \op{X}_{\varphi}} U_{\varphi \mid \psi} e^{-it \op{X}_{\varphi}}.
\eeq
}
\beq \label{RN for Dual Weights}
    U_{\varphi \mid \psi}^{-1} e^{\op{X}_{\varphi}/2} \ket{\psi,0} = U_{\varphi \mid \psi}^{-1} \ket{\psi,-i/2} = u^{\varphi \mid \psi}_{-i/2} \ket{\psi,-i/2} = \ket{\varphi,-i/2} = e^{\op{X}_{\varphi}/2} \ket{\varphi,0},
\eeq
and thus
\beq \label{Equality of traces}
    \tau_{\tilde{\psi}}(a) = \bra{\psi,0} e^{\op{X}_{\varphi}/2} U_{\varphi \mid \psi} a U_{\varphi \mid \psi}^{-1} e^{\op{X}_{\varphi}/2} \ket{\psi,0} = \bra{\varphi,0} e^{\op{X}_{\varphi}/2} a e^{\op{X}_{\varphi}/2} \ket{\varphi,0} = \tau_{\tilde{\varphi}}(a). 
\eeq
Eqn. \eqref{Equality of traces} again tells us that the choice of reference weight $\psi$ or $\varphi$ has no bearing on the structure of the overall crossed product algebra. Henceforward we will simply denote the trace by $\tau$. 

We now arrive at the important point. Suppose that $\psi$ is an unnormalizable weight and $\varphi$ is a normalizable state with $\varphi(\op{1}) = N_{\varphi} < \infty$. Let $P^{\psi}_-$ and $P^{\varphi}_-$ denote the projection on $\mathscr{K} \otimes L^2(\mathbb{R})$ onto the negative part of the spectrum of $\op{X}_{\psi}$ and $\op{X}_{\varphi}$, respectively. Then, we can easily compute:
\beq
    \tau(P^{\psi}_{-}) = \braket{\psi|\psi} \int_{\mathbb{R}_-} \frac{dx_{\psi}}{2\pi} e^{x_{\psi}}, \qquad \tau(P^{\varphi}_-) = \braket{\varphi|\varphi} \int_{\mathbb{R}_-} \frac{dx_{\varphi}}{2\pi} e^{x_{\varphi}} < \infty.
\eeq
The trace of $P^{\psi}_-$ remains infinite due to the non-normalizability of $\ket{\psi}$, but the trace of $P^{\varphi}_-$ is finite! We therefore conclude that the algebra $P^{\psi}_- A P^{\psi}_-$ remains type II$_{\infty}$, while the algebra $P^{\varphi}_- A P^{\varphi}_-$ is Type II$_1$. 

The upshot is the following: Although by \eqref{Relation between generators} the eigenvalues of $\op{X}_{\varphi}$ and $\op{X}_{\psi}$ agree, the projections onto the positive part of their spectra are sensitive to their eigenvectors. It \emph{is} possible to arrive at a Type II$_1$ subalgebra by bounding the spectrum of the generator of `time translations' in the modular crossed product, provided the associated weight is normalizable. Mathematically, we can always just choose to bound the generator of modular translations for a state. From a physical perspective, however, the bound on $\op{X}$ is only justified if this operator can be interpreted as generating genuine time translations. We must, therefore, determine physically which weight to reference our bound to.

In the case $B = \mathfrak{A}$ is the algebra of quantum fields (e.g. in the absence of an additional constraint group), the modular crossed product comes about by implementing the Hamiltonian constraint. That is, the algebra of observables must commute with $\op{X} + \op{\mathcal{H}}_{QFT}$, viewed as an operator on $\mathscr{H} \otimes L^2(\mathbb{R})$. By the commutation theorem \cite{vandaele1978continuous}, the algebra of observables commuting with this operator defines a crossed product $\mathfrak{A} \times_{\alpha} \mathbb{R}$, where $\alpha_t(a) = e^{-it \op{\mathcal{H}_{QFT}}} a e^{it\op{\mathcal{H}_{QFT}}}$. The realization that this algebra is a \emph{modular} crossed product arises from observation that there exists a state, $\omega \in P(\mathfrak{A})$, whose modular flow is equivalent to `time evolution': $\sigma^{\omega}_t = \alpha_t$. In the algebra $\mathfrak{A} \times_{\alpha} \mathbb{R}$, the automorphism $\alpha$ is implemented unitarily by $e^{it \op{X}}$, and thus $\op{X} = \op{X}_{\omega}$ acquires an interpretation as the generator of time translations in the crossed product. In this regard, bounding the spectrum of $\op{X}_{\omega}$ constitutes a bound on the true physical energy when the modular Hamiltonian of $\omega$ is the generator of time translations on the algebra before implementing the Hamiltonian constraint. 

Now, let us move to the case $B = \mathfrak{A} \times_{\alpha^H, \nu^H} H$, with the overall gauge invariant algebra being the modular crossed product of $B$. In this case, the Hamiltonian constraint will generically require that the physical algebra commute with an operator $\op{X} + \op{\mathcal{H}}_{QFT} + \op{\mathcal{H}}_{H}$ acting on $\mathscr{H} \otimes L^2(H,d\nu_H) \otimes L^2(\mathbb{R})$. Here, $\op{\mathcal{H}}_{QFT}$ remains the generator of time translations on the QFT algebra, but now there is possibly an additional contribution from $\op{\mathcal{H}}_H$, the generator of time translations on the operators $\ell_{\nu_H}(h)$. Taking $\omega \in P(\mathfrak{A})$ to be the state on the QFT algebra whose modular flow (on $\mathfrak{A}$) agrees with the time evolution generated by $\op{\mathcal{H}}_{QFT}$, the classical-quantum state $\ket{\omega} \otimes e^{-\op{\mathcal{H}}_H/2} \ket{e_{\nu_H}}$ will have a modular flow (on $\mathfrak{A} \times_{\alpha_H,\nu_H} H$) agreeing with the automorphism generated by $\op{\mathcal{H}}_{QFT} + \op{\mathcal{H}}_H$. Thus, it is physically natural to bound the operator $\op{X} = \op{X}_{\varphi}$ associated with the weight $\varphi$ whose vector representative is $\ket{\varphi} = \ket{\omega} \otimes e^{-\op{\mathcal{H}}_H/2} \ket{e_{\nu_H}}$. If the vector $e^{-\op{\mathcal{H}}_H/2} \ket{e_{\nu_H}}$ is normalizable, then bounding $\op{X}_{\varphi}$ will indeed land on a finite subalgebra. This was the case in subsections \ref{subsec:deSitter} and \ref{sec:dSII1} in which $\ket{\varphi}$ is the state of a compact observer moving along its worldtube in the static patch of a de Sitter spacetime. 

\section{The Infinite Tensor Product and its States} \label{app: ITP}

In section \ref{sec:fulldressedalg}, we employ the notion of an infinite tensor product algebra to conclude that the `neutral' weight for the boost supertranslation group is well defined, normal and semifinite. In section \ref{sec:dSII1}, we use the infinite tensor product to establish the existence of a physically motivated type II$_1$ subalgebra of the gravitational algebra of an observer in the de Sitter static patch. In this appendix, we introduce some aspects of this construction. For a more complete introduction, we refer the reader to \cite{Takesaki2003VolIII}. 

An inductive sequence of $C^*$ algebras is a sequence of $C^*$ algebras, $\{A_n\}_{n \in \mathbb{N}}$, along with a sequence of $*$-homomorphisms $\{\pi_n: A_n \rightarrow A_{n+1}\}_{n \in \mathbb{N}}$. For each $n,m \in \mathbb{N}$ we define
\beq
	\pi_{n+m,n} \equiv \pi_{n+m-1} \circ \pi_{n+m-2} \circ ... \circ \pi_{n+1} \circ \pi_n: A_n \rightarrow A_{n+m}.
\eeq
By construction $\pi_{l,m} \circ \pi_{m,n} = \pi_{l,n}$ provided $l > m > n$. Treating each $A_n$ as disjoint, we can define the set 
\beq
	X \equiv \bigsqcup_{n = 1}^{\infty} A_n.
\eeq
Given any $a \in A_n$ and $b \in A_m$ we then introduce the equivalence relation
\beq
	a \sim b \iff \exists N \text{ s.t. } \pi_{l,n}(a) = \pi_{l,m}(b) \; \forall l > N. 
\eeq
We will denote by $[a] \equiv \pi_{\infty,n}(a)$ the equivalence class of $a \in X$ under the relation $\sim$. Using this equivalence relation we can define the set $A_{\infty} \equiv X/\sim$. This set can be made into an involutive algebra by endowing it with the following operations:
\beq
	\lambda [a] \equiv [\lambda a], \;\; [a] + [b] \equiv [\pi_{l,n}(a) + \pi_{l,m}(b)], \;\; [a][b] \equiv [\pi_{l,n}(a)\pi_{l,m}(b)], \;\; [a]^* \equiv [a^*]. 
\eeq
In this way, the assignment $a \in A_n \mapsto \pi_{\infty,n}(a) \in A_{\infty}$ can be viewed as a homomorphism. 

If the set $\pi_{\infty,n}^{-1}(0)$ is closed for sufficiently large $n$, the inductive sequence $\{A_n,\pi_n\}_{n \in \mathbb{N}}$ is called proper. Each algebra $A_n$ is embedded homomorphically into $A_{\infty}$ with image $\overline{A}_n \simeq A_n/\pi_{\infty,n}^{-1}(0)$. Thus, for a proper inductive sequence of $C^*$ algebras, the algebra $\overline{A}_n$ becomes a $C^*$ algebra for sufficiently large $n$. In this way, the $C^*$ norm of $A_n$ can be used to induce a $C^*$ norm on $A_{\infty}$ giving it the structure of a pre-$C^*$ algebra. Closing $A_{\infty}$ in the topology induced by this norm results in a $C^*$ algebra, $A$, which is called the inductive limit of the sequence $\{A_n,\pi_n\}$. We will sometimes use the standard notation $A \equiv \text{Lim}_{n \rightarrow \infty} \{A_n,\pi_n\}$ to denote the inductive limit. 

A standard example of an inductive limit is the infinite tensor product of $C^*$ algebras. Let $\{A_n\}_{n \in \mathbb{N}}$ be a sequence of unital $C^*$ algebras and denote by
\beq
	\overline{A}_n \equiv A_1 \otimes A_2 \otimes ... \otimes A_n
\eeq	
the $n$-fold finite $C^*$ tensor product. There is a natural isomorphism between $\overline{A}_n$ and $\overline{A}_{n+1}$ in terms of the projection $\pi_n(a) \equiv a \otimes \op{1}_{A_{n+1}}$. As each $\pi_n$ is injective, the inductive sequence $\{\overline{A}_n, \pi_n\}_{n \in \mathbb{N}}$ is proper. The inductive limit of this sequence is the infinite tensor product which we denote by
\beq
	\text{Lim}_{n \rightarrow \infty} \overline{A}_n = \bigotimes_{n = 1}^{\infty} A_n. 
\eeq	

Now, suppose that $\{M_n\}_{n \in \mathbb{N}}$ is a sequence of von Neumann algebras. Let $\bigotimes_{n = 1}^{\infty} M_n$ be the $C^*$-infinite tensor product of these algebras. Given a collection $\{\omega_n\}_{n \in \mathbb{N}}$ where each $\omega_n$ is a state on $M_n$, we obtain a state $\omega \equiv \bigotimes_{n = 1}^{\infty} \omega_n$ on the infinite tensor product which is defined by the condition that
\beq
	\omega(x_1 \otimes x_2 \otimes ... \otimes x_N \otimes \op{1} \otimes \op{1} \otimes ...) = \prod_{n = 1}^N \omega_n(x_n). 
\eeq
Performing the GNS construction relative to the state $\omega$, we obtain a Hilbert space representation of $\bigotimes_{n = 1}^{\infty} M_n$ which we may use to close this $C^*$ algebra to a von Neumann algebra, $M$, which we call the von Neumann infinite tensor product of $\{M_n\}_{n \in \mathbb{N}}$ with respect to the state $\omega$. 

The case that is of interest to us involves decomposing the group algebra of $\mathcal{S}_{0}$ as an infinite tensor product. This follows from the observation that the space of functions on $\mathbb{S}^2$ can be decomposed into a basis of spherical harmonics labeled by $\ell,m$. In this regard, the group $\mathcal{S}_{0}$ is equivalent to the infinite Cartesian product of the group $\mathbb{R}$, one for each $\ell,m$. More rigorously, in the language described above, the group $C^*$ algebra of $\mathcal{S}_{0}$ is a $C^*$ infinite tensor product of group algebras $\mathfrak{A}(\mathbb{R})$: $A \equiv \bigotimes_{\ell > 0, m} \mathfrak{A}(\mathbb{R})$. 

To promote $A$ to a von Neumann algebra, we must introduce a preferred state. Each $\mathfrak{A}(\mathbb{R})$ is represented standardly on the Hilbert space $L^2(\mathbb{R},dt_{\ell,m})$. Thus, per the construction above, we can obtain a state on $A$ by specifying a collection of normalized vectors $\ket{\psi_{\ell,m}} \in L^2(\mathbb{R},dt_{\ell,m})$. To reproduce our construction from section \ref{subsubsec:HS}, we can choose each of these vectors to be a Gaussian wavepacket $\ket{\Omega_{\ell,m}}$. More to the point, we may regard each $\mathfrak{A}(\mathbb{R})$ as acting on the Hilbert space $L^2(\mathbb{R},d\gamma_{\ell,m})$ where $\gamma_{\ell,m}$ is a Gaussian measure for each $\ell,m$. Then, $\ket{\Omega_{\ell,m}}$ is the state in $L^2(\mathbb{R},d\gamma_{\ell,m})$ with wavefunction $1$ with respect to the chosen measure, as in \eqref{eq:gaussST}. Let $\omega_{\ell,m}$ be the state on $\mathfrak{A}(\mathbb{R})$ induced by the vector $\ket{\Omega_{\ell,m}}$. Then, $\omega \equiv \bigotimes_{l > 0, m} \omega_{\ell,m}$ can be interpreted as the state induced by the infinite Gaussian measure, $\gamma$, on $\mathcal{S}_{0}$ obtained as the `product' of the individual Gaussian measure $\gamma_{\ell,m}$ for each $\ell,m$. The group von Neumann algebra $\mathfrak{A}(\mathcal{S}_{0})$ is isomorphic to the weak closure of $\bigotimes_{\ell > 0, m} \mathfrak{A}(\mathbb{R})$ in the GNS representation induced by this state, which can be identified with the Hilbert space $L^2(\mathcal{S}_{0},d\gamma)$. 

At the same time, we can also introduce the improper state $\ket{e_{\gamma_{\ell,m}}}$ for each $\ell,m$, which gives rise to the neutral weight with respect to the chosen measure. The weight induced by $\ket{e_{\gamma_{\ell,m}}}$ on the algebra $\mathfrak{A}(\mathbb{R})$ is faithful, semifinite, and normal. Moreover, the overlap $\braket{e_{\gamma_{\ell,m}}|\Omega_{\ell,m}} = 1$. Thus, invoking the construction introduced in \cite{Blackadar1977Infinite}, the infinite tensor product
\beq \label{eq:ITPNeutralWeight}
    \ket{e_{\gamma}} \equiv \bigotimes_{\ell>1,m} \ket{e_{\gamma_{\ell,m}}},
\eeq
induces a semifinite, normal weight on the group von Neumann algebra $\mathfrak{A}(\mathcal{S}_0)$.

For each $\ell,m$, let $H_{\ell,m}$ be a positive, self-adjoint operator on $L^2(\mathbb{R},d\gamma_{\ell,m})$ such that
\beq
	Z_{\ell,m} \equiv \bra{e_{\gamma_{\ell,m}}} e^{-H_{\ell,m}} \ket{e_{\gamma_{\ell,m}}} < \infty, 
\eeq
and that the state induced by the normalized vector $\ket{\psi_{\ell,m}} \equiv e^{-(H_{\ell,m} + \log Z_{\ell,m})/2} \ket{e_{\gamma_{\ell,m}}}$ belongs to the folium of the state induced by $\ket{\Omega_{\ell,m}}$. Let us use the notation $\psi_{\ell,m}(\cdot) \equiv \bra{\psi_{\ell,m}} \cdot \ket{\psi_{\ell,m}}$ for the state on the algebra $\mathfrak{A}(\mathbb{R})$ induced by the vector $\ket{\psi_{\ell,m}}$. Then, by construction, $\psi \equiv \bigotimes_{\ell > 0,m} \psi_{\ell,m}$ is a state on the von Neumann infinite tensor product $\mathfrak{A}(\mathcal{S}_{0})$. This state is defined by the property that, for any finite $N$ and $\{x_{\ell,m}\; | \; 1\leq \ell \leq N, |m|\leq \ell\} \in \mathfrak{A}(\mathbb{R})$
\begin{flalign}
	\psi&\bigg(\bigotimes_{\substack{1 \leq \ell \leq N \\ |m| \leq \ell}} x_{\ell,m} \otimes \op{1} \otimes \op{1} \otimes ...\bigg) \nonumber \\
	&= \bigotimes_{\substack{1 \leq \ell \leq N \\ |m| \leq \ell}} \bra{e_{\gamma_{\ell,m}}} e^{-\frac{1}{2} \sum\limits_{\substack{1 \leq \ell \leq N \\ |m| \leq \ell}} (H_{\ell,m} + \log Z_{\ell,m})} \bigg(\bigotimes_{\substack{1 \leq \ell \leq N \\ |m| \leq \ell}} x_{\ell,m}\bigg) e^{-\frac{1}{2} \sum\limits_{\substack{1 \leq \ell \leq N \\ |m| \leq \ell}} (H_{\ell,m} + \log Z_{\ell,m})} \bigotimes_{\substack{1 \leq \ell \leq N \\ |m| \leq \ell}} \ket{e_{\gamma_{\ell,m}}}. 
\end{flalign}
In this sense, we may write
\beq
	\ket{\psi} = \lim_{N \rightarrow \infty} \bigg(e^{-\frac{1}{2} \sum\limits_{\substack{1 \leq \ell \leq N \\ |m| \leq \ell}} (H_{\ell,m} + \log Z_{\ell,m})} \bigotimes_{\substack{1 \leq \ell \leq N \\ |m| \leq \ell}} \ket{e_{\gamma_{\ell,m}}}\bigg) \equiv e^{-K/2} \ket{e_{\gamma}}.
\eeq
Here, 
\beq
	K \equiv \lim_{N \rightarrow \infty} \sum\limits_{\substack{1 \leq \ell \leq N \\ |m| \leq \ell}} (H_{\ell,m} + \log Z_{\ell,m}), 
\eeq
and $\ket{e_{\gamma}}$ is as defined in \eqref{eq:ITPNeutralWeight}. By construction, $\ket{\psi}$ belongs to the folium of $\omega$ and its modular Hamiltonian is given by $K$.

\providecommand{\href}[2]{#2}\begingroup\raggedright\endgroup

\end{document}